\documentclass[11pt]{article}
\newcommand{\Comment}[1]{{}}
\usepackage[textwidth = 430 pt, textheight = 630 pt]{geometry}
\usepackage{amsmath,euscript,amssymb,amsfonts,graphicx,bbm,upgreek,esvect,subcaption,mathtools}

\usepackage{color}
\usepackage{bbold}
\definecolor{MyDarkBlue}{rgb}{0.15,0.15,0.45}
\usepackage[linktocpage=true]{hyperref}
\hypersetup{
colorlinks=true,
citecolor=blue,
linkcolor=blue,
urlcolor=blue,
pdfauthor={Andrew~B.~ Royston},
pdftitle={Monopole-defect simulations},
pdfsubject={hep-th}
}

\DeclareFontFamily{OT1}{pzc}{}
\DeclareFontShape{OT1}{pzc}{m}{it}{<-> s * [1.10] pzcmi7t}{}
\DeclareMathAlphabet{\mathpzc}{OT1}{pzc}{m}{it}

\makeatletter
\DeclareFontFamily{OMX}{MnSymbolE}{}
\DeclareSymbolFont{MnLargeSymbols}{OMX}{MnSymbolE}{m}{n}
\SetSymbolFont{MnLargeSymbols}{bold}{OMX}{MnSymbolE}{b}{n}
\DeclareFontShape{OMX}{MnSymbolE}{m}{n}{
    <-6>  MnSymbolE5
   <6-7>  MnSymbolE6
   <7-8>  MnSymbolE7
   <8-9>  MnSymbolE8
   <9-10> MnSymbolE9
  <10-12> MnSymbolE10
  <12->   MnSymbolE12
}{}
\DeclareFontShape{OMX}{MnSymbolE}{b}{n}{
    <-6>  MnSymbolE-Bold5
   <6-7>  MnSymbolE-Bold6
   <7-8>  MnSymbolE-Bold7
   <8-9>  MnSymbolE-Bold8
   <9-10> MnSymbolE-Bold9
  <10-12> MnSymbolE-Bold10
  <12->   MnSymbolE-Bold12
}{}

\let\llangle\@undefined
\let\rrangle\@undefined
\DeclareMathDelimiter{\llangle}{\mathopen}%
                     {MnLargeSymbols}{'164}{MnLargeSymbols}{'164}
\DeclareMathDelimiter{\rrangle}{\mathclose}%
                     {MnLargeSymbols}{'171}{MnLargeSymbols}{'171}
\makeatother
\newcommand{\overbar}[1]{\mkern 1.5mu\overline{\mkern-1.5mu#1\mkern-1.5mu}\mkern 1.5mu}

\DeclarePairedDelimiterX\braket[2]{\langle}{\rangle}{#1\,\delimsize\vert\,\mathopen{}#2}
\def\sech{{\operatorname{sech}}}
\def\RS{{\operatorname{RS}}}
\def\sn{{\, \operatorname{sn}}}
\def\am{{\, \operatorname{am}}}
\def\cn{{\, \operatorname{cn}}}
\def\dn{{\, \operatorname{dn}}}
\def\Es{{\, \operatorname{Es}}}
\def\Ec{{\, \operatorname{Ec}}}

\def\ed{\, \textrm{d}}

\def\ie{{\it i.e.}~}

\def\pd{\partial}
\def\diag{ \, \textrm{diag}}
\def\ii{{\mathrm{i}}}

\def\varphibar{\overbar{\varphi}}

\def\half{\frac{1}{2}}

\def\BB{\mathcal{B}}

\def\HH{\mathcal{H}}
\def\MM{\mathcal{M}}
\def\PP{\mathcal{P}}

\def\SS{\mathcal{S}}

\def\JJ{\mathcal{J}}

\def\Re{ \, \mathrm{Re} \, }

\def\MM{\mathcal{M}}
\def\KK{\mathcal{K}}

\def\FF{\mathcal{F}}
\def\QQ{\mathcal{Q}}

\def\rmk{{\mathrm{k}}}

\def\ii{{\mathrm{i}}}

\begin{document}

%\rightline{}
 
   \vspace*{3truecm}

\centerline{\LARGE \bf {\sc Framework for the Forced Soliton Equation:}}
\centerline{\large \bf {\sc Regularization, Numerical Solutions, and Perturbation Theory}} 
% \vspace{.5cm}\centerline{\LARGE \bf {\sc }} 
\vspace{1.5truecm}
\centerline{ {\large {\bf{\sc
        Zachary J.~Allamon,}}}\footnote{E-mail address:
    \href{mailto:Zachary Allamon <zba5114@psu.eduu>}{\tt
      zba5114@psu.edu}
      Current Affiliation:  Department of Physics, The Pennsylvania State University, University Park, PA 16802, USA }   {\large {\bf{\sc Quentin A.~Hales,}}}\footnote{E-mail address:
    \href{mailto:Quentin Hales
      <quentin.hales@gmail.com>}{\tt
      quentin.hales@gmail.com} } {\large {\bf{\sc
        Andrew~B.~Royston,}}}\footnote{E-mail address:
    \href{mailto:Andy Royston <abr84@psu.eduu>}{\tt
      abr84@psu.edu} } }     
\centerline{ {\large {\bf{\sc Douglas L.~Rutledge}}}\footnote{E-mail address:
    \href{mailto:Douglas Rutledge
      <dlr223@psu.edu>}{\tt
      dlr223@psu.edu} }   {and}  {\large {\bf{\sc Erica A.~Yozie}}}\footnote{E-mail address:
    \href{mailto:Erica Yozie
      <ejy5118@psu.edu>}{\tt
      ejy5118@psu.edu}. 
      Current Affiliation:  Department of Physics, The Pennsylvania State University, University Park, PA 16802, USA}  }

\vspace{1cm}
\centerline{{\it Department of Physics, Penn State Fayette, The Eberly Campus}}
\centerline{{\it 2201 University Drive, Lemont Furnace, PA 15456, USA}}
%\vspace{.5cm}
%\centerline{${}^b${\it George~P.~\& Cynthia Woods Mitchell Institute}}
%\centerline{{\it for Fundamental Physics and Astronomy}}
%\centerline{{\it Texas A\&M University, College Station, TX 77843, USA}}

\vspace{1.5truecm}

%%%%%%%%%%%%%%%%%
\thispagestyle{empty}

\centerline{\sc Abstract}
\vspace{0.4truecm}
\begin{center}
  \begin{minipage}[c]{380pt}{The forced soliton equation is the starting point for semiclassical computations with solitons away from the small momentum transfer regime.  This paper develops necessary analytical and numerical tools for analyzing solutions to the forced soliton equation in the context of two-dimensional models with kinks.  Results include a finite degree of freedom regularization of soliton sector physics based on periodic and anti-periodic lattice models, a detailed analysis of numerical solutions, and the development of perturbation theory in the soliton momentum transfer to mass ratio $\Delta P/M$.  Numerical solutions at large transfer $\Delta P \gtrsim M$ are capable of exhibiting, in a smooth and controlled fashion, extreme phenomena such as soliton-antisoliton pair creation and superluminal collective coordinate velocities, which we investigate.}
\end{minipage}
\end{center}

\vspace{.4truecm}

\noindent

\vspace{.5cm}

\setcounter{page}{0}

\newpage

\renewcommand{\thefootnote}{\arabic{footnote}}
\setcounter{footnote}{0}

\linespread{1.1}
\parskip 7pt

{}~
{}~

\makeatletter
\@addtoreset{equation}{section}
\makeatother
\renewcommand{\theequation}{\thesection.\arabic{equation}}

\tableofcontents

%%%%%%%%%%%%%%%%
%%%%%%%%%%%%%%%%
\section{Introduction and Summary}
%%%%%%%%%%%%%%%%
%%%%%%%%%%%%%%%%

This paper is motivated by a simple sounding question from quantum field theory with solitons: What is the leading semiclassical behavior of soliton form factors?  Form factors are Fourier transforms of matrix elements of local operators between soliton states \cite{Dashen:1974ci,Dashen:1974cj,Goldstone:1974gf,Gervais:1974dc,Callan:1975yy,Christ:1975wt,Gervais:1975pa,Tomboulis:1975gf}.  Suppose the in-state consists of a soliton with momentum $P_i$ and the out-state a soliton with momentum $P_f$.  Very little is known about these form factors, not even their leading semiclassical behavior, when the transfer $P_f - P_i$ is comparable to the soliton mass $M$.  The exception is in integrable models where many impressive exact results---exact in both the coupling and the transfer to mass ratio---are available.  See, for example, \cite{Weisz:1977ii,Karowski:1978vz,Babujian:1998uw}.  

Through crossing symmetry, the answer to the question we started with could teach us about the answer to another fundamental question: What role do soliton-antisoliton pairs play in quantum corrections to processes involving ordinary particles?  However, in order to apply crossing symmetry to address this question one has to understand the form factor at momentum transfers above the threshold for pair creation, $\Delta P \geq 2M$.

Although the ultimate goal is to be able to address these questions in more realistic $(3+1)$-dimensional models with solitons, for the time being and the remainder of this paper we confine ourselves to $(1+1)$-dimensional models of a single real scalar field $\phi = \phi(t,x)$ with kink solitons.  For simplicity we assume the interactions of the model are controlled by a single coupling $g$, and the potential $V(\phi)$ has the scaling property $V(\phi) = \frac{1}{g^2} \widetilde{V}(g\phi)$, where $\widetilde{V}(\widetilde{\phi})$ is independent of $g$.  This allows $1/g^2$ to be pulled out in front of the action so that $g^2$ plays the role of $\hbar$ in the semiclassical expansion.  It also implies that the soliton mass scales like $1/g^2$ in units of the perturbative mass scale $m$ of the theory.

There has been significant recent progress in the computation of quantum ($g$-)corrections to observables involving kink solitons at small $\Delta P/M$, spurred in large part by the development of a new streamlined approach to such computations in \cite{Evslin:2019xte,Evslin:2020qow,Evslin:2020azr,Evslin:2021gxs}.  These computations tie the small parameter $\Delta P/M$ to $g$, typically assuming $P/M = O(g^2)$, or equivalently $P/m = O(1)$, although some work has considered $P/M = O(g)$ such that the transfer is relativistic with respect to the meson mass scale $m$ but still non-relativistic compared to the soliton's mass \cite{Evslin:2022xvs}.   

In contrast, there has been relatively little progress in understanding soliton sector observables away from the small transfer regime, where no scaling relationship is assumed between $g$ and $\epsilon \equiv \Delta P/M$.  The main reason for this is that the starting point for such computations---the classical saddle---must change when $\epsilon \neq 0$.  In \cite{Melnikov:2020ret,Melnikov:2020iol} a new saddle point equation was identified for this purpose: the forced soliton equation (FSE).  As will be reviewed in this paper, the FSE arises when one couples the collective coordinate degree of freedom, and only that degree of freedom, to an external force $F$.  The collective coordinate degree of freedom is distinguished by the form of the semiclassical expansion in the soliton sector of the theory.  It is defined as a canonically conjugate variable to the Noether momentum $P$ whose eigenvalues label the soliton one-particle states.

This paper is concerned with the construction of solutions to the FSE, both numerical ones and perturbative ones in $\epsilon$.  Before describing the main results, however, we should explain three additional facts about the FSE.  First, while \cite{Melnikov:2020ret,Melnikov:2020iol} described the FSE in the context of a general driving force $F(t)$ that couples to the collective coordinate, here we focus on one specific case---an instantaneous impact with impulse equal to the transfer: $F(t) = (P_f - P_i)\delta(t-t_\ast)$.  As it turns out, this is exactly the case that is relevant for form factors of a single local operator between soliton states.  Momentum conservation, which follows from translation invariance the theory, holds except at the time of the insertion $t_\ast$, and hence the momentum must be constant except at that time.

Second, the relevant solution to the FSE for the form factor problem is the solution to an associated (temporal) boundary value problem (BVP), where the position field $\phi$ is fixed to be that of pure kinks with momenta $P_i$ and $P_f$ at initial and final times, while the momentum field is whatever it needs to be to fit these boundary conditions.  Here, we instead solve an initial value problem (IVP), taking both the position and momentum fields before the impact to be those of a moving kink with velocity $\beta_i$ corresponding to the initial momentum $P_i$.  We view this as a warm-up for the more difficult boundary value problem.  Numerically speaking, we first need to develop a fast IVP solver that can be employed in a shooting method approach for the BVP.

Third, the FSE is not a Lorentz-covariant equation.  This is clear from the fact that a Lorentz transformation sends one degree of freedom into an admixture of all degrees of freedom.  The FSE approach to the form factor, like most approaches to soliton sector observables, breaks manifest Lorentz covariance.  Therefore, solutions to the FSE for different initial momenta $P_i$ are not related by a symmetry transformation, and it is important to consider the equation for generic initial momenta $P_i$ and transfers $\epsilon$.  The Lorentz covariance properties of the form factor within the FSE approach is a subject of work in progress.  

Now we turn to an overview of results.  

%%%%%%%%%%%%%%%%%%%%
\subsection{Regularization}\label{ssec:introreg}
%%%%%%%%%%%%%%%%%%%%

The first set of results concerns regularization, by which we mean reducing the degrees of freedom of the theory to a finite number.  The direct motivation for regularization is practical.  Numerical methods require it.  Extracting continuum and infinite volume results from such numerics in turn requires an understanding of the errors incurred by the regularization.  

Still, there are many regularizations one can consider from a numerical perspective, and the choices we make here are driven by a secondary motivation.  Our choices are best suited for implementing the standard regularization approach taken, or at least implied, by most quantum field theory computations in these theories, where continuum integrals over infinite volume are replaced by finite mode sums.  Hence we regularize the infinite 1D spatial length by putting the theory on a circle of circumference $L$, and we regularize the continuum by introducing a lattice.  While all work in this paper is with classical field theory and its regularization, the intent is to set up a framework well suited for implementation in quantum field theory.

In Section \ref{sec:FSE} we describe the continuum theory on a circle, using our two primary examples of $\phi^4$ theory and the sine-Gordon model.  Much of the discussion applies to any $V(\phi)$ possessing either a reflection symmetry (like $\phi^4$ theory) or a discrete translation symmetry (like sine-Gordon), while results specific to these two theories are clearly indicated as such.  We apply the classical canonical transformation of \cite{Gervais:1975pa,Tomboulis:1975gf} in the soliton sector to identify the collective coordinate degree of freedom and the remaining oscillator degrees of freedom for fluctuations orthogonal to the zero mode.  New results include a complete discussion of the normal modes for these oscillators around a moving kink.  For our primary examples of sine-Gordon and $\phi^4$ theory, this reduces to an eigenvalue problem for the $\nu_{\rm L} = 1$ and $2$ Lam\'e equations with twisted periodicity conditions, where the twisting is related to the kink velocity.  We use a recent result from the mathematics literature \cite{MR4717595} to characterize the error in approximating the eigenvalues by solutions to a transcendental equation that arises from imposing twisted periodicity conditions on the infinite volume scattering modes, which is a typical approach in the physics literature.  The approximation is exponentially accurate in $mL$ as expected.

In Section \ref{sec:Discretize} we introduce lattice models for any $V(\phi)$ with either a reflection or discrete translation symmetry.  These models are not new.  They were first discussed in \cite{Drell:1976bq,Drell:1976mj} and studied further in the Ph.D. thesis of Pearson \cite{Pearsonthesis}.  They contain two key ingredients:  1) a spatial derivative operator---the ``SLAC'' derivative or, rather, finite periodic/anti-periodic versions of it---whose spectrum agrees with the exact derivative operator up to a the UV cutoff; and 2) an interaction potential built from the underlying lattice degrees of freedom using an interpolation field based on the discrete Fourier transform.  The key point of these ingredients is that, when combined, they equip the lattice theory with a \emph{continuous} spatial translation symmetry.  

The existence of a continuous translation symmetry is what enables us to introduce a collective coordinate degree of freedom and perform a finite lattice version of the classical soliton sector canonical transformation.  From a numerical perspective this result justifies the particular discretization of the FSE we utilize.  From a physics perspective it lays the groundwork for filling an important logical gap that exists in much of the QFT soliton literature, going back to the original work of \cite{Dashen:1974cj}.  Computations of quantum corrections to soliton observables like the soliton mass require delicate cancellations between quantities in the soliton and vacuum sectors that each become infinite when regulators are removed.  It is always tacitly assumed that the regulators are the same in the two sectors, but finite modifications to the map of regulators can change the answers for physical quantities.  Without a fully IR and UV regularized definition of the theory and transformation of field variables, how can one be sure of this map?  This point has been emphasized in recent work \cite{Guo:2019hiu,Evslin:2022opz}.  Exhibiting a quantum version of the classical transformation constructed here should fill this gap and is work in progress.

We finish Section \ref{sec:Discretize} by describing our numerical implementation of the initial value problem for the FSE using fourth-order Runge--Kutta to march forward in time.

%%%%%%%%%%%%%%%%%%%%
\subsection{Numerical Solutions}
%%%%%%%%%%%%%%%%%%%%

The bulk of our numerical results are presented in Subsection \ref{ssec:numresults}.  We consider field histories, collective coordinate trajectories, and energy distributions for various solutions to the FSE initial value problem.  The problem is parameterized by the initial velocity $\beta_i$ of the incoming kink, the momentum transfer $\epsilon$ delivered by the instantaneous impact force, and the dimensionless circle size $mL$.  We always choose $mL$ large enough and run times short enough such that radiation does not have time to wrap around the circle and self-interact.  

One of the most surprising outcomes from runs with large transfers $\epsilon \gtrsim1$ is that the kink's collective coordinate velocity $\beta(t)$ can become superluminal for a brief period after the impact.\footnote{This can even happen with small to moderate transfers that are in the same direction as the initial velocity of a highly boosted kink.}  See for example the high impact scenario of Figures \ref{fig:worldsheets} and \ref{fig:ccvelocities}.  As we explain in Subsection \ref{ssec:Superluminal}, while this phenomenon is certainly interesting, there is no need for alarm.  The collective coordinate velocity is not associated to any energy transport, and the energy flow velocity remains subluminal in such solutions as required by special relativity.  We provide a toy model for understanding the mechanism that leads to $|\beta(t)| > 1$.  

We also push the transfer to the absolute limit to see when the solution breaks down.  The physics here is quite interesting too.  At large enough transfers, a kink-antikink pair is formed in addition to the original kink.  See Figures \ref{fig:BigSmack} and \ref{fig:BigPush}.  This does not necessarily lead to a breakdown of the solution however.  The solution only breaks down when the antikink of the newly formed pair is imparted with enough kinetic energy to catch up to the original kink and annihilate it.  When this happens, the coordinate system we are using on configuration space breaks down.  It breaks down because the kink zero mode can have zero overlap with the dynamical field, and this overlap appears in a denominator in the soliton sector canonical transformation.  In a global coordinate system, such as one that uses the original field variable, one could presumably track the solution through the annihilation.\footnote{However setting up the impact force would be challenging in this variable since the collective coordinate is a spatially nonlocal functional of it.}

%%%%%%%%%%%%%%%%%%%%
\subsection{Perturbation Theory}
%%%%%%%%%%%%%%%%%%%%

Our final set of results considers the opposite regime of small momentum transfer.  In Subsection \ref{ssec:pert} we lay out the general framework of classical perturbation theory in $\epsilon$.  This is for the solution to the FSE initial value problem with general forcing $F(t)$ and an arbitrary initial velocity pure kink solution as starting point.  Then, in Subsection \ref{ssec:pertcheck} and the accompanying Appendix \ref{app:pert}, we carry out detailed calculations for the case of a delta function impact to second order in the amplitude and first order in the frequency.

After the impact, we find at first order that the field oscillates around a configuration that is different from the initial kink.  This first order time averaged configuration coincides with the fields of a $\beta_f$ boosted kink expanded to first order in $\epsilon$, where $\beta_f$ is the velocity associated to the final momentum $P_f = P_i + M_0 \epsilon$ of a pure kink configuration.  In other words, at first order and under time averaging, all of the Noether momentum $P_f$ is carried by the kink.  This is perhaps not surprising since the impact force couples only to the collective coordinate, and at first order we do not see the effects of interactions distributing this momentum  to the other degrees of freedom.  Similarly, the first order corrections to the oscillator frequencies are consistent with the oscillator frequencies around a $\beta_f$-boosted kink.

At second order we see the first deviation of the time averaged fields from those of a $\beta_f$-boosted kink, and we obtain a neat expression for the corresponding deviation of the time-averaged collective coordinate velocity.  It takes the form of a sum over the spectrum involving the first-order corrected frequencies.

We compare our perturbative results against numerical solutions to the FSE for a kink with a relatively small initial velocity and for a highly relativistic initial kink, finding beautiful agreement in both cases.  See Figure \ref{fig:Pertpics}.

We conclude in Section \ref{sec:Conclusions} by highlighting some of the many directions for future work.

%%%%%%%%%%%%%%%%%%%%
%%%%%%%%%%%%%%%%%%%%
\section{Kinks on a Circle}\label{sec:FSE}
%%%%%%%%%%%%%%%%%%%%
%%%%%%%%%%%%%%%%%%%%

We start by recalling the models of interest and their formulation on a circle.  We review how the forced soliton equation arises as the equation of motion in the co-moving frame of the soliton, discuss its static solutions, and analyze several aspects of time-dependent solutions.

%%%%%%%%%%%%%%%%%%%%
\subsection{Classical Actions and Kinks on $\mathbbm{R}^{1,1}$}
%%%%%%%%%%%%%%%%%%%%

We consider two-dimensional theories of a single real scalar field on $\mathbbm{R}^{1,1}$ with action
\begin{equation}\label{Scl}
S = \int \ed^2 x \left( -\half \pd_\mu \phi \pd^\mu \phi - V(m^2;\phi) \right)~.
\end{equation}
Spacetime points are denoted $x^\mu = (t,x)$, and we work in signature $(-,+)$.  We take the potential $V$ to have a reflection symmetry or to be periodic, and to have neighboring degenerate isolated minima at $\phi = \phi_{{\rm vac},\pm}$.  Without loss of generality we assume $V(m^2;\phi_{{\rm vac},\pm}) = 0$.  The parameter $m$ controls the mass associated with perturbative particle fluctuations around $\phi_{{\rm vac},\pm}$, and we use the notation $V(m^2;\phi)$ when we want to emphasize the dependence on this parameter.  

Under these conditions there exist time-independent solutions to the equations of motion,
\begin{equation}
\frac{\pd^2 \phi}{\pd x^2} - V^{(1)}(\phi) = 0~,
\end{equation}
where $V^{(n)}(\phi) \equiv \frac{\ed^n V}{\ed \phi^n}$, that interpolate from one vacuum to the other as $x \to \pm \infty$.  They are called kinks and antikinks.  We take the kink to interpolate from $\phi_{{\rm vac},-}$ to $\phi_{{\rm vac},+}$ and denote the corresponding solution
\begin{equation}
\phi = \phi_0(x - X)~.
\end{equation}
The free parameter $X$ is the center-of-mass position of the kink.

Canonical examples, along with their static classical kink solutions, are \emph{c.f.} \cite{Rajaraman:1982is}
\begin{align}\label{models}
\textrm{$\phi^4$ theory:} \quad &  V = \frac{1}{g^2} \left( g^2 \phi^2 - \frac{1}{4} m^2 \right)^2~, \quad &   \phi_0 = \frac{m }{2 g}\tanh\left( \frac{m}{\sqrt 2} (x- X)\right)~, \cr
\textrm{sine-Gordon:} \quad &  V = \frac{m^2}{g^2} \left( 1 - \cos{(g \phi)} \right)~,  \quad & \phi_0 = \frac{4}{g} \arctan\left( e^{m (x- X)} \right) ~.
\end{align}
In the $\phi^4$ model the two minima are at $\phi_{{\rm vac},\pm} = \pm m/2g$ while in the sine-Gordon model there is an infinite sequence at $\phi = \frac{2\pi n}{g}$, $n \in \mathbbm{Z}$, and we have chosen $\phi_{{\rm vac},-}$ to correspond to $n = 0$ and $\phi_{{\rm vac},+}$ to correspond to $n=1$.  

Kinks are examples of topological solitons.  The space of finite-energy field configurations has disconnected components labeled by a topological charge, which is in turn specified by the pair of vacua that the field asymptotes to as $x \to \pm \infty$.  There is no homotopy from a field configuration in one sector to a field configuration in another that passes through finite-energy configurations only.  The kink and antikink are energy minimizers within their components of field configuration space.    

In the two models considered above there is a single additional parameter in the potential, the coupling constant $g$, that controls interactions.  One can of course consider more complicated models with multiple couplings and coupling constants.  We highlight these two because the $\phi^4$ model is possibly the simplest Lorentz-covariant field theory with topological solitons, while the sine-Gordon model stands out due to its classical and quantum integrability \cite{Faddeev:1977rm,ZAMOLODCHIKOV1978525,PhysRevD.11.2088}.  Furthermore, in the quantum field theory context, degenerate isolated minima---essential for the existence of solitons---are  protected from quantum corrections only if they are the result of a spontaneously broken discrete symmetry.  This must be a translation or reflection symmetry for the class of linear $\sigma$ models under consideration.  The sine-Gordon and $\phi^4$ models represent the simplest models of each type.

The classical mass of the kink is the energy in the field configuration $\phi_0$:
\begin{equation}\label{Mcl}
M_0 = \int \ed x \left\{ \frac{1}{2} (\pd_{x} \phi_0)^2 + V(m^2;\phi_0) \right\}~.
\end{equation}
On $\mathbbm{R}^{1,1}$ Derrick's theorem \cite{Derrick:1964ww} guarantees that each term in the integrand contributes equally, such that $M_0 = \int \ed x (\pd_x \phi_0)^2$.  For $\phi^4$-theory and the sine-Gordon model we have
\begin{equation}
\textrm{$\phi^4$-theory:} \quad M_0 = \frac{\sqrt{2} m^3}{6 g^2}~, \qquad \textrm{sine-Gordon:} \quad M_0 = \frac{8 m}{g^2} ~.
\end{equation}

An expansion in fluctuations around the soliton profile leads to the eigenvalue problem
\begin{equation}\label{evproblem}
\left( - \frac{\ed^2}{\ed x^2} + V^{(2)}(\phi_0)  \right) \uppsi = \upomega^2 \uppsi ~.
\end{equation}
For the theory on $\mathbbm{R}^{1,1}$ there will be a discrete spectrum with eigenvalues $0 \leq \upomega_a < V^{(2)}(\phi_{\rm vac}) \equiv \mu$ followed by continuum $\upomega_{\underline{k}} = \sqrt{k^2 + \mu^2}$ labeled by wave number $k \in \mathbbm{R}$.  Here $\mu$ is the mass of the perturbative particle created by $\phi$; $\mu = 2m$ in $\phi^4$-theory while $\mu = m$ in the sine-Gordon model.  The notation $\underline{k}$ is meant to distinguish a continuum label from a discrete one.

The discrete spectrum always includes the zero mode, $\uppsi_0 \propto \pd_{x} \phi$, and might or might not contain additional $L^2$ modes depending on the model.  In the $\phi^4$ model there is one additional discrete mode beyond the zero mode, while in sine-Gordon the only discrete mode is the zero mode.  Physically, the zero mode generates translational motion of the kink, the other discrete modes represent possible distortions in the shape of the kink, and the continuum modes represent radiation above the kink.  The radiation modes have the usual relativistic dispersion relation of a particle with mass $\mu$.  

We introduce some notation that will be convenient in the following.  Let $\rho = x - X$ be the soliton-centered or co-moving coordinate.  Let $f' \equiv \pd_\rho f$ and $\dot{f} \equiv \pd_t f$.  We denote the usual $L^2$ inner product for functions of $\rho$ with a bra-ket:
\begin{equation}
\langle f | g \rangle = \int \ed \rho f(t,\rho)^\ast g(t,\rho)~.
\end{equation}
Then the normalized zero mode is
\begin{equation}
\uppsi_0(\rho) = \frac{1}{\sqrt{\langle \phi_{0}' | \phi_{0}' \rangle }} \phi_{0}'(\rho)  = \frac{1}{\sqrt{M_0}} \phi_{0}'(\rho) ~,
\end{equation}
where the last equality is only valid for kinks on $\mathbbm{R}$.

For $\phi^4$ theory the eigenvalue problem \eqref{evproblem} is
\begin{equation}\label{fluceigenequation}
\left( - \pd_{z}^2 + 4 - 6 \sech^2(z)\right) \uppsi = h \uppsi~, \qquad z = \frac{m \rho}{\sqrt{2}}~, \qquad h = \frac{2\upomega^2}{m^2} ~,
\end{equation}
which has the form of the time-independent Schrodinger equation for the $n=2$ member of the Poschl--Teller family of potentials.  The eigenmodes are
\begin{align}\label{flucspectrum}
\upomega_0 = 0: \quad & \uppsi_0 = \sqrt{ \frac{3m}{4\sqrt{2}}} \sech^2(\tfrac{m \rho}{\sqrt{2}})~, \cr
\upomega_{1} = \frac{\sqrt{3} m}{\sqrt{2}} : \quad & \uppsi_{1} = \sqrt{ \frac{3 m}{2 \sqrt{2}}} \tanh(\tfrac{m \rho}{\sqrt{2}}) \sech(\tfrac{m \rho}{\sqrt{2}})~, \cr
\upomega_{\underline{k}} = \sqrt{k^2 + 2m^2} : \quad & \uppsi_{\underline{k}} = e^{\ii k \rho} \frac{ \left( 3m^2 \tanh^2(\tfrac{m \rho}{\sqrt{2}}) - 3\sqrt{2} \ii m k \tanh(\tfrac{m \rho}{\sqrt{2}}) - (2k^2 + m^2) \right) }{\sqrt{4\pi (2k^2 + m^2)(k^2 + 2m^2)}}~, \cr
\end{align}
These modes are orthogonal and normalized such that $\langle \uppsi_a | \uppsi_b \rangle = \delta_{ab}$, $\langle \uppsi_a | \uppsi_{\underline{k}} \rangle = 0$, and $\langle \uppsi_{\underline{k}} , \uppsi_{\underline{l}} \rangle = \delta(k - l)$.

For sine-Gordon theory the eigenvalue problem is the $n=1$ Poschl--Teller problem:
\begin{equation}
\left( -\pd_{z}^2 + 1 - 2 \sech^2(z) \right) \uppsi = h \uppsi~, \qquad z = m \rho~, \qquad h = \frac{\upomega^2}{m^2} ~.
\end{equation}
The normalized zero mode and continuum modes are
\begin{align}\label{sGstaticmodes}
\upomega_0 = 0: \quad & \uppsi_0(\rho) = \sqrt{\frac{m}{2}} \sech(m\rho) ~, \cr
\upomega_{\underline{k}} = \sqrt{k^2 + m^2} : \quad &  \uppsi_{\underline{k}}(\rho) = e^{\ii k \rho} \frac{\left(\ii k - m \tanh(m \rho) \right)}{\sqrt{2\pi (k^2 + m^2)}} ~.
\end{align}

%%%%%%%%%%%%%%%%
\subsection{Kinks at Finite $L$}\label{ssec:finiteLkinks}
%%%%%%%%%%%%%%%%

When dealing with numerical solutions it is necessary to reduce the infinitely many field-theoretic degrees of freedom to a finite set.  This can be achieved by putting the theory on an interval of length $L$ and then discretizing.  The spatial discretization will be discussed in Section \ref{sec:Discretize}.  Here we consider the continuous finite-$L$ theory with action
\begin{equation}\label{SclfinL}
S = \int \ed t \int_{S_{L}^1} \ed x \left( -\half \pd_\mu \phi \pd^\mu \phi - V(m^2;\phi) \right)~.
\end{equation}
The IR regularization of models with kink solitons has been discussed extensively, since both UV and IR regularization are generally required to extract quantum field theoretic corrections to \emph{e.g.}~the kink mass.  We follow \cite{Sakamoto:1999yk,Mussardo:2004zn,Mussardo:2005dx,Pawellek:2008st,Pawellek:2008gs}.

% % % % % % % % % % % %
\subsubsection{Anti-periodic $\phi^4$ Kinks}
% % % % % % % % % % % %

In $\phi^4$ theory the kink is an odd function about $x = X$, and a natural boundary condition at finite $L$ is to impose anti-periodicity.  This will have the added advantage of enabling us to employ spectral methods based on the discrete Fourier transform when it comes to the spatial discretization.  Vacuum (or kink) sector field configurations are defined to be those satisfying periodic (or anti-periodic) identifications:
\begin{align}\label{sectors}
\textrm{vacuum sector:} \quad &  \phi(t,x+L) = \phi(t,x) ~, \cr
\textrm{kink sector:} \quad & \phi(t,x+L) = - \phi(t,x) ~. 
\end{align}
Both imply boundary conditions that are consistent with the variational principle for \eqref{SclfinL}.  The Lagrangian density will be periodic in either case, given our $\mathbbm{Z}_2$ symmetry assumption on $V$.

There is no notion of a topological charge for finite $L$ with the definition \eqref{sectors}, since one can deform a configuration in one sector to a configuration in the other by passing through $\phi = 0$, which now has finite energy.  What is true is that the energy barrier between the minimizing configurations in the different sectors \eqref{sectors} becomes infinite as $L \to \infty$.

The static kink profile will be anti-periodic and $C^{\infty}$ provided we impose anti-periodicity on both $\phi_0$ and its first derivative.  Doing so determines a unique solution to the equations of motion:\footnote{Note the $\mathrm{k}$ appearing here is unrelated to the wave number $k$ that appeared earlier.}
\begin{equation}\label{twistedkink}
\phi_0(\rho) = \frac{m}{2g} \sqrt{\frac{2 \rmk^2}{\rmk^2 + 1}} \, \operatorname{sn}\left(\left. \frac{m \rho}{\sqrt{\rmk^2 + 1}} \right| \rmk^2 \right)~,
\end{equation}
where $\operatorname{sn}(z \,| \, \mathrm{k}^2)$ is a Jacobi elliptic function.  The elliptic modulus $\rmk$ is determined by $L$ as follows.  The double periodicity of $\operatorname{sn}(z \, |\,\rmk^2)$ is
\begin{equation}
\operatorname{sn}(z + 2\alpha \mathbf{K}(\rmk^2) + 2i \beta \mathbf{K}(1-\rmk^2) \,| \, \rmk^2) = (-1)^\alpha \operatorname{sn}(z\,| \, \rmk^2)~,
\end{equation}
for $\alpha,\beta$ integer and where $\mathbf{K}(\rmk^2)$ is the complete elliptic integral of the first kind.  Hence as a function of real $z$, $\operatorname{sn}(z \,| \, \rmk^2)$ is anti-periodic with period $2 \mathbf{K}(\rmk^2)$.  Therefore the condition $\phi_0(x+L) = - \phi_0(x)$ is equivalent to
\begin{equation}\label{ktoL}
m L = 2 \sqrt{1 + \rmk^2} \, \mathbf{K}(\rmk^2)~.
\end{equation}

The classical mass of the kink can be computed from \eqref{Mcl}.  The gradient and potential terms no longer contribute equally and must be computed separately.  One finds
\begin{equation}
M_0 = \frac{2 m^3}{3 g^2 (1 + \rmk^2)^{3/2}} \left(I_1(\rmk^2) + I_2(\rmk^2) \right)~,
\end{equation}
with
\begin{align}\label{I12def}
I_1(\rmk^2) :=&~ \frac{3 \rmk^2}{8} \int_{-\mathbf{K}(\rmk^2)}^{\mathbf{K}(\rmk^2)} \ed z (\pd_z \operatorname{sn}(z \,| \, \rmk^2) )^2 \cr
=&~ \frac{1}{4} \left[ (1 + \rmk^2) \mathbf{E}(\rmk^2) - (1 - \rmk^2) \mathbf{K}(\rmk^2) \right] ~, \quad \textrm{and} \cr
I_2(\rmk^2) :=&~ \frac{3}{8} \int_{-\mathbf{K}(\rmk^2)}^{\mathbf{K}(\rmk^2)} \ed z \left( \rmk^2 \operatorname{sn}^2(z \,| \, \rmk^2) - \frac{(1+ \rmk^2)}{2} \right)^2 \cr
=&~ \frac{1}{16} \left[ 4 (1 + \rmk^2) \mathbf{E}(\rmk^2) - (1 - \rm k^2)(1 + 3 \rmk^2) \mathbf{K}(\rmk^2) \right]~,
\end{align}
where $\mathbf{E}$ is the complete elliptic integral of the second kind.  The total result is
\begin{equation}\label{twistedM0}
M_0 = \frac{m^3}{24 g^2} \frac{1}{(1 + \rmk^2)^{3/2}} \left[8 (1 + \rmk^2) \mathbf{E}(\rmk^2) - (1 - \rmk^2 ) (5 + 3 \rmk^2) \mathbf{K}(\rmk^2)  \right] ~.
\end{equation}

The $m L \to \infty$ limit corresponds to $\rmk \to 1$, where we indeed have $\phi_0(x) \to \frac{m}{2g} \tanh(\frac{m}{\sqrt{2}} x)$.  Meanwhile $I_{1,2} \to \half$, and therefore $M_0 \to \sqrt{2} m^3/(6 g^2)$.  When $\rmk \to 0$ we see that $\phi_0$ is forced to vanish.  At the same time, $M_0 \to (m^4/16 g^2) L_0$ where $L_0 = \pi/m$.  This energy agrees with that of the zero field configuration on an interval of size $L_0$.  Thus for $L < L_0$ there is no kink.  Since we are ultimately interested in $m L \to \infty$, we will always assume $L > L_0$.

The equation for fluctuations around the static kink, \eqref{evproblem}, can be recast into the Jacobian form of the $\nu_{\rm L}=2$ Lam\'e equation:
\begin{align}\label{finiteLevproblem}
& \left[ - \pd_{z}^2 + 2 (2+1) \rmk^2 \operatorname{sn}^2(z \,| \, \rmk^2) \right] \uppsi(z) = h  \uppsi(z) ~, \quad \textrm{with} \cr
& z = \frac{m \rho}{\sqrt{1 + \rmk^2}}~, \quad h = \left( \frac{\upomega^2}{m^2} + 1 \right)(1 + \rmk^2)~.
\end{align}
The full spectrum, which is now discrete, consists of a finite number of ``algebraic'' eigenvalues and an infinite set of ``transcendental'' eigenvalues.  The former are the finite $mL$ analog of the zero mode and shape mode eigenvalues, and the latter are the discrete analog of the radiation modes.  The algebraic eigenvalues can be obtained explicitly for low $\nu_{\rm L}$.  Restricting to anti-periodic eigenfunctions, the two for $\nu_{\rm L} = 2$ are 
\begin{equation}\label{algebraicphi4}
\upomega_0 = 0~, \qquad \upomega_{1} = \sqrt{\frac{3 \rmk^2 m^2}{1 + \rmk^2}} ~.
\end{equation}
The corresponding eigenfunctions are Lam\'e polynomials of the form
\begin{align}\label{twistedzm}
& \uppsi_0 = N_0 \operatorname{cn}(z \,| \, \rmk^2) \operatorname{dn}(z \,| \, \rmk^2)~, \qquad \uppsi_{1} = N_{\rm b} \operatorname{sn}(z \,| \, \rmk^2) \operatorname{dn}(z \,| \, \rmk^2)~,
\end{align}
with normalization constants
\begin{align}
N_{0}^2 =&~ \frac{3 \rmk^2 m}{2 \sqrt{\rmk^2 + 1} ( (1 + \rmk^2) \mathbf{E}(\rmk^2) - (1 - \rmk^2) \mathbf{K}(\rmk^2))} ~, \cr
N_{\rm 1}^2 =&~ \frac{3 \rmk^2 m}{2\sqrt{\rmk^2 + 1} ( (1 - \rmk^2) \mathbf{K}(\rmk^2) + (2\rmk^2-1) \mathbf{E}(\rmk^2) )} ~.
\end{align}
The zero mode is, as usual, the derivative of the static solution up to normalization: $\uppsi_0 \propto \pd_\rho \phi_0$.  

The transcendental eigenvalues are doubly degenerate so that the exact spectrum has the form $\upomega_0 < \upomega_1 < \upomega_2 = \upomega_3 < \upomega_4 = \upomega_5 < \cdots$.  The $\upomega_{a}$, for $a \geq 2$ must be computed numerically.  We discuss relevant background material and our approach to this computation in Appendix \ref{app:Lame}.

When $mL$ is large, the approximate solution to \eqref{ktoL} is
\begin{equation}\label{approxksoln}
\rmk = 1 - 8 e^{-m L/\sqrt{2}} \left( 1 + O(e^{-m L/\sqrt{2}}) \right)~.
\end{equation}
For $m L \gtrsim 56$ we have $1 - \rmk^2 < 10^{-16}$, meaning that $\rmk^2$ is numerically indistinguishable from $1$ to within machine precision.  Thus, when taking $m L >56$ in our numerical simulations, we will use the $mL \to \infty$ results for the static kink, its mass, the zero mode profile, and the shape mode eigenvalue and profile.

\begin{figure}[t!]
 \centering
\begin{subfigure}{.45\textwidth}
  \centering
  \includegraphics[width=\linewidth]{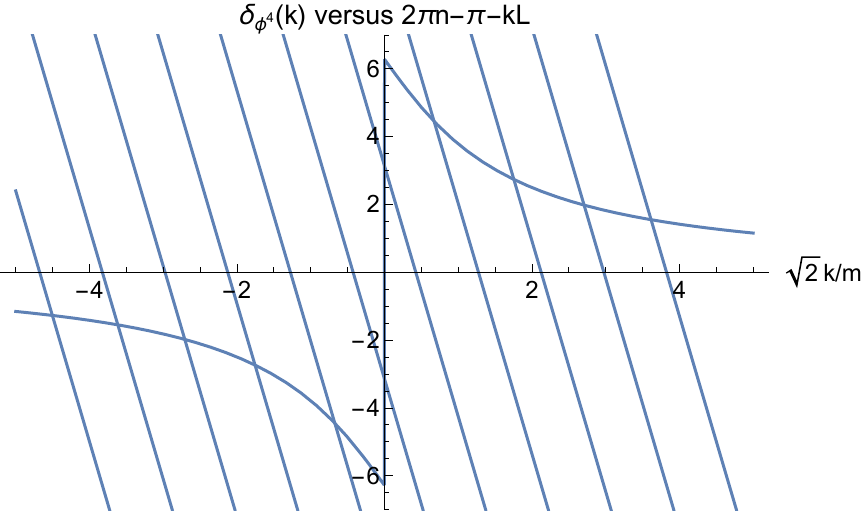}
  \caption{}
  \label{fig:quantcon1}
\end{subfigure}  \qquad %
\begin{subfigure}{.45\textwidth}
  \centering
  \includegraphics[width=\linewidth]{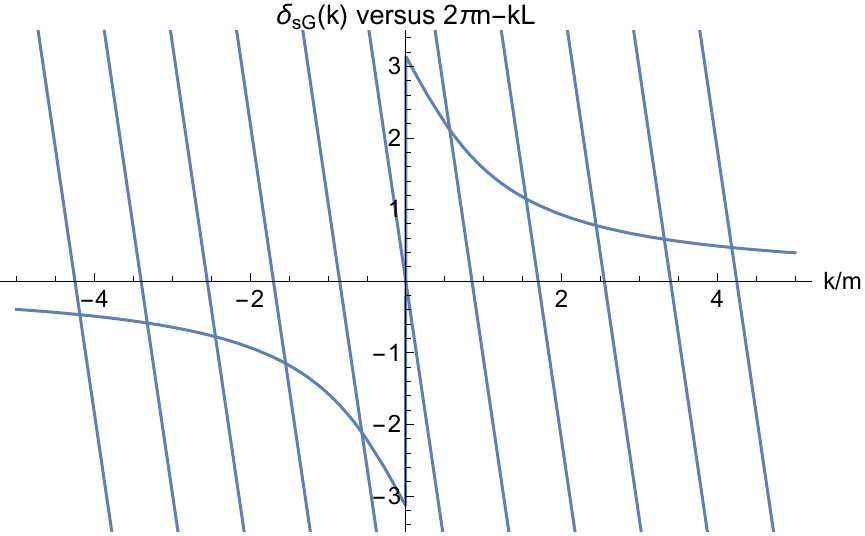}
  \caption{}
  \label{fig:quantcon2}
\end{subfigure}
\caption{Graphs of the quantization conditions for the approximate transcendental Lam\'e eigenvalues.  (a) is for $\phi^4$-theory, and (b) is for sine-Gordon.}
\label{fig:quantcon}
\end{figure}

Similarly, when $mL$ is large, a standard approach \cite{Dashen:1974cj} to obtaining the spectrum of radiation modes is to impose appropriate periodicity conditions on the continuum $mL \to \infty$ modes \eqref{flucspectrum}, as if they were the correct form of the modes at finite $m L$.  Requiring $\uppsi_{\underline{k}}(L/2) = - \uppsi_{\underline{k}}(-L/2)$ leads to the quantization condition $e^{\ii (k L+\delta(k))} = -1$, up to exponentially small corrections, with a phase shift that satisfies $\tan(\delta_{\phi^4}(k)/2) = 3k m/ (\sqrt{2} (k^2 + m^2))$.  This can be brought to the form
\begin{equation}\label{staticquantcon}
k L + \delta_{\phi^4}(k) = (2n - 1) \pi~, \qquad n \in \mathbbm{Z}~,
\end{equation}
with
\begin{equation}\label{phi4phase}
\delta_{\phi^4}(k) = -2\pi + 4\pi \Theta(\tfrac{k}{m}) -2 \left( \arctan( \tfrac{k}{\sqrt{2} m}) + \arctan(\tfrac{\sqrt{2} k}{m}) \right)~,
\end{equation}
where $\Theta$ is the Heaviside step function.  The solutions to this equation give a set of allowed wave numbers $k(n)$ and are shown graphically in Figure \ref{fig:quantcon1}.  Notice there is no solution when $n=0,1$.  These ``missing'' modes can be viewed as the two boundstate modes that are captured by the potential.

%%%%%%%%%%%%%%%%% 
%%%%%%%%%%%%%%%%%%

Taking $\upomega_{\underline{k(n)}} = \sqrt{k(n)^2 + 2m^2}$ does in fact give an exponentially accurate approximation (in $1/(mL)$) to the exact transcendental spectrum.  In Appendix \ref{app:Lame} we use a recent result \cite{MR4717595} on the $\rmk \to 1$ asymptotics of the Hill discriminant of the Lam\'e equation to determine the leading error in the approximation.  We note that $k(-n+1) = -k(n)$, for $n \geq 2$, which leads to the two-fold degeneracy of these approximate eigenvalues.  

When labeling the eigenvalues and corresponding eigenfunctions of the discrete transcendental spectrum, it is convenient to continue the values of the $a,b$-type index beyond the last algebraic mode, ordered by increasing energy.  The relationship between eigenvalue index $a$ and its corresponding quantum number $n_a$ for the approximate eigenvalues is taken to be
\begin{equation}
n_{0,1} = 0,1~, \quad n_{2j} = j+1~, \quad n_{2j+1} = -j~, \quad j \in \mathbbm{N}~.
\end{equation}
Then we denote $k(n_a) \equiv k_a$ and the associated approximate eigenvalue by $\upomega_{a}^{\approx} \equiv \upomega_{\underline{k(n_a)}}$.  Appendix \ref{app:Lame} shows that the difference $\upomega_a - \upomega_{a}^{\approx}$ is exponentially small in $1/(mL)$.

We also expect that the continuum scattering modes in \eqref{flucspectrum} with $k = k_a$ give an exponentially accurate approximation to the exact transcendental eigenfunctions.  In this context, however, the normalization must be modified if we want the new $\uppsi_a \propto \uppsi_{\underline{k_a}}$ to satisfy $\langle \uppsi_a | \uppsi_b \rangle = \delta_{ab}$ approximately. Specifically, we should multiply the continuum modes by $\sqrt{\frac{\ed k}{\ed n} |_{n_a}}$ so that
\begin{align}\label{DiracKronecker}
\sqrt{ \frac{\ed k}{\ed n} \bigg|_{n_a} \frac{\ed k}{\ed n} \bigg|_{n_b}} \int_{-L/2}^{L/2} \ed \rho \uppsi_{\underline{k_a}}^\ast(\rho) \uppsi_{\underline{k_b}}(\rho) \approx &~  \sqrt{ \frac{\ed k}{\ed n} \bigg|_{n_a} \frac{\ed k}{\ed n} \bigg|_{n_b}}  \int_{-\infty}^{\infty} \ed \rho \uppsi_{\underline{k_a}}^\ast(\rho) \uppsi_{\underline{k_b}}(\rho) \cr
=&~ \frac{\ed k}{\ed n} \bigg|_{n_a} \delta(k_a - k_b) \cr
=&~ \delta(n_a - n_b)~.
\end{align}
In the first line we approximated the integral over the period $L$ by an integral over $\mathbbm{R}$.  This approximation is exponentially accurate in the distributional sense, \ie\ when acting on an appropriate class of test functions.  The Dirac distribution we obtained in the last line acts on functions of a continuous $n$ via $f \mapsto \int \ed n' f(n') \delta(n - n') = f(n)$.  This is the appropriate continuum limit of $\{f_{n'}\} \mapsto \sum_{n'} \delta_{n n'} f_{n'} = f_n$, which is what we are after.

In the perturbative sector, the quantization condition for the allowed wave numbers is $k = 2\pi n/L$ so that $\sqrt{ \frac{\ed k}{\ed n}} = \sqrt{\frac{2\pi}{L}}$.   This is the usual rescaling from a plane-wave normalized mode to a discrete normalized mode.  However in the soliton sector we must take the derivative of \eqref{staticquantcon} with respect to $n$ which results in
\begin{equation}
\frac{\ed k}{\ed n}\bigg|_{n_a} = \frac{2\pi}{L + \delta'(k_a)} ~,
\end{equation}
where $\delta'(k_a) \equiv \frac{\ed \delta}{\ed k}(k_a)$.  The correction from the phase shift is only $O(1/L)$ suppressed compared to the leading $2\pi/L$ behavior and should be kept.  When considering finite but large $mL$ numerics it makes a noticeable difference.  Explicitly, for $\phi^4$-theory we have
\begin{equation}
\delta_{\phi^4}'(k) = - \frac{6\sqrt{2} m (k^2 + m^2)}{(2k^2 + m^2)(k^2 + 2m^2)}~,
\end{equation}
and our approximate transcendental eigenmodes are 
\begin{equation}\label{phi4approxmodes}
\uppsi_{a}^{\approx}(\rho) = \sqrt{\frac{2\pi}{L + \delta_{\phi^4}'(k_a)}} \uppsi_{\underline{k_a}}(\rho)~.
\end{equation}
One can confirm by direct integration over $\rho \in [-L/2,L/2]$ that these satisfy $\langle \uppsi_a | \uppsi_b \rangle = \delta_{ab} + O(e^{-m L/(2\sqrt{2})})$.

% % % % % % % % % % % % 
\subsubsection{Quasi-periodic sine-Gordon Kinks}
% % % % % % % % % % % %

In the sine-Gordon model the field $\phi$ is usually considered to be circle-valued with periodicity $\phi \sim \phi + 2\pi/g$.  The action \eqref{SclfinL} is invariant under shifts $\phi \to \phi + \frac{2\pi \nu}{g}$ for $\nu \in \mathbbm{Z}$.  If we take the one-dimensional space to be a circle with periodicity $L$, then at any fixed time, $x \mapsto \phi(t,x)$ is a map from $S^1$ to $S^1$.  The configuration space of such smooth maps is decomposed into an infinite sequence of disjoint components labeled by the winding number $\nu \in \mathbbm{Z}$.  The $\nu = 0$ sector is the perturbative sector while the $\nu = \pm 1$ sectors are the kink and antikink sectors.  Higher $|\nu|$ corresponds to multi-kink sectors, but there do not exist time-independent solutions to the equations of motion in these sectors.  Physically, sine-Gordon kinks apply attractive forces to each other.

The kink solution at finite $L$ is
\begin{equation}\label{sGLkink}
\phi_0(\rho) = \frac{\pi}{g} + \frac{2}{g} \operatorname{am}\left( \left. \frac{m \rho}{\rmk} \right| \rmk^2\right)~,
\end{equation}
where the elliptic modulus is related to the circumference via
\begin{equation}\label{sGLrmk}
m L = 2 \rmk \mathbf{K}(\rmk^2)~.
\end{equation}
Here $\operatorname{am}(z \, | \, \rmk^2)$ is the Jacobi amplitude function, which is the inverse of the incomplete elliptic integral of the first kind.
It is related to some of the other Jacobi elliptic functions introduced earlier by
\begin{equation}
\operatorname{sn}(z \,| \, \rmk^2) = \sin(\operatorname{am}(z \,| \, \rmk^2))~, \quad \operatorname{cn}(z \,| \, \rmk^2) = \cos(\operatorname{am}(z \,| \,\rmk^2))~, \quad \operatorname{dn}(z \,| \, \rmk^2) = \frac{\ed}{\ed z} \operatorname{am}(z \,| \, \rmk^2)~.
\end{equation}
The Jacobi amplitude function has quasi-periodicity
\begin{equation}
\operatorname{am}(z+2\mathbf{K}(\rmk^2) \,| \, \rmk^2) = \operatorname{am}(z \,| \, \rmk^2) + \pi~.
\end{equation}
This with \eqref{sGLrmk} implies $\phi_0 \to \phi_0 + 2\pi/g$ as $\rho \to \rho + L$.  

The classical kink mass is given by
\begin{align}\label{sGfinLmass}
M_0 =&~ \frac{2m \rmk}{g^2} \int_{-\mathbf{K}(\rmk^2)}^{\mathbf{K}(\rmk^2)} \ed z \left\{ \frac{1}{\rmk^2} \operatorname{dn}^2(z \,| \, \rmk^2) + \operatorname{cn}^2(z \,| \, \rmk^2) \right\}   \cr
=&~  \frac{4m}{g^2 \rmk} \left[ (\rmk^2 - 1) \mathbf{K}(\rmk^2) + 2 \mathbf{E}(\rmk^2) \right]~.
\end{align}
In the limit $L \to \infty$, we have $\rmk \to 1$, $\phi_0(\rho) \to \frac{4}{g} \operatorname{arctan}(e^{m\rho})$ and $M_0 \to 8m/g^2$.

The eigenvalue problem \eqref{evproblem} for fluctuations around the kink takes the form of the $\nu_{\rm L} = 1$ Lam\'e equation:
\begin{equation}\label{lame1standard}
\left[ - \frac{\ed^2}{\ed z^2} + 2\rmk^2 \operatorname{sn}^2(z \,| \, \rmk^2) \right] \uppsi = h \uppsi~, \qquad z =\frac{m \rho}{\rmk} ~, \qquad h = \rmk^2 \left(\frac{\upomega^2}{m^2} + 1 \right) ~.
\end{equation}
The normalized zero-mode is
\begin{equation}\label{sGfinLzm}
\uppsi_0(\rho) = \sqrt{\frac{m}{2 \rmk \mathbf{E}(\rmk^2)}} \operatorname{dn}\left(\left. \frac{m \rho}{\rmk} \right|  \rmk^2\right) ~.
\end{equation}
Since $\lim_{\rmk \to 1} \operatorname{dn}(z \,| \, \rmk^2) = \sech(z)$, we see that this has limit $\uppsi_0 \to \sqrt{\frac{m}{2}} \sech(m \rho)$ as $L \to \infty$.  At large $mL$, the solution to \eqref{sGLrmk} is $\rmk \approx 1 - 8 e^{-mL}$.  There are no further algebraic eigenvalues corresponding to periodic eigenfunctions.  The transcendental eigenvalues are doubly degenerate and described in Appendix \ref{app:Lame}.  

At large $mL$, the transcendental eigenvalues are well approximated by $\upomega_{\underline{k(n)}}$, where the $k(n)$ are the solutions to the quantization condition
\begin{equation}\label{sGquantcon}
k L + \delta_{\rm sG}(k) = 2\pi n~, \qquad n \in\mathbbm{Z}~,
\end{equation}
with phase shift
\begin{equation}\label{sGphase}
\delta_{\rm sG}(k) = -2 \arctan(k/m) + 2\pi \Theta(k/m) - \pi ~.
\end{equation}
Here there is no solution when $n=0$.  This ``missing'' mode can be thought of as the zero mode.  See Figure \ref{fig:quantcon2}.  The condition \eqref{sGquantcon} is found by imposing the periodicity condition $\uppsi_{\underline{k}}(-L/2) = \uppsi_{\underline{k}}(L/2)$ on the scattering modes in \eqref{sGstaticmodes}.  We have $k(n) = k(-n)$ and so we define $n_a$ via
\begin{equation}
n_0 = 0~, \quad n_{2j-1} = j~, \quad n_{2j} = -j~, \quad j \in \mathbbm{N}~.
\end{equation}
Then we set $k_a \equiv k(n_a)$, and the approximate transcendental spectrum is $\upomega_{a}^{\approx} = \upomega_{\underline{k_a}}$ for $a \geq 1$.  

As in the case of $\phi^4$-theory, the eigenfunctions can be approximated by setting $k = k_a$ in the scattering modes and modifying the normalization.  This leads to
\begin{align}\label{sGapproxmodes}
\uppsi_{a}^{\approx}(\rho) =& \sqrt{\frac{2\pi}{L + \delta_{\rm sG}'(k_a)}} \uppsi_{\underline{k_a}}(\rho)~, \qquad \textrm{where} \quad \delta_{\rm sG}'(k) = -\frac{2m}{k^2 + m^2}  ~.
\end{align}
%

%%%%%%%%%%%%%%%%
\subsection{Canonical Transformation in the Soliton Sector}
%%%%%%%%%%%%%%%%

The classical action in Hamiltonian form is
\begin{align}
S =&~ \int \ed t \left( \int \ed x \pi \dot{\phi} - H \right)~, \qquad \textrm{where} \cr
H =&~ \int \ed x \left( \half \pi^2 + \half (\pd_x \phi)^2 + V(m^2;\phi) \right)~.
\end{align}
The canonical transformation of Gervais--Jevicki--Sakita \cite{Gervais:1975pa} and Tomboulis \cite{Tomboulis:1975gf} is a change of variables on phase space, $(\pi ; \phi) \mapsto (P,\varpi; X, \chi)$, in the soliton sector of the theory, in which one of the new coordinates is taken to be the position parameter of the kink, while the remaining coordinate degrees of freedom are packaged into a field $\chi$ that is constrained to be orthogonal to the zero mode of the kink: $\langle \uppsi_0 | \chi \rangle =0$.  $P$ and $\varpi$ are the canonically conjugate momenta.  The change of variables for the position field is
\begin{equation}\label{phi2chi}
\phi(t,x) = \phi_0(t,x-X(t)) + \chi(t,x-X(t))~.
\end{equation}
The original motivation for this transformation was the desire to develop a perturbation theory for QFT in the presence of the soliton which is free from divergences associated with the zero eigenvalue of the linear operator for small fluctuations.  The degree of freedom associated with the zero mode $\uppsi_0$ is replaced by the collective coordinate $X(t)$, so that motion along the energy-minimizing trough in field configuration space parameterized by $X$ is treated fully nonlinearly.

The momentum field transformation is determined by the requirement that the full phase space transformation be canonical and is
\begin{equation}\label{pi2varpi}
\pi(t,x) = - \frac{ (P(t) + \langle \chi' | \varpi \rangle )}{\langle \uppsi_0 | \phi_{0}' + \chi' \rangle} \uppsi_0(x- X(t)) + \varpi(t,x-X(t)) ~.
\end{equation}
From the QFT perspective, this ensures that the measure of the phase space path integral is preserved,\footnote{More precisely, the measure is not quite invariant under this nonlinear point canonical transformation, but only so to leading order in the semiclassical expansion.  A known correction is generated that can be incorporated into the action as a two-loop quantum correction to the potential \cite{Gervais:1976ws}.} and from the classical perspective it ensures that Hamilton's equations maintain their form.  

It is convenient to make a further canonical transformation $(P,\varpi; X,\chi) \mapsto (P,\varpi;X,\varphi)$ with $\varphi(t,\rho) = \phi_0(t,\rho) + \chi(t,\rho)$, so that the transformation from the original fields is
\begin{align}\label{canonicaltrans}
\phi(t,x) =&~ \varphi(t,x-X(t))~, \cr
\pi(t,x) =&~ - \frac{( P(t) + \langle \varphi' | \varpi \rangle )}{\langle \uppsi_0 | \varphi' \rangle} \uppsi_0(x-X(t)) + \varpi(t,x-X(t))~.
\end{align}

This form of the canonical transformation \eqref{canonicaltrans} is valid for the finite-$L$ theory, taking $\phi_0$ and $\uppsi_0$ to be the the appropriate kink and zero mode, as described above for the $\phi^4$ and sine-Gordon models.  The new position field satisfies the constraint $\langle \uppsi_0 | \varphi \rangle = \langle \uppsi_0 | \phi_{0} \rangle$ while the momentum field still satisfies $\langle \uppsi_0 | \varpi \rangle = 0$.  

There are two situations in which the canonical transformation \eqref{canonicaltrans} can break down.  First, it can happen that there is not a unique solution for the collective coordinate.  Given a field configuration $\phi(t,x)$ in the one-soliton sector, the collective coordinate $X = X[\phi]$ is determined by imposing the constraint:
\begin{equation}\label{Xofphi}
\int \ed \rho \uppsi_0(\rho) \left( \phi(t, \rho + X(t)) - \phi_0(\rho) \right) = 0~.
\end{equation}
For a smooth $\phi$ satisfying soliton sector boundary conditions, one can argue that this equation always has at least one solution, but may in general have an odd number of solutions.  The physical interpretation of having multiple solutions is that $\phi$ has the appearance of a train of alternating kinks and antikinks.  

The second situation in which the transformation can break down is when the overlap $\langle \uppsi_0 | \varphi' \rangle$ vanishes.  Going back to \eqref{pi2varpi}, we see that this requires the fluctuation field $\chi$ around the static kink to be $O(1/g)$, which again signals the appearance of kink-antikink pairs since this is the amplitude required to traverse the potential barrier between vacua.  Hence the coordinates $(X,\varphi)$ are not good in such regions of configuration space.  This is hardly surprising since they are predicated on decomposing the field into ``kink plus fluctuation.''  

The action and Hamiltonian in the new variables are
\begin{align}\label{ssS}
S =&~ \int \ed t \left( P \dot{X} + \langle \varpi | \dot{\varphi} \rangle - H_T \right)~, \qquad \textrm{where} \cr
H_T =&~ \frac{ (P + \langle \varpi | \varphi' \rangle)^2}{2 \langle \uppsi_0 | \varphi' \rangle^2} + \int_{-L/2}^{L/2} \ed\rho \left\{ \tfrac{1}{2} \varpi^2 + \tfrac{1}{2} \varphi^{\prime 2} + V(m^2;\varphi) \right\} + \lambda \langle \uppsi_0 | \varphi \rangle + \nu \langle \uppsi_0 | \varpi \rangle \cr
\equiv &~ H +  \lambda \langle \uppsi_0 | \varphi - \phi_0 \rangle + \nu \langle \uppsi_0 | \varpi \rangle ~.
\end{align}
The last two terms in the total Hamiltonian $H_T$ enforce the constraints; $\lambda(t), \nu(t)$ are the associated Lagrange multipliers.

Additionally, we note it follows from \eqref{canonicaltrans} that
\begin{equation}\label{PNoether}
P_{\rm Noether} = - \int \ed x \pi \phi = P~,
\end{equation}
In other words, the collective coordinate momentum is the Noether charge associated with spatial translations.  Thus, $P$ plays a dual role as the canonically conjugate momentum to the collective coordinate and as the total momentum in the fields.

%%%%%%%%%%%%%%%%%%%%%%
\subsection{The Forced Soliton Equation}
%%%%%%%%%%%%%%%%%%%%%%
 
 The forced soliton equation arises as the classical equation of motion for $\varphi$ in the presence of an arbitrary $P(t)$.  Hamilton's equations with respect to $H_T$ are
 \begin{align}
 \dot{\varpi} =&~ - \frac{\delta H_T}{\delta \varphi} = - \frac{\delta H}{\delta \varphi} - \nu \uppsi_0 ~, \cr
 \dot{\varphi} =&~ \frac{\delta H_T}{\delta \varpi} = \frac{\delta H}{\delta \varpi} + \lambda \uppsi_0 ~.
 \end{align}
 We can solve for the Lagrange multipliers by taking the inner product of these equations with $\uppsi_0$.  Plugging the result back in, we find
 \begin{align}
 \dot{\varpi} =&~ - \left( \frac{\delta H}{\delta \varphi} \right)^\perp ~, \qquad \dot{\varphi} = \left( \frac{\delta H}{\delta \varpi} \right)^\perp ~,
 \end{align}
 where we have introduced the notation
 \begin{equation}
 f^\perp := f - \langle \uppsi_0 | f \rangle \uppsi_0~,
 \end{equation}
 for the projection of $f$ onto the orthogonal complement of $\uppsi_0$.  Here we have used that $\dot{\varpi},\dot{\varphi}$ are orthogonal to $\uppsi_0$.  This follows from the constraint equations and the fact that $\uppsi_0$ is time independent.
 
 The first variations of the Hamiltonian are
 \begin{align}\label{dHdphi}
\frac{\delta H}{\delta \varpi} =&~ \frac{ (P + \langle \varpi | \varphi' \rangle)}{\langle \uppsi_0 | \varphi' \rangle^2} \varphi' + \varpi  \cr
=&~ \beta \varphi' + \varpi~, \cr
\frac{\delta H}{\delta \varphi} =&~ - \frac{ (P + \langle \varpi | \varphi' \rangle )}{\langle \uppsi_0 | \varphi' \rangle^2} \varpi'  + \frac{ (P + \langle \varpi | \varphi' \rangle)^2}{\langle \uppsi_0 |\varphi' \rangle^3} \uppsi_0'  - \varphi'' + V^{(1)}(\varphi)  \cr
=&~ - \beta \varpi' + \beta^2 \langle \uppsi_0 | \varphi' \rangle \uppsi_0' - \varphi'' + V^{(1)}(\varphi) ~.
\end{align}
In the second steps we used the definition of the velocity functional,
\begin{equation}\label{betafunc}
\beta[P,\varpi,\varphi] := \frac{P + \langle \varpi | \varphi' \rangle}{\langle \uppsi_{0} |\varphi' \rangle^2}~.
\end{equation}
The name is appropriate since Hamilton's equation for $X$ gives $\dot{X} = \beta$.  Taking the orthogonal projection, the equations of motion are then
\begin{align}\label{HamiltonFSE}
\dot{\varphi} =&~ \frac{ (P + \langle \varpi | \varphi' \rangle )}{\langle \uppsi_0 |\varphi' \rangle^2} \left( \varphi' - \langle \uppsi_0 |\varphi' \rangle \uppsi_0 \right) + \varpi  \cr
=&~ \beta (\varphi' - \langle \uppsi_0 |\varphi' \rangle \uppsi_0 ) + \varpi~, \cr
\dot{\varpi} =&~   \frac{ (P + \langle \varpi | \varphi' \rangle )}{\langle \uppsi_0 |\varphi' \rangle^2} (\varpi' - \langle \uppsi_0 | \varpi' \rangle \uppsi_0) - \frac{ (P + \langle \varpi | \varphi' \rangle )^2}{\langle \uppsi_0 |\varphi' \rangle^3} \uppsi_0' + \varphi'' - V^{(1)}(\varphi) + \cr
&~ - \left\langle \uppsi_0 \bigg| \left( \varphi'' - V^{(1)}(\varphi) \right) \right\rangle \uppsi_0 \cr
=&~  \beta (\varpi' - \langle \uppsi_0 |\varpi' \rangle \uppsi_0 ) - \beta^2 \langle \uppsi_0 |\varphi' \rangle \uppsi_{0}' + \left(  \varphi'' - V^{(1)}(\varphi) \right)^\perp~,
\end{align}
where we used that $\varpi, \uppsi_0'$ are orthogonal to $\uppsi_0$.  We refer to these equations as the Hamiltonian form of the forced soliton equation.

We can use the first equation to eliminate $\varpi$ and obtain a second order equation for $\varphi$ as in \cite{Melnikov:2020ret}.  First, taking the inner product with $\varphi'$ one finds an expression for $\langle \varpi |\varphi' \rangle$, which in turn yields
\begin{equation}\label{beta2}
\beta[P,\varphi] = \frac{P + \langle \dot{\varphi} | \varphi ' \rangle}{\langle \varphi' | \varphi' \rangle} ~.
\end{equation}
Then substituting
\begin{equation}\label{varpisol}
\varpi = \dot{\varphi} - \beta (\varphi' - \langle \uppsi_0 | \varphi' \rangle \uppsi_0 )
\end{equation}
into the second equation leads to the forced soliton equation:
\begin{equation}\label{FSE}
\left( \pd_t - \beta[\varphi;P] \pd_\rho \right)^2 \varphi - \varphi'' + V^{(1)}(\varphi) + \frac{\dot{P}(t) \uppsi_0(\rho)}{\langle \uppsi_0 | \varphi' \rangle} = 0~.
\end{equation}
%

%%%%%%%%%%%%%%%%%%%
\subsection{Static Solutions}\label{ssec:boostedkinks}
%%%%%%%%%%%%%%%%%%%

We first discuss time-independent solutions to \eqref{FSE} in the finite-$L$ theory.  These play a role in setting up initial conditions for the time-dependent numerical solutions below.  Additionally, we will use these solutions to define the instantaneous (mechanical) momentum and energy of the kink for a general time-dependent solution.

Assuming time-independence for $P$ and $\varphi$, the velocity functional $\beta$ \eqref{beta2}, is a constant.  With the $\dot{P}$ term vanishing, the FSE reduces to the classical field equation in the new variable $\tilde{\rho} = \rho/\sqrt{1-\beta^2}$.  Hence the boosted kink solution $\varphi(\rho) = \phi_{\beta}(\rho)$ is
\begin{align}\label{boostedAPkink}
\textrm{$\phi^4$-theory:} \qquad \phi_{\beta}(\rho) :=&~ \frac{m}{2g} \sqrt{\frac{2 \rmk_{\beta}^2}{\rmk_{\beta}^2 + 1}} \, \operatorname{sn}\left(\left. \frac{m \rho}{\sqrt{(1-\beta^2)(\rmk_{\beta}^2 + 1)}} \right| \rmk_{\beta}^2 \right)~. \cr
\textrm{sine-Gordon:} \qquad \phi_{\beta}(\rho) :=&~ \frac{\pi}{g} + \frac{2}{g} \operatorname{am}\left( \left. \frac{m \rho}{\rmk_\beta \sqrt{1-\beta^2}} \right| \rmk_{\beta}^2 \right)~.
\end{align}
We must still impose anti-periodicity or quasi-periodicity under $\rho \to \rho + L$, which makes the elliptic parameter $\beta$-dependent.  It is now the solution to
\begin{align}\label{kbetatoL}
\textrm{$\phi^4$-theory:} \qquad m L =&~ 2 \sqrt{ (1-\beta^2) (\rmk_{\beta}^2 + 1)} \, \mathbf{K}(\rmk_{\beta}^2)~, \cr
\textrm{sine-Gordon:} \qquad m L =&~ 2 \rmk_{\beta} \sqrt{1-\beta^2} \mathbf{K}(\rmk_{\beta}^2)~.
\end{align}
Note that \eqref{kbetatoL} has the same form as \eqref{ktoL} and \eqref{sGLrmk} with $L \to L/\sqrt{1-\beta^2}$.  Indeed, if the static kink solution is denoted $\phi_0(\rho;L)$ then the boosted kink profile is 
\begin{equation}\label{boostedviastatic}
\phi_\beta(\rho) = \phi_0(\gamma\rho;\gamma L)~,
\end{equation}
where $\gamma = (1-\beta^2)^{-1/2}$ is the usual Lorentz contraction factor.  This result is general for the relationship between any static and boosted kink profile at finite $L$, except that we should in general allow for an offset $\rho_0$ in the boosted solution, $\phi_{\beta}(\rho) = \phi_0(\gamma(\rho-\rho_0);\gamma L)$, which is to be fixed by the requirement that the boosted solution satisfy the constraint $\langle \uppsi_0 | \phi_{\beta} - \phi_0 \rangle = 0$.  In the case of $\phi^4$-theory and sine-Gordon, the solution to the constraint is $\rho_0 = 0$.  We also note that the momentum field on this solution is not zero.  From \eqref{HamiltonFSE} we have
\begin{equation}\label{boostedAPkinkmom}
\varpi_{\beta}(\rho) = - \beta (\phi_{\beta}' - \langle \uppsi_0 | \phi_{\beta}' \rangle \uppsi_0)~,
\end{equation}
with $\uppsi_0$ given in \eqref{twistedzm} or \eqref{sGfinLzm} for $\phi^4$-theory and sine-Gordon respectively.

The velocity $\beta$ must be determined in terms of $P$ through \eqref{beta2}.  With $\varphi = \phi_\beta$ we have
\begin{align}\label{finiteLrelmass}
\textrm{$\phi^4$ theory:} \qquad \langle \phi_{\beta}' | \phi_{\beta}' \rangle =&~ \frac{4 m^3 \gamma}{3 g^2 (1 + \rmk_{\beta}^2)^{3/2}} I_1(\rmk_{\beta}^2)  \cr
=&~   \frac{\gamma m^3}{3 g^2 (1 + \rmk_{\beta}^2)^{3/2}} \left[ (1 + \rmk_{\beta}^2) \mathbf{E}(\rmk_{\beta}^2) - (1- \rmk_{\beta}^2) \mathbf{K}(\rmk_{\beta}^2) \right] \cr
\equiv &~ \gamma M_{\beta}' ~,   \cr
\textrm{sine-Gordon:} \qquad  \langle \phi_{\beta}' | \phi_{\beta}' \rangle =&~ \frac{4 m^2}{g^2 \rmk_{\beta}^2 (1-\beta^2)} \int_{-L/2}^{L/2} \ed \rho  \operatorname{dn}^2\left(\left. \frac{m \rho}{\rmk_{\beta} \sqrt{1-\beta^2}} \right| \rmk_{\beta}^2 \right) \cr 
=&~ \frac{8 m \mathbf{E}(\rmk_{\beta}^2)}{g^2 \rmk_{\beta} \sqrt{1-\beta^2}} \cr
\equiv &~ \gamma M_{\beta}'~,
\end{align}
where in the last steps we defined a ``mass'' which depends on $\beta$ only through the dependence of $\rmk_\beta$ on $\beta$.  The relationship \eqref{beta2} then says $P = p_{\beta}$ with
\begin{equation}\label{LorentzL}
p_{\beta} := \langle \phi_{\beta}' | \phi_{\beta}' \rangle \beta  = \gamma M_{\beta}' \beta ~.
\end{equation}
This is not the usual relativistic relationship between momentum and velocity because $M_{\beta}' \neq M_0$ as given in \eqref{twistedM0} or \eqref{sGfinLmass}, for any $\beta$.  This is to be expected since the finite $L$ boundary conditions break Lorentz symmetry.  In the limit $L \to \infty$, however, we have $M_{\beta}' \to M_0$ for any physical $\beta$, (\emph{i.e.}~$|\beta| < 1$).

The total energy of the solution \eqref{boostedAPkink} also differs from the relativistic expression, $\gamma M_0$, at finite $L$.  This energy can be computed from the Hamiltonian \eqref{ssS} evaluated on the solution.  Using \eqref{betafunc}, and \eqref{boostedAPkinkmom}, we find
\begin{align}\label{LorentzH}
E_{\beta} := H[\varpi_{\beta},\phi_{\beta}] =&~ \int d\rho \left\{ \half (1 + \beta^2) \phi_{\beta}^{\prime 2} + V(m^2;\phi_{\beta}) \right\}~.
\end{align}
Evaluating for the two cases,
\begin{align}
\textrm{$\phi^4$-theory:} \qquad E_{\beta} =&~ \frac{2 m^3 \gamma}{3 g^2 (1 + \rmk_{\beta}^2)^{3/2}} \left[ (1 + \beta^2) I_1(\rmk_{\beta}^2) + (1 - \beta^2) I_2(\rmk_{\beta}^2) \right] \cr
\equiv &~ \gamma M_{\beta} ~, \cr
\textrm{s-G:} \qquad E_{\beta} 
=&~ \frac{2 m}{g^2 \sqrt{1-\beta^2}} \left[ \frac{2}{\rmk_{\beta}} \mathbf{E}(\rmk_{\beta}^2) (1 + \beta^2) + 2( \mathbf{E}(\rmk_{\beta}^2) + (\rmk_{\beta}^2 - 1) \mathbf{K}(\rmk_{\beta}^2) ) (1-\beta^2) \right] \cr
\equiv &~ \gamma M_\beta~. 
\end{align}
In the last steps we defined another $\beta$-dependent mass, which is different from $M_0$ for finite $L$ and $\beta \neq 0$.  This mass agrees with $M_0$ when $\beta = 0$.  This is as it should be since, when $\beta = 0$, the solution \eqref{boostedAPkink} reduces to \eqref{twistedkink}.  Furthermore, when $L \to \infty$ we have $M_\beta \to M_0$ for any $\beta$, and we recover the usual relativistic relations among $E_{\beta}, p_{\beta}, \beta$.

Recalling the discussion around \eqref{approxksoln} and noting the comment under \eqref{kbetatoL}, we see that if $\gamma mL$ is large the difference between $\rmk_{\beta}$ and $1$ is exponentially small in this quantity.  Thus the differences between $M_0, M_{\beta}$ and $M_{\beta}'$, and therefore the violations of the usual relativistic relations between $E_{\beta} ,p_{\beta} ,\beta$, are exponentially small:
\begin{equation}\label{largeLrelativistic}
E_{\beta} \approx \sqrt{p_{\beta}^2 + M_{0}^2} \approx \gamma M_0~, \qquad (\textrm{large $\gamma mL$})~.
\end{equation}

We point out that the results \eqref{LorentzL}, \eqref{LorentzH}, and \eqref{largeLrelativistic} are always valid for $p_\beta$ and $E_\beta$ as defined in \eqref{LorentzL}, \eqref{LorentzH}, even on a time-dependent solution with a time-dependent $\beta$, as long as $|\beta| < 1$.  What is different on a time-dependent solution is that $H = H[\varpi,\varphi] \neq E_\beta$ and $P \neq p_\beta$.  $H$ and $P$ are the total energy and momentum in the system.  (Recall that $P = P_{\rm Noether}$.)  For a general time-dependent solution, they include contributions both from the collective coordinate degree of freedom and from the remaining field theory degrees of freedom.  This is why we introduced the notation $p_{\beta}, E_{\beta}$ for the values of $P, H$ on the special family of configurations labeled by $\beta$ and determined from \eqref{boostedAPkink} and \eqref{boostedAPkinkmom}.  This point is important and can be confusing, especially because $P$ is the conjugate momentum to the soliton's collective coordinate.  The difference between $p_{\beta}$ and $P$ is somewhat analogous to the difference between the mechanical momentum and canonical momentum of a charged particle coupled to an electromagnetic field.

We will use \eqref{LorentzL} and \eqref{LorentzH} with \eqref{boostedAPkink} to define the soliton's mechanical momentum and energy for a general time-dependent solution with a time-dependent $\beta$.  In other words, we will compute $\beta$ from the solution via \eqref{beta2}, and then use this $\beta$ in \eqref{boostedAPkink}, which is then used to compute $p_{\beta},E_{\beta}$.  This means that the configuration \eqref{boostedAPkink} will be time-dependent, and hence so will $p_{\beta},E_\beta$, but note this configuration is not the solution to the FSE.

%%%%%%%%%%%%%%%%
\subsection{Motivation for Time-dependent Solutions to the FSE}
%%%%%%%%%%%%%%%%

Returning to the action functional \eqref{ssS} in the soliton sector, the equations of motion for $P,X$ are
\begin{equation}\label{PXeqns}
\dot{P} = - \frac{\pd H}{\pd X} ~, \qquad \dot{X} = \frac{\pd H}{\pd P} = \beta~.
\end{equation}
In particular, since $H$ is explicitly $X$-independent, the first equation implies $P$ must be constant.  This is a consequence of the translation invariance of the theory.  As we pointed out in \eqref{PNoether}, $P$ is the Noether charge associated with translations.  It is therefore conserved on a solution to the full equations of motion.  One might reasonably ask: Why is it interesting to consider solutions to the FSE with a time-dependent $P(t)$?  We provide two answers.

First, in the context of quantum field theory, classical solutions to equations of motion play a role in the semiclassical expansion.  However, one is typically interested in the classical solutions that follow from the action modified by source terms: $S \to S_J = S + \int J \phi$.  Evaluating the action on a classical solution to the sourced equation of motion is part of the leading saddle-point approximation to the generator of correlation functions in the QFT.  

Furthermore, recent work \cite{Melnikov:2020ret,Melnikov:2020iol}, combined with the observation that $P$ must be constant away from insertions, shows that the leading semiclassical approximation to soliton form factors are controlled by solutions to the FSE with a specific time-dependent profile for $P(t)$.  For a single operator insertion at $(t,x) = (t_\ast,x_\ast)$, the relevant profile for $P(t)$ is a step function,
\begin{align}\label{deltasmack}
P(t) =&~ \left\{ \begin{array}{l l} P_i~, & t < t_\ast~, \\  P_f~, & t > t_\ast~, \end{array} \right.
\end{align}
corresponding to a delta function force at the time $t_\ast$ with strength equal to the momentum transfer, $P_f - P_i$, where $P_{i,f}$ are the momenta of the initial and final soliton states.

The relevant boundary conditions in time for this application of the FSE, however, are not the initial value formulation considered in this paper, but rather a boundary value formulation in which $\varphi$ is fixed at initial and final times $t_{i,f}$ while $\varpi$ is unconstrained.  Nevertheless, it is interesting to consider the case of a delta-function impulse in the initial value formulation.  This will be our main example below.

The second reason it is interesting to consider time-dependent solutions to the FSE lies fully within the realm of classical field theory.  A solution to the FSE describes a soliton being driven by an external force that couples directly to the collective coordinate.  Consider adding the following term to the action \eqref{ssS}:
\begin{equation}\label{SFX}
S_F = S + \int \ed t F X~.
\end{equation}
where $F(t)$ is the driving force.  Then the first of \eqref{PXeqns} is modified to $\dot{P} = F$.  The exact soliton trajectory and field response are found by solving the FSE with the corresponding $P(t)$, inserting this solution into the velocity functional $\beta$, and integrating again to get the trajectory $X(t)$.  The trajectory $X(t)$ together with the solution $\varphi(t,\rho)$ describe both the soliton's motion and the radiation created by its acceleration.

Studies of kink dynamics in the presence of external forces go back to the early 80's, where most were motivated as toy models for condensed matter systems.  (See \cite{Fernandez:1986bt} and references therein.)  Our work is the first to couple an external force directly and solely to the collective coordinate.  A solution to the FSE is a soliton analog of the Lienard--Weichert solution for the fields of an arbitrarily accelerating point charge in electromagnetism.  

A key difference is that the soliton is not a point particle but rather a classically extended object.  On the one hand this means that the field describing the soliton and its radiation is everywhere smooth.  On the other hand this also means that one cannot draw some of the usual conclusions of relativistic point particle mechanics for the collective coordinate trajectory.  For example, the collective coordinate velocity can in fact exceed the speed of light, in much the same way that the location of the maximum of energy density can exceed the speed of light in certain electromagnetic wave pulses.\footnote{The simplest example is an X-wave: two plane waves crossing each other at an angle.  While the energy flow velocity never exceeds the speed of light anywhere, the speed of the intersection locus, where the energy density is maximum, moves faster than the speed of light.}  We will see this in explicit solutions below, and we will devote considerable time to understanding the physics of this phenomenon.

%%%%%%%%%%%%%%%%
\subsection{Energy and Power of a Solution to the FSE}\label{ssec:EnergyandPower}
%%%%%%%%%%%%%%%%

The Hamiltonian evaluated on a general time-dependent solution to the FSE is
\begin{align}\label{Hbar}
H =&~ \frac{(P + \langle \varpi | \varphi' \rangle)^2}{2 \langle \uppsi_0 | \varphi' \rangle^2} + \int_{-L/2}^{L/2} \ed \rho \left\{ \tfrac{1}{2} \varpi^2 + \tfrac{1}{2} \varphi^{\prime 2} + V(m^2; \varphi) \right\}  \cr
=&~ \int_{-L/2}^{L/2} d\rho \left\{ \half (\dot{\varphi} - \beta \varphi')^2 + \half \varphi^{\prime 2} + V(m^2,\varphi) \right\}~.
\end{align}
One can compute the time derivative of this quantity using the equations of motion \eqref{HamiltonFSE}.  After some algebra, one finds the satisfying result
\begin{equation}\label{Power}
\dot{H} = \dot{P} \beta~,
\end{equation}
which says that the power delivered to the system is the external force acting on it times velocity.  It is natural that the collective coordinate velocity should appear here:  The collective coordinate is a center of mass coordinate in the sense that it is canonically conjugate to the total momentum, and the external force couples directly to the collective coordinate.  As is well known, there is no uniquely defined center of mass in relativistic theories \cite{Pryce:1948pf}.

Equation \eqref{Power} shows that the total energy in the system is constant whenever $\dot{P} = 0$, whether the field configuration itself is constant or not.  For example, if we consider a $P(t)$ that ramps from $P_i = P(t_i)$ to $P_f = P(t_f)$ and is constant for $t >t_f$, then $H$ must also be constant for $t > t_f$ even though the field configuration will not be.

Once energy has been transferred to the system through the driving force, it does not stay in the collective coordinate degree of freedom.  The coupling of the collective coordinate to the remaining field theory degrees of freedom (through the coupling of $P$ to the fields in \eqref{ssS}) allows some of the energy to be transferred into oscillations of the normal modes of the field.  Recall that these normal modes include both shape modes of the kink and radiation.  These normal modes can then interact with the collective coordinate and transfer energy back to it, etc.  In the following we will identity three time-dependent components of the total energy: 1) the energy associated with the soliton's collective coordinate, which we will identify with the soliton's rest energy plus kinetic energy; 2) the energy in the normal modes, which includes their kinetic energy and the interaction energy between different normal modes; and 3) the interaction energy between the collective coordinate and normal modes.

The energy associated with the soliton's collective coordinate is taken to be $E_{\beta}$, the energy \eqref{LorentzH} based on the boosted soliton profile \eqref{boostedAPkink} with instantaneous velocity $\beta$ determined by the solution to the FSE.  In order to separate out the remaining two components of the energy we introduce a second energy functional.  

Viewing a solution to the FSE as a functional of $P =P(t)$, the Hamiltonian \eqref{ssS} evaluated on a solution likewise becomes a functional of $P$, $H[P]$.  It is not this Hamiltonian, however, that one would then vary with respect to $P$ to find the equation of motion for $X$.  Rather, returning to the action \eqref{ssS}, one would vary
\begin{equation}
H_{\rm eff}[P] := H[P] - \langle \varpi | \dot{\varphi} \rangle~,
\end{equation}
where dependence of the second term on $P$ is through dependence of the solution to the FSE on $P$.  It is in terms of $H_{\rm eff}$ that the action has the standard form $S = \int (P \dot{X} - H_{\rm eff})$.  We will refer to this Hamiltonian as the soliton effective Hamiltonian.  It is the leading (tree-level) approximation to the quantum soliton effective Hamiltonian defined in \cite{Gervais:1975pa,Melnikov:2020ret,Melnikov:2020iol}.  It is $H_{\rm eff}$ that naturally accounts for all energy associated with the soliton: its rest energy, kinetic energy, and effective potential energy due to its interaction with the normal modes.  Thus we define the difference $H - H_{\rm eff} = \langle \varpi | \dot{\varphi} \rangle$ to be the energy in the normal modes, and we define the difference $H_{\rm eff} - E_{\beta}$ to be the interaction energy between the soliton and normal modes:
\begin{equation}\label{Eintdef}
E_{\rm nm} := \langle \varpi | \dot{\varphi} \rangle~,  \qquad E_{\rm int} := H - E_{\rm nm} - E_{\beta} ~,
\end{equation}
so that $H = E_{\beta} + E_{\rm int} + E_{\rm nm} = H_{\rm eff} + E_{\rm nm}$.

While the decomposition $H = H_{\rm eff} + E_{\rm nm}$ is well-defined on any solution to the FSE, the decomposition $H_{\rm eff} = E_{\beta} + E_{\rm int}$ need not be.  This decomposition relies on evaluating the boosted kink configuration \eqref{boostedAPkink} on the instantaneous velocity $\beta = \beta(t)$ of the solution.  It can be the case, however, that $|\beta| > 1$.  While there is no natural definition of the boosted profile \eqref{boostedAPkink} when $\beta > 1$, and hence no natural definition of $E_{\beta}$, the solution to the FSE is still well-defined and well-behaved.  We will analyze such solutions in detail below.

%%%%%%%%%%%%%%%%
\subsection{Continuity and Jumping Conditions Through a Kick}\label{ssec:Jump}
%%%%%%%%%%%%%%%%

Our main example of a driving force $F$ for the collective coordinate, \eqref{SFX}, is
\begin{equation}\label{deltaimpact}
F(t) = (P_f - P_i) \delta(t - t_\ast)~,
\end{equation}
so that the collective coordinate momentum has the step function profile \eqref{deltasmack}.  Here we obtain some useful results for the behavior of a solution to the FSE through an instantaneous impulse.\footnote{We thank I.~Melnikov for contributions to the analysis of this subsection.}

We begin by considering continuity conditions on the fields at $t_{\ast}$.  We assume the solution $\varphi$ and all of its $\rho$ derivatives are continuous across $t = t_\ast$ and that the discontinuity in its first time derivative is finite.  Then we can derive this discontinuity by integrating \eqref{FSE} over a small neighborhood $[t_\ast - \varepsilon, t_\ast + \varepsilon]$ of $t_\ast$.  In the limit $\varepsilon \to 0$, the only terms that can contribute are $\dot{P}$ terms and terms containing $\ddot{\varphi}$.  There are two terms of each type, since additional ones arise when the $\pd_t$ hits $\beta$ in \eqref{FSE}.  We find that
\begin{align}\label{FSEjump}
0 =&~ \lim_{\varepsilon \to 0_+} \int_{t_\ast - \varepsilon}^{t_\ast + \varepsilon} dt \left\{ \ddot{\varphi} - \frac{\langle \varphi' | \ddot{\varphi} \rangle}{\langle \varphi' | \varphi' \rangle} \varphi' + \left( \frac{\uppsi_0}{\langle \uppsi_0 | \varphi' \rangle } - \frac{\varphi'}{\langle \varphi' | \varphi' \rangle}\right) \dot{P} \right\}  \cr
=&~ \lim_{\varepsilon \to 0_+} \int_{t_\ast - \varepsilon}^{t_\ast + \varepsilon} \bigg\{ dt \partial_t \left( \dot{\varphi} - \frac{\langle \varphi' |\dot{\varphi} \rangle}{\langle \varphi' | \varphi' \rangle } \varphi' \right) + \left( \frac{\uppsi_0}{\langle \uppsi_0 | \varphi' \rangle } - \frac{\varphi'}{\langle \varphi' | \varphi' \rangle}\right) (P_f - P_i) \delta(t-t_\ast) \bigg\} \cr
=&~ \left( \mathbbm{1} - \frac{|\varphi'(t_\ast) \rangle \langle \varphi'(t_\ast) |}{\langle \varphi'(t_\ast) | \varphi'(t_\ast) \rangle} \right) \left| \dot{\varphi}_+(t_{\ast}) - \dot{\varphi}_-(t_{\ast}) +  \frac{(P_f - P_i) \uppsi_0}{\langle \uppsi_0 | \varphi'(t_\ast) \rangle}  \right\rangle ~.
\end{align}
In the second line we used the assumption that $\varphi'$ is continuous.  In the last line we emphasize the time-dependence of the various kets, so that a ket $|f(t) \rangle$ corresponds to a function $f(t,\rho)$.  The factor on the far left in the last line is a projector onto the orthogonal complement of $\varphi'(t_\ast,\rho)$, and we have defined
\begin{equation}
f_{\pm}(\rho) = \lim_{\varepsilon \to 0_+} f(t_\ast \pm \varepsilon, \rho)~,
\end{equation}
for any quantity that is not continuous across the jump.

The general solution to \eqref{FSEjump} for the discontinuity in $\dot{\varphi}$ is
\begin{equation}
\dot{\varphi}_+(\rho) - \dot{\varphi}_-(\rho) = - \frac{(P_f - P_i) \uppsi_0(\rho)}{\langle \uppsi_0 | \varphi'(t_\ast) \rangle} + \alpha \varphi'(t_\ast,\rho)~,
\end{equation}
where the second term lies in the kernel of the projector.  The coefficient $\alpha$ is determined by requiring that the discontinuity be orthogonal to $\uppsi_0$, so that the constraint $\langle \uppsi_0 | \varphi \rangle = 0$ continues to hold for $t > t_\ast$.  Hitting both sides with $\langle \uppsi_0 |$ and setting the left-hand side to zero, we find $\alpha = (P_f - P_i) (\langle \uppsi_0 | \varphi'(t_\ast))^{-2}$ so that the solution is
\begin{equation}
\dot{\varphi}_+(\rho) - \dot{\varphi}_-(\rho) = \frac{(P_f - P_i)}{\langle \uppsi_0 | \varphi'(t_\ast) \rangle^2} \left( \varphi'(t_\ast, \rho) - \langle \uppsi_0 | \varphi'(t_\ast) \rangle \uppsi_0 \right)~.
\end{equation}

Now let us use this result to consider the jump in other quantities, starting with the jump in the velocity functional, $\beta$, \eqref{beta2}.  We find
\begin{align}\label{betajump}
\Delta \beta \equiv \beta_+ - \beta_- =&~ \frac{(P_f - P_i) + \langle \dot{\varphi}_+ - \dot{\varphi}_- | \varphi'(t_\ast) \rangle}{\langle \varphi'(t_\ast) | \varphi'(t_\ast) \rangle} = \frac{(P_f - P_i)}{\langle \uppsi_0 | \varphi'(t_\ast) \rangle^2}~.
\end{align}
Using this and \eqref{varpisol} we can compute the jump in the momentum field, $\varpi$:
\begin{align}
\varpi_+(\rho) - \varpi_-(\rho) =&~ \frac{(P_f - P_i)}{\langle \uppsi_0 | \varphi'(t_\ast) \rangle^2} \left( \varphi'(t_\ast, \rho) - \langle \uppsi_0 | \varphi'(t_\ast) \rangle \uppsi_0 \right) + \cr
&~ - \Delta \beta \left( \varphi'(t_\ast,\rho) - \langle \uppsi_0 |\varphi'(t_\ast) \rangle \uppsi_0(\rho) \right) \cr
=&~ 0~.
\end{align}
Hence the momentum field is continuous across the jump.  This is a crucial result for the numerics since it means the standard RK4 scheme applied to the Hamiltonian FSE  is able to march right through the impulse.

Having obtained the jump in the velocity, $\Delta \beta$, we can also obtain the jump in the total energy of the system, $H$, using \eqref{Power}.  We set 
\begin{equation}
\beta(t) = \beta_- + \Delta\beta \Theta(t-t_\ast) + O((t-t_\ast)^1)~,
\end{equation}
and use the distribution identity $\Theta(\tau) \delta(\tau) = \half \delta(\tau)$ to get
\begin{align}\label{Hbarjump1}
\Delta H =&~ \lim_{\varepsilon \to 0_+} \int_{t_\ast - \varepsilon}^{t_\ast + \varepsilon} \dot{H} dt  =  \lim_{\varepsilon \to 0_+}  \int_{t_\ast - \varepsilon}^{t_\ast + \varepsilon}  dt \left[ \beta_- + \Delta\beta \Theta(t-t_\ast) \right] (P_f - P_i) \delta(t-t_\ast) \cr
=&~ (P_f - P_i) \beta_- + \frac{(P_f - P_i)}{2} \Delta \beta \cr
=&~  (P_f - P_i) \left( \beta_- + \frac{(P_f - P_i)}{2 \langle \uppsi_0 | \varphi'(t_\ast) \rangle^2} \right) ~.
\end{align}

If we choose initial conditions before the impulse such that the solution is a constant velocity solution, \eqref{boostedAPkink}, with $\beta = \beta_i$ determined by $P_i$ through \eqref{LorentzL}, then we can evaluate this jump explicitly.  In this case $\beta_- = \beta_i$, $H_- = E_{\beta_i}$, and $\varphi'(t_\ast,\rho) = \phi_{\beta_i}'(\rho)$.  Hence the final total energy, $H_f  = H_+$ is
\begin{equation}\label{Hbarjump2}
H_f = E_{\beta_i} + (P_f - P_i) \beta_i + \frac{(P_f - P_i)^2}{2 \langle \uppsi_0 | \phi_{\beta_i}' \rangle^2} ~.
\end{equation}
The integral appearing in the last term, which is essentially the overlap of the static and boosted zero modes, does not have any simple closed-form analytic expression as far as we are aware.  Hence, this is an interesting result for the total energy in the system after a delta function impulse, which we will test numerically.

It is tempting to contemplate the possible Lorentz covariance properties of \eqref{Hbarjump1} and \eqref{Hbarjump2} since they relate the change in total energy of the system to the change in total momentum.  However it is not useful to do so because the kink is not a point particle colliding elastically with a brick wall.  Rather, the ``particle'' in this context is the entire field theory which has ``internal'' degrees of freedom that can be excited.  Even though we might choose initial conditions so that, before the collision, the system behaves as a particle, after the collision the internal degrees of freedom are excited leading to a change in this particle's ``mass.''    

%%%%%%%%%%%%%%%%
\subsection{Classical Perturbation Theory for Small Driving Force}\label{ssec:pert}
%%%%%%%%%%%%%%%%

If the force $F$ that couples to the collective coordinate is small we can solve the FSE perturbatively.  We start with the full action
\begin{align}
S_F = \int \ed t \left\{ P \dot{X} + \langle \varpi | \dot{\varphi} \rangle - H[P,\varpi,\varphi] - \lambda \langle \uppsi_0 | \varphi - \phi_0 \rangle - \nu \langle \uppsi_0 | \varpi \rangle + F X \right\}~,
\end{align}
and we assume $F = O(\epsilon)$, where $\epsilon$ is a small quantity.  More precisely, we assume that when $F$ acts over a finite time $t$ in units of the meson mass, it gives rise to a change in momentum that is $O(\epsilon)$ in units of the soliton mass.  Hence we write $F(t) = \epsilon M_0 m f(m t)$, where the dimensionless function $f(\tau)$ specifies the force profile.  In the case of \eqref{deltaimpact} for example, we have $\epsilon = (P_f - P_i)/M_0$ and $f(\tau) = \delta(\tau - mt_\ast)$.

After eliminating the Lagrange multipliers, the exact equations of motion are
\begin{align}\label{exactEOMs}
& \dot{P} = F ~, \qquad \dot{X} = \beta[P,\varpi,\varphi] ~, \cr
& |\dot{\varpi} \rangle = - \PP_{\uppsi_0}^\perp \left| \frac{\delta H}{\delta \varpi} \right\rangle ~, \qquad |\dot{\varphi} \rangle = \PP_{\uppsi_0}^\perp \left| \frac{\delta H}{\delta \varpi} \right\rangle ~,
\end{align}
with $\PP_{\uppsi_0}^{\perp} := \mathbbm{1} - |\uppsi_0 \rangle \langle \uppsi_0|$.  The solution to the first of these is
\begin{equation}
P = P_i + \epsilon M_0 \int_{m t_i}^{m t} \ed \tau f(\tau) \equiv P_i + P^{(1)}(t)~,
\end{equation}
where $P^{(1)}$ is $O(\epsilon)$ in units of the soliton mass.  One can then check that the ansatz
\begin{equation}
|\varpi \rangle = \sum_{n=0}^{\infty} |\varpi^{(n)} \rangle ~, \qquad | \varphi \rangle = \sum_{n = 0}^{\infty} | \varphi^{(n)} \rangle~, \qquad X = \sum_{n = 0}^{\infty} X^{(n)}~,
\end{equation}
where $\varpi^{(n)}, \varphi^{(n)}$, and $X^{(n)}$ are $O(\epsilon^n)$ suppressed relative to $\varpi^{(0)}, \varphi^{(0)}$ and $X^{(0)}$, leads to a consistent $\epsilon$-expansion of the remaining equations of motion, provided the leading order configuration is the boosted kink,
\begin{equation}
\varpi^{(0)}(t,\rho) = \varpi_{\beta_i}(\rho)~, \qquad \varphi^{(0)}(t,\rho) = \phi_{\beta_i}(\rho) ~, \qquad X^{(0)}(t) = \beta_i t + X_i~,
\end{equation}
and all higher order perturbations are orthogonal to the zero mode:
\begin{equation}
\langle \uppsi_0 | \varpi^{(n)} \rangle = 0 = \langle \uppsi_0 | \varphi^{(n)} \rangle~, \qquad n \geq 1~.
\end{equation}

Let
\begin{equation}
\JJ = \left( \begin{array}{c c} 0 & - \mathbbm{1}  \\ \mathbbm{1} & 0 \end{array} \right)
\end{equation}
denote the symplectic form on $(\varpi,\varphi)$ phase space, and let
\begin{align}
\HH_{12} = \left( (\PP_{\uppsi_0}^{\perp})_{13} \otimes \mathbbm{1}_2 \right) \left( \begin{array}{c c} \frac{\delta^2 H}{\delta \varpi_3 \delta \varpi_4} & \frac{\delta^2 H}{\delta \varpi_3 \delta \varphi_4} \\[1ex] \frac{\delta^2 H}{\delta \varphi_3 \varpi_4} & \frac{\delta^2 H}{\delta \varphi_3 \delta \varphi_4} \end{array} \right) \left( (\PP_{\uppsi_0}^{\perp})_{42} \otimes \mathbbm{1}_2 \right)
\end{align}
denote the quadratic form for fluctuations.  Here we use a shorthand $\varpi_1 = \varpi(t,\rho_1)$, and a repeated index in an expression like $\frac{\delta^2 H}{\delta \varpi_1 \delta \varpi_2} | \varpi^{(1)}\rangle_2$ means integrate over $\rho_2$.  Then the $\varpi,\varphi$ equations of motion at $O(\epsilon^n)$ take the form
\begin{align}\label{linearizedFSE}
\pd_t \left( \begin{array}{c} | \varpi^{(n)} \rangle_1 \\ |\varphi^{(n)} \rangle_1 \end{array} \right) =&~ \JJ (\HH_i)_{12} \left( \begin{array}{c} | \varpi^{(n)} \rangle_2 \\ |\varphi^{(n)} \rangle_2 \end{array} \right) + \JJ \left( \begin{array}{c} | \SS_{\varpi}^{(n)} \rangle_1 \\ | \SS_{\varphi}^{(n)} \rangle_1 \end{array} \right) ~, 
\end{align}
where the source terms $\SS_{\varpi,\varphi}^{(n)}$ depend only on the lower-order solution and the notation $(\HH_i)_{12}$ means evaluate $\HH_{12}$ on the leading order configuration $(P, \varpi, \varphi) = (P_i, \varpi_{\beta_i}, \phi_{\beta_i})$.  We will also utilize the notation $|\zeta^{(n)} \rangle = (|\varpi^{(n)} \rangle , |\varphi^{(n)} \rangle )^T$ for our phase space fluctuation field so that \eqref{linearizedFSE} takes the form
\begin{equation}\label{linearizedFSEzeta}
\pd_t | \zeta^{(n)} \rangle = \JJ \HH_i | \zeta^{(n)} \rangle + \JJ | \SS^{(n)} \rangle~.
\end{equation}
In terms of derivatives of $H$, the source terms at the first two orders are
\begin{align}\label{firstordersource}
& | \SS_{\varpi}^{(1)} \rangle_1 = P^{(1)} (\PP_{\uppsi_0}^{\perp})_{12} \left| \left. \tfrac{\delta \pd H}{\delta \varpi \pd P} \right|_i \right\rangle_2~, \qquad | \SS_{\varphi}^{(1)} \rangle = P^{(1)} (\PP_{\uppsi_0}^{\perp})_{12} \left| \left. \tfrac{\delta \pd H}{\delta \varphi \pd P} \right|_i \right\rangle_2 ~,
\end{align}
and
\begin{align}\label{secondordersource}
| \SS_{\varpi}^{(2)} \rangle_1 =&~  (\PP_{\uppsi_0}^{\perp})_{12} \bigg\{ P^{(1)}  \left. \tfrac{ \delta^2 \pd  H}{\delta \varpi_2 \delta \varphi_3 \pd P} \right|_i |\varphi^{(1)} \rangle_3 +  \left.  \tfrac{\delta^3 H}{\delta \varpi_2 \delta \varpi_3  \delta \varphi_4} \right|_i |\varpi^{(1)} \rangle_3 | \varphi^{(1)} \rangle_4 +  \cr
&~ \qquad  \qquad + \half  \left.  \tfrac{\delta^3 H}{\delta \varpi_2 \delta \varphi_3  \delta \varphi_4} \right|_i |\varphi^{(1)} \rangle_3 | \varphi^{(1)} \rangle_4 \bigg\} ~, \cr
| \SS_{\varphi}^{(2)} \rangle_1 =&~  (\PP_{\uppsi_0}^{\perp})_{12}  \bigg\{  \half P^{(1)2}  \left| \left. \tfrac{\delta \pd^2 H}{\delta \varphi \pd^2 P} \right|_{i} \right\rangle_2 + P^{(1)} \left[  \left. \tfrac{ \delta^2 \pd H}{\delta \varphi_2 \delta \varpi_3 \pd P} \right|_i | \varpi^{(1)} \rangle_3 + \left. \tfrac{ \delta^2 \pd H}{\delta \varphi_2 \delta \varphi_3 \pd P} \right|_i | \varphi^{(1)} \rangle_3 \right] + \cr
&~ \qquad \qquad + \half \left. \tfrac{ \delta^3 H}{\delta \varphi_2 \delta \varpi_3 \delta \varpi_4} \right|_i |  \varpi^{(1)} \rangle_3 | \varpi^{(1)} \rangle_4 + \left. \tfrac{ \delta^3 H}{\delta \varphi_2 \delta \varpi_3 \delta \varphi_4} \right|_i | \varpi^{(1)} \rangle_3 | \varphi^{(1)} \rangle_4 + \cr
&~ \qquad \qquad + \half \left. \tfrac{\delta^3H}{\delta \varphi_2 \delta \varphi_3 \delta \varphi_4} \right|_i | \varphi^{(1)} \rangle_3 | \varphi^{(1)} \rangle_4 \bigg\} ~. 
\end{align}
Here we used that $H$ is quadratic in the momenta $(P,\varpi)$ to eliminate some terms from $| \SS_{\varpi}^{(2)} \rangle$.  

Finally, if the solution to \eqref{linearizedFSE} can be found order by order, then we can insert it back into $\beta[P,\varpi,\varphi]$ and integrate to obtain the collective coordinate trajectory.  For the expansion of the collective coordinate velocity to $O(\epsilon^2)$ we obtain
\begin{align}\label{pertccvel}
\beta =&~ \beta_i + \beta^{(1)} + \beta^{(2)} + O(\epsilon^3)~, \qquad \textrm{with} \cr
\beta^{(1)} =&~  \frac{ P^{(1)} + \langle \phi_{\beta_i}' | \varpi^{(1)} \rangle + \beta_i  \braket{\phi_{\beta_i}'' +\langle \uppsi_0 | \phi_{\beta_i}' \rangle \uppsi_0' }{ \varphi^{(1)} } }{\langle \uppsi_0 | \phi_{\beta_i}' \rangle^2} ~, \cr
\beta^{(2)} =&~ \frac{  \langle \phi_{\beta_i}' | \varpi^{(2)} \rangle + \beta_i  \braket{\phi_{\beta_i}'' +\langle \uppsi_0 | \phi_{\beta_i}' \rangle \uppsi_0' }{ \varphi^{(2)} } + \langle \varpi^{(1)} | \varphi^{(1)\prime} \rangle - \beta_i \langle \uppsi_{0}' | \varphi^{(1)} \rangle^2 }{\langle \uppsi_0 | \phi_{\beta_i}' \rangle^2} + \cr
&~ + \frac{2 \langle \uppsi_0' | \varphi^{(1)} \rangle}{\langle \uppsi_0 | \phi_{\beta_i}' \rangle} \beta^{(1)} ~.
\end{align}

The case $\beta_i = 0$ is special.  Then, since $\phi_{0}' \propto \uppsi_0$ is orthogonal to $\varpi^{(1)}$, we find that $\beta^{(1)} \to P^{(1)} / \langle \uppsi_0 | \phi_{0}' \rangle^2$ while the $\varpi^{(2)},\varphi^{(2)}$ terms of $\beta^{(2)}$ drop.  Hence it is sufficient to work to first order in the fields to obtain the collective coordinate velocity to second order, but it turns out that the first order sources vanish when $\beta_i = 0$.  Therefore the first correction to the fields starts at second order and the first time-dependent correction to $\beta$ starts at third order, coming from the last term of \eqref{pertccvel} with $\varphi^{(1)} \to \varphi^{(2)}$:
\begin{align}\label{betapertbetai0}
\beta_i = 0 \quad \Rightarrow \quad \beta = \frac{P^{(1)}}{\langle \uppsi_0 | \phi_{0}' \rangle^2} \left( 1 + \frac{2 \langle \uppsi_0' | \varphi^{(2)} \rangle}{\langle \uppsi_0' | \phi_{0}' \rangle} \right) +  O(\epsilon^4)~.
\end{align}

The main task at hand in implementing this perturbation theory is to solve the linearized FSE, \eqref{linearizedFSE}.  Reference \cite{Melnikov:2020ret} shows how the eigenvalue problem for $\HH_i$ can be reduced to one involving the static soliton fluctuation operator \eqref{evproblem} but with twisted periodicity conditions depending on the background velocity $\beta_i$.  We review key details from that analysis in Appendix \ref{app:boostedmodes} and extend it significantly to give a complete picture of the finite $L$ eigenvalue problem for sine-Gordon and $\phi^4$-theory.  The result is a set of phase space normal modes $|\eta_a \rangle$ and eigenvalues $\nu_a > 0$, for $a = 1,2,\ldots$ satisfying the eigenvalue problem
\begin{equation}\label{JHevproblem}
\ii \JJ \HH_i | \eta_a \rangle = \nu_a | \eta_a \rangle ~,
\end{equation}
and symplectic orthonormality conditions
\begin{equation}\label{symorth}
\langle \eta_a | \ii \JJ | \eta_b \rangle = \delta_{ab} ~, \qquad \langle \eta_{a}^\ast | \ii \JJ | \eta_b \rangle = 0 ~.
\end{equation}

The components $|\eta_{\varpi a} \rangle, |\eta_{\varphi a} \rangle$ of these phase space kets take the form
\begin{align}\label{etafrompsi}
|\eta_{\varpi a} \rangle =&~ - \PP_{\uppsi_0}^\perp \left[  (\beta_i \pd_\rho + \ii \nu_a) |\psi_a \rangle - \beta_i \frac{\langle \uppsi_0 | \psi_a \rangle}{\langle \uppsi_0 | \phi_{\beta_i}' \rangle} | \phi_{\beta_i}'' \rangle \right] ~, \cr
|\eta_{\varphi a} \rangle =&~ |\psi_a \rangle - \frac{\langle \uppsi_0 | \psi_a \rangle}{\langle \uppsi_0 | \phi_{\beta_i}' \rangle} |\phi_{\beta_i}' \rangle~
\end{align}
where the $\psi_a$ and $\nu_a$ are \emph{boosted} modes and eigenvalues solving
\begin{equation}\label{bmeqn}
\left\{ - (1 - \beta_{i}^2) \pd_{\rho}^2 + 2 \ii \nu_a \beta_i \pd_\rho + V^{(2)}(\phi_{\beta_i}) - \nu_{a}^2 \right\} | \psi_a \rangle = 0~,
\end{equation}
subject to the same periodicity conditions as the boosted zero mode:
\begin{align}\label{bmperiodicity}
\textrm{$\phi^4$-theory:} \quad \psi_a(\rho + L) = - \psi_a(\rho)~, \qquad \textrm{sine-Gordon:} \quad \psi_a(\rho + L) = \psi_a(\rho)~.
\end{align}
These conditions are what ensure symplectic orthogonality \eqref{symorth} and determine the eigenvalues $\nu_a$.  Symplectic orthonormality requires
\begin{equation}\label{bmnorm}
1 = \int \ed \rho \left\{ \ii \beta_i \left( \psi_a \psi_{a}^{\prime \ast} - \psi_{a}^\ast \psi_{a}' \right) + 2\nu_a \psi_{a}^\ast \psi_a \right\}~.
\end{equation}

Notice that the boosted zero mode, $\psi_0 \propto \phi_{\beta_i}'$, solves \eqref{bmeqn} with $\nu_0 = 0$.  However, this solution is projected out in \eqref{etafrompsi} where it corresponds to the trivial solution $|\eta_a \rangle = 0$.  Indeed, the expression for $|\eta_{\varphi a} \rangle$ has the form $|\eta_{\varphi a} \rangle = \PP | \psi_a \rangle$, where $\PP$ is a projector with kernel spanned by the boosted zero mode and co-kernel spanned by the static zero mode.

Equations \eqref{bmeqn} and \eqref{bmperiodicity} are equivalent to the fluctuation equations we encountered earlier around the static kink, but with twisted---or Floquet---periodicity conditions.  Inserting the ansatz
\begin{equation}
\psi_a(\rho) = e^{\ii \gamma_{i}^2 \beta_i \nu_a (\rho - \rho_0)} \widetilde{\uppsi}_a(\gamma(\rho - \rho_0))~,
\end{equation}
where $\rho_0$ is the constant that appears in the general relationship between the boosted and static kink profiles as discussed under \eqref{boostedviastatic}, we find the eigenvalue problem
\begin{equation}\label{twistedEVproblem1}
\left[ - \pd_{y}^2 + V^{(2)}(\phi_0(y;\gamma_i L)) \right] \widetilde{\uppsi}_{a}(y) = \widetilde{\upomega}_{a}^2 \widetilde{\uppsi}_a(y)~, \qquad \textrm{with} \quad\widetilde{\upomega}_a = \gamma_i \nu_a~,
\end{equation}
in terms of the Lorentz contracted variable $y = \gamma_i (\rho - \rho_0)$.  This equation reduces to the $\nu_{\rm L} = 1$ and $2$ Lam\'e equations for sine-Gordon and $\phi^4$-theory respectively, in terms of a new elliptic modulus $\rmk_{\beta_i}$ determined by $\gamma_i L$ in the same way that the original elliptic modulus $\rmk$ is determined by $L$.  The (anti-)periodicity conditions for $\psi_a$ become twisted periodicity conditions for $\widetilde{\uppsi}_a$:
\begin{align}\label{twistedEVproblem2}
 \widetilde{\uppsi}_a(y + \gamma_i L) =&~  e^{\ii \mu_a} \widetilde{\uppsi}_a(y) ~, \qquad \textrm{where} \quad e^{\ii \mu_a} = \left\{ \begin{array}{l l} -e^{- \ii \beta_i \gamma_i \widetilde{\upomega}_a L}~,  & \textrm{$\phi^4$-theory}~, \\ e^{-\ii \beta_i \gamma_i  \widetilde{\upomega}_a L}~, & \textrm{sine-Gordon}~. \end{array} \right.
\end{align}
The normalization condition \eqref{bmnorm} is
\begin{align}
1 =&~  \int_{-\gamma L/2}^{\gamma L/2} \ed y \left\{ 2 \widetilde{\upomega}_a \widetilde{\uppsi}_{a}^\ast \widetilde{\uppsi}_a + \ii \beta W(\widetilde{\uppsi}_a, \widetilde{\uppsi}_{a}^\ast) \right\}~,
\end{align}
where $W(f,g)(y) = f \pd_y g - g \pd_y f$ is the Wronskian.  When $\beta_i \to 0$ the eigenvalue problem \eqref{twistedEVproblem1}, \eqref{twistedEVproblem2} reduces to the static one, $\widetilde{\upomega}_a \to \upomega_a$, while $\widetilde{\uppsi}_a \to \frac{1}{\sqrt{2\upomega_a}} \uppsi_a$ due to the usual symplectic normalization condition on phase space modes.

The Floquet problem for the Lam\'e equation is fairly well studied and relevant results are summarized in Appendices \ref{app:Lame} and \ref{app:boostedmodes}.  In particular, exponentially accurate approximations for the $\widetilde{\upomega}_a$ and $\widetilde{\uppsi}_a$ at large $mL$ are described, following along similar lines as in the static case.  These expressions are analytic except for a transcendental equation that must be solved for the approximate eigenvalues.  As a check of our work, we also obtain the exact solutions by numerical integration and compare them in some examples.

Returning to our perturbation theory discussion, we describe the solution to \eqref{linearizedFSE} through the first two orders in broad strokes.  We assume that the external force is zero and hence $P^{(1)} = 0$ until some initial time $t_\ast$.  At first order we insert the ansatz
\begin{equation}\label{firstorderansatz}
|\zeta^{(1)} \rangle = \sum_{a =1}^{\infty} \left( A_{a}^{(1)} e^{-\ii \nu_a (t-t_\ast)} |\eta_a \rangle + A_{a}^{(1)\ast}e^{\ii \nu_a (t - t_\ast)}| \eta_{a}^\ast \rangle \right)~,
\end{equation}
where the $A_a = A_a(t)$ satisfy the initial conditions $A_a(t_\ast) = 0$.  Using \eqref{JHevproblem} and \eqref{symorth}, we find
\begin{equation}\label{A1eom}
\dot{A}_{a}^{(1)} = -\ii  e^{\ii \nu_a (t- t_\ast)} \langle \eta_a | \SS^{(1)} \rangle  \quad \Rightarrow \quad A_{a}^{(1)}(t) = -\ii  \int_{0}^{t - t_\ast} \ed t' e^{\ii \nu_a t'} \langle \eta_a  | \SS^{(1)} \rangle ~.
\end{equation}
Hence the first order solution is
\begin{equation}
|\zeta^{(1)} \rangle = \sum_{a=1}^{\infty} 2 \Re \left\{ -\ii e^{-\ii \nu_a (t- t_\ast)} |\eta_a \rangle \int_{0}^{t - t_\ast} \ed t' e^{\ii \nu_a t' } \langle \eta_a  | \SS^{(1)} \rangle \right\}~~, \qquad t\geq t_\ast~.
\end{equation}

The new issue that can arise at second order is resonance.  When we insert the first order solution into \eqref{secondordersource}, do any terms in $\SS^{(2)}$ oscillate at the same frequency as the normal modes?  Clearly the answer depends on the time-dependence of $P^{(1)}$, but for the case of primary interest where $P^{(1)}$ has a step function profile the answer is yes.  Those terms in $|\SS^{(2)} \rangle$ that are linear in $|\varpi^{(1)}\rangle$ or $|\varphi^{(1)} \rangle$ will give rise to source terms that oscillate at the same frequency as the normal modes.

If $\SS^{(2)}$ contains resonant terms then we must modify our ansatz for $\varpi,\varphi$ to allow both the amplitude and frequency to receive a series of corrections in $\epsilon$, an idea familiar from perturbation theory in quantum mechanics:
\begin{align}\label{modifiedepsexp}
|\zeta \rangle = \sum_{a=1}^{\infty} \left\{ (A_{a}^{(1)} + A_{a}^{(2)} + \cdots) e^{-\ii (\nu_a + \nu_{a}^{(1)} + \cdots)(t-t_\ast)} |\eta_a \rangle + \textrm{c.c.} \right\}~,
\end{align}
where the c.c.~denotes complex conjugate.  The $\epsilon$ expansion on the left-hand side of \eqref{linearizedFSE} is reorganized starting at second order, since the time derivative brings down the frequency series.  We find
\begin{align}
| \dot{\zeta}^{(2)}\rangle =&~ \sum_{a=1}^{\infty} \left\{ \left( \dot{A}_{a}^{(2)} - \ii \nu_{a}^{(1)} A_{a}^{(1)} \right) e^{-\ii (\nu_a + \cdots)(t-t_\ast)} |\eta_a \rangle + \textrm{c.c.} \right\}~.
\end{align}
Meanwhile the modified time-dependence of the ansatz does not affect the form of the $\epsilon$ expansion on the right-hand side of \eqref{linearizedFSE} since no time derivatives are involved.  Therefore the analog of equation \eqref{A1eom} at second order is
\begin{equation}\label{A2eom}
\dot{A}_{a}^{(2)} - \ii \nu_{a}^{(1)} A_{a}^{(1)} = -\ii e^{\ii (\nu_a + \cdots) (t-t_\ast)} \langle \eta_a | \SS^{(2)} \rangle ~.
\end{equation}
Having resonance means that there are terms on the right-hand side of this equation with the same time dependence as $A_{a}^{(1)}$.  We write
\begin{equation}\label{s2res}
e^{\ii (\nu_a + \cdots) (t-t_\ast)} \langle \eta_a  | \SS^{(2)}(t) \rangle = s_{a}^{(2,{\rm res})} + s_{a}^{(2)}(t) ~,
\end{equation}
where $s_{a}^{(2,{\rm res})}$ is time independent, while $s_{a}^{(2)}(t)$ is oscillatory.  We then fix $\nu_{a}^{(1)}$ by demanding time-independent terms on both sides match.  Inserting this solution back into \eqref{A2eom}, the equation is then integrated to obtain $A_{a}^{(2)}$.

In Appendix \ref{app:pert} we compute the sources $\SS^{(1)}$ and $\SS^{(2)}$ explicitly for the case of a step function profile, $P^{(1)}(t) = \epsilon M_0 \Theta(t - t_\ast)$ and we use them to obtain the fields $\varpi,\varphi$ through $O(\epsilon^2)$ in the amplitude expansion and $O(\epsilon)$ in the frequency expansion.  Then in Subsection \ref{ssec:pertcheck} we compute the first two corrections to the collective coordinate velocity \eqref{pertccvel} and compare these perturbative results with full fledged numerical solutions to the FSE.

%%%%%%%%%%%%%%%%
%%%%%%%%%%%%%%%%
\section{Kinks on a Finite Lattice}\label{sec:Discretize}
%%%%%%%%%%%%%%%%
%%%%%%%%%%%%%%%%

In this section we consider the discretization of our finite $L$ theory.  We utilize a spectral method based on the discrete Fourier transform to approximate spatial derivatives.  Our reasons for doing so are two-fold.  First, past studies of numerical soliton solutions to nonlinear wave equations have noted the sensitivity of such solutions to the accuracy of the spatial derivative approximation.  See, for example, \cite{MR1433936}.  Spectral methods often provide exponential accuracy for derivative operators.  

Second, the lattice derivative we employ is extremely natural from the QFT perspective.  It was first introduced in early studies of nonperturbative QFT on the lattice by a group at the Stanford Linear Accelerator Center \cite{Drell:1976bq,Drell:1976mj}, and has become known as the ``SLAC'' derivative.  It plays an essential role in obtaining lattice models that maintain a \emph{continuous} translation symmetry, allowing one to define a collective coordinate and implement a discrete analog of the soliton sector canonical transformation.  We demonstrate this at the classical level in Subsections \ref{ssec:latmodel} and \ref{ssec:LatticeCC} below, while a study of the regularized QFT based on these lattice models will appear elsewhere.

The lattice construction allows us to obtain a discrete form of the forced soliton equation.  Having reduced the system to a finite number of degrees of freedom, we employ standard fourth-order Runge--Kutta to march forward in time.

%%%%%%%%%%%%%%%%
\subsection{The SLAC Derivative on a Finite Lattice}\label{sec:SLAC}
%%%%%%%%%%%%%%%%

% % % % % % % % % % % % %
\subsubsection{Anti-periodic Lattice for $\phi^4$ Theory}
% % % % % % % % % % % % %

We begin by introducing a lattice spacing $a$ such that $L/a = 2N$, an even integer.  Our reason for restricting to an even number of sites for anti-periodic lattices is explained at the end of this subsection.  It is merely a technical convenience.  Let the lattice sites be at
\begin{equation}
x_j = \left(j - \tfrac{1}{2}\right) a~, \qquad j = -N +1 ~, \ldots, ~N~.
\end{equation}
Given an anti-periodic function $f(x + L) = -f(x)$, let $f_j = f(x_j)$.  Then the discrete Fourier transform $\{ f_j \} \mapsto \{ \breve{f}_n \} = \FF \{ f_j \}$ and its inverse are
\begin{align}\label{DFT}
\breve{f}_n = \frac{1}{\sqrt{2N}} \sum_{j = -N+1}^{N} f_j e^{-2\pi \ii (n- \frac{1}{2}) x_j/L} ~, \qquad  f_j = \frac{1}{\sqrt{2N}} \sum_{n = -N+1}^{N} \breve{f}_n e^{2\pi \ii (n-\frac{1}{2}) x_j/L} ~.
\end{align}
Note the latter implies $f_{j + 2N} = -f_j$, as required by the anti-periodicity of $f$, thanks to the one-half in the $(n- \frac{1}{2})$ factor of the exponential.  The relations \eqref{DFT} can be written in matrix notation as $\breve{f}_n = \sum_j (\FF_{{\rm A}_{2N}})_{nj} f_j$ and $f_j = \sum_n(\FF_{{\rm A}_{2N}}^{\dag})_{jn} \breve{f}_n$ where $\FF_{{\rm A}_{2N}}$ is the unitary matrix with elements 
\begin{equation}
(\FF_{{\rm A}_{2N}})_{nj} = \frac{1}{\sqrt{2N}} e^{-2\pi \ii (n-\half) x_j/L} = \frac{1}{\sqrt{2N}} e^{2\pi \ii (n - \half) (j - \half)/(2N)} ~.
\end{equation}

Starting from only the $\{f_j\}$, one can construct a smooth approximation, $f_{\rm approx}(x)$, to the original function $f(x)$, called an interpolating function, as follows.  First, compute the $\{ \breve{f}_n \}$ using the first of \eqref{DFT}, then set
\begin{equation}\label{interpolating}
f_{\rm approx}(x) = \frac{1}{\sqrt{2N}} \sum_{n= -N+1}^{N} \breve{f}_n e^{2\pi \ii (n - \half) x/L} ~.
\end{equation}
By construction, $f_{\rm approx}$ agrees with $f$ on the lattice: $f_{\rm approx}(x_j) = f(x_j)$.  For general $x$ not a lattice site, this approximation is exponentially accurate (in $1/N$) provided the original function $f(x)$ is analytic in a strip of the complex plane containing the real axis.  Examples of such functions are the family of anti-periodic kinks, \eqref{boostedAPkink}.  If $f$ is smooth $(C^{\infty})$ but not necessarily analytic, then the error in the approximation vanishes faster than any power of $1/N$ as $N \to \infty$.

To see that \eqref{interpolating} gives an exponentially accurate approximation to an analytic $f$, notice that the form of \eqref{interpolating} is the same as that of a partial sum Fourier series, except that the $\breve{f}_n$ are not quite the Fourier series coefficients.  The first point is that the $\breve{f}_n$ are exponentially close to the Fourier series coefficients, $c_n$, of $f$.  To see this, observe that the right-hand side of the first equation of \eqref{DFT} is a Riemann sum approximation to the integral that determines $c_n$.  It is by construction a midpoint Riemann sum, but the midpoint rule is equivalent to a trapezoidal rule for periodic integrands.  For general integrands the trapezoidal rule is only $O(1/N^2)$ accurate.  However, for periodic integrands analytic in a strip containing the real axis, it is well known that the trapezoidal rule for the Riemann sum approximation to the integral is exponentially accurate.  See \cite{MR3245858} for an overview of this phenomenon.  Thus, $\breve{f}_n$ is exponentially close to $c_n$ for all $n$.  The second point is that the Fourier series partial sum gives an exponentially accurate approximation to the original function if the original function is analytic.  This follows from the exponential decay of the Fourier coefficients $c_n$ as $n \to \infty$, which in turn is a consequence of the analyticity of $f$ and the Riemann--Lebesgue lemma.

We then obtain exponentially accurate approximations to derivatives of $f$ by taking derivatives of $f_{\rm approx}$ and evaluating the results at the lattice points.  This defines a set of derivative operators directly on the lattice data.  The $p^{\rm th}$ derivative operator, denoted $D_{{\rm A}_{2N}}^{(p)}$, is given by
\begin{align}\label{SLACDasOrdinaryD}
( D_{{\rm A}_{2N}}^{(p)} f)_j =&~ \frac{\ed^p f_{\rm approx}}{\ed x^p} \bigg|_{x_j} = \frac{1}{\sqrt{2N}} \sum_{n = -N+1}^{N} e^{2\pi \ii (n - \half) x_j/L} \left( \frac{2\pi \ii (n- \half)}{L} \right)^p \breve{f}_n \cr
=&~ \sum_{n = - N+1}^{N} (\FF_{{\rm A}_{2N}}^{\dag})_{jn} \left( \frac{2\pi \ii (n- \half)}{L} \right)^p \sum_{k = -N+1}^{N} (\FF_{{\rm A}_{2N}})_{nk} f_k ~,
\end{align}
or equivalently
\begin{equation}\label{Dpdiagonalized}
D_{{\rm A}_{2N}}^{(p)} = \FF_{{\rm A}_{2N}}^\dag (\Lambda_{{\rm A}_{2N}})^p \FF_{{\rm A}_{2N}} ~,
\end{equation}
where
\begin{equation}
 \Lambda_{{\rm A}_{2N}} = \diag(\lambda_{-N+1}, \ldots, \lambda_{N}) ~, \quad \lambda_n = \frac{2\pi \ii (n - \half)}{L} ~.
 \end{equation}
This formula makes it clear that $D_{{\rm A}_{2N}}^{(p)} D_{{\rm A}_{2N}}^{(q)} = D_{{\rm A}_{2N}}^{(p+q)}$.  Explicit expressions for the matrix elements of the first two derivatives are
\begin{align}\label{DA1}
& (D_{{\rm A}_{2N}})_{jj} = 0~, \qquad (D_{{\rm A}_{2N}})_{jk} = \frac{\pi}{L} (-1)^{j - k} \csc\left(\tfrac{\pi (j-k)}{2N}\right)~, \quad j \neq k~,
\end{align}
and
\begin{align}\label{DA2}
(D_{{\rm A}_{2N}}^{(2)})_{jj} =&~ - \frac{\pi^2}{3 L^2} ( (2N)^2 - 1)~, \cr
(D_{{\rm A}_{2N}}^{(2)})_{jk} =&~ \frac{2\pi^2}{L^2} (-1)^{j-k + 1} \cot\left(\tfrac{\pi (j-k)}{2N}\right) \csc\left(\tfrac{\pi (j-k)}{2N}\right)~, \quad j \neq k~.
\end{align}

The derivative operator $D_{{\rm A}_{2N}}$ constructed above can also be obtained from a rather different perspective.  Our approach thus far has been to first IR regularize by defining a continuum theory at finite $L$ and then UV regularize by introducing a lattice.  One could UV regularize first by discretizing the continuum theory on $\mathbbm{R}$ to an infinite lattice and then IR regularize by restricting to periodic or anti-periodic sequences depending on the sector.  On the infinite lattice the direct analog of the derivative operator we constructed above is known as the SLAC derivative \cite{Drell:1976bq,Drell:1976mj}.  The action of the SLAC derivative, $D$, is
\begin{align}\label{slacDrealine}
(D f)_j :=&~ \frac{1}{a} \bigg\{ \cdots + - \frac{1}{4} f(x_j + 4a) + \frac{1}{3} f(x_j + 3a) - \frac{1}{2} f(x_j + 2a) + \cr
&~ \qquad \qquad + f(x_j + a) - f(x_j - a) + \cr
&~ \qquad +  \frac{1}{2} f(x_j -2a) - \frac{1}{3} f(x_j -3a) + \frac{1}{4} f(x_j -4a) - + \cdots \bigg\} \cr
=&~ \frac{1}{a} \sum_{k=1}^{\infty} \frac{(-1)^{k+1}}{k} \left( f(x_j + k a) - f(x_j - k a) \right)~.
\end{align}
If $f$ is representable by a Fourier transform, $f(x) = \int \frac{dp}{\sqrt{2\pi}} e^{\ii p \rho} \hat{f}(p)$, one finds that
\begin{align}\label{Dfft}
(D f)_j =&~ \frac{2 \ii}{a} \int_{-\infty}^{\infty} \frac{dp}{\sqrt{2\pi}} \left( \sum_{k=1}^{\infty} \frac{(-1)^{k+1}}{k} \sin(kpa) \right) e^{\ii p x_j} \hat{f}(p) \cr
=&~ \int_{-\infty}^{\infty} \frac{dp}{\sqrt{2\pi}} \left(\ii p - \frac{2\pi \ii}{a} \left\lfloor \frac{pa + \pi}{2\pi} \right\rfloor \right) e^{\ii p x_j} \hat{f}(p) ~,
\end{align}
where $\lfloor \cdot \rfloor$ denotes the Floor function.  The function of $p$ in parentheses in the last line is a sawtooth that agrees with $\ii p$ in the fundamental domain $|p| < \pi/a$.  Thus $D$ has exactly the same spectrum as the ordinary derivative in the first Brillouin zone but not beyond it, and this is the point of the construction.  The spectrum is continuous because the spatial lattice is infinite.

If one now restricts to $f_j$ that are anti-periodic, $f_{j + 2N} = -f_j$, the sum in \eqref{slacDrealine} can be broken up into $2N$ separate sums, each one an infinite sum determining the coefficient of $f_{-N + j}$ for $j = 1, \ldots 2N$.  One can show that this approach leads to the same $D_{{\rm A}_{2N}}$ we found above.  This approach is discussed in the context of periodic sequences in \cite{Costella:2002js}.

By restricting to lattices with an even number of sites in the above discussion we have ensured that we do not have a lattice point on the boundary of the first Brillouin zone, $|p| = \pi/a$.  For odd lattices there will be a sample point on the boundary, and a modification to the definition of $D$ should be made.  This is analogous to what one does for periodic functions with even lattices, as described in \cite{Costella:2004re}.  Here, since we deal exclusively with field configurations in the kink sector, \eqref{sectors}, it is convenient for $\phi^4$-theory to restrict consideration to even lattices.

% % % % % % % % % % % % % % %
\subsubsection{Periodic Lattice for the sine-Gordon Model}
% % % % % % % % % % % % % % %

In the winding number $\nu$ sector of the sine-Gordon model we can work with periodic fields $\phi^{(\nu)}(t,x), \pi^{(\nu)}(t,x)$ related to the original quasi-periodic fields by
\begin{equation}\label{winding}
\phi(t,x) = \phi^{(\nu)}(t,x) + \frac{2\pi \nu x}{gL} ~, \qquad \pi(t,x) = \pi^{(\nu)}(t,x)~.
\end{equation}
Therefore when discretizing the theory it is natural to consider a periodic lattice, and for the reasons mentioned above it is convenient to restrict to the case of an odd number of lattice sites.  Therefore we take the lattice sites to be at
\begin{equation}
x_j = j a~, \qquad j = -N~, \ldots, ~ N~, \qquad a = \frac{L}{2N+1}~.
\end{equation}

Given a set of periodic lattice data $\{ f_j \}$, the discrete Fourier transform and its inverse are now
\begin{equation}
\breve{f}_n = \sum_{j = -N}^N (\FF_{{\rm P}_{2N+1}})_{nj} f_j ~, \qquad f_{j} = \sum_{n = -N}^N (\FF_{{\rm P}_{2N+1}}^\dag)_{jn} \breve{f}_n ~,
\end{equation}
with
\begin{equation}
(\FF_{{\rm P}_{2N+1}})_{jn} = \frac{1}{\sqrt{2N+1}} e^{-2\pi \ii n j/(2N+1)} ~.
\end{equation}
These can be used to construct an interpolating function
\begin{equation}
f_{\rm approx}(x) = \frac{1}{\sqrt{2N+1}} \sum_{n = -N}^N e^{2\pi \ii n x/L} \breve{f}_n ~,
\end{equation}
and a set of derivative operators
\begin{equation}
D_{{\rm P}_{2N+1}}^{(p)} = \FF_{{\rm P}_{2N+1}}^\dag (\Lambda_{{\rm P}_{2N+1}})^p \FF_{{\rm P}_{2N+1}} ~,
\end{equation}
with
\begin{equation}
\Lambda_{{\rm P}_{2N+1}} = \diag(\lambda_{-N}~, \ldots,~ \lambda_N)~, \qquad \lambda_n = \frac{2\pi \ii n}{L} ~,
\end{equation}
which give the derivatives of the interpolating function at the lattice points:
\begin{equation}
\sum_{k = -N}^N (D_{{\rm P}_{2N+1}}^{(p)})_{jk} f_k = \frac{\ed^p}{\ed x^p} f_{\rm approx}(x) \bigg|_{x = x_j} ~.
\end{equation}
The matrix elements of these derivative operators have the same form as \eqref{DA1}, \eqref{DA2} with $2N \to 2N+1$.  The first two are
\begin{align}\label{DP1}
& (D_{{\rm P}_{2N+1}})_{jj} = 0~, \qquad (D_{{\rm P}_{2N+1}})_{jk} = \frac{\pi}{L} (-1)^{j - k} \csc\left(\tfrac{\pi (j-k)}{2N+1}\right)~, \quad j \neq k~,
\end{align}
and
\begin{align}\label{DP2}
(D_{{\rm P}_{2N}}^{(2)})_{jj} =&~ - \frac{\pi^2}{3 L^2} ( (2N+1)^2 - 1)~, \cr
(D_{{\rm P}_{2N+1}}^{(2)})_{jk} =&~ \frac{2\pi^2}{L^2} (-1)^{j-k + 1} \cot\left(\tfrac{\pi (j-k)}{2N+1}\right) \csc\left(\tfrac{\pi (j-k)}{2N+1}\right)~, \quad j \neq k~.
\end{align}
We see from these expressions that the $D_{{\rm A}}$ are anti-periodic under a translation by the full lattice, $(D_{{\rm A}_{2N}}^{(p)})_{j+2N,k} = - (D_{{\rm A}_{2N}}^{(p)})_{jk}$, while the $D_{\rm P}$ are periodic, $(D_{{\rm P}_{2N+1}}^{(p)})_{j+2N+1,k} = (D_{{\rm P}_{2N+1}}^{(p)})_{jk}$.

\begin{figure}[t!]
 \centering
\begin{subfigure}{.50\textwidth}
  \centering
  \includegraphics[width=\linewidth]{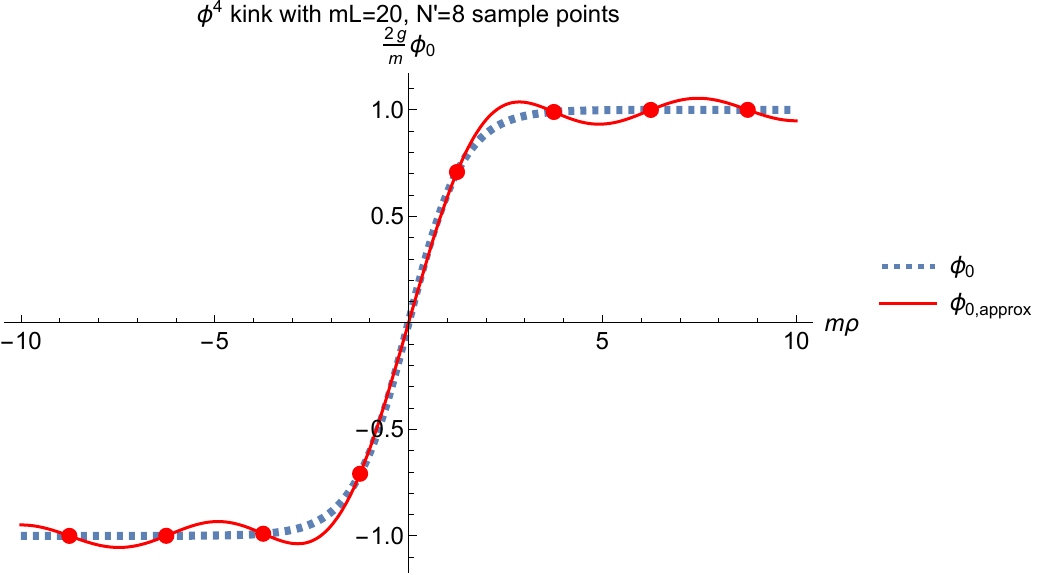}
  \caption{}
  \label{fig:InterpSmallLlowN}
\end{subfigure}  \quad%
\begin{subfigure}{.4\textwidth}
  \centering
  \includegraphics[width=\linewidth]{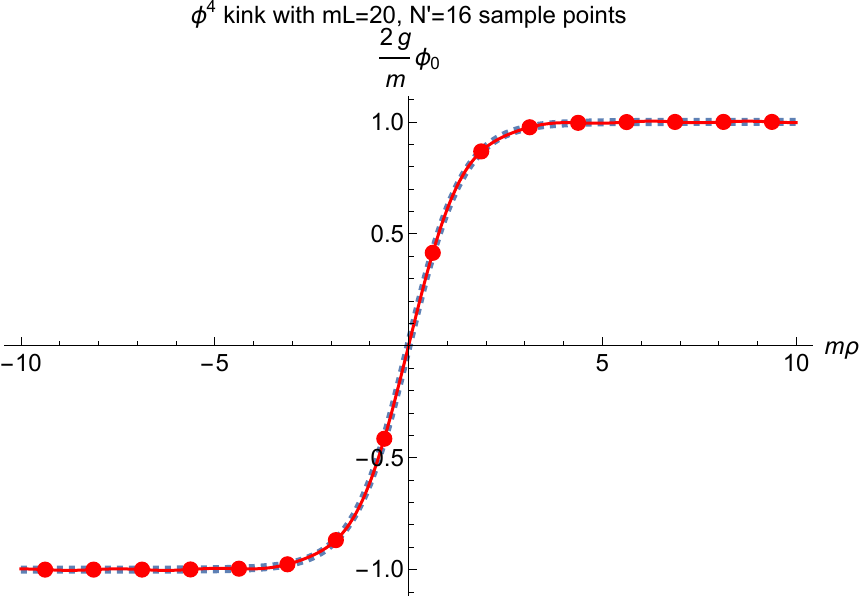}
  \caption{}
  \label{fig:InterpSmallLhighN}
\end{subfigure}  \\[1ex]
\begin{subfigure}{.50\textwidth}
  \centering
  \includegraphics[width=\linewidth]{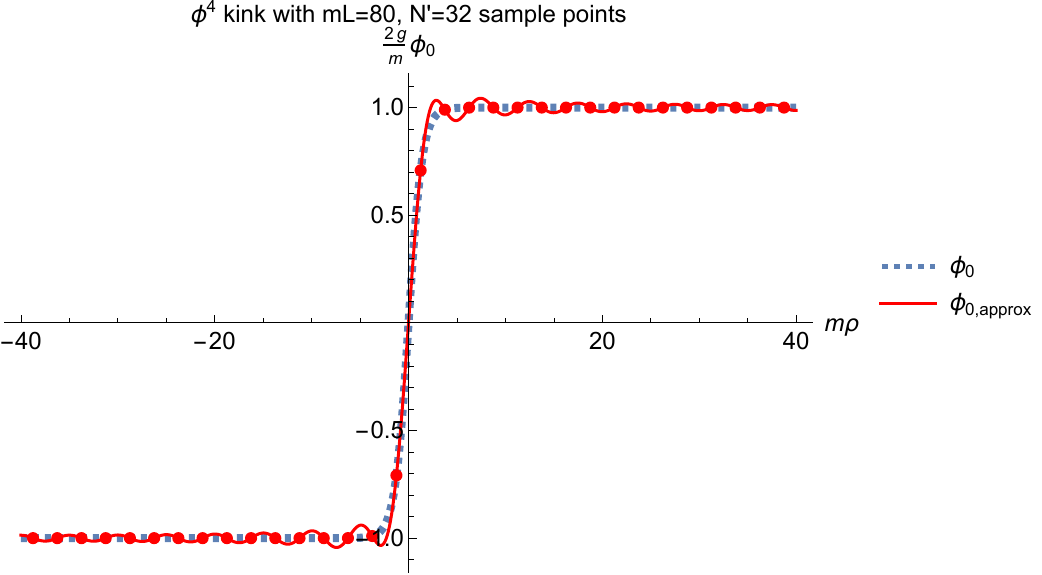}
  \caption{}
  \label{fig:InterpLargeLlowN}
\end{subfigure}  \quad%
\begin{subfigure}{.4\textwidth}
  \centering
  \includegraphics[width=\linewidth]{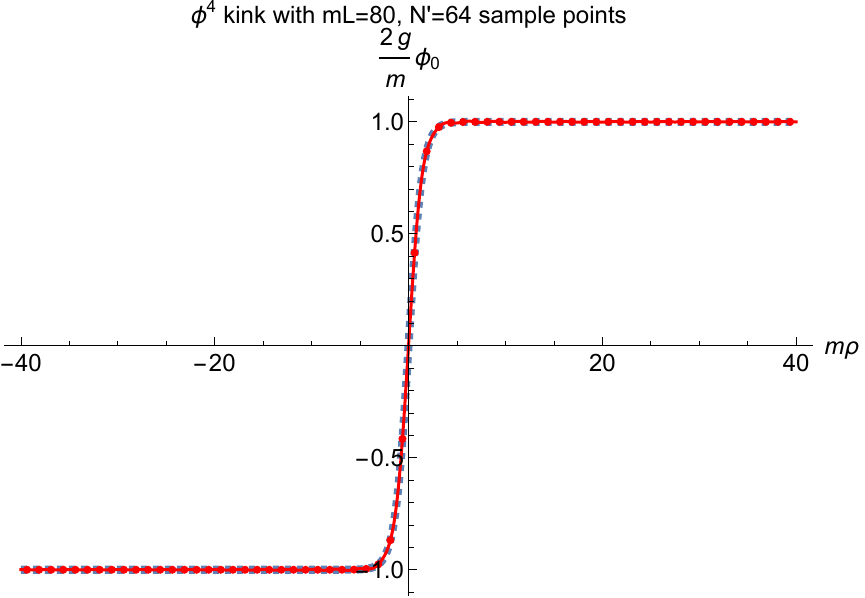}
  \caption{}
  \label{fig:InterpLargeLhighN}
\end{subfigure}
\caption{Comparison of the interpolating function with the exact profile for the anti-periodic $\phi^4$ kink for two choices of $mL$.}
\label{fig:Interpolation}
\end{figure}

Finally we note that the interpolating function can be expressed directly in terms of the lattice data.  In both cases discussed---anti-periodic even lattice and periodic odd lattice---the expression takes the same form:
\begin{equation}\label{DirichletKernel}
f_{\rm approx}(x) = \sum_j \frac{\sin(\frac{N' \pi (x- x_j)}{L}) }{N' \sin(\tfrac{\pi (x- x_j)}{L})} f_j = \sum_j \frac{a \sin(\tfrac{\pi (x-x_j)}{a})}{L \sin(\tfrac{\pi (x-x_j)}{L})} f_j~,
\end{equation}
where $N'$ is the total number of lattice points, $2N$ or $2N+1$ in the two cases respectively.  Although the error in the approximation $f_{\rm approx}(x)$ for a smooth analytic $f(x)$ decreases exponentially fast with increasing $N_{\sigma}'$, the absolute value of this error depends on how rapidly $f$ varies, and more sample points will be required for an $f$ that varies on smaller length scales.  We illustrate this in Figure \ref{fig:Interpolation} by taking $f = \phi_0$, the anti-periodic kink solution \eqref{twistedkink} in $\phi^4$-theory, for a relatively small $mL = 20$ and a relatively large $mL = 80$.

  %%%%%%%%%%%%%%%%% 
%%%%%%%%%%%%%%%%%%

%%%%%%%%%%%%%%%%
\subsection{Lattice Theory with a Continuous Translation Symmetry}\label{ssec:latmodel}
%%%%%%%%%%%%%%%%

Now we use the above technology to define a lattice version of \eqref{SclfinL} with a continuous translation symmetry.  The idea is simple.  We replace the continuum fields in \eqref{SclfinL} with interpolating functions based on the lattice degrees of freedom:
\begin{equation}\label{latticemodel1}
S[\pi,\phi] \to S[\pi_{\rm approx},\phi_{\rm approx}] \equiv S[\pi_j, \phi_j] ~.
\end{equation}
This idea is not new.  It goes back to \cite{Drell:1976bq} and was further studied in the doctoral thesis of Pearson \cite{Pearsonthesis} shortly after.  

Note that the form of $S[\pi_j,\phi_j]$ will depend on the sector of configuration space in which we work.  In $\phi^4$ theory, following \eqref{sectors}, we would use interpolating functions based on the periodic and anti-periodic DFT for the vacuum and kink sector respectively.  We denote the corresponding lattice fields by $\pi_{j}^{({\rm v})},\phi_{j}^{({\rm v})}$ and $\pi_{j}^{(\rm k)},\phi_{j}^{(\rmk)}$.   In sine-Gordon theory we would first make the change of variables \eqref{winding}, which introduces explicit $\nu$ dependence, and then we would replace $\pi^{(\nu)},\phi^{(\nu)}$ with interpolating functions built on the periodic DFT.

Since our interest here is purely in the kink and $\nu = 1$ sectors of these theories, we will simply denote the lattices fields in these sectors by $\pi_j,\phi_j$ in the following.  The appropriate DFT's and SLAC derivatives, $\FF_{{\rm A}_{2N}}, D_{{\rm A}_{2N}}$ and $F_{{\rm P}_{2N+!}}, D_{{\rm P}_{2N+1}}$ will be denoted $\FF,D$.  The total number of lattice sites will be denoted $N'$, so that $N' = 2N$ for $\phi^4$-theory and $N' = 2N+1$ for sine-Gordon.  In either case $N' = L/a$.  We will distinguish the two cases of the $\phi^4$ kink sector and sine-Gordon $\nu = 1$ sector as the need arises.

We also note that nothing in the previous subsection was special to $\phi^4$ theory or sine-Gordon in particular.  The anti-periodic lattice would be appropriate for any $V(\phi)$ with a reflection symmetry, and the periodic lattice would be appropriate for any periodic $V(\phi)$.

The lattice degrees of freedom are taken as canonical variables satisfying equal time Poisson brackets
\begin{equation}\label{latticePB}
\{ \phi_j, \pi_k \} = a^{-1} \delta_{jk} ~.
\end{equation}
The $a^{-1}$ is needed on dimensional grounds and to obtain the correct continuum limit for the interpolating fields, as we now explain.  The mode coefficients $\breve{\phi}_n = \sum_j \FF_{nj} \phi_j$, $\breve{\pi}_n = \sum_j \FF_{nj} \pi_j$ satisfy $\{ \breve{\phi}_m , \breve{\pi}_{n}^\ast \} = a^{-1} \delta_{mn}$ or equivalently $\{ \breve{\phi}_m , \breve{\pi}_{-n} \} = a^{-1} \delta_{mn}$.  From here we obtain the brackets for the interpolating fields:
\begin{equation}\label{interpbracket}
\{ \phi_{\rm approx}(t,x), \pi_{\rm approx}(t,y) \} = \frac{\sin(\tfrac{\pi (x - y)}{a}) }{L \sin(\tfrac{\pi (x-y)}{L})} ~.
\end{equation}
Integrating the right-hand side of this expression against a periodic test function $f(y)$ with period $L$ and taking the limit $a/L \to 0$ agrees with the action of the Dirac delta distribution on $f$:
\begin{align}
\lim_{a/L \to 0} \int_{-L/2}^{L/2} \ed y \frac{\sin(\tfrac{\pi (x - y)}{a}) }{L \sin(\tfrac{\pi (x-y)}{L})} f(y) =&~ \lim_{a/L \to 0} \int_{\frac{2x - L}{2a}}^{\frac{2x+L}{2a}} \ed z \frac{\sin{z}}{\frac{\pi L}{a} \sin(\tfrac{a}{L} z)} f(x - \tfrac{a}{\pi} z) \cr
=&~ f(x) \int_{-\infty}^{\infty} \ed z \frac{\sin(z)}{\pi z} \cr
=&~ f(x)~.
\end{align}
Hence the continuum limit of \eqref{interpbracket} comes out correctly: $\{ \phi(x),\pi(y)\} = \delta(x-y)$.

The equations of motion for the lattice theory can be obtained with respect to $\pi_j,\phi_j$ or with respect to the mode coefficients $\breve{\pi}_n, \breve{\phi}_n$.  With respect to the mode coefficients we have
\begin{align}
\frac{\delta S}{\delta \breve{\phi}_n} =&~ \int \ed y \frac{\delta S}{\delta \phi_{\rm approx}(y)} \frac{ \delta \phi_{\rm approx}(y)}{\delta \breve{\phi}_n} =  \frac{1}{\sqrt{N'}} \int \ed y e^{ 2\pi \ii n y/L} \frac{\delta S}{\delta \phi_{\rm approx}(y)} ~,
\end{align}
for the odd lattice of the sine-Gordon model or the same with $e^{2\pi \ii n y/L} \to e^{2\pi \ii (n - \half) y/L}$ for the even lattice of $\phi^4$-theory.  The point is that the equations of motion, when expressed in terms of the interpolating fields, take the same form as in the continuum theory, except that we only demand they hold when projected against the lowest $N'$ Fourier modes.  

By taking a continuum solution and evaluating it on the lattice we obtain an exponentially accurate approximate solution to the lattice theory.  It is not a true solution since the interpolating field built from the lattice sample will not contain the higher Fourier modes of the continuum solution.  In the continuum theory these higher field modes contribute to the low $n$ modes of the equation of motion through nonlinear terms.  These contributions, which are exponentially small based on decay estimates for Fourier mode coefficients of analytic functions, are missing when we use the interpolating field, and therefore we do not get an exact solution.

We now show that the lattice model defined by using interpolating fields in place of continuum fields in the continuum action, e.g.~\eqref{latticemodel1}, has a continuous translation symmetry.  The basic idea is to find a transformation of lattice data that implements the ordinary translation action on the interpolating field. More precisely, the translation action with parameter $h$ on the soliton sector interpolating fields should be 
\begin{equation}
\pi_{\rm approx}(t,x) \to {\rm T}_h[\pi_{\rm approx}](t,x) = \pi_{\rm approx}(t,x+h)~,
\end{equation}
and $\phi_{\rm approx} \to {\rm T}_h[\phi_{\rm appxox}]$ with
\begin{align}\label{interpolatingtrans}
\textrm{$\phi^4$-theory:} \qquad & {\rm T}_h[\phi_{\rm approx}](t,x) = \phi_{\rm approx}(t,x+h) ~, \cr
\textrm{sine-Gordon:} \qquad & {\rm T}_h[\phi_{\rm approx}](t,x) = \phi_{\rm approx}(t,x+h) + \frac{2\pi h}{g L}~,
\end{align}
where the inhomogeneous term in the sine-Gordon case is due to the explicit $x$-dependence in the $\nu = 1$ case of the change of variables \eqref{winding}.  

To find the lattice transformation that implements \eqref{interpolatingtrans} on the interpolating fields, we exponentiate the appropriate SLAC derivative operator.  Consider the ansatz
\begin{align}\label{latticetrans}
\textrm{$\phi^4$-theory:} \qquad {\rm T}_h[\phi]_j(t) =&~ \sum_{k = -N}^N (e^{h D})_{jk} \phi_{k}(t)  ~, \cr
\textrm{sine-Gordon:} \qquad {\rm T}_h[\phi]_j(t) =&~ \sum_{k = -N}^N (e^{h D})_{jk} \phi_{k}(t) + \frac{2\pi  h}{gL} \mathbf{1}_j~,
\end{align}
where $\mathbf{1}$ is the constant function with value $1$ at every lattice point.  Then in the sine-Gordon case, for example,
\begin{align}\label{latticetransdemo}
{\rm T}_h[\phi_{\rm approx}](t,x) =&~ \frac{1}{2N+1} \sum_{j = -N}^N \sum_{n_1 =-N}^N e^{2\pi \ii n (x - x_j)/L} {\rm T}_h[\phi]_j(t) \cr
=&~ \frac{1}{2N+1} \sum_{j = -N}^N \sum_{n_1 =-N}^N e^{2\pi \ii n_1 (x - x_j)/L} \times \cr
&~ \times  \left\{ \frac{1}{2N+1} \sum_{k = -N}^N  \sum_{n_2 = -N}^N  e^{2\pi \ii n_2 (x_j - x_k + h)/L} \phi_{k}(t) + \frac{2\pi  h}{L} \mathbf{1}_j \right\} \cr
=&~ \frac{1}{2N+1} \sum_{k = -N}^N \sum_{n = -N}^N e^{2\pi \ii n (x - x_k + h)/L} \phi_{k}(t) + \frac{2\pi h}{L} \cr
=&~ \phi_{\rm approx}(t,x+h) + \frac{2\pi \nu h}{g L} 
\end{align}
as desired.  In the second-last step we used
\begin{equation}\label{Diraccomb}
\sum_{j = -N}^N e^{2\pi \ii m j/(2N+1)} = (2N+1) \sum_{\ell \in \mathbbm{Z}} \delta_{m,\ell (2N+1)} ~.
\end{equation}
The right-hand side of \eqref{Diraccomb} is a discrete Dirac comb.  In our case we only picked up the $m=0$ term since either $m = n_2 - n_1$ or $m = -n_1$ when applied to \eqref{latticetransdemo}, and in either case $|m| < 2N+1$.  In $\phi^4$ theory we drop the inhomogeneous shift, and the computation works similarly.  Likewise, the $\pi$ transformation for either theory only requires the action of $e^{h D}$.

One can check that the symmetry \eqref{latticetrans} is generated by
\begin{align}\label{latticePNoether}
\textrm{$\phi^4$-theory:} \qquad   P_{\rm Noether} =&~ - a \sum_j \pi_j (D\phi)_j ~, \cr
\textrm{sine-Gordon:} \qquad P_{\rm Noether} =&~ -a \sum_j \pi_j \left( (D\phi)_j + \frac{2\pi}{gL} \mathbf{1}_j\right) ~,
\end{align}
via the action of the Poisson bracket.  This is a lattice version of the Noether momentum for the continuum theory, \eqref{PNoether}.  

The existence of this symmetry implies, for example, that the lattice version of the static kink solution comes in a one-parameter family just like the continuum solution.  We will use it in the next subsection to introduce a collective coordinate and implement a lattice version of the the soliton sector canonical transformation.

Before doing that, however, let us discuss to what extent the action can be directly expressed in terms of the lattice data $\pi_j,\phi_j$.  There is no difficulty with the standard quadratic terms.  In particular, the gradient energy term evaluates to a quadratic form that utilizes the SLAC second derivative operator:
\begin{align}
\int \ed x \,\pi_{\rm approx} \dot{\phi}_{\rm approx} =&~ a \sum_j \pi_{j} \dot{\phi}_j ~, \cr
\int \ed x \, \pi_{\rm approx}^2 =&~ a \sum_j \pi_{j}^2 ~, \cr
\textrm{$\phi^4$-theory:} \quad \int \ed x (\pd_x \phi_{\rm approx})^2 =&~ - a \sum_{j,k} \phi_j(\FF \Lambda^2 \FF)_{jk} \phi_k = - a \sum_{j,k} \phi_j D^{(2)}_{jk} \phi_k  \cr
\textrm{sine-Gordon:} \quad \int \ed x \left(\frac{2\pi}{gL} +  \pd_x \phi_{\rm approx} \right)^2 =&~ \frac{4\pi^2}{g^2 L} - a \sum_{j,k} \phi_j D^{(2)}_{jk} \phi_k ~.
\end{align}
The appearance of $a$ in front of the $\sum_j \pi_j \dot{\phi}_j$ is consistent with the $a^{-1}$ in the Poisson bracket \eqref{latticePB}.  The constant shift in the sine-Gordon expression for the gradient energy is due to \eqref{winding}.  

The interaction terms are more involved.  We can define potential functions $V_{\rm ti}(\vv{\phi})$ of the lattice data implicitly through
\begin{align}\label{VtiImplicit}
\textrm{$\phi^4$-theory:} \qquad V_{\rm ti}(\vv{\phi}) =&~ \frac{1}{g^2} \int \ed x \left( g^2 \phi_{\rm approx}^{2} - \frac{m^2}{4} \right)^2 ~,  \cr
\textrm{sine-Gordon theory:} \qquad  V_{\rm ti}(\vv{\phi}) =&~ \frac{m^2}{g^2} \int \ed x \left[ 1- \cos\left(\frac{2\pi x}{L} + g\phi_{\rm approx}\right) \right]~,
\end{align}
The subscript ``ti'' refers to ``translation invariant,'' and here begin using the notation $\vv{\phi}$ to refer to the collection of lattice fields $\{\phi_j\}$.  To get explicit expressions for $V_{\rm ti}$ in terms of the $\phi_j$ we have to insert \eqref{DirichletKernel} and evaluate the integral over $x$.  This can be done order by order in principle.  We are able to do it explicitly for low orders in practice.

We evaluate $V_{\rm ti}(\phi_j)$ completely for $\phi^4$-theory in Appendix \ref{app:latquartic}.  The result can be neatly summarized as follows:
\begin{align}\label{VtiVloc1}
\textrm{$\phi^4$-theory:} \quad V_{\rm ti}(\vv{\phi}) =&~ V_{\rm loc}(\vv{\phi}) + \frac{g^2 a^3}{\pi^2} \sum_{j} \left[ \phi_{j}^3 (D^{(2)}\phi)_j + 3 \phi_{j}^2 (D\phi)_j (D\phi_j) \right]~,
\end{align}
where $V_{\rm loc}(\vv{\phi})$ is the ultra-local part of the interaction potential obtained by simply replacing the continuum field $\phi$ with $\phi_j$ and the continuum integral with a Riemann sum:
\begin{equation}\label{Vlocphi4}
V_{\rm loc}(\vv{\phi}) = \frac{a}{g^2} \sum_j \left( g^2 \phi_{j}^2 - \frac{m^2}{4} \right)^2 ~.
\end{equation}
$V_{\rm loc}$ by itself would break translation invariance of the lattice model.  One way to see this is to express it in terms of the Fourier modes by plugging in $\phi_j = \sum_n (\FF^\dag)_{jn} \breve{\phi}_n$.  On the quartic terms the sum over $j$ implements a version of the Dirac comb \eqref{Diraccomb} with $m \to n_1 + n_2 +n_3 + n_4$.  In this case there are contributions to the potential from some $\ell \neq 0$ teeth of the comb.  These contributions, called ``umklapps,'' were discussed in \cite{Drell:1976bq}.  They break total momentum conservation.  The additional set of terms in \eqref{VtiVloc1} provide just the right couplings to restore translation invariance.  This does not appear obvious from the lattice form of the expression but is clear when we consider the origin of this expression in terms of interpolating functions \eqref{VtiImplicit}.

The correction terms relating the two potentials in \eqref{VtiVloc1} have the appearance of higher-derivative operators weighted by inverse powers of the UV cutoff $\Lambda \propto 1/a$.  For lattice configurations $\vv{\phi}$ that are obtained by sampling a periodic analytic function, such as the continuum kink solution, we can argue on general grounds that the difference between $V_{\rm ti}$ and $V_{\rm loc}$ must be exponentially small in $N'$.  $V_{\rm ti}$ and $V_{\rm loc}$ give two different approximations to the continuum interaction energy $\int \ed x V(m^2;\phi)$ based on interpolating functions.  $V_{\rm ti}$ uses an interpolating function for the field itself while $V_{\rm loc}$ uses an interpolating function for the interaction energy density.  Therefore both should be exponentially close to the continuum interaction energy and hence exponentially close to each other.

With the explicit expression \eqref{VtiVloc1} in hand for $\phi^4$-theory, we are also able to see this from a different point of view.  Recalling the relationship between the SLAC derivative and the ordinary derivative of the interpolating function, \eqref{SLACDasOrdinaryD}, we see that the difference in potentials is a Riemann sum approximation to an integral of a total derivative of a periodic function:
\begin{align}
\textrm{$\phi^4$-theory:} \quad V_{\rm ti} - V_{\rm loc} =&~ g^2 a \sum_j \left[ \frac{a^2}{4\pi^2} \pd_{x}^2\phi_{\rm approx}^4 \right]_{x = x_j}~.
\end{align}
The Riemann sum is exponentially close to the integral for periodic analytic functions, and the integral in this case is zero.  Hence the difference in potentials is exponentially small.

We cannot be as explicit in the case of sine-Gordon theory since there is an infinite set of higher and higher order interactions, and we have not been able to obtain a closed form expression for the interaction at arbitrary order.  We have evaluated $V_{\rm ti}$ through quartic order in interactions, for arbitrary sector $\nu$, and find similar results to the $\phi^4$ case.  These results are consistent with the general argument that
\begin{equation}
V_{\rm ti} = V_{\rm loc} + \textrm{exponentially small}~,
\end{equation}
where in this case
\begin{equation}\label{VlocsG}
\textrm{sine-Gordon:} \quad V_{\rm loc}(\vv{\phi}) = \frac{m^2 a}{g^2} \sum_j \left[ 1 - \cos\left( \frac{2\pi x_j}{L} + g \phi_j\right)\right]~.
\end{equation}

Summarizing, we can write the translation-invariant lattice action, restricted to the soliton sector, in the form
\begin{align}\label{latticeSandH}
S[\vv{\pi},\vv{\phi}] =&~ \int \ed t \left( a \sum_j \pi_j \dot{\phi_j} - H[\vv{\pi},\vv{\phi}] \right)~, \qquad \textrm{with} \cr
H[\vv{\pi}, \vv{\phi}] =&~ \frac{a}{2} \sum_j \pi_{j}^2 - \frac{a}{2} \sum_{j,k} \phi_j D_{jk}^{(2)} \phi_k + V_{\rm ti}(\phi_j) + c ~,
\end{align}
where $c = 0$ in $\phi^4$-theory and $c = 2\pi^2/(g^2 L)$ for sine-Gordon.  

We introduce two more quantities in preparation for the next subsection: $\phi_{0j}$ and $\uppsi_{0j}$.  Time-independent solutions to the equations of motion have $\vv{\pi} = 0$, and $\vv{\phi}$ solving
\begin{equation}\label{latticeEOM}
-a \sum_k D_{jk}^{(2)} \phi_k + \frac{\pd V_{\rm ti}}{\pd \phi_j} = 0~, \quad \forall j~.
\end{equation}
As discussed above, we expect a one-parameter family of exact solutions corresponding to the one-parameter family of kinks in the continuum theory, and furthermore sampling the continuum solution on the lattice should give an exponentially accurate approximation to the solution to \eqref{latticeEOM}.  Let us denote the exact lattice kink solution to \eqref{latticeEOM} centered at $x = 0$ by $\vv{\phi}_0$ with components $\phi_{0j}$ in analogy with the continuum configurations \eqref{twistedkink} and \eqref{sGLkink}.  More precisely, in the case of sine-Gordon, $\phi_{0j}$ is well-approximated by the $x = x_j$ value of the periodic function
\begin{equation}\label{periodicsGkink}
\phi_{0}^{(\nu =1)}(x) = \frac{\pi}{g}\left(1-\frac{2x}{L}\right) + \frac{2}{g} \operatorname{am}\left( \left. \frac{m x}{\rmk} \right| \rmk^2\right)~.
\end{equation}
where \eqref{sGLrmk} gives the relation between $L$ and the elliptic modulus $\rmk$.

Since we have a one-parameter family of exact solutions, the linearized equation of motion around the background $\{\phi_{0i}\}$,
\begin{equation}
\sum_k \left(-a D_{jk}^{(2)}  + \frac{\pd V_{\rm ti}}{\pd \phi_j \pd \phi_k}(\phi_{0i}) \right) \uppsi_k = 0~, \quad \forall j~,
\end{equation}
must have a zero mode, $\vv{\uppsi}_0$.  Using the explicit translation action discussed above and working infinitesimally in $h$, we can say exactly what that zero mode is in terms of $\{ \phi_{0 j} \}$:
\begin{align}\label{latticezm}
\textrm{$\phi^4$-theory:} \quad \uppsi_{0j} =&~ N_0 (D \phi_0)_j , \quad N_{0}^{-2} =   a \sum_{j} (D \phi_0)_j (D\phi_0)_j, \cr
\textrm{sine-Gordon:} \quad  \uppsi_{0j} =&~ N_0 \left( (D \phi_0)_j + \frac{2\pi}{gL} \mathbf{1}_j \right) , \quad N_{0}^{-2} = a \sum_j \left( (D_0 \phi_0)_j + \frac{2\pi}{gL} \mathbf{1}_j \right)^2 . \qquad
\end{align}
$N_0$ is a normalization factor such that $a \sum_j \uppsi_{0j}^2 = 1$.  These lattice zero modes are exponentially well approximated by the lattice values of the continuum zero modes \eqref{twistedzm} and \eqref{sGfinLzm}.  In particular, the constant shift in the sine-Gordon lattice zero mode \eqref{latticezm} cancels the derivative of the linear-in-$x$ shift in \eqref{periodicsGkink}.

%%%%%%%%%%%%%%%%
\subsection{Lattice Canonical Transformation}\label{ssec:LatticeCC}
%%%%%%%%%%%%%%%%

Our goal in this subsection is to construct a lattice analog of the soliton sector canonical transformation \eqref{canonicaltrans} for the translation-invariant lattice models discussed above.  The key feature of the canonical transformation is the collective coordinate degree of freedom.  We implement this degree of freedom on the lattice by utilizing the translation symmetry \eqref{latticetrans} with $h \to -X(t)$.  Hence our ansatz for the coordinate part of the phase space transformation $(\vv{\pi}; \vv{\phi}) \mapsto (P,\vv{\varpi}; X,\vv{\varphi})$ is
\begin{align}\label{phij2varphij}
\textrm{$\phi^4$-theory:} \qquad \phi_j(t) =&~ \sum_k (e^{-X(t) D})_{jk} \varphi_k(t) ~, \cr
\textrm{sine-Gordon:} \qquad \phi_j(t) =&~ \sum_k (e^{-X(t) D})_{jk} \varphi_k(t) - \frac{2\pi}{gL} X(t) \mathbf{1}_j ~.
\end{align}
Again we comment on the extra shift in the sine-Gordon case.  Remember that $\phi_j(t)$ is a lattice version of the field $\phi^{(\nu = 1)}(t,x)$, as defined in \eqref{winding}, so that it is periodic when evaluated on the kink solution rather than quasi-periodic.  Therefore a lattice version of the original quasi-periodic field would be obtained by adding $2\pi x_j/(gL)$ to $\phi_j(t)$.  Adding this term to both sides of \eqref{phij2varphij} leads to the appearance of the expected combination $x_j - X \mathbf{1}_j$.  Hence, $\varphi_j$ is a co-moving version of $\phi_j$.

In order to preserve coordinate degrees of freedom $\vv{\varphi}$ must satisfy a constraint.  We impose the lattice version of the continuum constraint $\langle \uppsi_0 | \varphi - \phi_0 \rangle = 0$:
\begin{equation}
\Psi_{\varphi} := a \sum_{j} \uppsi_{0j} ( \varphi_j - \phi_{0j}) = 0~.
\end{equation}
In the $\phi^4$ lattice theory, as in the $\phi^4$ continuum theory, the $L^2$ inner product of $\vv{\uppsi}_0$ and $\vv{\phi}_0$ is zero.  This follows from \eqref{latticezm} and the fact that $D_{jk}$ is antisymmetric.  Hence the constraint simplifies to $\sum_j \uppsi_{0j} \varphi_j = 0$.  In sine-Gordon theory we instead have $\sum_j \uppsi_{0j} \phi_{0j} = N_0 \frac{2\pi}{gL} \sum_j \phi_{0j}$.  While not zero, the key point is that this quantity is a constant, independent of time.

We make the following ansatz for the transformation of momenta:
\begin{equation}\label{pij2varpij}
\pi_j(t) = \Pi_0[P,\vv{\varpi};\vv{\varphi}] \sum_k (e^{-X(t) D})_{jk} \uppsi_{0k} + \sum_k (e^{-X(t) D})_{jk} \varpi_k(t) ~,
\end{equation}
where $\Pi_0$ is a functional to be determined and $\vv{\varpi}$ is constrained to be orthogonal to the zero mode:
\begin{equation}
\Psi_{\varpi} := a \sum_j \uppsi_{0j} \varpi_j  = 0~.
\end{equation}
We assume $X,P$ and $\varphi_j,\varpi_j$ are canonically conjugate pairs on the extended phase space with brackets $\{~,~\}_{\rm P}$: $\{X,P\}_{\rm P} = 1$ and $\{ \varphi_j, \varpi_k \}_{\rm P} = a^{-1} \delta_{jk}$.  We use Dirac brackets on the physical phase space to account for the second class constraints, $\Psi_{\alpha} = (\Psi_{\varphi},\Psi_{\varpi})$.  The relationship is
\begin{equation}
\{F,G\} = \{F,G\}_{\rm P} - \sum_{\alpha,\beta} \{ F, \Psi_{\alpha} \}_{\rm P} ( \{ \Psi_{\alpha}, \Psi_{\beta} \}_{\rm P} )^{-1}  \{ \Psi_{\beta}, G \}_{\rm P} ~,
\end{equation}
Using $a \sum_j \uppsi_{0j} \uppsi_{0j} = 1$ we find $\{ \Psi_{\varphi}, \Psi_{\varpi} \}_{\rm P} = 1$ while $\{ \Psi_{\varphi} , \Psi_{\varphi} \}_{\rm P} = 0 =  \{ \Psi_{\varpi}, \Psi_{\varpi} \}_{\rm P}$.  Hence we have
\begin{equation}
\{ X, P \} = 1~, \qquad \{ \varphi_j , \varpi_k \} = a^{-1} \delta_{jk} - \uppsi_{0j} \uppsi_{0k} ~,
\end{equation}
with all other brackets vanishing.  

Now we solve for the functional $\Pi_0[P,\vv{\varpi};\vv{\varphi}]$ by demanding that the transformation $(\vv{\pi}; \vv{\phi}) \mapsto (P,\vv{\varpi}; X,\vv{\varphi})$ given by \eqref{phij2varphij} and \eqref{pij2varpij} is canonical---that the Dirac brackets of $\vv{\phi}, \vv{\pi}$ satisfy
\begin{equation}
\{ \phi_j, \phi_k \} = 0~, \qquad \{ \phi_j, \pi_k \} = a^{-1} \delta_{jk}~, \qquad \{\pi_j, \pi_k \} = 0~.
\end{equation}
The first of these is automatic since the transformation in question is a point transformation: the old coordinates are a function of the new coordinates only.

Analyzing the remaining two brackets, we find that they hold for the $\phi^4$-theory transformation if and only if $\Pi_0$ satisfies the following system:
\begin{align}
0 =&~ 1 + \frac{\pd \Pi_0}{\pd P} a \sum_j \uppsi_{0j} (D \varphi)_j ~, \cr
\textrm{$\phi^4$-theory:} \qquad 0 =&~ a^{-1} \frac{\pd \Pi_0}{\pd \varpi_j} - \frac{\pd \Pi_0}{\pd P} (D\varphi)_j ~, \cr
0 =&~ \Pi_0 \frac{\pd \Pi_0}{\pd P} (D \uppsi_0)_j + \frac{\pd \Pi_0}{\pd P} (D \varpi)_j + a^{-1} \frac{\pd \Pi_0}{\pd \varphi_j} ~.
\end{align}
In the sine-Gordon case, the resulting equations can be obtained from these by sending $(D \varphi)_j \to (D \varphi_j) + \frac{2\pi}{gL} \mathbf{1}_j$.

These  equations are integrable and the solution is
\begin{align}\label{Pi0solution}
\textrm{$\phi^4$-theory:} \qquad \Pi_0 =&~ - \frac{ (P + a \sum_j \varpi_j (D \varphi)_j ) }{a \sum_j \uppsi_{0j} (D \varphi)_j } ~, \cr
\textrm{sine-Gordon:} \qquad  \Pi_0 =&~ - \frac{\left( P + a \sum_j \varpi_j ( (D\varphi)_j + \frac{2\pi}{gL} \mathbf{1}_j) \right) }{a \sum_j \uppsi_{0j} ( (D \varphi)_j + \frac{2\pi}{gL} \mathbf{1}_j)} ~,
\end{align}
resulting in the time-independent canonical transformations
\begin{align}
\textrm{$\phi^4$-theory:} \quad    \pi_j =&~ -  \frac{ (P + a \sum_i \varpi_i (D \varphi)_i ) }{a \sum_i \uppsi_{0i} (D \varphi)_i } \sum_k (e^{-X D})_{jk} \uppsi_{0k} + \sum_k (e^{- X D})_{jk} \varpi_k ~, \cr
\phi_j =&~ \sum_k (e^{-X D})_{jk} \varphi_k ~,
\end{align}
and
\begin{align}
\textrm{sine-Gordon:} & \cr
\pi_j =&~ -  \frac{ ( P + a \sum_i \varpi_i ( (D \varphi)_i + \tfrac{2\pi}{gL} \mathbf{1}_i ) ) }{a \sum_i \uppsi_{0i} ( (D \varphi)_i + \tfrac{2\pi}{gL} \mathbf{1}_i ) } \sum_k (e^{-X D})_{jk} \uppsi_{0k} + \sum_k (e^{- X D})_{jk} \varpi_k ~, \cr
\phi_j =&~ \sum_k (e^{-X D})_{jk} \varphi_k - \frac{2\pi}{gL} X \mathbf{1}_j ~.  \raisetag{24pt}
\end{align}
An arbitrary integration constant that shifts the value of $P$ was set to zero in \eqref{Pi0solution}.  This choice ensures that the collective coordinate momentum $P$ agrees with the Noether momentum $P_{\rm Noether}$ in \eqref{latticePNoether}.  To see this, contract both sides of the $\pi$ equation with $\sum_k (e^{-X D})_{jk} (D\varphi)_k$ or $\sum_k (e^{-X D})_{jk} ((D \varphi)_k + \frac{2\pi}{gL} \mathbf{1}_k)$ and then use the $\phi$ equation.

The lattice action and Hamiltonian \eqref{latticeSandH} can be expressed in the new variables.  First, one can check that
\begin{equation}
a \sum_j \pi_j \dot{\phi}_j = P \dot{X} + a \sum_j \varpi_j \dot{\varphi}_j ~,
\end{equation}
as required by the canonical transformation.  Then using $(e^{-X D})^T = e^{X D}$, $\sum_k D_{jk} \mathbf{1}_k = 0$, $e^{-X D} D = D e^{-X D}$, and $a \sum_j \uppsi_{0j}^2 = 1$, we quickly find
\begin{align}
\textrm{$\phi^4$-theory:} \quad \frac{a}{2} \sum_{j} \pi_j \pi_j =&~ \frac{ (P + a \sum_i \varpi_i (D\varphi)_i)^2}{2 ( a \sum_i \uppsi_{0i} (D\varphi)_i )^2} + a \sum_j \varpi_{j}^2~,
\end{align}
and the same with $(D\varphi)_i \to (D\varphi_i) + \frac{2\pi}{gL} \mathbf{1}_i$ for sine-Gordon, and 
\begin{equation}
\frac{a}{2} \sum_{j,k} \phi_j D_{jk}^{(2)} \phi_k = \frac{a}{2} \sum_{j,k} \varphi_j D_{jk}^{(2)} \varphi_k ~,
\end{equation}
for both cases.

Now consider \eqref{VtiImplicit} for the interaction potentials.  We insert the canonical transformation into the interpolating function construction.  Since the coordinate part of the transformation is given by the action of the translation operator \eqref{latticetrans},
\begin{equation}
\phi_j = {\rm T}_{-X}[ \varphi]_j ~,
\end{equation}
and the translation operator implements the action \eqref{interpolatingtrans} on the interpolating function, it follows that
\begin{align}
\textrm{$\phi^4$-theory:} \quad \phi_{\rm approx}(x) =&~ \varphi_{\rm approx}(x-X) ~, \cr
\textrm{sine-Gordon:} \quad \phi_{\rm approx}(x) =&~ \varphi_{\rm approx}(x-X) - \frac{2\pi}{gL} X ~,
\end{align}
where $\varphi_{\rm approx}(x)$ is the interpolating function built on the lattice data $\varphi_j$.  Inserting these into \eqref{VtiImplicit}, we find that all $X$ dependence can be removed by a change of integration variables.  In the sine-Gordon case, for example,
\begin{align}
V_{\rm ti}(\vv{\phi}) =&~ \frac{m^2}{g^2} \int_{S^1} \ed x \left[ 1 - \cos \left( \frac{2\pi (x - X)}{L} + g \varphi_{\rm approx}(x - X)\right) \right] \cr
=&~ \frac{m^2}{g^2} \int_{S^1} \ed \rho \left[ 1- \cos\left( \frac{2\pi \rho}{L} + g \varphi_{\rm approx}(\rho) \right) \right] \cr
=&~ V_{\rm ti}(\vv{\varphi}) ~.
\end{align}
The $\phi^4$ case works the same: $V_{\rm ti}(\vv{\phi}) = V_{\rm ti}(\vv{\varphi})$.

The soliton sector action and Hamiltonian \eqref{latticeSandH} in the new variables are thus
\begin{align}\label{ssSlattice}
S[P, \vv{\varpi},X,\vv{\varphi},\nu,\lambda] =&~ \int \ed t \left\{ P \dot{X} + a \sum_{j} \varpi_j \dot{\varphi}_j - H_T[P,\vv{\varpi} , \vv{\varphi},\nu,\lambda] \right\}~, \qquad  \cr
H_{T}P,\vv{\varpi} , \vv{\varphi},\nu,\lambda] =&~ H[P,\vv{\varpi}, \vv{\varphi}] + \lambda a \sum_j \uppsi_{0j} (\varphi_j - \phi_{0j}) + \nu a \sum_j \uppsi_{0j} \varpi_j ~,
\end{align}
where
\begin{align}\label{ssHlattice}
\textrm{$\phi^4$-theory:} & \cr
H =&~ \frac{ (P + a \sum_i \varpi_i (D\varphi)_i)^2}{2 ( a \sum_i \uppsi_{0i} (D\varphi)_i )^2}  + \frac{a}{2} \sum_j \varpi_{j}^2 - \frac{a}{2} \sum_{j,k} \varphi_j D_{jk}^{(2)} \varphi_k + V_{\rm ti}(\vv{\varphi}) ~, \cr
\textrm{sine-Gordon:} & \cr
H =&~ \frac{ (P + a \sum_i \varpi_i ((D\varphi)_i + \frac{2\pi}{gL} \mathbf{1}_i))^2}{2 ( a \sum_i \uppsi_{0i} ((D\varphi)_i + \frac{2\pi}{gL} \mathbf{1}_i) )^2}  + \frac{a}{2} \sum_j \varpi_{j}^2 - \frac{a}{2} \sum_{j,k} \varphi_j D_{jk}^{(2)} \varphi_k + V_{\rm ti}(\vv{\varphi}) + c~. \quad  \cr
\end{align}
The Lagrange multipliers $\nu,\lambda$ in the total Hamiltonian enforce the constraints $\Psi_{\varphi} = 0 = \Psi_{\varpi}$.  Recall that in the sine-Gordon case $c = 2\pi^2/(g^2 L)$.  Since $D^{(2)} = D^2$, we can rewrite the gradient energy term as
\begin{equation}
-\frac{a}{2} \sum_{j,k} \varphi_j D_{jk}^{(2)} \varphi_k + c  = \frac{a}{2} \sum_j \left( (D\varphi)_j + \frac{2\pi}{gL}\mathbf{1}_j \right)^2  ~.
\end{equation}

The lattice action and Hamiltonian \eqref{ssSlattice} are a natural spatial discretization of \eqref{ssS}, and we could have arrived at them directly from \eqref{ssS} by making the replacements
\begin{equation}
 \varphi' \to \left\{ \begin{array}{l l} (D\varphi)_j ~, & \textrm{$\phi^4$-theory}, \\ (D\varphi)_j + \tfrac{2\pi}{gL} \mathbf{1}_j ~, & \textrm{sine-Gordon}~, \end{array} \right.  \qquad \varpi \to \varpi_j ~, \qquad \uppsi_0 \to \uppsi_{0j} ~, \nonumber
 \end{equation}
 \begin{equation}
 \int \ed \rho \to a \sum_j ~,   \qquad V(m^2; \varphi) \to V_{\rm ti}(\vv{\varphi}) ~.
\end{equation}
If our only goal was to discretize the soliton-sector Hamiltonian and its equations of motion, then this approach would be satisfactory.  Instead, we obtained \eqref{ssSlattice} via canonical transformation from a fully regularized lattice model of the original $1+1$-dimensional field theory, \eqref{latticemodel1}.  As we discussed in Subsection \ref{ssec:introreg}, this result lays the groundwork for resolving a long-standing issue in the quantum field theory of solitons which we intend to return to in the future.

In the remainder of this paper we focus on the discretized equations of motion for \eqref{ssSlattice} coupled to an external forcing term for the collective coordinate as in \eqref{SFX}.  This leads to a discretized form of the FSE whose initial value problem is amenable to direct numerical integration via standard Runge--Kutta methods.  In fact, as far as numerical solutions go, we will argue that nothing is lost by replacing the potential $V_{\rm ti}(\vv{\varphi})$ in the soliton-sector Hamiltonian \eqref{ssHlattice} with the simpler $V_{\rm loc}(\vv{\varphi})$ introduced above in \eqref{Vlocphi4} and \eqref{VlocsG}.

%%%%%%%%%%%%%%%%%%%
\subsection{Discretized FSE}
%%%%%%%%%%%%%%%%%%%

We begin with \eqref{ssSlattice} coupled to an external force for the collective coordinate:
\begin{align}
S_F[P, \vv{\varpi},X,\vv{\varphi},\nu,\lambda] =&~ \int \ed t \left\{ P \dot{X} + a \sum_{j} \varpi_j \dot{\varphi}_j - H_T[P,\vv{\varpi} , \vv{\varphi},\nu,\lambda] + X F \right\} ~.
\end{align}
The equations of motion in the collective coordinate sector are simple thanks to the $X$-independence of $H$.  We have Newton's Second Law for the $X$ equation of motion,
\begin{equation}\label{Newton2}
\dot{P} = F~,
\end{equation}
while the $P$ equation of motion gives the collective coordinate velocity in terms of the collective coordinate momentum and the other degrees of freedom.  We have
\begin{equation}
\dot{X} = \beta[P,\vv{\varpi},\vv{\varphi}] = \frac{\pd H_T}{\pd P} = \frac{P + a \sum_i \varpi_i (D\varphi)_i }{ ( a \sum_i \uppsi_{0i} (D\varphi)_i )^2}~, \quad \textrm{for $\phi^4$-theory}~,
\end{equation}
and the same with $(D\varphi)_i \to (D\varphi)_i + \frac{2\pi}{gL} \mathbf{1}_i$ for sine-Gordon.

The Hamiltonian FSE is Hamilton's equations for $\vv{\varpi},\vv{\varphi}$ in the presence of the $P(t)$ specified by \eqref{Newton2}.  The Lagrange multipliers are solved for by contracting the equations with $\vec{\uppsi}_0$.  Inserting their solution back in leads to
\begin{align}\label{projectedform}
a \dot{\varpi}_j = - \sum_k (\delta_{jk} - a \uppsi_{0j} \uppsi_{0k}) \frac{\pd H}{\pd \varphi_k} ~, \qquad a \dot{\varphi}_j = \sum_k (\delta_{jk} - a \uppsi_{0j} \uppsi_{0k}) \frac{\pd H}{\pd \varpi_k} ~,
\end{align}
where we used the facts that $\sum_j \uppsi_{0j} \dot{\varphi}_j = 0 = \sum_j \uppsi_{0j} \dot{\varpi}_j$, by the constraint equations.  It is clear from the form of \eqref{projectedform} that the constraints are preserved under time evolution if initial data $\varpi_j(t_i), \varphi_j(t_i)$ satisfies them.  We then straightforwardly compute the following equations of motion:
\begin{align}\label{latticeFSE1}
& \textrm{$\phi^4$-theory:}  \cr
& \dot{\varphi}_j = \varpi_j + \beta (D \varphi)_j - \beta \uppsi_{0j} \left( a \textstyle\sum_k \uppsi_{0k} (D \varphi)_k \right) ~, \cr
& \dot{\varpi}_j = \beta  \left[ (D\varpi)_j - \uppsi_{0j} \left( a \textstyle\sum_k  \uppsi_{0k} (D \varpi)_k \right) \right] - \beta^2 (D \uppsi_0)_j \left( a \textstyle\sum_k \uppsi_{0k} (D\varphi)_k \right) + \cr
& \qquad \quad  + (D^{(2)} \varphi)_j - \frac{1}{a} \frac{\pd V_{\rm ti}}{\pd \varphi_j} - \uppsi_{0j} \left[ a \sum_k \uppsi_{0k} \left( (D^{(2)}\varphi)_k - \frac{1}{a} \frac{\pd V_{\rm ti}}{\pd \varphi_k} \right) \right] ~, \cr
& \textrm{sine-Gordon:}  \cr
& \dot{\varphi}_j = \varpi_j + \beta \left( (D \varphi)_j + \tfrac{2\pi}{gL} \mathbf{1}_j \right)  - \beta \uppsi_{0j} \left[ a \textstyle\sum_k \uppsi_{0k} \left(  (D \varphi)_k + \tfrac{2\pi}{gL} \mathbf{1}_k \right) \right] ~, \cr
& \dot{\varpi}_j = \beta  \left[ (D\varpi)_j - \uppsi_{0j} \left( a \textstyle\sum_k  \uppsi_{0k} (D \varpi)_k \right) \right] - \beta^2 (D \uppsi_0)_j \left[ a \textstyle\sum_k \uppsi_{0k} \left( (D\varphi)_k + \tfrac{2\pi}{gL} \mathbf{1}_k \right)  \right] + \cr
& \qquad \quad  + (D^{(2)} \varphi)_j - \frac{1}{a} \frac{\pd V_{\rm ti}}{\pd \varphi_j} - \uppsi_{0j} \left[ a \sum_k \uppsi_{0k} \left( (D^{(2)}\varphi)_k - \frac{1}{a} \frac{\pd V_{\rm ti}}{\pd \varphi_k} \right) \right]  ~.
\end{align}

These equations are a discrete analog of the Hamiltonian FSE \eqref{HamiltonFSE}.  For an analytic (anti)-periodic solution, we know that the SLAC derivative operator is an exponentially accurate approximation to the true derivative and the Riemann sums are exponentially accurate approximations to to the $L^2$ inner products appearing in \eqref{HamiltonFSE}.  Additionally the potential $\pd_j V_{\rm ti}$ is an exponentially accurate approximation to $V^{(1)}$ in \eqref{HamiltonFSE}.  As we discussed in the paragraphs following \eqref{Vlocphi4}, however, it is not the only exponentially accurate approximation.  We could alternatively use $V_{\rm loc}$ in place of $V_{\rm ti}$ without sacrificing any accuracy.  Doing so improves computation time.  For the purposes of numerics, therefore, we make this replacement in \eqref{latticeFSE1}.

%%%%%%%%%%%%%%%%
\subsection{Numerical Implementation}\label{ssec:TimeEvolution}
%%%%%%%%%%%%%%%%

We view \eqref{latticeFSE1} as a finite-dimensional first-order system of coupled ODE's.  We can apply fourth-order Runge--Kutta (RK4) to integrate forward in time.  In this subsection we do some rescalings to bring the system to as simple a form as possible, set the initial conditions, and review the RK4 routine.

To begin we briefly return to the continuum finite-$L$ theories in Subsections \ref{ssec:finiteLkinks} through \ref{ssec:boostedkinks} and describe rescaled versions of the static kink profiles, zero modes, and boosted kink profiles.  We do this because we will sample the kink profiles on the lattice to form initial conditions, and we will sample the zero mode on the lattice to approximate the lattice zero mode $\uppsi_{0j}$.  

All dependence on the parameters $m,g$ can be removed from the kink configurations and the equations of motion by rescaling spacetime coordinates and fields appropriately.  We set
\begin{align}
& \tau = m t~, \qquad \widetilde{x} = m x~, \qquad \widetilde{X} = m X~, \qquad \sigma = m \rho = \widetilde{x} - \widetilde{X} ~, \qquad \widetilde{L} = m L~,
\end{align}
and we rescale the original fields $\pi, \phi \to \widetilde{\pi}, \widetilde{\phi}$ via
\begin{equation}
\widetilde{\phi} = \left\{ \begin{array}{l l}  \tfrac{2g}{m} \phi~, &  \textrm{$\phi^4$-theory} \\ g \phi~, & \textrm{sine-Gordon}~,  \end{array} \right.  \qquad \widetilde{\pi} = \left\{ \begin{array}{l l} \tfrac{2g}{m^2} \pi ~, & \textrm{$\phi^4$ -theory} \\ \tfrac{g}{m} \pi ~, & \textrm{sine-Gordon}~.  \end{array} \right.
\end{equation}
The Hamiltonian becomes
\begin{align}
H[\pi,\phi] =&~ \left\{ \begin{array}{l l} \tfrac{m^3}{4g^2} \widetilde{H}[\widetilde{\pi},\widetilde{\phi}] ~,  & \textrm{$\phi^4$-theory}~, \\ \tfrac{m}{g^2} \widetilde{H}[\widetilde{\pi},\widetilde{\phi}]~, & \textrm{sine-Gordon}~, \end{array} \right.
\end{align}
with
\begin{equation}
\widetilde{H} = \int \ed \widetilde{x} \left\{ \half \widetilde{\pi}^2 + \half (\pd_{\widetilde{x}} \widetilde{\phi})^2 + \widetilde{V}(\widetilde{\phi}) \right\}~, 
\end{equation}
where the potentials are
\begin{align}
\textrm{$\phi^4$-theory:} \quad \widetilde{V} = \frac{1}{4} (\widetilde{\phi}^2 - 1)^2~, \qquad \textrm{sine-Gordon:} \quad \widetilde{V} = 1 - \cos(\widetilde{\phi}) ~.
\end{align}

The static kinks are
\begin{align}\label{rescaledkink}
\textrm{$\phi^4$-theory:} \qquad \widetilde{\phi}_0(\sigma) =&~  \sqrt{\frac{2 \rmk^2}{\rmk^2 + 1}} \sn\left(\left. \tfrac{\sigma}{\sqrt{\rmk^2 + 1}} \right|  \rmk^2\right)~, \cr
\textrm{sine-Gordon:} \qquad \widetilde{\phi}_0(\sigma) =&~ \pi + 2 \am\left(\left. \tfrac{\sigma}{\rmk} \right| \rmk^2\right)~,
\end{align}
where the elliptic modulus $\rmk$ is determined by $\widetilde{L}$ according to
\begin{equation}
\textrm{$\phi^4$-theory:} \quad \widetilde{L} = 2 \sqrt{\rm k^2 + 1} \mathbf{K}(\rmk^2)~, \qquad \textrm{sine-Gordon:} \quad \widetilde{L} = 2 \rmk \mathbf{K}(\rmk^2)~.
\end{equation}
Note in sine-Gordon theory the more relevant quantity for the lattice is the periodic configuration
\begin{equation}\label{prescaledkink}
\widetilde{\phi}_{0}^{(\nu = 1)}(\sigma) = \widetilde{\phi}_0(\sigma) - \frac{2\pi \sigma}{\widetilde{L}}~.
\end{equation}
The rescaled normal modes are related to the originals by $\widetilde{\uppsi}_n(\sigma) = \frac{1}{\sqrt{m}} \uppsi_n(\sigma/m)$.  In particular the normalized zero modes are
\begin{align}\label{rescaledzm}
& \textrm{$\phi^4$-theory:}  \cr
& \widetilde{\uppsi}_0(\sigma) =  \sqrt{ \frac{3 \rmk^2}{2 \sqrt{\rmk^2 + 1} \left[ (\rmk^2 + 1) \mathbf{E}(\rmk^2) - (1 - \rmk^2) \mathbf{K}(\rmk^2)\right] } } \cn\left( \left. \tfrac{\sigma}{\sqrt{\rmk^2 + 1}} \right| \rmk^2\right) \dn\left(\left.  \tfrac{\sigma}{\sqrt{\rmk^2 + 1}} \right| \rmk^2\right)~, \cr
& \textrm{sine-Gordon:} & \cr
& \widetilde{\uppsi}_0(\sigma) = \frac{1}{\sqrt{2 \rmk \mathbf{E}(\rmk^2)}} \dn\left( \left.  \tfrac{\sigma}{\rmk} \right| \rmk^2\right)~.
\end{align}

The continuum canonical transformation in the soliton sector maintains its form if we define
\begin{align}
\textrm{$\phi^4$-theory:} \qquad & \widetilde{\varphi} = \frac{2g}{m} \varphi ~, \quad \widetilde{\varpi} = \frac{2g}{m^2} \varpi~, \qquad \widetilde{P} = \frac{4 g^2}{m^3} P ~, \cr
\textrm{sine-Gordon:} \qquad  & \widetilde{\varphi} = g \varphi~, \quad \widetilde{\varpi} = \frac{g}{m} \varpi~, \quad \widetilde{P} = \frac{g^2}{m} P ~.
\end{align}
Then
\begin{align}
\widetilde{\pi}(\tau,\widetilde{x}) =&~ - \frac{( \widetilde{P} + \langle \widetilde{\varpi} \tilde{|} \pd_\sigma \widetilde{\varphi} \rangle ) }{\langle \widetilde{\uppsi}_0 \tilde{|} \pd_{\sigma} \widetilde{\varphi} \rangle} \widetilde{\uppsi}_0(\widetilde{x} - \widetilde{X}(\tau)) + \widetilde{\varpi}(\tau,\widetilde{x} - \widetilde{X}(\tau)) ~, \cr
\widetilde{\phi}(\tau,\widetilde{x}) =&~ \widetilde{\varphi}(\tau, \widetilde{x} - \widetilde{X}(\tau)) ~,
\end{align}
where $\langle f \tilde{|} g\rangle := \int \ed \sigma f(\tau,\sigma)^\ast g(\tau,\sigma)$.  Hence the Hamiltonian $\widetilde{H}$ in the new variables has the same form as $H$ in the old, and the Hamiltonian form of the FSE \eqref{HamiltonFSE} maintains its form with all quantities replaced by there rescaled counterparts, including $t$ and $\rho$ derivatives replaced with $\tau$ and $\sigma$ derivatives. 

The boosted kink configurations are $\widetilde{\varphi}(\sigma) = \widetilde{\phi}_\beta(\sigma)$ with
\begin{align}
\textrm{$\phi^4$-theory:} \qquad \widetilde{\phi}_{\beta}(\sigma) :=&~ \sqrt{\frac{2 \rmk_{\beta}^2}{\rmk_{\beta}^2 + 1}}\sn\left(\left. \tfrac{\sigma}{\sqrt{(1-\beta^2)(\rmk_{\beta}^2 + 1)}} \right| \rmk_{\beta}^2 \right)~. \cr
\textrm{sine-Gordon:} \qquad \widetilde{\phi}_{\beta}(\sigma) :=&~ \pi + 2\am\left(\left. \tfrac{\sigma}{\rmk_\beta \sqrt{1-\beta^2}} \right| \rmk_{\beta}^2 \right)~,
\end{align}
where the $\beta$-dependent elliptic modulus is the solution to
\begin{equation}
\textrm{$\phi^4$-theory:} \quad \frac{\widetilde{L}}{\sqrt{1-\beta^2}} = 2 \sqrt{\rm k_{\beta}^2 + 1} \mathbf{K}(\rmk_{\beta}^2)~, \qquad \textrm{sine-Gordon:} \quad \frac{\widetilde{L}}{\sqrt{1-\beta^2}} = 2 \rmk_{\beta} \mathbf{K}(\rmk_{\beta}^2)~.
\end{equation}
The periodic boosted kink configuration for sine-Gordon is $\widetilde{\phi}_{\beta}^{(\nu = 1)}(\sigma) = \widetilde{\phi}_{\beta}(\sigma) - \frac{2\pi \sigma}{\widetilde{L}}$.  The co-moving momentum field is nonzero on these configurations and given by
\begin{equation}
\widetilde{\varpi}_{\beta}(\sigma) = - \beta \left( \pd_{\sigma} \widetilde{\phi}_\beta - \langle \widetilde{\uppsi}_0 \tilde{|} \pd_{\sigma} \widetilde{\phi}_\beta \rangle \widetilde{\uppsi}_0(\sigma) \right)~.
\end{equation}

The boosted kink configurations carry total momentum $\widetilde{P} = \widetilde{p}_\beta$ and total energy $\widetilde{H} = \widetilde{E}_\beta$ given by
\begin{equation}\label{rescaledpE}
\widetilde{p}_\beta = \frac{\widetilde{M}_{\beta}' \beta}{\sqrt{1-\beta^2}} ~, \qquad \widetilde{E}_{\beta} = \frac{\widetilde{M}_\beta}{\sqrt{1- \beta^2}} ~,
\end{equation}
where the two $\beta$-dependent masses are
\begin{align}
\textrm{$\phi^4$-theory:} \qquad \widetilde{M}_{\beta}' =&~ \frac{4}{3 (\rmk_{\beta}^2 + 1)^{3/2}} \left[ (\rmk_{\beta}^2 + 1) \mathbf{E}(\rmk_{\beta}^2) - (1 - \rmk_{\beta}^2) \mathbf{K}(\rmk_{\beta}^2) \right] ~, \cr
\textrm{sine-Gordon:} \qquad \widetilde{M}_{\beta}' =&~ \frac{8}{\rmk_{\beta}} \mathbf{E}(\rmk_{\beta}^2) ~,
\end{align}
and
\begin{align}
\textrm{$\phi^4$-theory:} \qquad \widetilde{M}_{\beta} =&~ \frac{8}{3 (\rmk^2 + 1)^{3/2}} \left[ (1 + \beta^2) I_1(\rmk_{\beta}^2) + (1 - \beta^2) I_2(\rmk_{\beta}^2) \right], \cr
\textrm{sine-Gordon:} \qquad \widetilde{M}_{\beta} =&~ (1 + \beta^2) \frac{4}{\rmk_{\beta}} \mathbf{E}(\rmk_{\beta}^2) + (1 - \beta^2) 4 \left[ \mathbf{E}(\rmk_{\beta}^2) - (1 - \rmk_{\beta}^2) \mathbf{K}(\rmk_{\beta}^2) \right] , \qquad
\end{align}
with $I_{1,2}$ defined in \eqref{I12def}.  In the $\widetilde{L} \to \infty$ limit these masses become $\beta$-independent and agree with the $\widetilde{L} \to \infty$ limit of the kink mass
\begin{equation}
\lim_{\widetilde{L} \to \infty} \widetilde{M}_{\beta}'  = \lim_{\widetilde{L} \to \infty} \widetilde{M}_{\beta} = \lim_{\widetilde{L} \to \infty} \widetilde{M}_0  ~,
\end{equation}
which, in these rescaled variables, is
\begin{equation}
\textrm{$\phi^4$-theory:} \quad \lim_{\widetilde{L} \to \infty} \widetilde{M}_0 = \frac{2\sqrt{2}}{3}~, \qquad \textrm{sine-Gordon:} \quad \lim_{\widetilde{L} \to \infty} \widetilde{M}_0 = 8~.
\end{equation}

Now we return to the lattice and introduce a rescaled lattice spacing $\tilde{a} = ma$.  For $\phi^4$-theory and sine-Gordon we consider an odd and even number of lattice points respectively,
\begin{align}
\textrm{$\phi^4$-theory:} \qquad \widetilde{x}_j =&~ \left( j - \tfrac{1}{2} \right) \tilde{a} ~, \qquad j = -N + 1, \ldots, N~, \cr
\textrm{sine-Gordon:} \qquad \widetilde{x}_j =&~ j \tilde{a} ~, \qquad j = -N~, \ldots, N~,
\end{align}
with $N' = \widetilde{L}/\tilde{a}$ the total number of sites: $N' = 2N$ or $N' = 2N+1$ respectively.  We introduce lattice degrees of freedom $\widetilde{\pi}_j, \widetilde{\phi}_j$ and carry out the lattice canonical transformation $( \widetilde{\pi}_j; \widetilde{\phi}_j) \mapsto (\widetilde{P}, \widetilde{\varpi}_j ; \widetilde{X}, \widetilde{\varphi}_j)$ just as before.   All of these quantities are rescaled in exactly the same way as their continuum counterparts.  The rescaled lattice action and Hamiltonian will be of the same form as \eqref{ssSlattice} and \eqref{ssHlattice} in terms of the rescaled fields.  

Note that the rescaled lattice kink configuration $\widetilde{\phi}_{0j}$ is the solution to the rescaled version of \eqref{latticeEOM}:
\begin{equation}
- \tilde{a} \sum_k \widetilde{D}_{jk}^{(2)} \widetilde{\phi}_{k} + \frac{\pd \widetilde{V}_{\rm ti}}{\pd \widetilde{\phi}_j} = 0~, \qquad \forall j~,
\end{equation}
where the rescaled translation-invariant potential is
\begin{align}\label{VtiImplicit}
\textrm{$\phi^4$-theory:} \qquad \widetilde{V}_{\rm ti} =&~ \frac{1}{4} \int \ed \widetilde{x} \left( \widetilde{\phi}_{\rm approx}^{2} - 1  \right)^2 ~,  \cr
\textrm{sine-Gordon theory:} \qquad  \widetilde{V}_{\rm ti} =&~  \int \ed \widetilde{x} \left[ 1- \cos\left(\frac{2\pi \widetilde{x}}{\widetilde{L}} + \widetilde{\phi}_{\rm approx}\right) \right]~,
\end{align}
and the interpolating functions are built from the lattice data $\widetilde{\phi}_j$ in the same way as \eqref{DirichletKernel} with $x,x_j a,L \to \widetilde{x}, \widetilde{x}_j, \tilde{a},\widetilde{L}$.  The SLAC derivative operator is dimensionful, and the rescaled $\widetilde{D}$ uses $\widetilde{L}$ in place of $L$.  The configuration $\widetilde{\phi}_{0j}$ is well-approximated by sampling the kink configuration \eqref{rescaledkink} on the lattice for $\phi^4$-theory and the periodic kink configuration \eqref{prescaledkink} on the lattice for sine-Gordon.

The rescaled lattice zero mode $\widetilde{\uppsi}_{0j}$ can be obtained from $\widetilde{\phi}_{0j}$ by a rescaled version of \eqref{latticezm}.  In particular, for sine-Gordon the shift term $\frac{2\pi}{gL} \mathbf{1}_j$ is replaced by $\frac{2\pi}{\widetilde{L}} \mathbf{1}_j$.  However, exponentially accurate approximations to these zero modes are more easily obtained by sampling the continuum zero modes \eqref{rescaledzm} on the lattice.

Now we are ready to give a rescaled form of the discretized Hamiltonian FSE \eqref{latticeFSE1} that is suitable for numerical implementation.  We introduce a shorthand for the four distinct Riemann sums that appear:
\begin{align}
\RS_1 :=&~  \left\{ \begin{array}{l l}  \tilde{a} \sum_j \widetilde{\varpi}_j (\widetilde{D} \widetilde{\varphi})_j~, & \textrm{$\phi^4$-theory} \\  \tilde{a} \sum_j \widetilde{\varpi}_j \left( (\widetilde{D} \widetilde{\varphi})_j + \frac{2\pi}{\widetilde{L}} \mathbf{1}_j \right) ~, & \textrm{sine-Gordon} ~, \end{array} \right.    \cr
\RS_2 :=&~ \left\{ \begin{array}{l l} \tilde{a} \sum_j \widetilde{\uppsi}_{0j} (\widetilde{D} \widetilde{\varphi})_j ~, & \textrm{$\phi^4$-theory}~, \\ \tilde{a} \sum_j \widetilde{\uppsi}_{0j} \left( (\widetilde{D} \widetilde{\varphi})_j + \frac{2\pi}{\widetilde{L}} \mathbf{1}_j \right)~, & \textrm{sine-Gordon}~, \end{array} \right. \cr
\RS_3 :=&~ \tilde{a} \sum_j \widetilde{\uppsi}_{0j} \left( (\widetilde{D}^{(2)} \widetilde{\varphi})_j - \frac{1}{\tilde{a}} \frac{\pd \widetilde{V}_{\rm loc}}{\pd \widetilde{\varphi}_j} \right)~,  \cr
\RS_4 :=&~ \tilde{a} \sum_j \widetilde{\uppsi}_{0j} (\widetilde{D} \widetilde{\varpi})_j ~.
\end{align}
Then the discretized Hamiltonian FSE is
\begin{align}\label{disHFSE}
\textrm{$\phi^4$-theory:} \qquad  \pd_{\tau} \widetilde{\varphi}_j =&~ \frac{(\widetilde{P} + \RS_1)}{\RS_{2}^2} \left( (\widetilde{D} \widetilde{\varphi})_j - \RS_2  \widetilde{\uppsi}_{0j} \right) + \widetilde{\varpi}_j ~, \cr
\textrm{sine-Gordon: } \qquad   \pd_{\tau} \widetilde{\varphi}_j =&~ \frac{(\widetilde{P} + \RS_1)}{\RS_{2}^2}  \left(  (\widetilde{D} \widetilde{\varphi})_j + \tfrac{2\pi}{\widetilde{L}} \mathbf{1}_j  - \RS_2  \widetilde{\uppsi}_{0j} \right) + \widetilde{\varpi}_j ~, \cr
\textrm{both:} \qquad \pd_{\tau} \widetilde{\varpi}_j =&~ \frac{ (\widetilde{P} + \RS_1)}{\RS_{2}^2} \left( (\widetilde{D} \widetilde{\varpi})_j - \RS_4 \widetilde{\uppsi}_{0j} \right) - \frac{ (\widetilde{P} + \RS_{1})^2}{\RS_{2}^3} (\widetilde{D} \widetilde{\uppsi}_0)_j + \cr
&~ + (\widetilde{D}^2 \widetilde{\varphi})_j - \frac{1}{\tilde{a}} \frac{\pd \widetilde{V}_{\rm loc}}{\pd \widetilde{\varphi}_j} -\RS_3  \widetilde{\uppsi}_{0j} ~.
\end{align}
In the last two expressions we have exchanged $\widetilde{V}_{\rm ti}$ for $\widetilde{V}_{\rm loc}$.  As discussed above this improves run time and does not sacrifice accuracy.  The local potentials are
\begin{equation}
\textrm{$\phi^4$-theory:} \quad \widetilde{V}_{\rm loc} = \frac{\tilde{a}}{4} (\widetilde{\varphi}_{j}^2 - 1)^2 ~, \qquad \textrm{sine-Gordon:} \quad \widetilde{V}_{\rm loc} = \tilde{a} \left[ 1 - \cos \left( \frac{2\pi \sigma_j}{\widetilde{L}} + \widetilde{\varphi}_j \right) \right] ~.
\end{equation}

Combining the momenta and coordinates into a single phase space vector, $\vv{\zeta} = (\vv{\widetilde{\varpi}},\vv{\widetilde{\varphi}})$, \eqref{disHFSE} takes the form of a coupled first order system of ordinary differential equations:
\begin{equation}
\pd_{\tau} \vv{\zeta} = \vv{F}_{\zeta}(\tau,\vv{\zeta})~.
\end{equation}
The explicit time dependence of the right-hand side arises solely through the explicit time dependence of $\widetilde{P}$.  The collective coordinate momentum $\widetilde{P}$ is viewed as a given function of time, determined by a given external force that couples directly to $\widetilde{X}$.  We will focus entirely on instantaneous impact forces that deliver an impulse at some time $\tau_\ast$:
\begin{equation}
\widetilde{P}(\tau) = \widetilde{P}_i +  (\widetilde{P}_f - \widetilde{P}_i) \Theta(\tau - \tau_\ast)~.
\end{equation}

We discretize the time variable into steps of size $\Delta \tau$ and apply a standard RK4 scheme to march forward in time.  This is an explicit scheme, meaning that the numerical solution at a subsequent time step is determined entirely by knowledge of the previous step.  Over a finite time interval $\tau \in [0,T]$, this scheme will be accurate to $O(\Delta \tau^4)$ with a coefficient that depends on $T$.  Notions of long-time stability, which probe the $T \to \infty$ behavior of the numerical solution at fixed $\Delta \tau$, can be defined for Hamiltonian systems of the form we study.  Achieving this form a stability generally requires an implicit scheme.  See, for example, \cite{sanz-serna_1992,Stuart1994ModelPI}.  It will be sufficient for our purposes to work with the standard RK4 scheme, and we will verify the energy relation \eqref{Power} numerically as a test of reliability.  In particular, we will find that the total energy is conserved to excellent precision when $\dot{P} = 0$.

The RK4 scheme is implemented as follows.  Let $\vv{\zeta}(\tau_0) \equiv \vv{\zeta}_0 = (\vv{\widetilde{\varpi}}(\tau_0),\vv{\widetilde{\varphi}}(\tau_0))$, satisfying the constraints $\sum_j \widetilde{\uppsi}_{0j} \widetilde{\varpi}_j = 0 = \sum_j \widetilde{\uppsi}_{0j} (\widetilde{\varphi}_j - \widetilde{\phi}_{0j})$, be given.  In this work we will restrict consideration to initial conditions that are a pure boosted kink with initial velocity $\beta_i$ corresponding to the initial momentum $\widetilde{P}_i$ via the relation \eqref{rescaledpE}.  These configurations satisfy the constraints.  Then the numerical solution at time $\tau_{n+1} = \tau_n + \Delta \tau$ is given by
\begin{equation}
\vv{\zeta}_{n+1} = \vv{\zeta}_{n} + \frac{1}{6} ( \vv{\zeta}_{n}^{(1)} + 2 \vv{\zeta}_{n}^{(2)} + 2 \vv{\zeta}_{n}^{(3)} + \vv{\zeta}_{n}^{(4)} )~,
\end{equation}
where 
\begin{align}
 \vv{\zeta}_{n}^{(1)} =&~  \Delta \tau \, \vv{F}_{\zeta} \left( \tau_n , \vv{\zeta}_n \right)~, \cr
 \vv{\zeta}_{n}^{(2)} =&~  \Delta \tau \,  \vv{F}_{\zeta} \left( \tau_n + \frac{\Delta \tau}{2}, \vv{\zeta}_n + \half \vv{\zeta}_{n}^{(1)} \right)~, \cr
 \vv{\zeta}_{n}^{(3)} =&~ \Delta \tau \,  \vv{F}_{\zeta} \left( \tau_n + \frac{\Delta \tau}{2}, \vv{\zeta}_n + \half \vv{\zeta}_{n}^{(2)} \right)~, \cr
 \vv{\zeta}_{n}^{(4)} =&~  \Delta \tau  \, \vv{F}_{\zeta} \left( \tau_n + \Delta\tau, \vv{\zeta}_n + \vv{\zeta}_{n}^{(3)} \right)~.
 \end{align}
 In the code implementing this routine we refer to the solution at time $\tau_n$ as $\vv{\zeta}_{n} = \vv{\zeta}_{n}^{(0)}$.
 
 The numerical solution to the FSE given by the collection of $\{\vv{\zeta}_n\}$ can be used to visualize the kink and radiation in the co-moving frame, where the kink position is always at $\sigma = 0$.  In order to view the system from the lab frame we must also solve
 \begin{equation}\label{Xtildeeom}
 \pd_{\tau} \widetilde{X} = \beta = \frac{ \widetilde{P} + \RS_1}{\RS_{2}^2}
 \end{equation}
to find the motion of the collective coordinate.  This is easily done by computing $\beta_n = \beta(\tau_n)$ at each time step from the solution $\vv{\zeta}_n$ and the given $\widetilde{P}$, and then using the same RK4 procedure to integrate \eqref{Xtildeeom}.  With the solution for $\widetilde{X}(\tau)$ in hand, we then get the original lab frame position field via
 \begin{equation}
 \widetilde{\phi}(\tau,\widetilde{x}) = \varphi(\tau,\sigma + \widetilde{X}(\tau))~.
 \end{equation}
We can also evaluate any quantity of interest on the solution $\{ \vv{\zeta}_{n} \}$, such as the energy functionals introduced in Subsection \ref{ssec:EnergyandPower}.  

The Mathematica notebook included as supplemental material with this submission has detailed comments explaining the implementation of this approach.

%%%%%%%%%%%%%%%%
%%%%%%%%%%%%%%%%
\section{Results and Analysis}
%%%%%%%%%%%%%%%%
%%%%%%%%%%%%%%%%

In this section we present numerical results, investigate the phenomenon of superluminal collective coordinate velocities, and compare numerical results with perturbation theory calculations for the case of small momentum transfer.

%%%%%%%%%%%%%%%%
\subsection{Results}\label{ssec:numresults}
%%%%%%%%%%%%%%%%

The Mathematica code that generates solutions to the FSE for initial conditions given by a boosted kink with velocity $\beta_i$ and an instantaneous impact force with momentum transfer $\widetilde{P}_f - \widetilde{P}_i$ depends on seven inputs.  These are the initial kink velocity $\beta_i$; the time of impact $\tau_\ast$; the momentum transfer in units of the kink mass, denoted $\epsilon = (\widetilde{P}_f - \widetilde{P}_i)/\widetilde{M}_0 = (P_f - P_i)/M_0$; the size of the circle $\widetilde{L}$; the number $N_\sigma$ determining the number of spatial lattice points $N_{\sigma}'$ via $N_{\sigma}' = 2N_\sigma$ for $\phi^4$-theory and $N_{\sigma}' = 2N_\sigma + 1$ for sine-Gordon; the temporal step size $\Delta \tau$; and the total number of time steps $N_\tau$.  The code starts at $\tau_0 = 0$.  The initial momentum $\widetilde{P}_i = \widetilde{p}_{\beta_i}$ is determined from $\beta_i$ using \eqref{rescaledpE}, so specifying $\epsilon$ is equivalent to specifying $\widetilde{P}_f$.  The two versions of the main code are called ``phi4FSEivp'' and ``sGFSEivp.''

The output is a list of 12 items, some of which are themselves lists.  The first four are the initial and final total momentum $\widetilde{P}_{i,f}$ and the (theoretically computed) initial and final total energy $\widetilde{H}_{i,f}$.  The initial energy $\widetilde{H}_i = \widetilde{E}_{\beta_i}$ is computed from \eqref{rescaledpE}, while the final energy is obtained from our analysis of the jumping conditions in Subsection \ref{ssec:Jump}.  See equation \eqref{Hbarjump2}.  Similarly, the fifth item is the jump in collective coordinate velocity at the kick, computed via \eqref{betajump}.  The next four items are the co-moving position field history $\{ \widetilde{\varphi}_{j}(\tau_n) \}$, the co-moving momentum field history $\{ \widetilde{\varpi}_j(\tau_n) \}$, the collective coordinate trajectory $\{ \widetilde{X}(\tau_n) \}$, and the collective coordinate velocity $\{ \beta(\tau_n) \}$.  The first two of these are lists of lists with the structure
\begin{equation}
\{ \widetilde{\varphi}_{j}(\tau_n) \} \to \left\{ \{ \{\sigma_j, \widetilde{\varphi}_j(\tau_n)\}~|~j = 1,\ldots, N_{\sigma}' \} ~|~ n = 0~, \ldots, N_t -1 \right\}~,
\end{equation}
and similarly for $\{ \widetilde{\varpi}_j(\tau_n) \}$.  These can readily be converted to a table of list plots which can then be animated.  The next two are lists of the form $\{ \{\tau_n, \widetilde{X}(\tau_n)\}~|~ n = 0 \ldots N_\tau - 1 \}$, and similarly for $\{ \beta(\tau_n) \}$.

The last three output items are the time development of the (numerically computed) total Hamiltonian $\{ \widetilde{H}(\tau_n)/\widetilde{M}_0  = H(\tau_n)/M_0 \}$, the kink effective Hamiltonian $\{ H_{\rm eff}(\tau_n)/M_0 \}$, and the instantaneous mechanical kink energy $\{ E_{\beta}(\tau_n)/M_0 \}$, all in units of the kink mass.  Each is given in the same list structure as the collective coordinate trajectory and velocity.  These three quantities allow us to separate the total energy of the configuration into the three components discussed in Subsection \ref{ssec:EnergyandPower}: the mechanical energy of the kink $E_{\beta}$, the interaction energy between the kink and the normal modes $E_{\rm int} = H_{\rm eff} - E_{\beta}$, and the energy in the normal modes $E_{\rm nm} = H - H_{\rm eff}$.  The mechanical energy is only defined for $|\beta| < 1$, so when the collective coordinate velocity is superluminal we do not separate $H_{\rm eff}$ into mechanical and interaction components.  Therefore the list structure of the mechanical energy values is $\{ \{ \tau_n, E_{\beta}(\tau_n)/M_0 \}~|~ |\beta(\tau_n)| < 1~, \textrm{with } n = 0,\ldots N_t - 1\}$.

 %%%%%%%%%%%%%%%%% 
 \begin{figure}[th!]
 \centering
\begin{subfigure}{.40\textwidth}
  \centering
  \includegraphics[width=\linewidth]{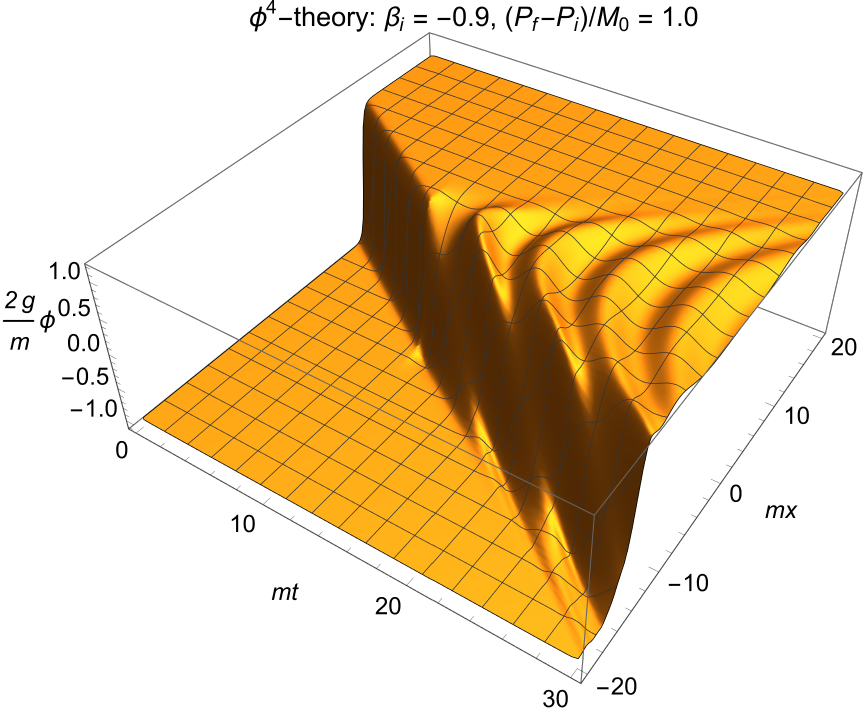}
  \caption{}
  \label{fig:worldsheetphi4low}
\end{subfigure}  \qquad%
\begin{subfigure}{.40\textwidth}
  \centering
  \includegraphics[width=\linewidth]{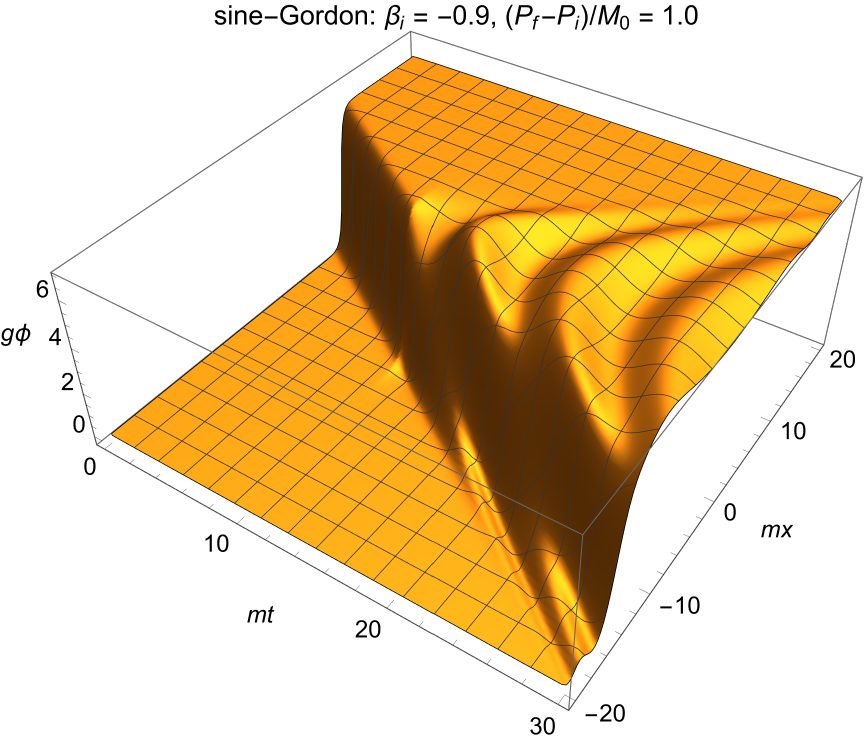}
  \caption{}
  \label{fig:worldsheetsGlow}
\end{subfigure}  \\[1ex] 
\begin{subfigure}{.40\textwidth}
  \centering
  \includegraphics[width=\linewidth]{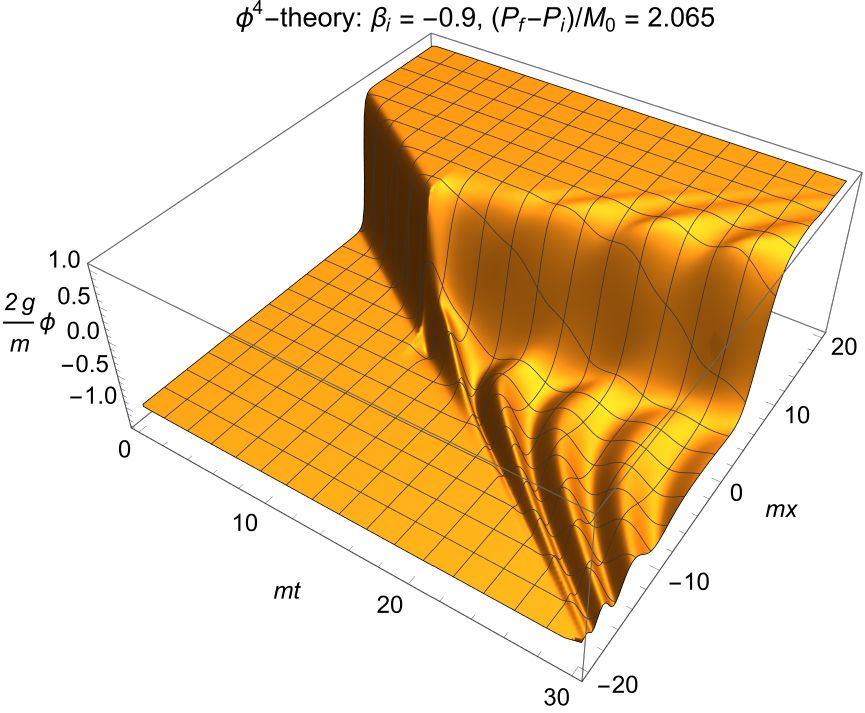}
  \caption{}
  \label{fig:worldsheetphi4mid}
\end{subfigure}  \qquad %
\begin{subfigure}{.40\textwidth}
  \centering
  \includegraphics[width=\linewidth]{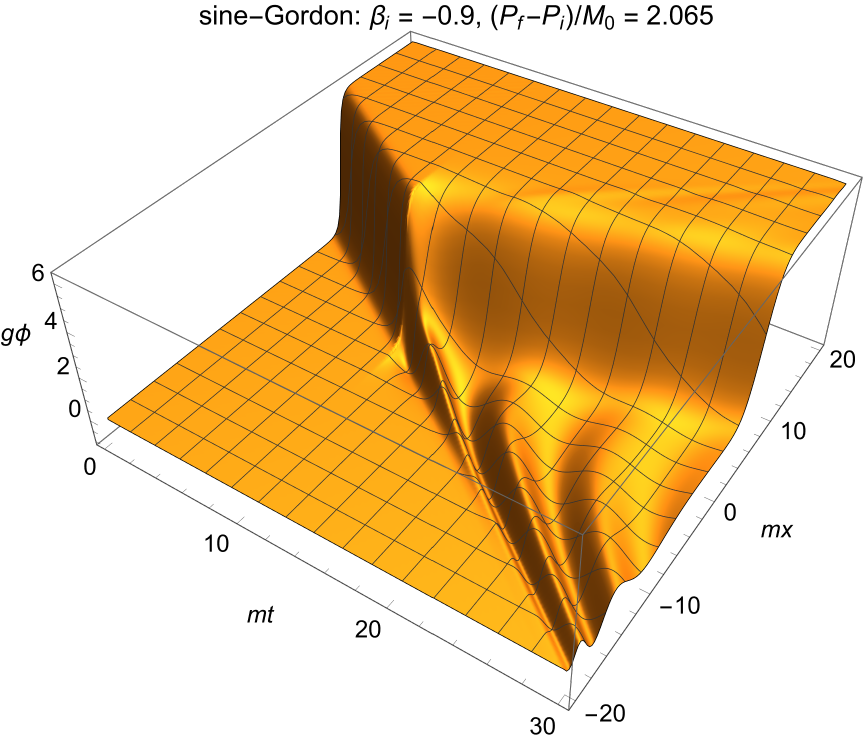}
  \caption{}
  \label{fig:worldsheetsGmid}
\end{subfigure}  \\[1ex]
\begin{subfigure}{.40\textwidth}
  \centering
  \includegraphics[width=\linewidth]{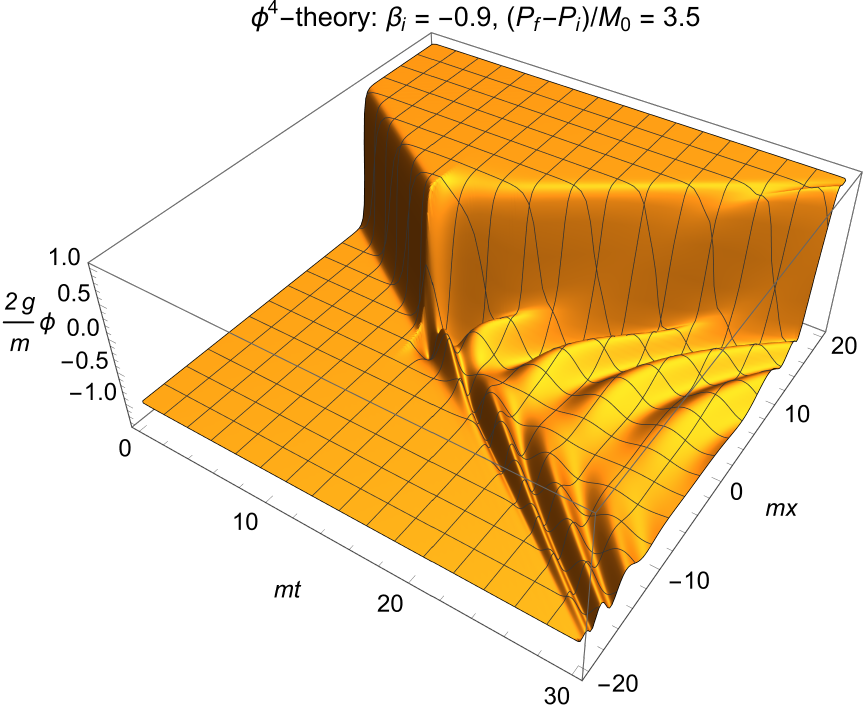}
  \caption{}
  \label{fig:worldsheetphi4high}
\end{subfigure}  \qquad %
\begin{subfigure}{.40\textwidth}
  \centering
  \includegraphics[width=\linewidth]{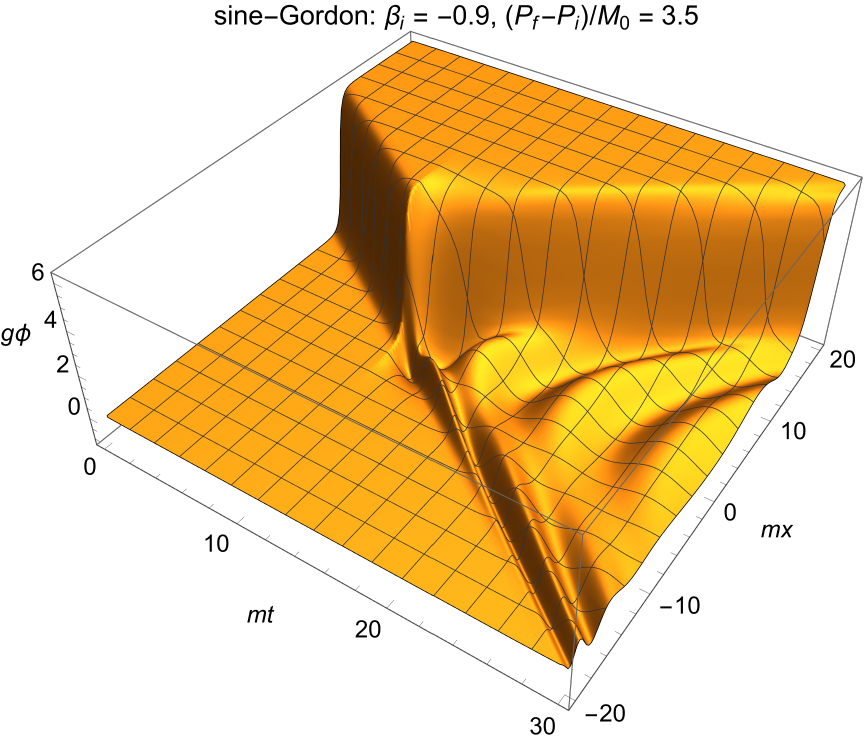}
  \caption{}
  \label{fig:worldsheetsGhigh}
\end{subfigure}
\caption{Spacetime evolution of the scalar field $\phi(t,x)$, with the kink coming in from the left and the impulse applied at $mt = 10$.  These runs use a box size of $mL = 80$ with $N_{\sigma}' = 240$ or $241$ lattice points and a time step of $m \Delta t = 0.04$.  The initial position of the kink is chosen so that it is at $x =0$ when the impact occurs, and we restrict the viewing range to $-20 \leq m x \leq 20$ so that (anti-)periodicity of the configuration is not observable.}
\label{fig:worldsheets}
\end{figure}
%%%%%%%%%%%%%%%%%%

 In Figure \ref{fig:worldsheets} we show the spacetime history of the scalar field in the lab frame, $\phi(t,x) = \varphi(t, x + X(t))$, in three different scenarios.  In all three cases we start with a kink traveling to the left with speed $\beta_i = 0.9$ corresponding to an initial momentum of $P_i \approx -2.605 M_0$.  The delta function force is applied at $m t_\ast = 10$.  The three rows of the figure correspond to increasing momentum transfers of $(P_f - P_i)/M_0 = 1.0$, $2.605$, and $3.5$.  The left column is the $\phi^4$-theory kink, and the right column is the sine-Gordon kink.  Note that the final momentum $P_f$ is distributed between the kink's collective coordinate and the radiation, and the latter receives a non-negligible fraction.  For example, the final total momentum in the second case is approximately zero, but the kink is clearly moving off to the right.  The radiation carries an equal and opposite momentum left.  

The behavior of the $\phi^4$ kink and sine-Gordon kink is qualitatively very similar with the only noticeable difference being the frequency of field oscillations after the kick.  This can be attributed to the fact that the $\phi^4$-potential well is steeper, having radius of curvature $\widetilde{V}^{(2)}(\widetilde{\phi}_{\rm vac}) = 2$ for $\phi^4$-theory versus $\widetilde{V}^{(2)}(\widetilde{\phi}_{\rm vac}) = 1$ for sine-Gordon.

  %%%%%%%%%%%%%%%%% 
 \begin{figure}[t!]
 \centering
\begin{subfigure}{.48\textwidth}
  \centering
  \includegraphics[width=\linewidth]{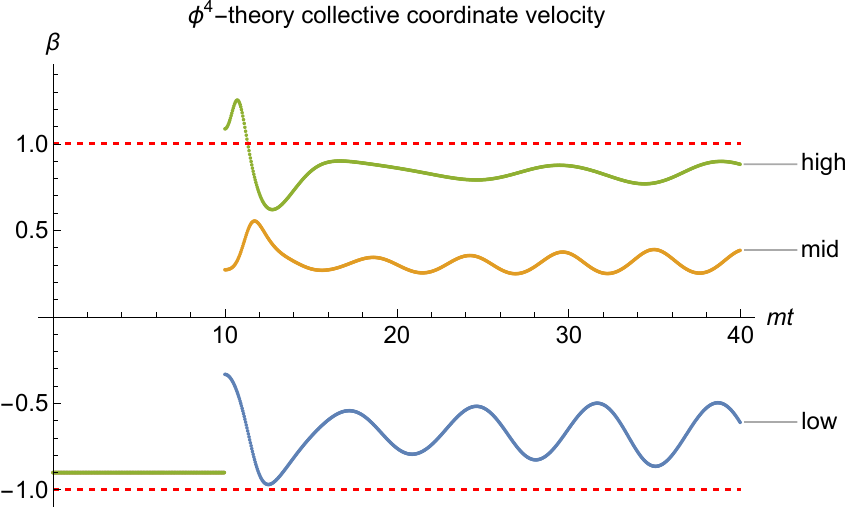}
  \caption{}
  \label{fig:ccvelocitiesphi4}
\end{subfigure}  \quad%
\begin{subfigure}{.48\textwidth}
  \centering
  \includegraphics[width=\linewidth]{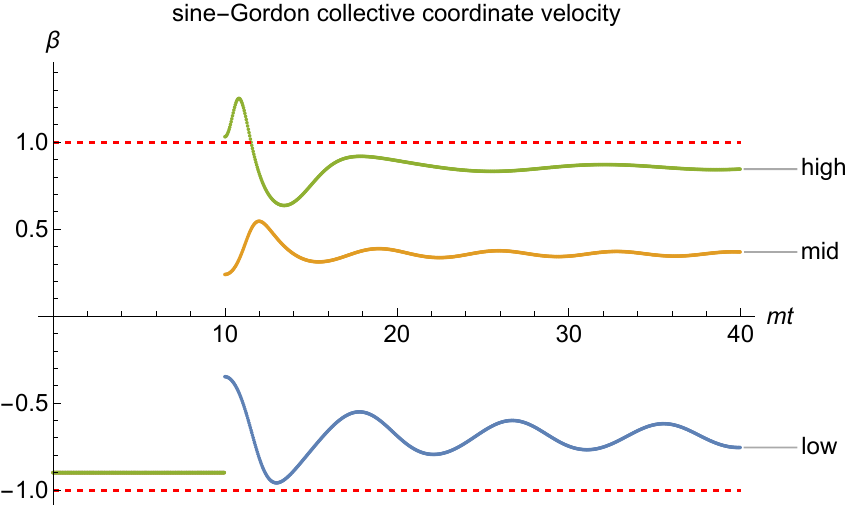}
  \caption{}
  \label{fig:ccvelocitiessG}
\end{subfigure} 
\caption{Collective coordinate velocities for the scenarios depicted in Figure \ref{fig:worldsheets}, with $\phi^4$-theory on the left and sine-Gordon on the right.  Low, mid, and high refer to the transfers $(P_f - P_i)/M_0 = 1.0$, $2.065$, and $3.5$ respectively, and dotted lines at $\beta = \pm 1$ are shown for reference.  These runs are stopped at $m t = 40$ before the left- and right-moving radiation wrap around to the other side of the circle and start interacting.  When this happens we can no longer assume the $mL \to \infty$ limiting behavior is faithfully represented.}
\label{fig:ccvelocities}
\end{figure}
%%%%%%%%%%%%%%%%%%

 In Figure \ref{fig:ccvelocities} we compare the collective coordinate velocities for the six scenarios of Figure \ref{fig:worldsheets}.  The discontinuities in the velocities at the kick, $\Delta \beta$, agree with the predictions from \eqref{betajump}, which are $\Delta \beta = 0.5678$, $1.172$, and $1.987$ for $\phi^4$-theory and $0.5517$, $1.139$, and $1.931$ for sine-Gordon.  As one can see, in the high momentum transfer case the collective coordinate velocities are superluminal after the impact.  We compare the collective coordinate velocity with other measures of ``kink velocity'' in Subsection \ref{ssec:Superluminal} below.  The punchline is that the collective coordinate velocity does not correspond to any physical energy transport velocity, so it can exceed the speed of light without any violation of special relativity.  When one looks at the energy flow velocity of the solution, it is always and everywhere subluminal as required by special relativity. 

  %%%%%%%%%%%%%%%%% 
 \begin{figure}[t!]
 \centering
\begin{subfigure}{.50\textwidth}
  \centering
  \includegraphics[width=\linewidth]{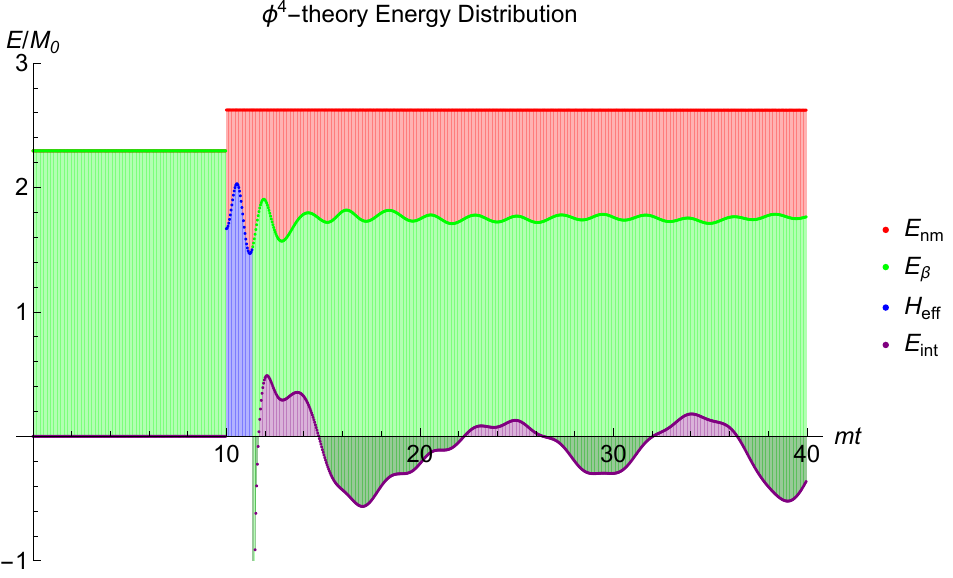}
  \caption{}
  \label{fig:EnergyDistributionphi4}
\end{subfigure}  \quad%
\begin{subfigure}{.46\textwidth}
  \centering
  \includegraphics[width=\linewidth]{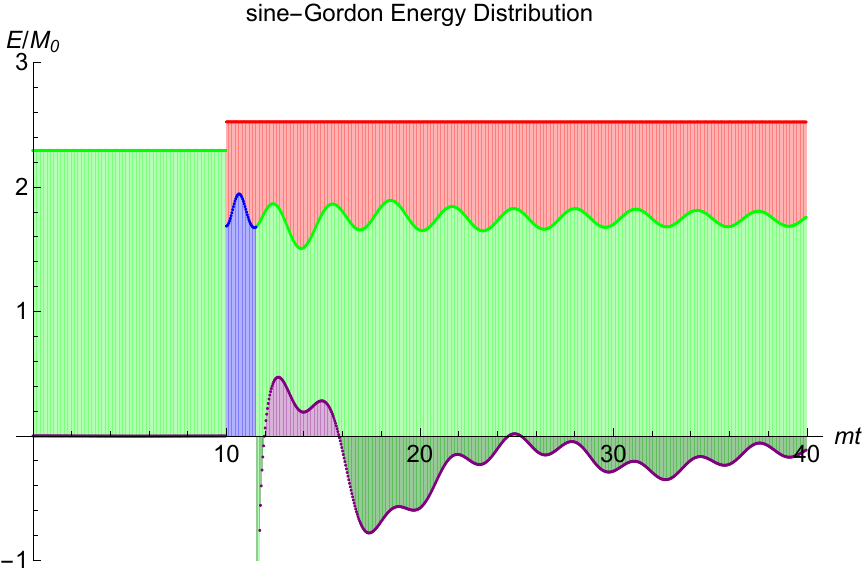}
  \caption{}
  \label{fig:EnergyDistributionsG}
\end{subfigure} 
\caption{Energy Distributions for the high impact scenarios depicted in Figures \ref{fig:worldsheetphi4high} and \ref{fig:worldsheetsGhigh}.  The total energy is composed of kink mechanical energy $E_{\beta}$ (green), interaction energy $E_{\rm int}$ between the kink and normal modes (purple), and energy in the normal modes $E_{\rm nm}$ (red).  Before the kick, all energy is in $E_{\beta}$.  For the brief period after the kick when the kink collective coordinate is superluminal, the decomposition of $H_{\rm eff}$ into mechanical and interaction energy is not possible, so we just indicate the value of $H_{\rm eff}$ directly (blue).}
\label{fig:EnergyDistributions}
\end{figure}
%%%%%%%%%%%%%%%%%%

In Figure \ref{fig:EnergyDistributions} we present the breakdown of total energy into normal mode energy, mechanical kink energy, and interaction energy as described in Subsection \ref{ssec:EnergyandPower} for the high momentum transfer scenarios from Figures \ref{fig:worldsheets} and \ref{fig:ccvelocities}.  The discontinuities in the total energies agree with the predictions from \eqref{Hbarjump2}, which are $\Delta H = 0.3276 M_0$ for the $\phi^4$ case and $\Delta H = 0.2293M_0$ for sine-Gordon.  When the kink collective coordinate is superluminal the decomposition of $H_{\rm eff}$ into kink mechanical energy and interaction energy is not possible.  Our physical interpretation of this is that the kink is not sufficiently particle-like at these times for this decomposition to make sense.

 %%%%%%%%%%%%%%%%% 
 \begin{figure}[t!]
 \centering
  \includegraphics[width=0.5\linewidth]{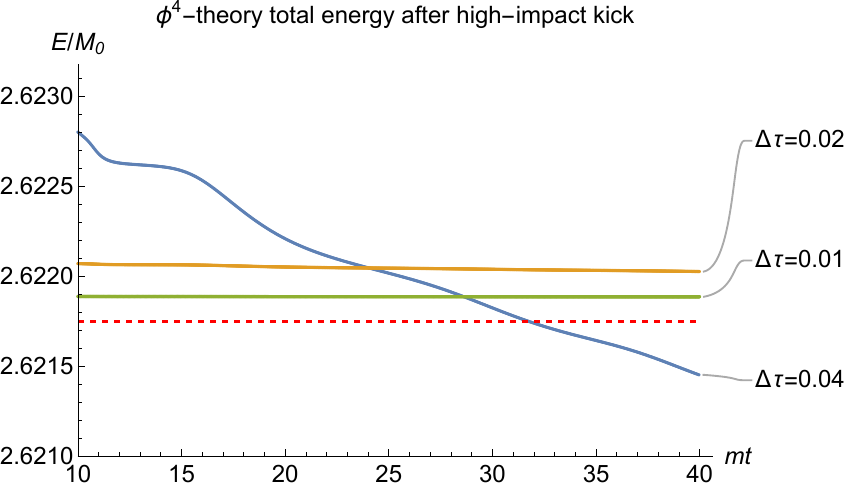}
 \caption{Total energy for the $\phi^4$-theory high impact scenario after the kick, computed with decreasing time step sizes.  Note the scale on the $y$-axis.  With smaller time steps the energy change becomes vanishingly small and the value approaches that predicted by the jumping condition from \eqref{Hbarjump2} (red dashed line).}
\label{fig:EnergyChange}
\end{figure}
%%%%%%%%%%%%%%%%%%

Aside from the instant of the impact, total energy must be conserved by an exact solution to the FSE.  The graphs in Figure \ref{fig:EnergyDistributions} appear to respect this well, however if we zoom in on the total energy there is a slight decrease over time.  As Figure \ref{fig:EnergyChange} shows, this energy change can be made arbitrarily small by decreasing the time step of the RK4 routine and hence can be associated with numerical error.  

A smaller $\Delta \tau$ means more steps if we hold the total time interval of the run fixed.  This increases run time proportionally.  Run time scales quadratically with the number of spatial lattice points since the SLAC derivative operator utilizes the entire lattice to determine the derivative at each lattice point.  Therefore in general we want to use as large of a spatial lattice size $\tilde{a} = \widetilde{L}/N_{\sigma}'$ as is reasonable.  Since the size of the kink is $O(1)$ independently of $\widetilde{L}$, we have to scale up $N_{\sigma}'$ proportionally to $\widetilde{L}$.  The choices for the above runs correspond to a lattice spacing of $\tilde{a} \approx 0.33$ which is sufficient to give a reasonably smooth profile when plot points are joined with straight lines.  We could use fewer lattice points if we joined plot points with the interpolating function as in Figure \ref{fig:Interpolation}, but it is more time consuming to generate the interpolating function.

Having fixed the lattice spacing, we then want to push the time step size as large as possible.  For the purposes of the numerics presented in this paper, we view the percentage energy change of the $\Delta \tau = 0.04$ step size in Figure \ref{fig:EnergyChange} as acceptable.  If we try to go much larger the RK4 routine will eventually diverge.  If one uses the same finite difference operator for both time and space derivatives, then one generally requires $\Delta \tau < \Delta \sigma$ for second order hyperbolic PDE's.  Since we treat spatial and temporal derivatives differently, our condition on the time step is more restrictive.  While we are not aware of theoretical results on this restriction for our particular implementation, experimentally we find that the RK4 routine diverges at $\Delta \tau \sim 0.15$ for our $\tilde{a} = 0.33$ lattice spacing, but the energy loss as we approach this time step is more significant.

The runs to generate each of the graphs shown in Figure \ref{fig:worldsheets} used $N_{\sigma}' = 240$ or $241$ spatial lattice points and 1000 time steps.  They each took about one minute on a 2022 Macbook Pro with a 2.6 GHz i7 Intel processor.  A smaller scale run, with $\widetilde{L} = 20$, $N_{\sigma}' = 60$, and 250 time steps for example, takes only six seconds.

An interesting physics question is how great of a momentum transfer can we deliver before the solution breaks down.  Based on the discussion under \eqref{canonicaltrans}, one might expect this occurs when enough energy is added to the system to create a kink-antikink pair in addition to the original kink.  For such configurations the change of coordinates $\phi \mapsto (X,\varphi)$ can break down because $X$ cannot be uniquely determined from $\phi$ or because the overlap $\langle \uppsi_0 | \varphi' \rangle$, which appears in a denominator in the momentum transformation, vanishes.

 %%%%%%%%%%%%%%%%% 
 \begin{figure}[h!]
 \centering
\begin{subfigure}{.40\textwidth}
  \centering
  \includegraphics[width=\linewidth]{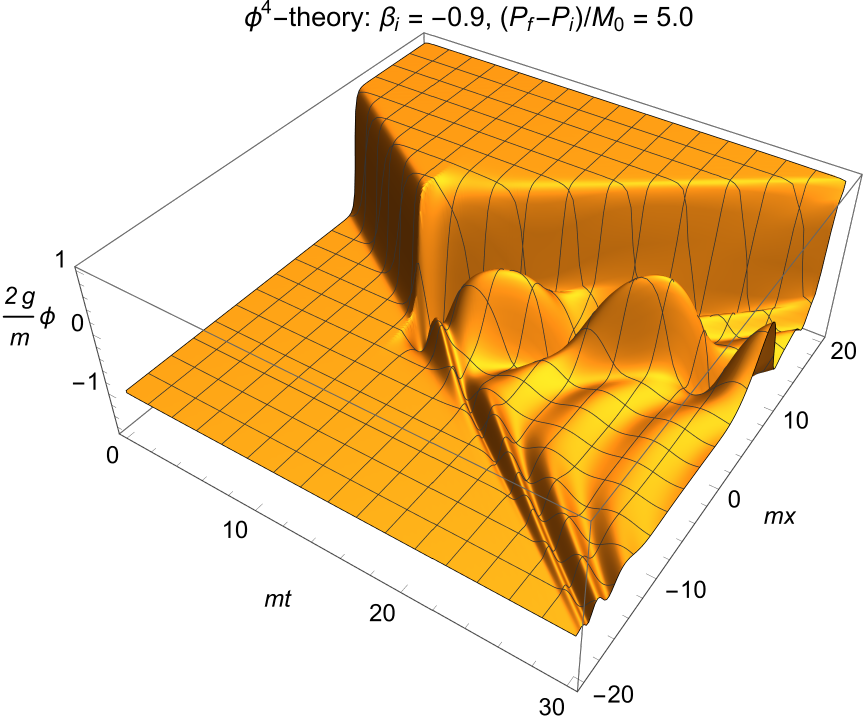}
  \caption{}
  \label{fig:worldsheetBigSmack1}
\end{subfigure}  \qquad \qquad%
\begin{subfigure}{.45\textwidth}
  \centering
  \includegraphics[width=\linewidth]{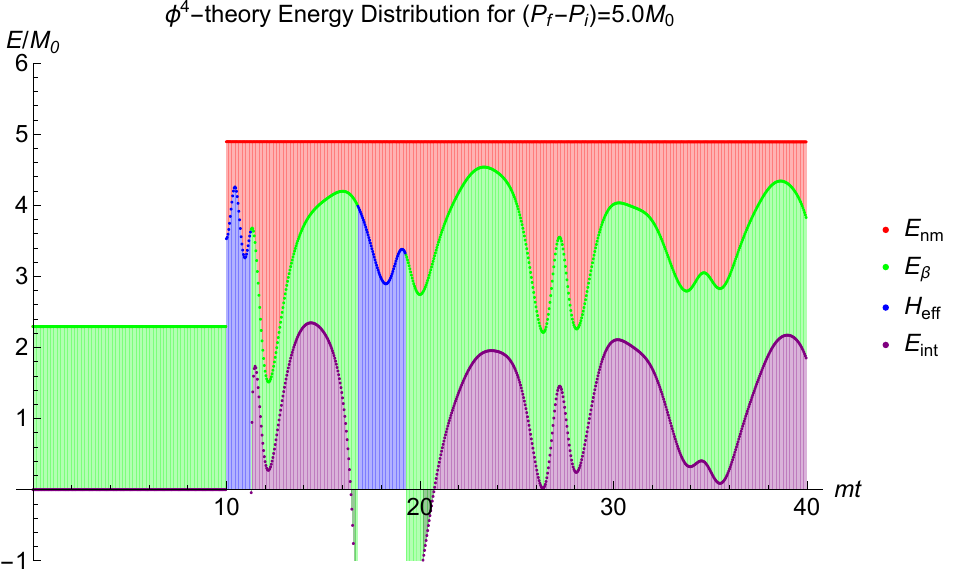}
  \caption{}
  \label{fig:EnergyDistributionBigSmack1}
\end{subfigure}  \\
\begin{subfigure}{.40\textwidth}
  \centering
  \includegraphics[width=\linewidth]{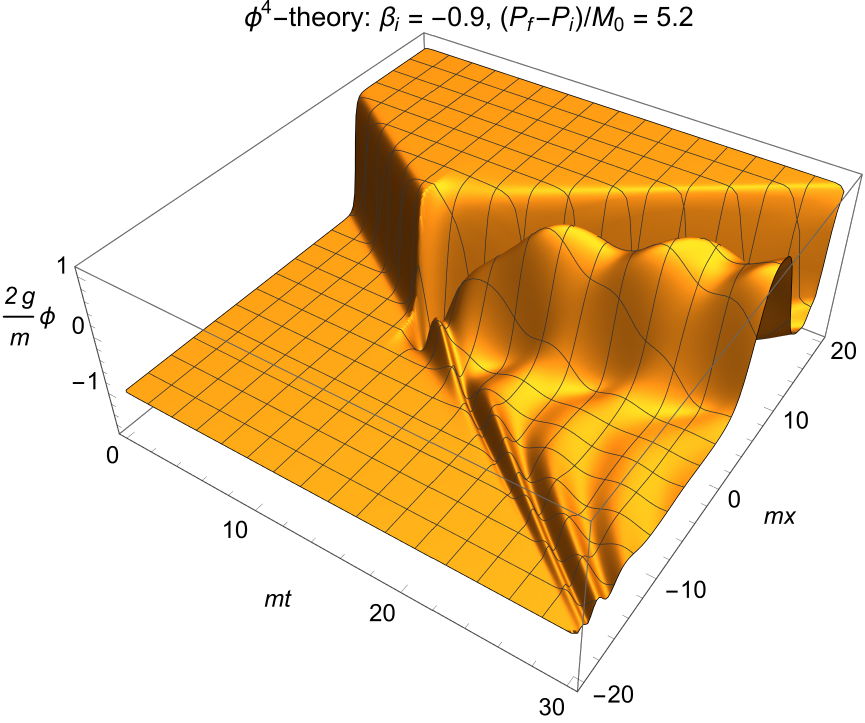}
  \caption{}
  \label{fig:worldsheetBigSmack2}
\end{subfigure}  \qquad \qquad %
\begin{subfigure}{.45\textwidth}
  \centering
  \includegraphics[width=\linewidth]{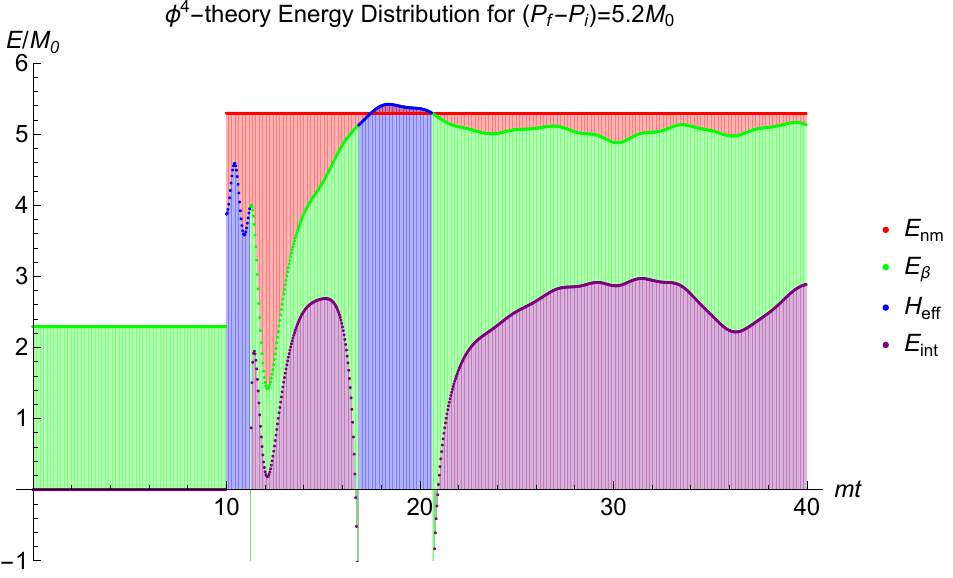}
  \caption{}
  \label{fig:EnergyDistributionBigSmack2}
\end{subfigure}  \\
\begin{subfigure}{.40\textwidth}
  \centering
  \includegraphics[width=\linewidth]{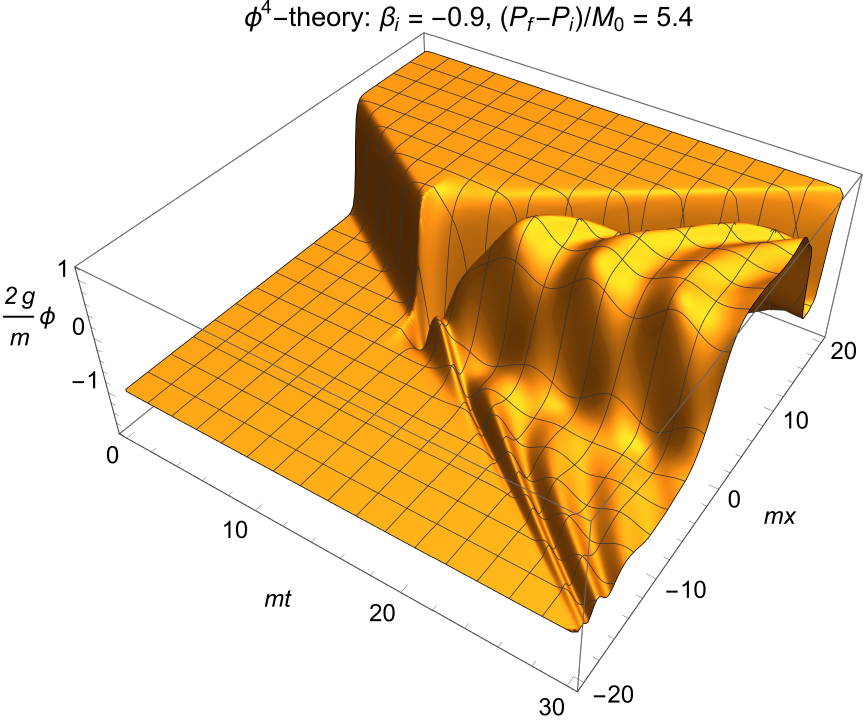}
  \caption{}
  \label{fig:worldsheetBigSmack3}
\end{subfigure}  \qquad \qquad %
\begin{subfigure}{.45\textwidth}
  \centering
  \includegraphics[width=\linewidth]{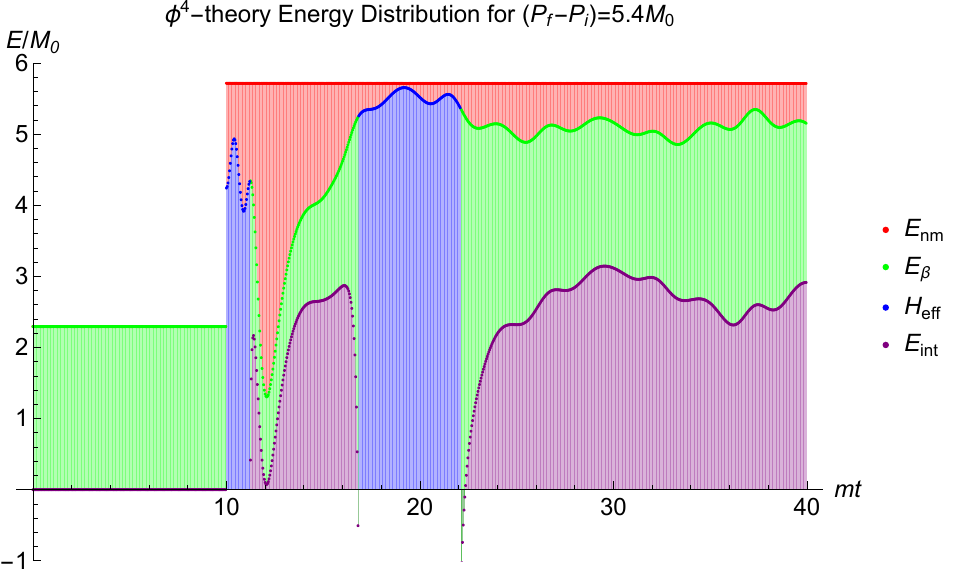}
  \caption{}
  \label{fig:EnergyDistributionBigSmack3}
\end{subfigure}
\caption{Spacetime evolution and energy distributions of ultra-high impulses directed oppositely to the incoming velocity of the kink.  We see the appearance of a kink-antikink pair when the interaction energy between the collective coordinate degree of freedom and normal modes (purple) exceeds roughly $2M_0$.  The numerical solution is able to track the kink-antikink pair as long as the antikink does not have sufficient speed to catch the original kink and annihilate it.  We find this happens at $P_f - P_i = 5.5M_0$ for the $\beta_i = -0.9$ kink, and the numerical solution breaks down.}
\label{fig:BigSmack}
\end{figure}
%%%%%%%%%%%%%%%%%%

 %%%%%%%%%%%%%%%%% 
 \begin{figure}[h!]
 \centering
\begin{subfigure}{.40\textwidth}
  \centering
  \includegraphics[width=\linewidth]{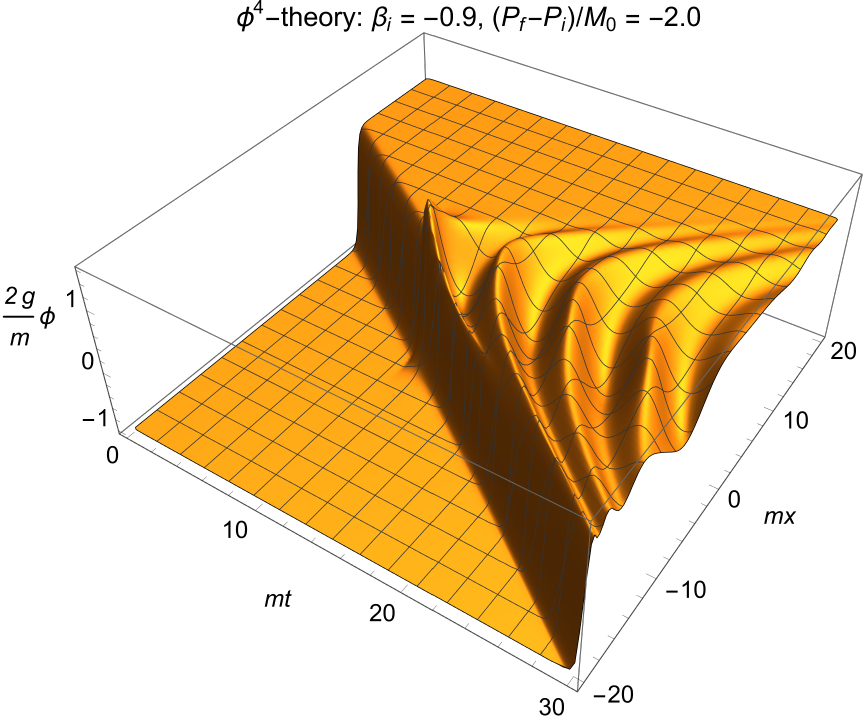}
  \caption{}
  \label{fig:worldsheetBigPush1}
\end{subfigure}  \qquad \qquad%
\begin{subfigure}{.45\textwidth}
  \centering
  \includegraphics[width=\linewidth]{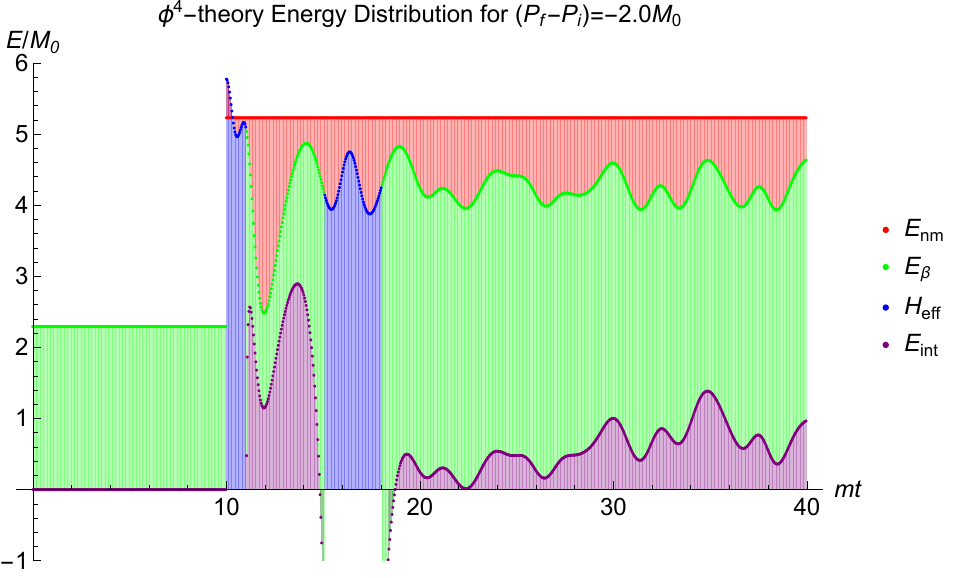}
  \caption{}
  \label{fig:EnergyDistributionBigPush1}
\end{subfigure}  \\
\begin{subfigure}{.40\textwidth}
  \centering
  \includegraphics[width=\linewidth]{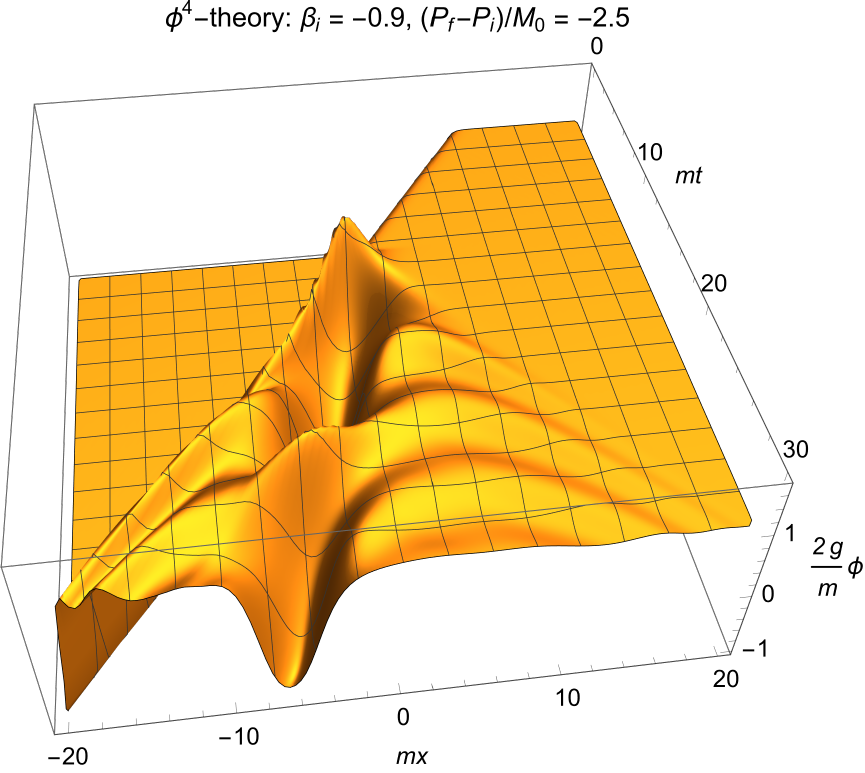}
  \caption{}
  \label{fig:worldsheetBigPush2}
\end{subfigure}  \qquad \qquad %
\begin{subfigure}{.45\textwidth}
  \centering
  \includegraphics[width=\linewidth]{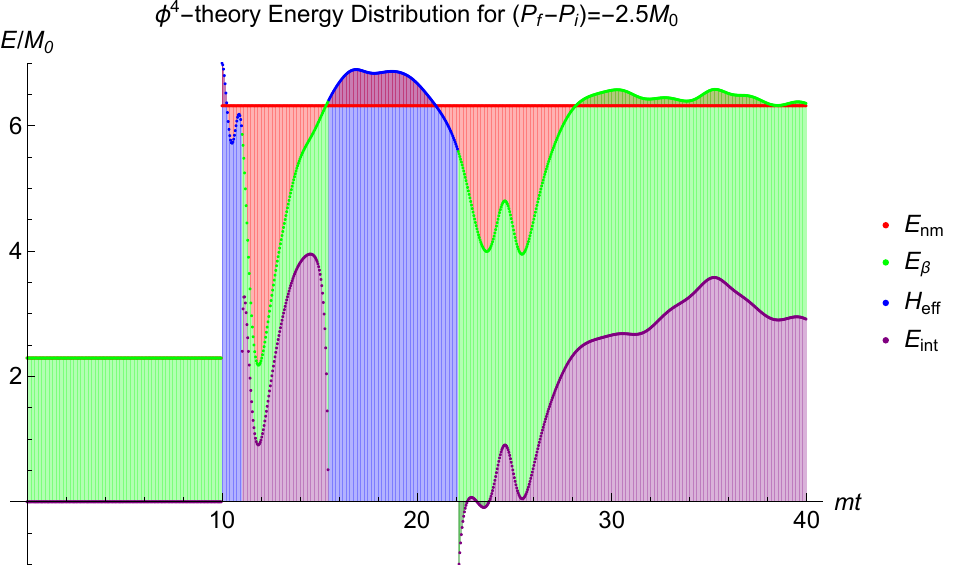}
  \caption{}
  \label{fig:EnergyDistributionBigPush2}
\end{subfigure}  \\
\begin{subfigure}{.40\textwidth}
  \centering
  \includegraphics[width=\linewidth]{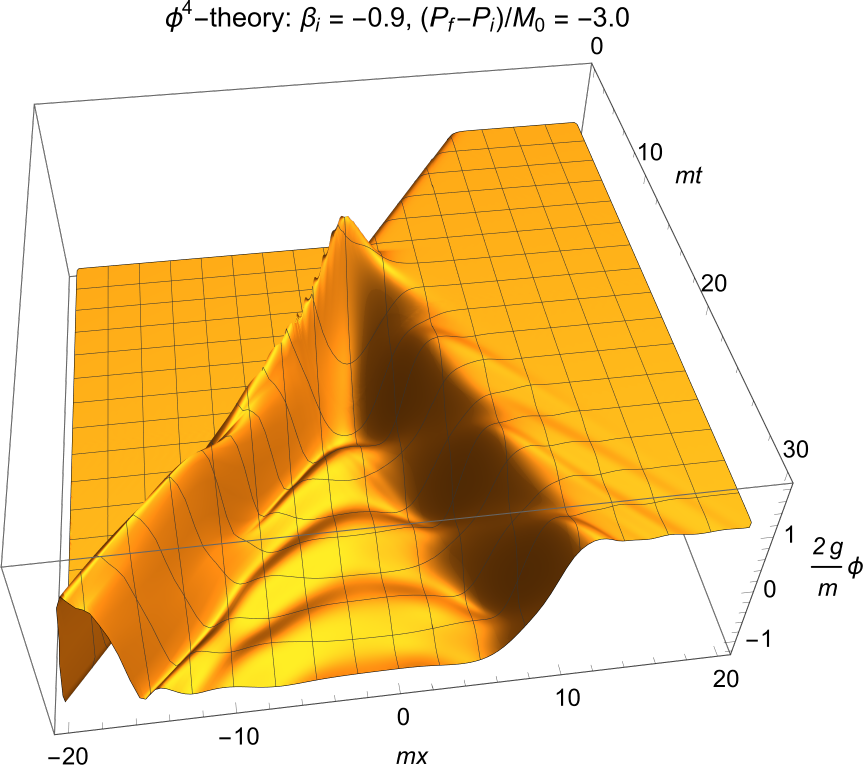}
  \caption{}
  \label{fig:worldsheetBigPusy3}
\end{subfigure}  \qquad \qquad %
\begin{subfigure}{.45\textwidth}
  \centering
  \includegraphics[width=\linewidth]{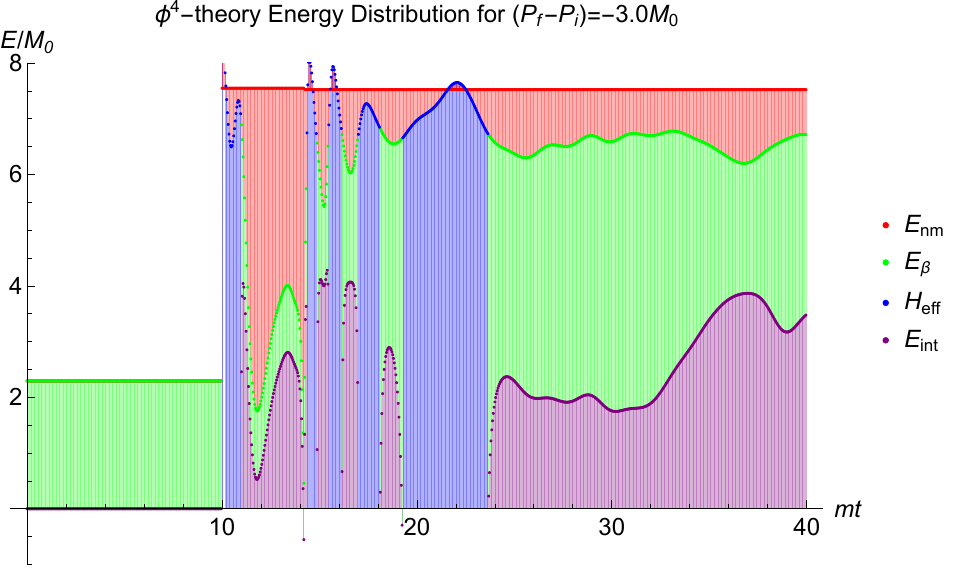}
  \caption{}
  \label{fig:EnergyDistributionBigPush3}
\end{subfigure}
\caption{Spacetime evolution and energy distributions of ultra-high impulses directed with the incoming velocity of the kink.  Again a kink-antikink pair appears when the interaction energy exceeds roughly $2M_0$.  The numerical solution is stable as long as the antikink does not have sufficient speed to catch the original kink and annihilate it.  The last solution displayed has $P_f - P_i = -3.0M_0$ and is just below the threshold for this to happen.}
\label{fig:BigPush}
\end{figure}
%%%%%%%%%%%%%%%%%%

Somewhat surprisingly, this expectation is not quite correct in two respects.  First, we find that it takes more energy than just twice the kink mass to create a kink-antikink pair.  This can be explained from the fact that a portion of the added energy goes into redirecting the original kink.  Second, even when we do create a kink-antikink pair, the numerical FSE solution can remain stable and track it, as long as the antikink from the newly created pair does not catch up to the original kink and annihilate it.  It appears that the annihilation of the original kink is what causes the solution to break down, and this will happen if there is enough energy imparted to the newly created kink-antikink pair.  

On the one hand this makes perfect sense.  When the original kink is annihilated, the field $\varphi$ has the form of small fluctuations around the vacuum at what used to be the position of the kink.  This is exactly what can lead to the vanishing of the overlap $\langle \uppsi_0 | \varphi' \rangle$.  On the other hand, it shows that entering regions of configuration space where \eqref{Xofphi} has multiple solutions is not necessarily a problem---\emph{i.e.}~it need not lead to a breakdown of the FSE solution.  The reason is that the ``correct'' solution to \eqref{Xofphi} in this region is determined, in the context of this paper, to be the one that is smoothly connected to a trajectory $X(t)$ that started in a region of configuration space where the solution to \eqref{Xofphi} is unique.

Figures \ref{fig:BigSmack} and \ref{fig:BigPush} illustrate this for $\beta_i = -0.9$ kink.  Figure \ref{fig:BigSmack} considers large momentum transfers in the opposite direction of the kink's initial velocity, while Figure \ref{fig:BigPush} considers large momentum transfers in the same direction of the kink's initial velocity.  In the first case, it takes a higher momentum transfer to create the kink-antikink pair and eventually break the solution because more of the impulse is used to redirect the original kink.  We find that the numerical solution breaks down at around $P_f - P_i \gtrsim 5.4 M_0$ in the first case and $P_f - P_i \lesssim -3.0M_0$ in the second case.

These figures also suggest that it is primarily the interaction energy between the collective coordinate degree of freedom and the normal mode degrees of freedom, $E_{\rm int}$ defined in \eqref{Eintdef}, that determines when a kink-antikink pair will form.  Heuristically this makes sense.  The impulse is delivered initially to the collective coordinate, and some of the added energy must be transferred to the normal mode degrees of freedom to create the kink-antikink pair, which can be thought of as a coherent state of normal modes.  The interaction energy measures this exchange and needs to exceed roughly twice the soliton mass in order for the pair to form.

We also observe in some of these scenarios that $H_{\rm eff}$ can exceed the total energy $H$ of the system.  Recall that $H_{\rm eff}$ is the total energy associated with the kink.  It is decomposed into kink mechanical energy (green) and kink-normal-mode interaction energy (purple) when this decomposition is possible---\emph{i.e.}~when the kink collective coordinate is subluminal.  When the decomposition is not possible, $H_{\rm eff}$ is still well-defined and color coded in blue.  Having $H_{\rm eff} > H$ means that the difference, which is the energy in the normal modes, $E_{\rm nm}$ (red), is negative.  This is possible since $E_{\rm nm}$ includes both kinetic normal mode energy and interaction energy among the normal modes.  Since the kink-antikink pair is a coherent bound configuration of normal modes, it makes sense that this interaction energy contributes negatively.

We have chosen to focus on the $\phi^4$ kink for the discussion of ulltra-high impulses.  The behavior of the sine-Gordon kink is qualitatively similar.

%%%%%%%%%%%%%%%%%%%%
 \subsection{Superluminal Collective Coordinate: Don't Worry, it's Okay}\label{ssec:Superluminal}
 %%%%%%%%%%%%%%%%%%%%
 
Let us return to what is perhaps the most startling feature of the solutions we exhibited in the previous subsection: the possibility of faster than light motion for the collective coordinate.  The common phrase ``Nothing travels faster than the speed of light,'' is inaccurate and not implied by the theory of special relativity.  What is true is that energy cannot be transported faster than the speed of light. 

The classic example is a point particle accelerated by a constant force.  In Newtonian mechanics its velocity will increase linearly forever, whereas in special relativity the velocity never exceeds c, the speed of light.  In this example, the location of the point particle is the location of energy, since mass is a form of energy in special relativity. Therefore, the speed limit on energy transport does imply a speed limit on the particle’s motion.

More interesting examples occur in electricity and magnetism.  Consider an X-wave formed by two plane wave crossing at an angle. Each plane wave travels at speed $c$ in a direction perpendicular to the wave front, but the location where they cross, which is the location where the energy density is maximum, travels faster than the speed of light. There is no contradiction with special relativity because, in field theories describing continuous systems, the quantity that is constrained to travel less than the speed of light is the energy flow velocity.  The energy flow velocity is defined in terms of components of the energy-momentum tensor as
\begin{equation}\label{energyflowvel}
v_i(t,\vec{x}) = \frac{T_{0i}(t,\vec{x})}{T_{00}(t,\vec{x})} ~.
\end{equation}
One can prove that the energy flow velocity is always less than $c$, even if the velocity of the maximum energy density is greater than $c$.  

Theories with solitons are an interesting case since the soliton possesses both particle-like and field-like properties.  The results of the previous subsection show that the collective coordinate of the soliton can exceed the speed of light---at least if an appropriate force is applied or, equivalently, appropriate initial conditions are considered.  One might argue that the required initial conditions are unphysical, but this is not the case.  It is possible to set up initial conditions at $t = 0$ with a subluminal collective coordinate which then becomes superluminal through time development.  See Figure \ref{fig:kinkvelocitiesA} below.

%%%%%%%%%%%%%%%%%%
  \begin{figure}[t!]
 \centering
\begin{subfigure}{.75\textwidth}
  \centering
  \includegraphics[width=\linewidth]{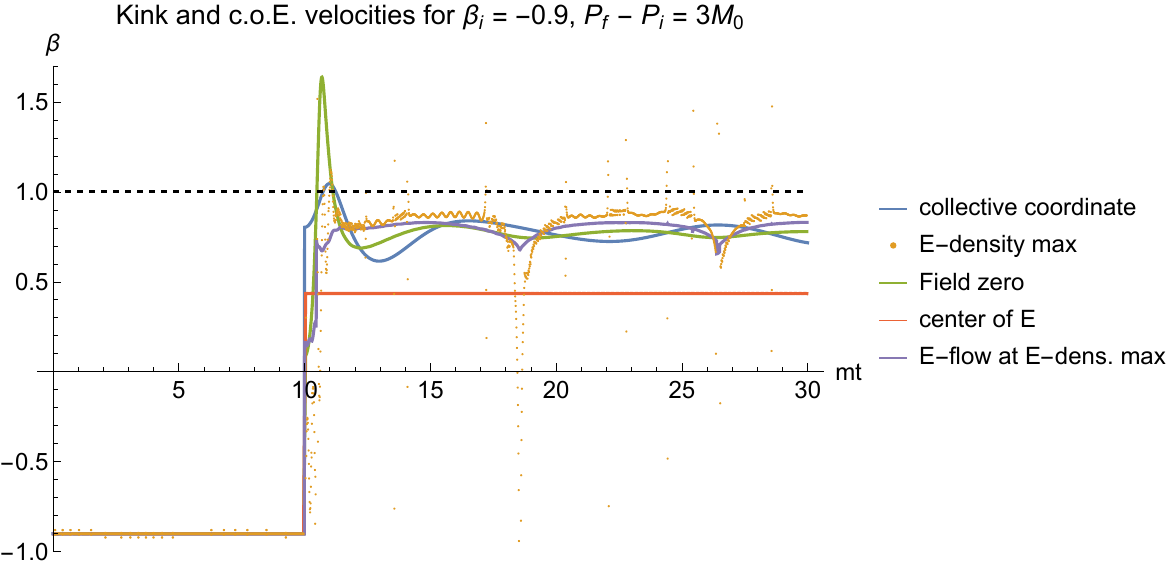}
  \caption{}
  \label{fig:kinkvelocities1}
\end{subfigure} \\
\begin{subfigure}{.5\textwidth}
  \centering
  \includegraphics[width=\linewidth]{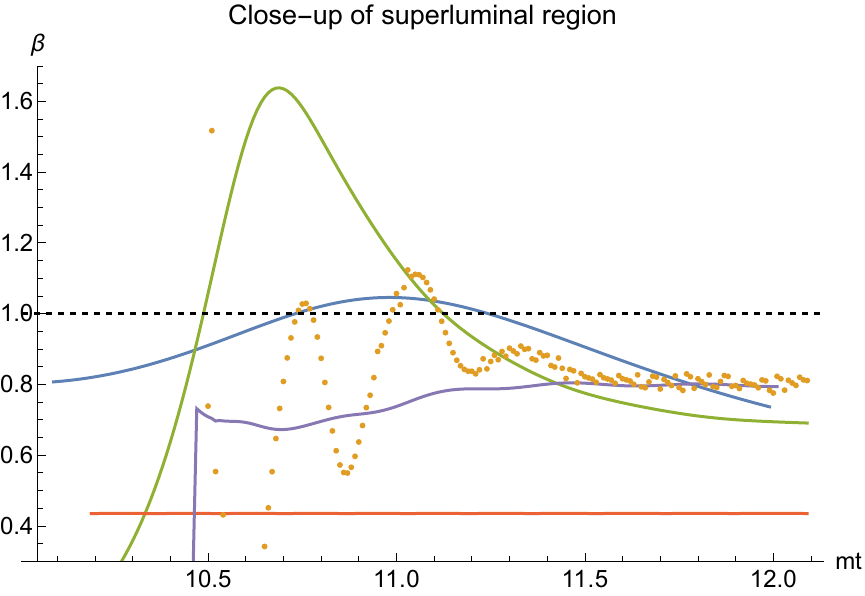}
  \caption{}
  \label{fig:kinkvelocities2}
\end{subfigure}  \\
\begin{subfigure}{.75\textwidth}
  \centering
  \includegraphics[width=\linewidth]{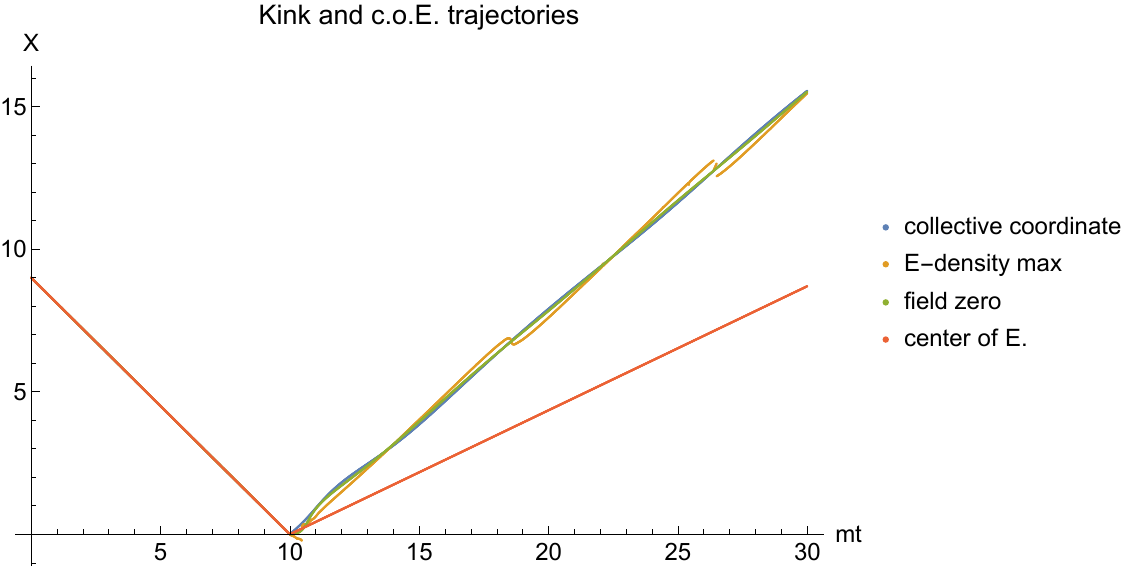}
  \caption{}
  \label{fig:kinktrajectories1}
\end{subfigure}%
\caption{Velocities (top, middle) and trajectories (bottom) for several measures of the kink position and center-of-energy position in a high impact scenario with $\beta_i = -0.9$ and $P_f - P_i = 3.0 M_0$.  Notice that the collective coordinate, maximum energy density, and field zero velocities are all less than one immediately after the kick and then increase above one.}
\label{fig:kinkvelocitiesA}
\end{figure}
%%%%%%%%%%%%%%%%%%

%%%%%%%%%%%%%%%%%%
 \begin{figure}[t!]
\centering
 \includegraphics[width=0.75\linewidth]{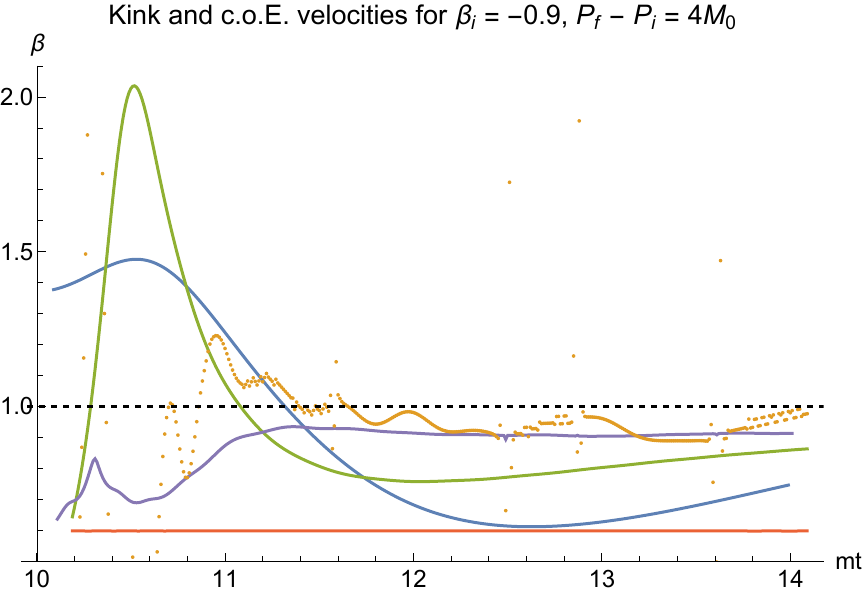}
  \caption{Kink and center-of-energy velocities at an even higher momentum transfer.  The color coding is the same as in Figure \ref{fig:kinkvelocitiesA}.}
\label{fig:kinkvelocitiesB}
\end{figure}
%%%%%%%%%%%%%%%%%%

It is well known from other contexts such as monopole scattering that the collective coordinate of a soliton, while being a perfectly good coordinate on the region of configuration space in question, need not represent the position of anything physical.  Therefore the question of whether a superluminal collective coordinate implies a superluminal soliton is one of semantics.  The more interesting questions are 1) how does the energy flow velocity behave when the collective coordinate is superluminal, 2) is there any better notion of ``the soliton's position'' when the collective coordinate is superluminal, and 3) can we understand the physical mechanism for a superluminal collective coordinate in more detail?

Again, regarding question 1), it is easy to prove that $v(t,x)$ as defined in \eqref{energyflowvel} is less than one in magnitude for the class of models described by the action \eqref{Scl}.  It boils down to the positivity of $(\pd_t \phi - \pd_x \phi)^2$ and $V(m^2;\phi)$.  Therefore the time development of the fields after the impact must have $|v(t,x)| < 1$ even if $|\beta| > 1$.  We verified this using our numerical solutions to the FSE in a couple high momentum transfer scenarios in $\phi^4$ theory.  The results for the collective coordinate velocity, the energy flow velocity evaluated at the maximum energy density, as well as several other measures of velocity, are illustrated in Figures \ref{fig:kinkvelocitiesA} and \ref{fig:kinkvelocitiesB}.

Specifically in these figures we plot the velocities and positions of the collective coordinate (blue), the maximum of the energy density (gold), the location where the field passes through zero (green), and the center of energy (red).  The trajectories of the first three, which can all be considered as measures of the kink's position, track closely with each other and are all superluminal for a time period after the kick.  The small jumps in the maximum energy density position are due to wave crests passing over the kink.  This, along with our numerical computation of the time derivative of the position of the maximum energy density, cause a fair amount of noise in the velocity graph for the maximum energy density.  Meanwhile, the center of energy position is a conserved quantity and is a relativistic analog of the center of mass position for the entire system.  It travels at a constant (subluminal) velocity when the force is not acting.  Its velocity is significantly less than the kink velocities because it accounts for the energy carried in the normal modes in addition to the energy carried by the collective coordinate.  

Finally, in the velocity graphs we also plot the energy flow velocity (purple) of equation \eqref{energyflowvel}, evaluated at the maximum of the energy density.  This shows the local energy flow rate in the vicinity of the kink, and we can see that it is always subluminal.

These graphs provide our answer to question 2): No, there is no good (\emph{a.k.a.}~subluminal) measure of the kink's position when its collective coordinate is superluminal.  We suggest there is a complementarity principle at play.  The soliton can be described as both a particle and field, with one description or the other more suitable in a given context.  In the context at hand, when the collective coordinate velocity is superluminal, the particle description is wholly inadequate.  We simply have an evolving field configuration.  We know a soliton is present by the topological charge, but there is no sense in which we can ascribe a meaningful position to it.  This is much like the situation in the vicinity of a head-on collision between two 't Hooft--Polyakov monopoles, though we might argue it is more severe.  Not only does the collective coordinate lack a good physical interpretation, it is particularly unphysical due to its superluminal speed.

\begin{figure}[t!]
 \centering
 \includegraphics[width=0.9\linewidth]{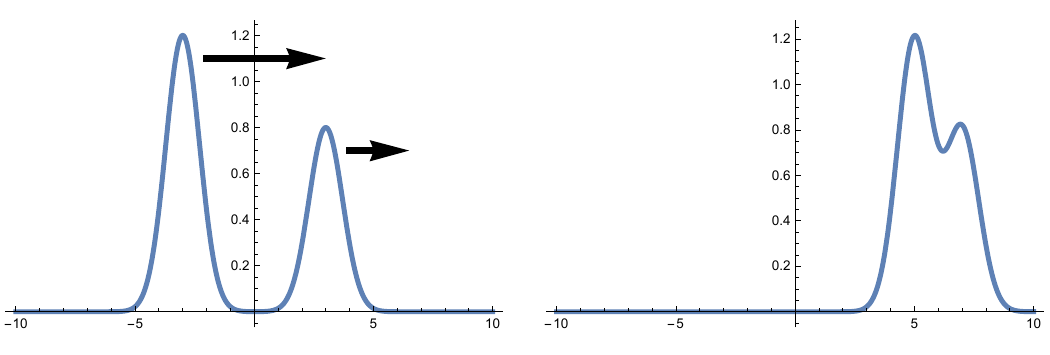}
\caption{A larger lump (representing the soliton) overtaking a smaller one (representing a wave crest).}
\label{fig:lumps}
\end{figure}

This leaves us with question 3).  Can we gain a better physical understanding of the mechanisms at play that can lead to a $|\beta| > 1$?  Since the collective coordinate tracks closely with the maximum of the energy density, our point of view is that the superluminal kink velocity is  essentially an overlapping wave phenomenon, and we will try to understand it from that perspective.  The basic idea is the following.  Consider the two lumps in Figure \ref{fig:lumps} with the bigger one moving faster.  As the larger lump passes over the smaller lump, the velocity of the maximum of the combination initially exceeds the velocity of the larger lump and then drops down below it, similarly to what we see in the collective coordinate graphs.  Here we are thinking of the larger lump as representing the kink and the smaller lump as representing the initial ripple created by the force.  (Although the shape of the kink does not appear to be well-represented by a lump, the shape of the energy density in the kink is.)  If the speed of the larger lump is near one but below it, we can easily arrange for the the speed of the maximum to exceed one during the time of overlap.

%%%%%%%%%%%%%%%%% 
%%%%%%%%%%%%%%%%%%

The above cartoon is simple and gets the main idea across, but it is rather ad hoc.  To construct a more detailed toy model, we need a single theory which admits lumps as wave packets.  These wave packets must be able to travel at different speeds, which means that the model must have a nontrivial dispersion relation.  We also require that the model is linear so that wave packets can be superimposed to form a new solution.  A natural candidate that meets all these requirements is Klein-Gordon theory.  There are no solitons in Klein-Gordon theory, but we can use a wave packet as a substitute.

Therefore our toy model involves the following steps:  
\begin{enumerate}
\item Start with the following configuration for the Klein-Gordon field $\phi(t,x)$ (with Klein-Gordon mass parameter $m$):
\begin{align}\label{KGconfig}
\phi(t,x) =&~ \phi_{\rm wp}(t,x) + \phi_{\rm par}(t,x)~,  \cr
\phi_{\rm wp}(t,x) =&~ \cos(k_0 x - \omega_0 t) \exp \left[ - \frac{\sigma_{k}^2}{2} \left(x - \tfrac{k_0}{\omega_0} t \right)^2 \right] ~, \cr
\phi_{\rm p}(t,x) =&~ F_0 \Theta(t) \sin(m t) x e^{- \sigma_{0}^2 x^2/2} ~. 
\end{align}
The wave packet $\phi_{\rm wp}(t,x)$ moves at group velocity
\begin{equation}
\beta_{\rm g} = \frac{k_0}{\omega_0} = \frac{k_0}{\sqrt{k_{0}^2 + m^2}}~,,
\end{equation}
and represents the soliton.  This is an approximate solution to the source-free Klein-Gordon equation that ignores wave-packet spreading.  The approximation is good when $k_0 \gg m$ so that $\beta_g$ is relativistic, the width $\sigma_k \sim m$, and for times $t$ that are not too large: $m t \ll 1/\beta_g$.  

The particular configuration $\phi_{\rm p}(t,x)$ comes about as follows.  We start with a static version of the wave-packet configuration given by $\phi_0 = \cos(mt) e^{-\sigma_{0}^2 x^2/2}$.  This is an approximate source-free solution if $\sigma_0 \ll m$.  We use this to build a forcing term $F(t,x) \propto \delta(t) \pd_x \phi_0$ in analogy with the forcing term $\delta(t) \phi_{0}'$ in the FSE \eqref{FSE}.  Finally, we construct a particular solution by integrating $F(t,x)$ against the retarded Greens function for the Klein-Gordon equation.  $\phi_{\rm p}(t,x)$ is an approximation to this particular solution that ignores spreading of the wave packet and is good as long as $\sigma_0 \ll m$.
\item Evaluate the Klein-Gordon energy density on the configuration \eqref{KGconfig}:
\begin{equation}
\HH(t,x) = \half (\pd_t \phi)^2 + \half (\pd_x \phi)^2 + \frac{m^2}{2} \phi^2 ~.
\end{equation}
\item Time-average the energy density in the co-moving frame of the wave packet.  The wave packet is supposed to represent the soliton, but its energy density oscillates due to the phase factor $\cos(k_0 x - \omega_0 t)$.  To eliminate this oscillation we time average the energy density over one cycle in the co-moving frame of the wave packet.  The co-moving frame is parameterized by coordinates $(t,\rho)$ with $\rho = x - \beta_{\rm g} t$.  This is analogous to $\rho = x - X(t)$ for the soliton, where we've taken the ``collective coordinate'' trajectory to be $X(t) = \beta_{\rm g} t$.  The time-averaging is then defined by integrating over $t$ for fixed $\rho$:
\begin{equation}
\langle \HH \rangle(t,\rho) := \frac{1}{T} \int_{t}^{t + T} \ed t' \HH(t',\rho)~,
\end{equation}
with period $T = 2\pi \gamma_{\rm g}/m$.  This is the period of $\cos(k_0 x - \omega_0 t) = m(\gamma_{\rm g} \beta_{\rm g} \rho - \gamma_{\rm g}^{-1} t)$ at fixed $\rho$.  While the averaging makes the $F_0$-independent part of $\langle \HH \rangle$ time independent, for $F_0$-dependent part of $\langle \HH \rangle$ still depends on $t$ because this part of the energy density is not $T$-periodic.
\item Find the maximum of $\langle \HH \rangle(t,\rho)$ and denote it $\rho = \rho_{\rm max}(t)$.
\item Evaluate the velocity of the maximum of the time-averaged energy density and transform back to the lab frame: $\beta_{\textrm{E-max}} := \beta_{\rm g} + \dot{\rho}_{\rm max}$.  
\end{enumerate}
We carry out these last three steps numerically since the integrals involved in the time averaging procedure are not simple.

%%%%%%%%%%%%%%%%% 
 \begin{figure}[t!]
 \centering
\begin{subfigure}{.45\textwidth}
  \centering
  \includegraphics[width=\linewidth]{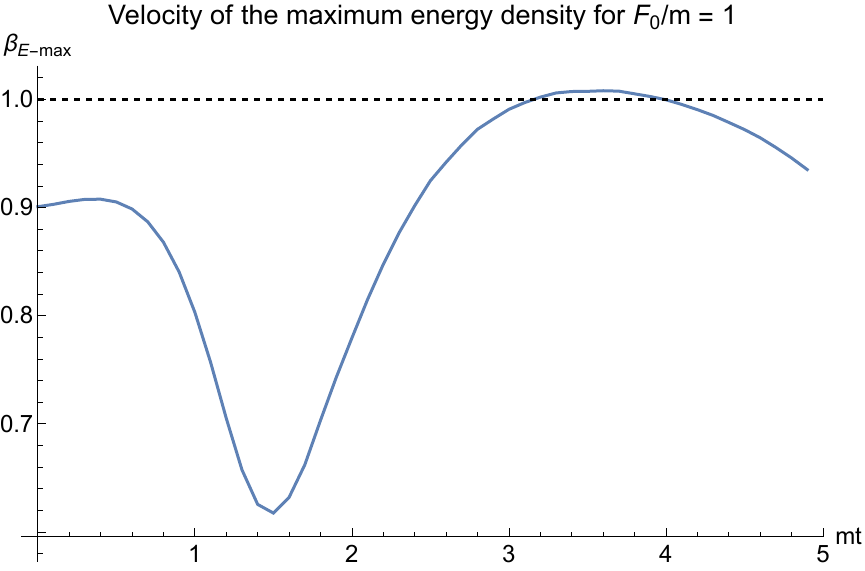}
  \caption{}
  \label{fig:KGanalogy1}
\end{subfigure}  \qquad %
\begin{subfigure}{.45\textwidth}
  \centering
  \includegraphics[width=\linewidth]{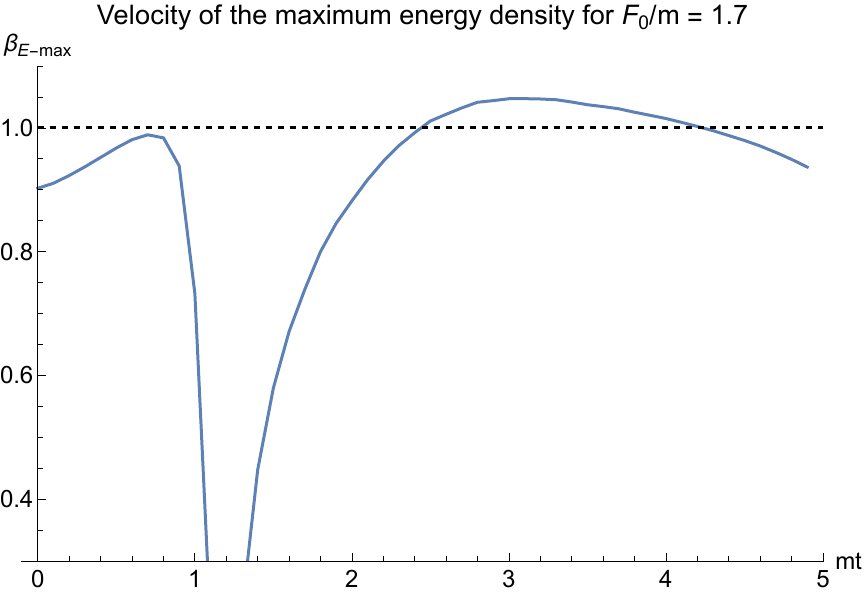}
    \caption{}
  \label{fig:KGanalogy2}
\end{subfigure}  \\[1ex]
\begin{subfigure}{.45\textwidth}
  \centering
  \includegraphics[width=\linewidth]{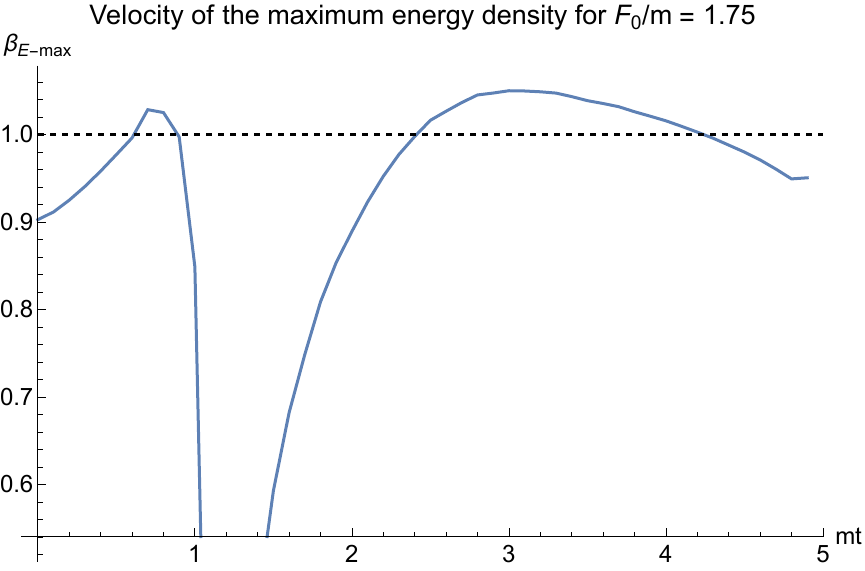}
  \caption{}
  \label{fig:KGanalogy3}
\end{subfigure}  \qquad %
\begin{subfigure}{.45\textwidth}
  \centering
  \includegraphics[width=\linewidth]{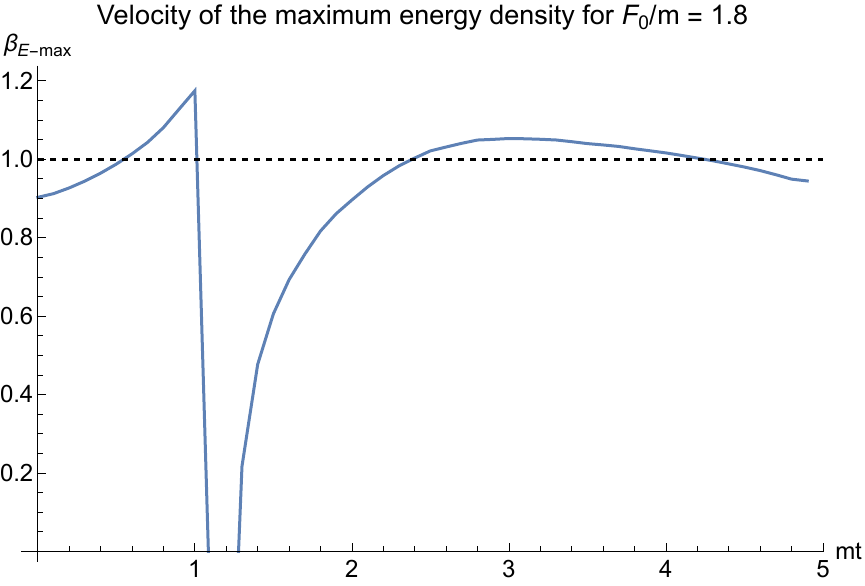}
  \caption{}
  \label{fig:KGanalogy4}
\end{subfigure}  
\caption{Graphs of the velocity of the maximum of the time-averaged energy density in our toy model for several different values of forcing coefficient.  We take the group velocity of the wave packet representing the ``soliton'' to be $\beta_{\rm g} = 0.9$, and the widths that appear in the configuration \eqref{KGconfig} to be $\sigma_k = m$ and $\sigma_0 = 0.3m$.}
\label{fig:KGanalogy}
\end{figure}
%%%%%%%%%%%%%%%%%%

In Figure \ref{fig:KGanalogy} we present the results for $\beta_{\textrm{E-max}}$ for several increasing choices of forcing coefficient $F_0$.  The group velocity of our wave-packet soliton is taken to be $\beta_{\rm g} = 0.9$, and we choose the widths $\sigma_k = m$, $\sigma_0 = 0.3m$.  When the forcing coefficient satisfies $F_0 \gtrsim 1.75m$, we see a discontinuity in $\beta_{\textrm{E-max}}$.  It is due to the same phenomonenon we observed in the graphs of the maximum energy density position for the FSE solution.  When a wave crest from the particular solution passes over the peak of the wave packet representing the soliton there can be a jump in position of the maximum energy density.  

The main point of Figure \ref{fig:KGanalogy} is that $\beta_{\textrm{E-max}}$ exhibits a qualitatively similar behavior to the kink velocities in Figures \ref{fig:kinkvelocitiesA} and \ref{fig:kinkvelocitiesB}.  Hence we conclude that the superluminal collective coordinate is an overlapping wave phenomenon and nothing to worry about.  It is an interesting feature of our solutions that clearly illustrates the possibility of non-particle-like behavior for a soliton.

%%%%%%%%%%%%%%%%%%%%
\subsection{Comparison with Perturbation Theory for Small $\Delta P/M_0$}\label{ssec:pertcheck}
%%%%%%%%%%%%%%%%%%%% 
  
In this final subsection we switch gears and compare our numerical results with perturbation theory in the case of an instantaneous impulse with small momentum transfer.  Thus we take our momentum profile to be
\begin{equation}\label{smallepsP}
P(t) = P_i + \epsilon M_0 \Theta(t-t_\ast) \equiv P_i + P^{(1)}~,
\end{equation}
where $\epsilon = (P_f - P_i)/M_0$ is assumed small.  Recalling the discussion in Subsection \ref{ssec:pert}, the initial value problem for the FSE admits a perturbative expansion in $\epsilon$ of the form
\begin{equation}\label{modifiedfieldexp}
|\zeta \rangle =  |\zeta_{\beta_i} \rangle + \sum_{a = 1}^{\infty} 2 \Re \left\{ (A_{a}^{(1)} + A_{a}^{(2)} + \cdots) e^{-\ii (\nu_a + \nu_{a}^{(1)} + \cdots)(t-t_\ast)} | \eta_a \rangle \right\}~,
\end{equation}
where $|\eta_a \rangle$ and $\nu_a > 0$ are a complete set of phase space normal modes and frequencies for the linearized problem around the boosted kink, and $A_{a}^{(n)}(t),\nu_{a}^{(n)}$ are the $O(\epsilon^n)$ corrections to the $a^{\rm th}$ normal mode amplitude and frequency.  We obtained the equations \eqref{A1eom} and \eqref{A2eom} that determine $A_{a}^{(1,2)}$ and $\nu_{a}^{(1)}$ in terms of certain source terms $\SS^{(1,2)}$ constructed from coefficients of the Hamiltonian and lower order solutions---see \eqref{secondordersource}.

In Appendix \ref{app:pert} we evaluate the expressions \eqref{secondordersource} for the case \eqref{smallepsP} and obtain the following results for $A_{a}^{(1,2)}$ and $\nu_{a}^{(1)}$.  First, $A_{a}^{(1)}$ is given by
\begin{align}\label{A1sol}
A_{a}^{(1)} =&~ - \frac{1}{\nu_a} \left(e^{\ii (\nu_a + \nu^{(1)} +  \cdots)(t-t_\ast)} - 1 \right) s_{a}^{(1)} ~, \qquad \textrm{with} \cr
s_{a}^{(1)} =&~ \frac{M_0 \epsilon}{\langle \uppsi_0 | \phi_{\beta_i}' \rangle^2} \left\{  \langle \eta_{\varpi a} | \phi_{\beta_i}' \rangle +  \beta_i \langle \eta_{\varphi a} | \phi_{\beta_i}'' + \langle \uppsi_0 | \phi_{\beta_i}' \rangle \uppsi_0' \rangle  \right\}~. 
\end{align}
Note if we want to consider times $t - t_\ast = O(\epsilon^{-1})$, then cannot drop the $\nu^{(1)}$ term in the phase of this expression.  Likewise, to identify the $O(\epsilon^n)$ contribution to the fields in \eqref{modifiedfieldexp}, we take the $A_{a}^{(n)}$ term from the amplitude expansion and keep the whole series of corrections in the phase.  Hence the first-order solution is
\begin{equation}\label{1storderfieldsol}
|\zeta^{(1)} \rangle = \sum_{a = 1}^\infty 2 \Re \left\{ \frac{1}{\nu_a} \left( e^{- \ii (\nu_a + \nu_{a}^{(1)} + \cdots)(t-t_\ast)} - 1 \right) s_{a}^{(1)} |\eta_a \rangle \right\} ~.
\end{equation}

A nice quantity to use for comparing perturbative and numerical FSE solutions is the collective coordinate velocity $\beta$.  On the perturbative side, we expanded $\beta$ to second order in terms of field fluctuations in \eqref{pertccvel}.  In this expression $( |\varpi^{(n)} \rangle, |\varphi^{(n)} \rangle )^T$ should be identified with the term in \eqref{modifiedfieldexp} proportional to $A^{(n)}$ but that includes all frequency corrections in the phase.  Hence, using \eqref{A1sol}, we obtain the following result for the first order correction to the collective coordinate velocity:
\begin{align}
\beta^{(1)} =&~ \frac{M_0 \epsilon}{\langle \uppsi_0 | \varphi_{\beta}' \rangle^2} + \sum_{a = 1}^{\infty} 2 \Re \left\{\frac{1}{\nu_a} \left( e^{- \ii (\nu_a + \nu_{a}^{(1)} + \cdots)(t-t_\ast)} - 1 \right) b_{a}^\ast s_{a}^{(1)} \right\} ~, \qquad \textrm{with} \quad \cr
b_a =&~ \frac{1}{\langle \uppsi_0 | \phi_{\beta_i}' \rangle^2} \left\{ \langle \eta_{\varpi a} | \phi_{\beta_i}' \rangle + \beta_i \left( \langle \eta_{\varphi a} |  \phi_{\beta_i}'' \rangle + \langle \uppsi_0 | \phi_{\beta_i}' \rangle \langle \eta_{\varphi a} |  \uppsi_0' \rangle \right) \right\} ~.
\end{align}
Notice that $s_{a}^{(1)}$ involves the same combination of terms as $b_a$: we have $s_{a}^{(1)} = M_0 \epsilon b_a$.  It  will be convenient to define the phase space ket
\begin{equation}\label{Bket}
| B_i \rangle = \frac{1}{\langle \uppsi_0 |\phi_{\beta_i}' \rangle^2} \left( \begin{array}{c} |\phi_{\beta_i}'\rangle \\[1ex] \beta_i ( |\phi_{\beta_i}'' \rangle + \langle \uppsi_0 |\phi_{\beta_i}' \rangle |\uppsi_0' \rangle ) \end{array} \right) ~,
\end{equation}
in terms of which the first order source is $|\SS^{(1)} \rangle = P^{(1)} \PP_{\uppsi_0}^\perp | B_i \rangle$.

The first order solution \eqref{1storderfieldsol} does not oscillate around zero, but rather
\begin{equation}
|\zeta_{\rm av}^{(1)} \rangle := - M_0 \epsilon \sum_{a=1}^{\infty} 2 \Re \left\{ \frac{1}{\nu_a} |\eta_a \rangle \langle \eta_a | B_i \rangle \right\} = - (M_0\epsilon) \HH_{i}^{-1} | B_i \rangle~.
\end{equation}
In the second step we noted that the sum is the spectral representation of $(\HH_i)^{-1}$.  In the appendix we show that $|\zeta_{\rm av}^{(1)} \rangle$ is precisely the $O(\epsilon)$ term in an expansion of the time-independent boosted kink profile $|\zeta_{\beta_f} \rangle$, where $\beta_f$ is the velocity associated to the final momentum $P_f = P_i + P^{(1)} = P_i + M_0 \epsilon$.  (See equation \eqref{HHzetaresult}.)  Therefore we write
\begin{equation}
|\zeta_{\rm av}^{(1)} \rangle = |\zeta_{\beta_f}^{(1)} \rangle ~.
\end{equation}
Likewise, the first order velocity oscillates around $\beta_{f}^{(1)}$, the $O(\epsilon)$ term in the expansion of $\beta_f$:
\begin{equation}\label{beta1oscdef}
\beta^{(1)} = \beta_{f}^{(1)} + 2 M_0 \epsilon \sum_{a=1}^{\infty} \frac{ |b_a|^2}{\nu_a} \cos\left((\nu_a + \nu_{a}^{(1)} + \cdots)(t-t_\ast)\right) \equiv \beta_{f}^{(1)} + \beta_{\rm osc}^{(1)}~.
\end{equation}
Physically, these results mean that at first order, all of the time-averaged momentum transfer imparted by the external force goes into the collective coordinate degree of freedom.  This makes sense as the force couples directly to this degree of freedom only.  We expect interactions to distribute the time-averaged net momentum to the other degrees of freedom, but this can only happen at higher order in perturbation theory.  Hence an interesting quantity to investigate is the difference $\beta^{(2)} - \beta_{f}^{(2)}$ and its time average.  Using \eqref{Bket} we note for future reference that
\begin{equation}\label{beta1Bket}
\beta_{f}^{(1)} = \frac{M_0 \epsilon}{\langle \uppsi_0 | \phi_{\beta_i}' \rangle^2} + \langle B_i | \zeta_{f}^{(1)} \rangle~, \qquad \beta_{\rm osc}^{(1)} = \langle B_i | \zeta_{\rm osc}^{(1)} \rangle ~,
\end{equation}
where $|\zeta_{\rm osc}^{(1)} \rangle$ is the oscillating part of the first order solution.  Explicitly,
\begin{equation}\label{1storderavplusosc}
|\zeta^{(1)} \rangle = |\zeta_{\beta_f}^{(1)} \rangle + |\zeta_{\rm osc}^{(1)} \rangle~, \qquad |\zeta_{\rm osc}^{(1)} \rangle =  M_0\epsilon \sum_{a=1}^{\infty} 2 \Re \left\{ \frac{b_a}{\nu_a} |\eta_a \rangle e^{-\ii (\nu_a + \cdots)(t-t_\ast)} \right\} ~.
\end{equation}

At second order there is a host of source terms \eqref{secondordersource} that appear in the linearized equation of motion \eqref{linearizedFSEzeta}.  Our physical interpretation of the first order result provides useful guidance on how to proceed.  We write the first order solution in the form \eqref{1storderavplusosc}, and we decompose the second-order source terms by the power of the oscillating part of the first order field that they contain, which can be zero, one, or two.  We denote the three contributions accordingly:
\begin{equation}
|\SS^{(2)} \rangle = |\SS^{(2,0)}\rangle + |\SS^{(2,1)}\rangle + |\SS^{(2,2)} \rangle ~,
\end{equation}
and each is computed in Appendix \ref{app:pert}.

The $|\SS^{(2,1)} \rangle$ contribution has the form of a phase space linear operator acting on $|\zeta_{\rm osc}^{(1)}\rangle$, and a little calculation shows that the linear operator in question has a very natural interpretation.  It is the $O(\epsilon)$ term in the expansion of $\HH_f$, the quadratic fluctuation operator around a boosted kink with velocity $\beta_f$:
\begin{equation}
|\SS^{(2,1)}\rangle = \HH_{f}^{(1)} | \zeta_{\rm osc}^{(1)} \rangle~.
\end{equation}
This term is precisely the contribution to $|\SS^{(2)}\rangle$ responsible for resonance.  As long as $mL$ is finite and/or $\beta_i$ is nonzero, there are no resonant terms from $|\SS^{(2,2)}\rangle$ because there is no choice of $a,b,c$ such that $\nu_a \pm \nu_b \pm \nu_c = 0$.\footnote{When $mL$ is large and $\beta_i = 0$ we can of course have $\nu_a \pm \nu_b \pm \nu_c \approx 0$, but this does not lead to any breakdown of perturbation theory at finite times.  Although we do get a small denominator from the frequency sum, the numerator for such terms is of the form $e^{\ii (\nu_a \pm \nu_b \pm \nu_c)(t-t_\ast)} - 1$ and has a canceling approximate zero, at least for $t-t_\ast = O(1)$.}  Hence going back to \eqref{s2res} we identify the resonant term in $A_{a}^{(2)}$'s equation of motion:
\begin{equation}
s_{a}^{(2,{\rm res})} =  e^{\ii (\nu_a + \cdots)(t-t_\ast)} \langle \eta_a | \SS^{(2,1)} \rangle \bigg|_{\textrm{time-ind}} =  \frac{ s_{a}^{(1)}}{\nu_a} \langle \eta_a | \HH_{f}^{(1)} | \eta_a \rangle ~,
\end{equation}
and therefore the first order shift in the frequencies is
\begin{equation}\label{1storderfreqshift}
\nu_{a}^{(1)} = \langle \eta_a | \HH_{f}^{(1)} | \eta_a \rangle~.
\end{equation}
Aside from its familiar form based on experience with perturbation theory in quantum mechanics, the result is physically reasonable given our interpretation of the time-averaged first order solution.

Taking \eqref{1storderfreqshift} into account, the $A^{(2)}$ equation of motion, \eqref{A2eom}, becomes
\begin{align}\label{A2eomexp}
\dot{A}_{a}^{(2)} =&~  -\ii \frac{s_{a}^{(1)} \nu_{a}^{(1)} }{\nu_a} e^{\ii (\nu_a + \cdots)(t-t_\ast)}  - \ii e^{\ii (\nu_a + \cdots)(t-t_\ast)} \left( \langle \eta_a |\SS^{(2,0)} \rangle + \langle \eta_{a} | \SS^{(2,2)} \rangle \right) + \cr
&~ -\ii   \sum_{\substack{b = 1 \\ b\neq a}}^\infty \frac{s_{b}^{(1)} \langle  \eta_a | \HH_{f}^{(1)} | \eta_b \rangle}{\nu_b} e^{\ii (\nu_a - \nu_b + \cdots)(t-t_\ast)} + \cr
&~ -\ii \sum_{b= 1}^{\infty} \frac{s_{b}^{(1)\ast} \langle \eta_a |\HH_{f}^{(1)} | \eta_{b}^\ast \rangle}{\nu_b} e^{\ii (\nu_a + \nu_b + \cdots)(t-t_\ast)} ~.  \raisetag{30pt}
\end{align}
All time dependence has been exposed in this expression except for that hiding in $|\SS^{(2,2)}\rangle$; $|\SS^{(2,0)}\rangle$ is time-independent.  The time dependence of $|\SS^{(2,2)}\rangle$ can be parameterized as
\begin{align}\label{JKkets}
|\SS^{(2,2)} \rangle =&~ \PP_{\uppsi_0}^{\perp} \sum_{b,c=1}^{\infty} \bigg\{ |J^{bc} \rangle e^{-\ii (\nu_b + \nu_c + \cdots)(t-t_\ast)} + |J^{bc \ast} \rangle e^{+\ii (\nu_b + \nu_c + \cdots)(t-t_\ast)} + \cr
&~ \qquad \qquad +  2 | K^{bc} \rangle e^{-\ii (\nu_b - \nu_c + \cdots)(t-t_\ast)}  \bigg\} ~,
\end{align}
where the time-independent phase space kets $|J^{bc}\rangle$ are complex and symmetric on $b,c$, while the time-independent phase space ket $|K^{bc} \rangle$ satisfy a Hermitian condition: $|K^{bc \ast} \rangle = |K^{cb} \rangle$.  They are given explicitly in the appendix.  See equation \eqref{JKexp}.

We can now integrate in time and insert the result back into \eqref{modifiedfieldexp} to get the second order fields.  The full result is not particularly illuminating so we focus on its time-independent---or equivalently, time-averaged---piece.  We write
\begin{equation}
|\zeta^{(2)} \rangle = |\zeta_{\rm av}^{(2)} \rangle + |\zeta_{\rm osc}^{(2)}\rangle~,
\end{equation}
and we find that $|\zeta_{\rm av}^{(2)}\rangle$ receives contributions from three places: the $|\SS^{(2,0)}\rangle$ term, the $\nu_{a}^{(1)}$ term, and the $|K^{bb}\rangle$ terms:
\begin{align}\label{zeta2av}
|\zeta_{\rm av}^{(2)} \rangle = - \sum_{a=1}^{\infty} 2 \Re \bigg\{\frac{1}{\nu_a} |\eta_a \rangle \langle \eta_a | \SS^{(2,0)} \rangle + \frac{s_{a}^{(1)} \nu_{a}^{(1)}}{\nu_{a}^2} |\eta_a \rangle  + \frac{2}{\nu_a} |\eta_a \rangle \sum_{b=1}^{\infty} \langle \eta_a | K^{bb} \rangle \bigg\}~.
\end{align}
The sum over $a$ can be carried out on the first term, and we show in the appendix that the result is exactly the second order term in the expansion of the $\beta_f$-boosted kink, $|\zeta_{\beta_f} \rangle$:
\begin{equation}
- \sum_{a=1}^{\infty} 2 \Re \left\{ \frac{1}{\nu_a} |\eta_a \rangle \langle \eta_a | \SS^{(2,0)} \rangle \right\} = - \HH_{i}^{-1} | \SS^{(2,0)}\rangle = | \zeta_{\beta_f}^{(2)} \rangle~.
\end{equation}
Therefore the remaining two terms in \eqref{zeta2av} represent the leading-in-$\epsilon$ deviation of the time-averaged fields from those of a $\beta_f$ boosted kink.  We write
\begin{align}\label{deltazeta2av}
|\zeta_{\rm av}^{(2)} \rangle =&~ |\zeta_{\beta_f}^{(2)} \rangle + | \delta \zeta_{\rm av}^{(2)} \rangle ~, \cr
|\delta \zeta_{\rm av}^{(2)}\rangle =&~  -  \sum_{a=1}^{\infty} 2 \Re \bigg\{ \frac{M_0\epsilon b_a \nu_{a}^{(1)}}{\nu_{a}^2} |\eta_a \rangle  + \frac{2}{\nu_a} |\eta_a \rangle \sum_{b=1}^{\infty} \langle \eta_a | K^{bb} \rangle \bigg\} ~.
\end{align}

Inserting the decompositions $|\zeta^{(1,2)}\rangle = |\zeta_{\rm av}^{(1,2)} \rangle + |\zeta_{\rm osc}^{(1,2)} \rangle$ into \eqref{pertccvel} and using our results for the average pieces we find the second order velocity
\begin{align}\label{beta2avosc}
& \beta^{(2)} = \beta_{f}^{(2)} + \langle B_i | \delta \zeta_{\rm av}^{(2)} \rangle + \langle B_i | \zeta_{\rm osc}^{(2)} \rangle + \cr
 & \quad + \frac{\langle \varpi_{\beta_f}^{(1)} |\varphi_{\rm osc}^{(1)\prime} \rangle + \langle \varpi_{\rm osc}^{(1)} | \phi_{\beta_f}^{(1)\prime} \rangle - 2 \beta_i \langle \uppsi_0 |  \phi_{\beta_f}^{(1)\prime} \rangle \langle \uppsi_0 |\varphi_{\rm osc}^{(1)\prime} \rangle }{\langle \uppsi_0 |\phi_{\beta_i}' \rangle^2}   - \frac{2 \langle \uppsi_0 | ( \beta_{f}^{(1)}  \varphi_{\rm osc}^{(1)\prime} +  \beta_{\rm osc}^{(1)}  \phi_{\beta_f}^{(1)\prime} ) \rangle}{\langle \uppsi_0 |\phi_{\beta_i}' \rangle}  + \cr
&  \quad + \frac{\langle \varpi_{\rm osc}^{(1)} | \varphi_{\rm osc}^{(1)\prime} \rangle - \beta_i \langle \uppsi_0 | \varphi_{\rm osc}^{(1)\prime} \rangle^2}{\langle \uppsi_0 |\phi_{\beta_i}' \rangle^2} - 2 \beta_{\rm osc}^{(1)} \frac{\langle \uppsi_0 |\varphi_{\rm osc}^{(1)\prime}\rangle}{\langle \uppsi_0 |\phi_{\beta_i}' \rangle} ~.
\end{align}
The second line of this expression is due to the terms in $\beta^{(2)}$ that are quadratic in the first order fields and where we've taken one of those fields to be a component of $|\zeta_{\rm av}^{(1)}\rangle = |\zeta_{\beta_f}^{(1)}\rangle$ and one to be a component of $|\zeta_{\rm osc}^{(1)} \rangle$.  They can be understood in the following way.  Let $|B_f\rangle$ be the ket analogous to \eqref{Bket} but corresponding to velocity $\beta_f$, and let $|B_{f}^{(1)}\rangle$ be the $O(\epsilon)$ term in its expansion around $\epsilon = 0$.  Then we find that this whole set of terms is exactly $\langle B_{f}^{(1)} | \zeta_{\rm osc}^{(1)} \rangle$.  The last line of \eqref{beta2avosc} comes from those terms quadratic in the oscillating part of the first order fields.  It can be written in the form $\langle \zeta_{\rm osc}^{(1)} | \QQ_1 | \zeta_{\rm osc}^{(1)} \rangle/\langle \uppsi_0 |\phi_{\beta_i}' \rangle^2$, where $\QQ_1$ is a quadratic form that appears in the analysis of $|\SS^{(2,2)}\rangle$.  Meanwhile those terms quadratic in the average part of the first order fields went into constructing the second-order contribution to $\beta_f$.  Hence we write
\begin{equation}\label{beta2simp1}
\beta^{(2)} = \beta_{f}^{(2)} + \langle B_i | \delta \zeta_{\rm av}^{(2)}\rangle + \langle B_i | \zeta_{\rm osc}^{(2)} \rangle + \langle B_{f}^{(1)} | \zeta_{\rm osc}^{(1)} \rangle + \frac{\langle \zeta_{\rm osc}^{(1)} | \QQ_1 | \zeta_{\rm osc}^{(1)} \rangle}{\langle \uppsi_0 | \phi_{\beta_i}' \rangle^2} ~.
\end{equation}

The final term of this expression contributes a piece to the time-independent part of $\beta^{(2)}$ when we take the same normal mode from both $\zeta_{\rm osc}^{(1)}$'s.  Hence the time-averaged second order velocity is
\begin{align}
\beta_{\rm av}^{(2)} =&~ \beta_{f}^{(2)} + \langle B_i | \delta \zeta_{\rm av}^{(2)} \rangle + \frac{2 (M_0 \epsilon)^2}{\langle \uppsi_0 | \phi_{\beta_i}' \rangle^2} \sum_{a=1}^{\infty}  \frac{|b_a|^2 \langle \eta_a | \QQ_1 | \eta_a \rangle}{\nu_{a}^2}  \cr
\equiv&~ \beta_{f}^{(2)} + \delta \beta_{\rm av}^{(2)}~,
\end{align}
with the latter two terms representing the deviation from $\beta_{f}^{(2)}$.  Using \eqref{deltazeta2av} and with the aid of $-(M_0 \epsilon) \HH_{i}^{-1} |B_i\rangle = |\zeta_{\beta_f}^{(1)}\rangle$, we see that the sum over $a$ can be carried out on the second term of $\langle B_i | \delta \zeta_{\rm av}^{(2)}\rangle$ involving $|K^{bb}\rangle$.  The result is related to the other two terms making up the deviation $\delta \beta_{\rm av}^{(2)} = \beta_{\rm av}^{(2)} - \beta_f$.  Specifically, in the appendix we show that
\begin{equation}
\frac{2}{M_0\epsilon} \langle \zeta_{\beta_f}^{(1)} | K^{aa} \rangle + \frac{2 (M_0\epsilon)^2 |b_{a}|^2}{\nu_{a}^2 \alpha_{i}^2} \langle \eta_a | \QQ_1 | \eta_a \rangle = \frac{ M_0\epsilon |b_a|^2 \nu_{a}^{(1)}}{\nu_{a}^2} ~,
\end{equation}
and therefore the deviation simplifies remarkably to
\begin{align}\label{beta2avdev}
\delta \beta_{\rm av}^{(2)} =&~ - M_0\epsilon \sum_{a=1}^{\infty} \frac{|b_a|^2 \nu_{a}^{(1)}}{\nu_{a}^2} ~.
\end{align}
%

%%%%%%%%%%%%%%%%% 
 \begin{figure}[t!]
 \centering
\begin{subfigure}{.45\textwidth}
  \centering
  \includegraphics[width=\linewidth]{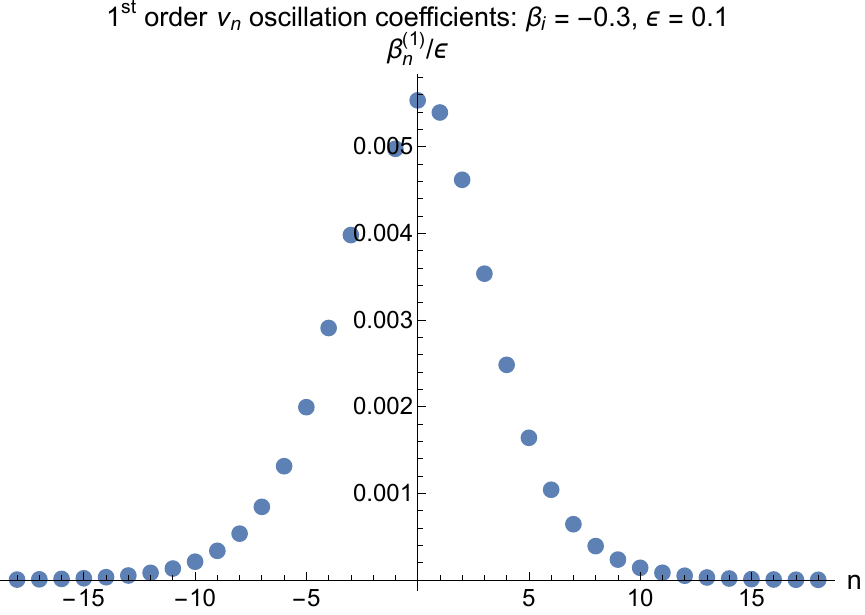}
  \caption{}
  \label{fig:Betaa1}
\end{subfigure}  \qquad %
\begin{subfigure}{.45\textwidth}
  \centering
  \includegraphics[width=\linewidth]{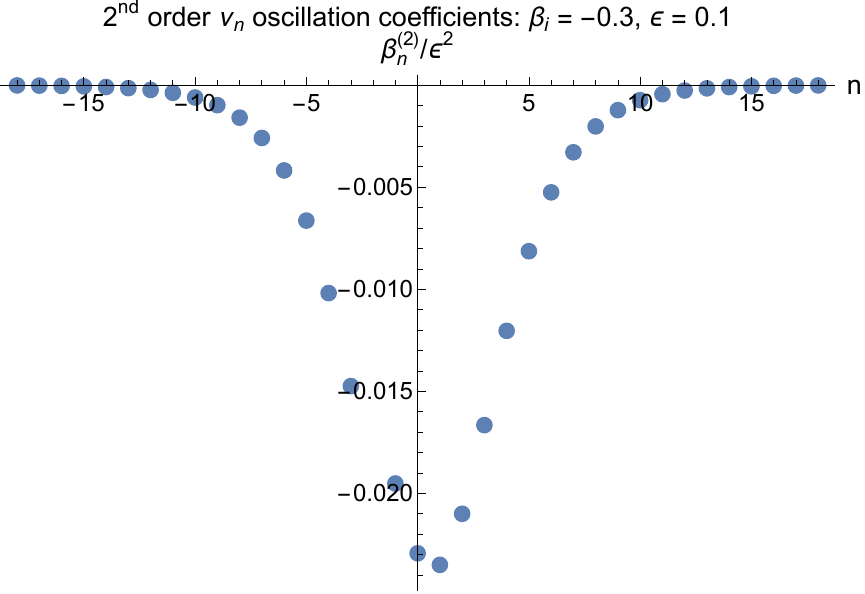}
  \caption{}
  \label{fig:Betaa2}
\end{subfigure}  \\[1ex]
\begin{subfigure}{.45\textwidth}
  \centering
  \includegraphics[width=\linewidth]{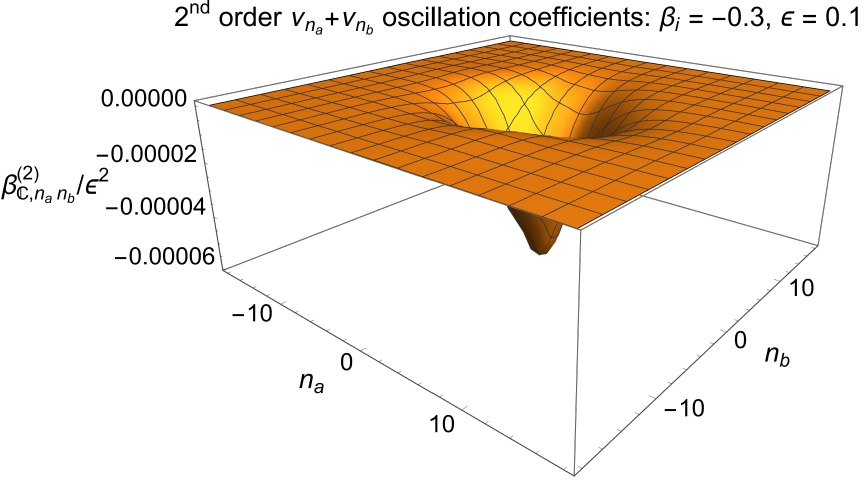}
  \caption{}
  \label{fig:BetaC2}
\end{subfigure}  \qquad %
\begin{subfigure}{.45\textwidth}
  \centering
  \includegraphics[width=\linewidth]{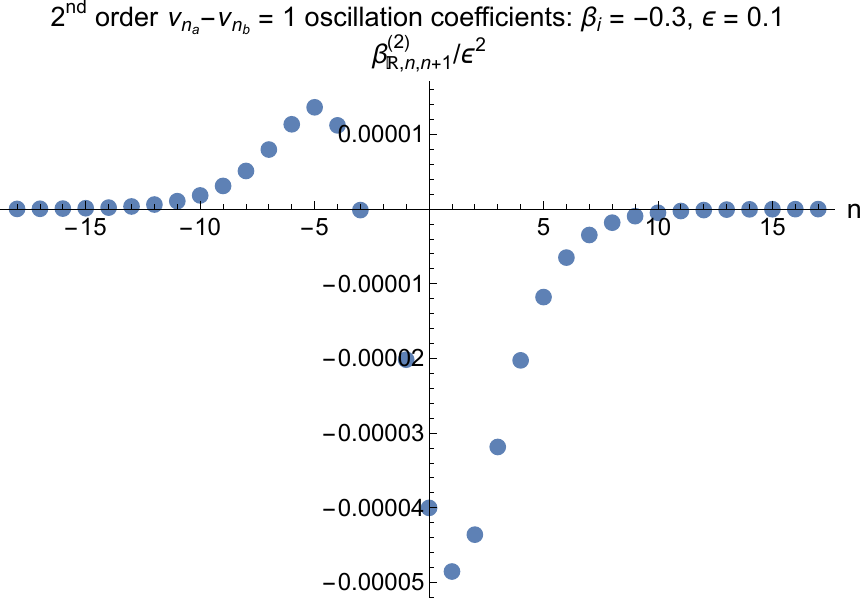}
  \caption{}
  \label{fig:BetaR2}
\end{subfigure}  
\caption{The first and second-order oscillation coefficients for the collective coordinate velocity as defined in the text, with the powers of $\epsilon$ factored out.  These correspond to a sine-Gordon kink with initial velocity $\beta_i = -0.3$ and an $mL$ of 40.  With this large value of $mL$ we use the approximate normal modes obtained from imposing periodic boundary conditions on the $mL \to \infty$ eigenfunctions as discussed around Equation \eqref{boostedapproxmodes}.  Due to parity symmetry the second order coefficients are all real.  In the case of $\beta_{\mathbbm{C},ab}^{(2)}$ we made a 3D list plot with points joined by flat surfaces for ease of visualization.  In the case of $\beta_{\mathbbm{R},ab}^{(2)}$, the coefficients are peaked near the diagonal and fall off quickly away from the diagonal so we graphed the case $b = a+1$.}
\label{fig:BetaCoefficients}
\end{figure}
%%%%%%%%%%%%%%%%%%

The remaining terms that go into $\beta^{(2)}$ involve oscillations with frequencies $\nu_a, \nu_a + \nu_b$, and $\nu_a - \nu_b$.  Thus we write
\begin{equation}
\beta^{(2)} = \beta_{\rm av}^{(2)} + \beta_{\rm osc}^{(2)} = \beta_{f}^{(2)} + \delta \beta_{\rm av}^{(2)} + \beta_{\rm osc}^{(2)}~,
\end{equation}
with
\begin{align}
\beta_{\rm osc}^{(2)} =&~ \sum_{a=1}^{\infty} 2 \Re \left\{ \beta_{a}^{(2)} e^{-\ii (\nu_a + \cdots)(t-t_\ast)} \right\} + \sum_{b,c=1}^{\infty} 2 \Re \left\{ \beta_{\mathbbm{C},bc}^{(2)} e^{-\ii (\nu_b + \nu_c + \cdots)(t-t_\ast)} \right\} + \cr
&~ + \sum_{\substack{b,c = 1 \\ b\neq c}}^{\infty} 2 \Re \left\{ \beta_{\mathbbm{R},bc}^{(2)} e^{-\ii (\nu_b - \nu_c + \cdots)(t-t_\ast)} \right\}~.
\end{align}
The coefficients $\beta_{a}^{(2)}$, $\beta_{\mathbbm{C},ab}^{(2)}$, and $\beta_{\mathbbm{R},ab}^{(2)}$ are recorded in the appendix in terms of phase space overlaps defined earlier.  See equations \eqref{betaa2final}-\eqref{betaRab2final}.   In particular, $\beta_{\mathbbm{C},ab}^{(2)}$ is complex symmetric in $ab$ while $\beta_{\mathbbm{R},ab}^{(2)}$ is Hermitian.  While we're at it, we also define $\beta_{a}^{(1)}$ to be the corresponding set of coefficients for $\beta_{\rm osc}^{(1)}$.  Looking at \eqref{beta1oscdef} we have $\beta_{a}^{(1)} = M_0 \epsilon |b_{a}|^2/\nu_a$.  We use this and the formulas of the appendix to compute these coefficients and the deviation \eqref{beta2avdev} numerically, and some results are presented in Figure \ref{fig:BetaCoefficients}.

As these graphs show, the coefficients are nontrivial in a localized region of mode number space and die out quickly away from this region.  The modes that contribute most significantly are those that are most shaped like the kink zero mode.  These correspond to the eigenmodes with the fewest oscillations.  Note that, especially at high $\beta_i$, this is not the same as the eigenmodes with the smallest eigenvalue.  Hence we find it convenient to plot these mode coefficients with respect to the kink rest frame wave number index $n$ rather than the lab frame energy index $a$.  See the discussion at the end of Appendix \ref{app:twistedEVs} and especially Figure \ref{fig:TwistedEVbetadep} for details.

%%%%%%%%%%%%%%%%% 
 \begin{figure}[t!]
 \centering
\begin{subfigure}{.99\textwidth}
  \centering
  \includegraphics[width=\linewidth]{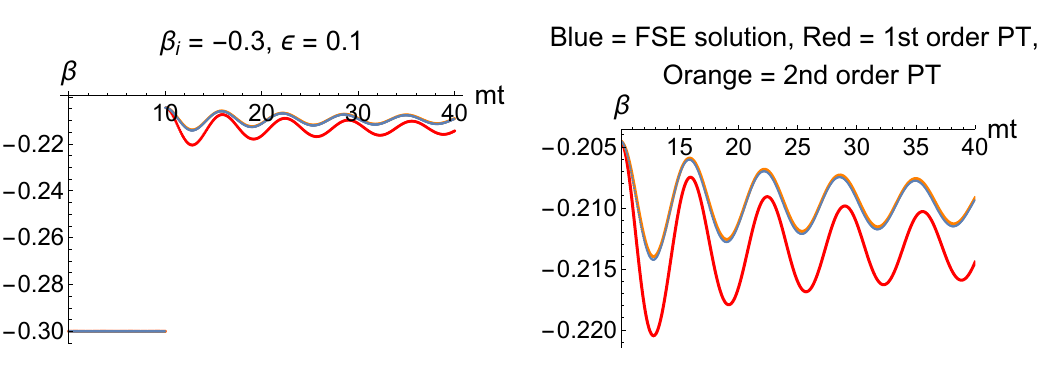}
  \caption{}
  \label{fig:Pertpicsmallbeta}
\end{subfigure}   \\[1ex]
\begin{subfigure}{.99\textwidth}
  \centering
  \includegraphics[width=\linewidth]{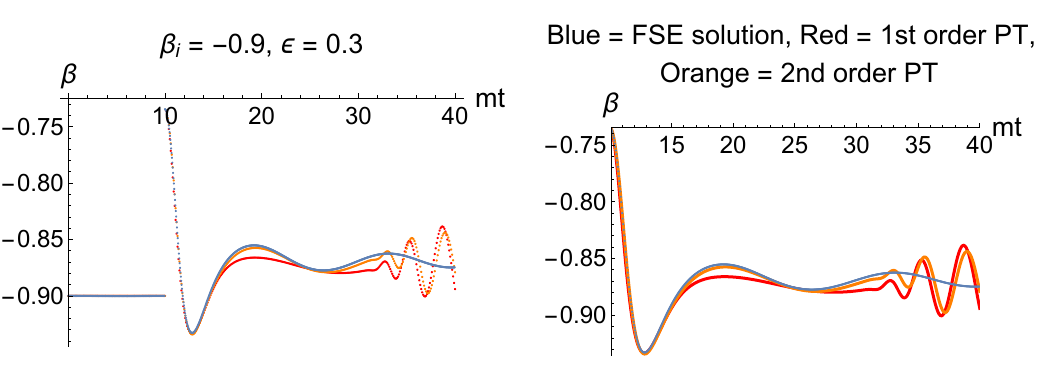}
  \caption{}
  \label{fig:Pertpiclargebeta}
\end{subfigure} 
\caption{Comparisons of the collective coordinate velocity $\beta(t)$ for the sine-Gordon kink as determined by a numerical solution to the FSE (blue) and the perturbative analysis in this subsection.  The top graphs have an initial kink velocity $\beta_i = -0.3$ and a momentum transfer of $0.1$.  The bottom graphs have $\beta = -0.9$ and $\epsilon = 0.3$.  The graphs on the right zoom in on the solution after the kick, which occurs at $m t_\ast = 10$.  The size of the circle is $mL = 40$.  The red graph represents the prediction from first order perturbation theory with the leading order frequencies $\nu_a$.  The orange graph is the prediction from second order perturbation theory with the first order correction to the frequencies included.  It is nearly indistinguishable from the FSE solution.  We can see the expected breakdown of the perturbative solution at large $t-t_\ast$ in the bottom case where higher order corrections to the frequencies are expected to be more significant.}
\label{fig:Pertpics}
\end{figure}
%%%%%%%%%%%%%%%%%%

Since the coefficients die off rapidly we can use truncated mode sums to approximate $\beta^{(1)}$ and $\beta^{(2)}$.  In Figure \ref{fig:Pertpics} we compare the results of our first- and second-order perturbation theory calculations with numerical solutions to the FSE for a couple values of $\beta_i$ and $\epsilon$ and find excellent agreement.  This agreement serves as a very nontrivial check on both our numerical approach to the FSE and the detailed calculations of Appendix \ref{app:pert}.

%%%%%%%%%%%%%%%%
%%%%%%%%%%%%%%%%
\section{Future Directions}\label{sec:Conclusions}
%%%%%%%%%%%%%%%%
%%%%%%%%%%%%%%%%

Going forward, the next steps of the program outlined in the Introduction are clear.  We need to tackle the boundary value problem for the FSE, and this is work in progress on both the numerical and perturbation theory fronts.  

The numerical BVP is exciting because it allows us to access the physically interesting regime of transfers $\epsilon \gtrsim 1$.  The basic idea is to employ a shooting method.  We perturb the initial conditions for the momentum field and try to find those that will land us on a pure kink position field in the final configuration.  We work in a finite-dimensional mode cut-off space and use a Newton method to find a minimum of an appropriate cost function.  This approach is amenable to parallel computing.  For the application to QFT form factors, however, we can expect potential difficulties in extracting the physically relevant $mL \to \infty$ and $T = t_f - t_i \to (1-\ii \epsilon)\infty$ limits.

Perturbation theory in $\epsilon$ for the BVP is a more straightforward modification of the work done here.  Extending these techniques to the QFT calculation of the form factor will then enable us to compare with known results from integrability for sine-Gordon \cite{Weisz:1977ii} and make new predictions in $\phi^4$ theory.  

Finally, we would like to extend the framework of the FSE beyond the realm of two-dimensional models with kink solitons.

%%%%%%%%%%%%%%%%
\section*{Acknowledgments}
%%%%%%%%%%%%%%%%

We thank Andrew Bowman for contributions in the early stages of this work and Jarah Evslin, Ilarion Melnikov, and Costis Papageorgakis for illuminating discussions.  We also thank Hans Volkmer for assistance in verifying our results for Lam\'e eigenvalues.  ABR is supported by NSF grant number PHY-2112781. The work of QAH and ZJA was supported by 2021 and 2023 Erickson Discovery Grants for Undergraduate Research through the Pennsylvania State University.  The work of EAY was supported by at 2023 Erickson Discovery Grant and a 2023 MC REU grant through the Penn State College of Engineering.  We also thank the Division of Faculty Affairs at Penn State Fayette for support in early stages of the project.

%%%%%%%%%%%%%%%%%%%%%%%%%%%%%%%%%%%%%%%%%%%%%%%%%%%%

\appendix

%%%%%%%%%%%%%%%%%%%
%%%%%%%%%%%%%%%%%%%
\section{Lam\'e Equation and Eigenfunctions}\label{app:Lame}
%%%%%%%%%%%%%%%%%%%
%%%%%%%%%%%%%%%%%%%

Most results quoted here can be found in \cite{MR2655369}, available as part of the online database \cite{NIST:DLMF}.  See in particular \cite[\href{https://dlmf.nist.gov/29}{Ch.~29}]{NIST:DLMF}.  A standard textbook reference is \cite{MR698781}, and we found the introductory section of  \cite{MR2377687} to be a helpful overview of results.  

The Lam\'e differential equation of order $\nu_{\rm L}$ is
\begin{equation}\label{Lamediffeq}
\pd_{z}^2 w + \left(h - \nu_{\rm L} (\nu_{\rm L} + 1) \rmk^2 \operatorname{sn}^2(z, \rmk^2) \right) w = 0~,
\end{equation}
with $h$ the eigenvalue.  As a Schr\"odinger equation with a periodic potential (of period $2 \mathbf{K}(\rmk^2) \equiv 2\mathbf{K}$), standard results apply.  See, \emph{e.g.}~\cite{MR559928}.  Equipping \eqref{Lamediffeq} with a twisted periodicity condition
\begin{equation}\label{twistedeigen}
w(z + 2\mathbf{K}) = e^{\ii \mu} w(z) \equiv \xi w(z)~,
\end{equation}
defines a self-adjoint boundary value problem.  For any real $\mu$ there will be an infinite discrete set of solutions.  Each such solution lies in one of the allowed zones, which are intervals delimited by $h$-values corresponding to periodic ($\xi = 1$) and anti-periodic ($\xi = -1$) solutions.  These form a sequence $h_0 < h_1 \leq h_2 < h_3 \leq h_4 < \cdots$ where $h_0$ is a periodic eigenvalue, followed by alternating pairs of anti-periodic and periodic eigenvalues.  The allowed zones are the intervals $[h_{2j}, h_{2j+1}]$, whose union gives the Bloch spectrum.   The complementary intervals $(h_{2j+1}, h_{2j})$ are the forbidden zones.  Solutions to \eqref{Lamediffeq} for $h$ in an allowed zone are everywhere bounded and have $\mu$ real.  Solutions for $h$ in a forbidden zone are unbounded.  The less than or equal signs allow for the possibility that forbidden zones can vanish, in which case there is a doubly degenerate eigenvalue.

In addition to periodicity, the Lam\'e equation has parity symmetries about both $z = \mathbf{K}$ and $z = 0$.  If $w(z)$ is a periodic solution (with either period $2\mathbf{K}$ or $4\mathbf{K}$), then reflection symmetry about $z = \mathbf{K}$ implies $w(2\mathbf{K}-z)$ is a periodic solution with the same periodicity.  By taking sums and differences we can restrict attention to solutions that are even about $z = \mathbf{K}$ and those that are odd about $z = \mathbf{K}$.  Periodic solutions that are even about $z = \mathbf{K}$ are denoted $\Ec_{\nu_{\rm L}}(z,\rmk^2)$, while periodic solutions that are odd about $z = \mathbf{K}$ are denoted $\Es_{\nu_{\rm L}}(z,\rmk^2)$.  If the periodicity of such a solution is $2p \mathbf{K}$, (where $p = 1$ or $2$), then $\Ec_{\nu_{\rm L}}^j$ for $j = 0,1,2,\ldots$ and $\Es_{\nu_{\rm L}}^j$, for $j = 1,2, \ldots$ denotes a solution with $p j$ zeros in the half-open interval $[0,2p \mathbf{K})$.  In either case of $p = 1,2$, periodicity plus reflection symmetry about $z = \mathbf{K}$ implies reflection symmetry about $z = -\mathbf{K}$.  Hence the $\Ec$'s satisfy Neumann conditions at $z = \pm \mathbf{K}$ while the $\Es$'s satisfy Dirichlet conditions at $z = \pm \mathbf{K}$.  The eigenvalues associated with $\Ec_{\nu_{\rm L}}^j$ and $\Es_{\nu_{\rm L}}^j$ are denoted $h = a_{\nu_{\rm L}}^j(\rmk^2)$ and $h = b_{\nu_{\rm L}}^j(\rmk^2)$ respectively.  

Since the differential equation and boundary conditions are invariant under $z \to -z$, the $\Ec$'s and $\Es$'s can be further divided into even and odd cases about $z = 0$.  $\Ec$'s that are even functions of $z$ have periodicity $2\mathbf{K}$ and correspond to even $j$'s, while those that are odd functions of $z$ have periodicity $4\bf{K}$ and correspond to odd $j$'s.  $\Es$'s that are even functions of $z$ have periodicity $4\mathbf{K}$ and correspond to odd $j$'s, while those that are odd functions of $z$ have periodicity $2\mathbf{K}$ and correspond to even $j$'s.  The periodicity and parity about $z = 0$ is summarized here:
\begin{align}\label{EcEsbreakdown}
 a_{\nu_{\rm L}}^{2m}(\rmk^2) = h_{4m} \qquad  & \Ec_{\nu_{\rm L}}^{2m}(z,\rmk^2) \qquad  \textrm{even} \qquad  \textrm{$2 \mathbf{K}$-periodic} ~, \cr
a_{\nu_{\rm L}}^{2m+1}(\rmk^2) = h_{4m+3} \qquad & \Ec_{\nu_{\rm L}}^{2m+1}(z,\rmk^2) \qquad \textrm{odd}  \qquad \textrm{$4 \mathbf{K}$-periodic}~, \cr
 b_{\nu_{\rm L}}^{2m+1}(\rmk^2) = h_{4m+1} \qquad  & \Es_{\nu_{\rm L}}^{2m+1}(z,\rmk^2) \qquad \textrm{even}  \qquad \textrm{$4 \mathbf{K}$-periodic}~, \cr
 b_{\nu_{\rm L}}^{2m+2}(\rmk^2) = h_{4m+2} \qquad & \Es_{\nu_{\rm L}}^{2m+2}(z,\rmk^2) \qquad \textrm{odd}  \qquad \textrm{$2\mathbf{K}$-periodic}~,
\end{align}
where $m = 0,1,2,\ldots$.  Each one of these corresponds to a unique solution up to normalization, and functions with $4\mathbf{K}$ periodicity have $2\mathbf{K}$ anti-periodicity.  We also see that the eigenvalues are ordered according to
\begin{equation}
a^0 < b^1 \leq a^1 < b^2 \leq a^2 < b^3 \leq a^3 < \cdots~.
\end{equation}

In fact, when $\nu_{\rm L}$ is a non-negative integer there is a coalescing of eigenvalues when $j$ gets large enough \cite{MR2400,MR1361127}:
\begin{equation}
b_{\nu_{\rm L}}^j(\rmk^2) = a_{\nu_{\rm L}}^j(\rmk^2) ~, \qquad j = \nu_{\rm L}+1~, \nu_{\rm L} +2~, \ldots~,
\end{equation}
Thus there are only $\nu_{\rm L}$ disallowed bands to the right of the first allowed band; the Bloch spectrum becomes 
\begin{equation}
\sigma(h) = [a_{\nu_{\rm L}}^0, b_{\nu_{\rm L}}^1] \cup [a_{\nu_{\rm L}}^1 , b_{\nu_{\rm L}}^2] \cup \cdots \cup [a_{\nu_{\rm L}}^{\nu_{\rm L}-1}, b_{\nu_{\rm L}}^{\nu_{\rm L}}] \cup [ a_{\nu_{\rm L}}^{\nu_{\rm L}}, \infty) ~.
\end{equation}

The two cases of interest for us are the $\nu_{\rm L} = 2$ equation with $2\mathbf{K}$-antiperiodic functions ($\phi^4$-theory) and the $\nu_{\rm L} = 1$ equation with periodic eigenfunctions (sine-Gordon).  Thus in $\phi^4$-theory we take the odd $j$ $\Ec$'s and $\Es$ only, while in sine-Gordon we take the even $j$ $\Ec$'s and $\Es$'s only.  The sine-Gordon spectrum is
\begin{equation}
\textrm{sine-Gordon:} \qquad a_{1}^0 < b_{1}^2 = a_{1}^2 < b_{1}^4 = a_{1}^4 < b_{1}^6 = a_{1}^6 < \cdots ~,
\end{equation}
where $a_{1}^0$ corresponds to the zero mode.  The $\phi^4$-theory spectrum is
\begin{equation}
\textrm{$\phi^4$-theory:} \qquad b_{2}^1 < a_{2}^1 < b_{2}^3 = a_{2}^3 < b_{2}^5 = a_{2}^5 < \cdots ~,
\end{equation}
with $b_{2}^1$ corresponding to the zero mode and $a_{2}^1$ the shape mode.  Hence, using \eqref{ktoL}, \eqref{finiteLevproblem} and \eqref{sGLrmk}, \eqref{lame1standard}, the relationship between $(h_j,\rmk)$ and $(\frac{\upomega_a}{m}, mL)$ is
\begin{align}\label{staticparametermap}
\textrm{sine-Gordon:} \quad mL =&~ 2\rmk \mathbf{K}(\rmk^2)~, \cr
 a_{1}^{2j} =&~ \rmk^2 \left( \frac{\upomega_{2j}^2}{m^2} + 1 \right)~, \quad b_{1}^{2j+2} =  \rmk^2 \left( \frac{\upomega_{2j+1}^2}{m^2} + 1 \right) \cr
\textrm{$\phi^4$-theory:} \quad mL =&~  2\sqrt{1 + \rmk^2} \mathbf{K}(\rmk^2)~, \cr
 b_{2}^{2j+1} =&~  (1 + \rmk^2) \left( \frac{\upomega_{2j}^2}{m^2} + 1 \right)~, \quad a_{2}^{2j+1} = (1 + \rmk^2) \left( \frac{\upomega_{2j+1}^2}{m^2} + 1 \right)~, \qquad 
\end{align}
for $j \in \mathbbm{N}_0$.

In general, the $2\nu_{\rm L} + 1$ band edges, $h_0,\ldots, h_{2\nu_{\rm L}}$ can be obtained from eigenvalues of certain finite-dimensional tri-diagonal matrices related to a Fourier series ansatz for the solutions.  They are algebraic functions of the $\rmk^2$ that can be computed explicitly for low $\nu_{\rm L}$.  For example we have
\begin{align}
\nu_{\rm L} = 1: \qquad & a_{1}^0 = \rmk^2~, \quad b_{1}^1 = 1~, \quad a_{1}^1 = 1 + \rmk^2~, \cr
\nu_{\rm L} = 2: \qquad & a_{2}^0 = 2 (1 + \rmk^2) - 2 \sqrt{ 1 - \rmk^2 + \rmk^4} ~, \quad b_{2}^1 = 1 + \rmk^2~, \quad a_{2}^1 = 1 + 4\rmk^2~, \cr
& b_{2}^2 = 4 + \rmk^2~, \quad a_{2}^2 = 2 (1 + \rmk^2) +2 \sqrt{1 - \rmk^2 + \rmk^4} ~.
\end{align}
The corresponding eigenfunctions are polynomials in $\sn, \cn$ and/or $\dn$.  Using \eqref{staticparametermap} we see that $a_{1}^0$ and $b_{2}^1$ correspond to the zero mode $\upomega_0 = 0$ for sine-Gordon and $\phi^4$-theory respectively.  Meanwhile, $a_{2}^1$ corresponds to the shape mode and gives the $\upomega_1$ of \eqref{algebraicphi4}.

In contrast, the doubly degenerate eigenvalues embedded in the topmost band $[h_{2\nu_{\rm L}}, \infty)$ are transcendental eigenvalues.  They are an infinite set of solutions to a transcendental equation, and the equation can be written down explicitly for low values of $\nu_{\rm L}$.  Reference \cite{MR2377687} gives this transcendental equation in the form of a dispersion relation expressing the twist parameter $\mu$ as a function of the eigenvalue $h$, $\mu = \mu_{\nu_{\rm L}}(h,\rmk^2)$.  We get the periodic or anti-periodic eigenvalues by solving this equation for $\mu = 0$ or $\pi$.  Explicitly, for $\nu_{\rm L} = 1$ the equation is 
\begin{equation}
\mu_{1}(h,\rmk^2) = -\ii Z(\alpha_0 \, | \, \rmk^2) + \frac{\pi}{2 \mathbf{K}(\rmk^2)} ~, \qquad\textrm{with} \quad \dn^2(\alpha_0 \, | \, \rmk^2) = h - \rmk^2~,
\end{equation}
where $Z$ is the Jacobi Zeta function.  There is an infinite sequence of purely imaginary $\alpha_0$'s that solve the latter relation for $h$ real and large enough.  The $\nu_{\rm L} = 2$ dispersion relation is quite a bit more complicated but also obtained explicitly in \cite{MR2377687}.

An alternative representation \cite{MR156022} of the eigenvalues is given in terms of the Hill discriminant, $\Delta(h)$, defined as follows.  Let $w_{1,2}(z,h)$ be the solutions to the differential equation with eigenvalue $h$ satisfying initial conditions $w_1(0) = 1 =  \pd_z w_2(0)$ and $\pd_z w_1(0) = 0 = w_2(0)$.  Then the Hill discriminant is
\begin{equation}
\Delta(h) = 2 \left( w_1(2\mathbf{K},h) + \pd_z w_2(2\mathbf{K},h) \right) ~.
\end{equation}
The Hill discriminant satisfies $|\Delta| \leq 2$ in the allowed zones and $|\Delta| > 2$ in the forbidden zones, with those $h$ satisfying $\Delta(h) = \pm 2$ corresponding to periodic and anti-periodic eigenvalues respectively.  More generally, the eigenvalues corresponding to the twisted periodicity condition \eqref{twistedeigen} are the solutions to \cite{MR3075381}
\begin{equation}
\Delta(h) = 2 \cos(\mu)~.
\end{equation}

Although a closed form expression for the Hill discriminant of the Lam\'e equation $\Delta(h) \equiv \Delta_{\nu_{\rm L}}(h,\rmk^2)$ is not known, some analytic results are available, and it is easily computed numerically from its definition.  Recently, Reference \cite{MR4717595} proved an asymptotic estimate for $\Delta_{\nu_{\rm L}}(h,\rmk^2)$ for any real $\nu_{\rm L} \geq -1/2$ in the limit $k \to 1$.  In the case of nonnegative integer $\nu_{\rm L}$ and $h > \nu_{\rm L} (\nu_{\rm L} + 1)$, set
\begin{align}\label{kappahB}
\kappa = \sqrt{ h - \nu_{\rm L} (\nu_{\rm L} + 1)} ~, \qquad B = \frac{ (\ii \kappa  -1) (\ii \kappa - 2) \cdots  (\ii \kappa - \nu_{\rm L})}{( \ii \kappa + 1 ) (\ii \kappa + 2) \cdots (\ii \kappa + \nu_{\rm L})} \equiv e^{\ii \delta(\kappa)}~,
\end{align}
where we noted that $B$ is a phase.  Then, with $\mathbf{K} = \mathbf{K}(\rmk^2)$ and $\mathbf{E} = \mathbf{E}(\rmk^2)$,
\begin{equation}\label{Deltaapprox}
\left| \Delta_{\nu_{\rm L}}(h,\rmk^2) - 2 \Re(B e^{2\ii \kappa \mathbf{K}} ) \right| \leq 8 h^{1/2} \kappa^{-2} \nu_{\rm L} (\nu_{\rm L} + 1) \left( \mathbf{E} + 1 - 2 \tanh\mathbf{K} \right)~.
\end{equation}
Suppose $\rmk = 1- e^{-\tau}$.  In the context of this paper, $\tau = mL/\sqrt{2} - \ln(8)$ for $\phi^4$ theory or $mL- \ln(8)$ for sine-Gordon, plus exponentially small corrections in $mL$.  Then $\mathbf{E} + 1 - 2\tanh{\mathbf{K}} = O(\tau \cdot e^{-\tau})$.

We show that this result is consistent with the quantization conditions \eqref{staticquantcon} and \eqref{sGquantcon} for large $mL$.  In the sine-Gordon case, setting $\Delta = 2$ gives
\begin{align}\label{sGDcon}
2 =&~ \Delta_{1}(h,\rmk^2) = 2 \Re(B e^{2\ii \kappa \mathbf{K}}) + O(mL e^{-mL})~. 
\end{align}
We have from \eqref{sGLrmk} that $2\mathbf{K} = mL/\mathrm{k} = mL(1 + O(e^{-mL}))$.  The condition $h > 2$ tells us we are in the semi-infinite allowed zone, so this approximation to $\Delta$ can only be useful for the transcendental eigenvalues.  $\kappa$ is a measure of the distance from the lower edge of this zone, so the fact that the error term in \eqref{Deltaapprox} goes like $\kappa^{-2}$ means that, even at large $mL$, the approximation might not be great for the lowest transcendental eigenvalues.   Since $B \to 1$ as $\kappa \to \infty$, we can choose the phase so that $\delta(\kappa) \to 0$ as $\kappa \to \infty$.  This choice is physically natural since at infinite energy the potential is irrelevant and there should be no phase shift.  The condition $\Delta = 2$ becomes $\cos(2\kappa \mathbf{K} + \delta(\kappa)) = 1$ or $2\kappa \mathbf{K} + \delta(\kappa) = 2\pi n$ for $n \in \mathbbm{Z}$.  Each solution is doubly degenerate, but $\kappa$ is restricted to be positive.  We can account for the degeneracy by extending $\kappa \in \mathbbm{R}$ if we extend the definition of the phase so that $\delta(-\kappa) = - \delta(\kappa)$.  This can be done by allowing for a discontinuity by $2\pi$ at the origin.  This leads to $\delta(\kappa) = \delta_{\rm sG}(\kappa)$ with
\begin{equation}\label{sGdeltaofkappa}
\delta_{\rm sG}(\kappa) = -2 \arctan(\kappa)  + 2\pi \Theta(\kappa) - \pi~.
\end{equation}
Then the condition \eqref{sGDcon} is
\begin{equation}\label{nu1quantcon}
\kappa m L + \delta_{\rm sG}(\kappa) = 2\pi n + O\left(|\kappa|^{-1} \cdot \sqrt{mL} e^{-mL/2}\right)~, \qquad n \in \mathbbm{Z}~,
\end{equation}
The leading error on the right-hand side of \eqref{nu1quantcon} is due to the error in \eqref{sGDcon}.  Equation \eqref{nu1quantcon} agrees with the form of \eqref{sGquantcon} if $\kappa = k/m$.  Using $h = ( \frac{\upomega^2}{m^2} + 1)(1 + O(e^{-mL}))$ from \eqref{lame1standard}, we see that this identification of $\kappa$ is consistent with $\upomega^2 = k^2 + m^2$.

%%%%%%%%%%%%%%%%% 
 \begin{figure}[t!]
 \centering
\begin{subfigure}{.45\textwidth}
  \centering
  \includegraphics[width=\linewidth]{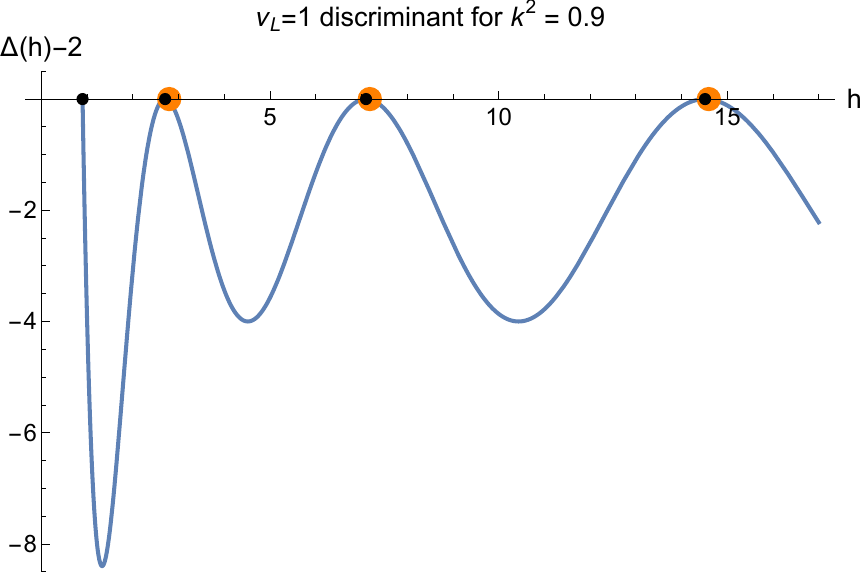}
  \caption{}
  \label{fig:LameEvs1}
\end{subfigure}  \qquad %
\begin{subfigure}{.45\textwidth}
  \centering
  \includegraphics[width=\linewidth]{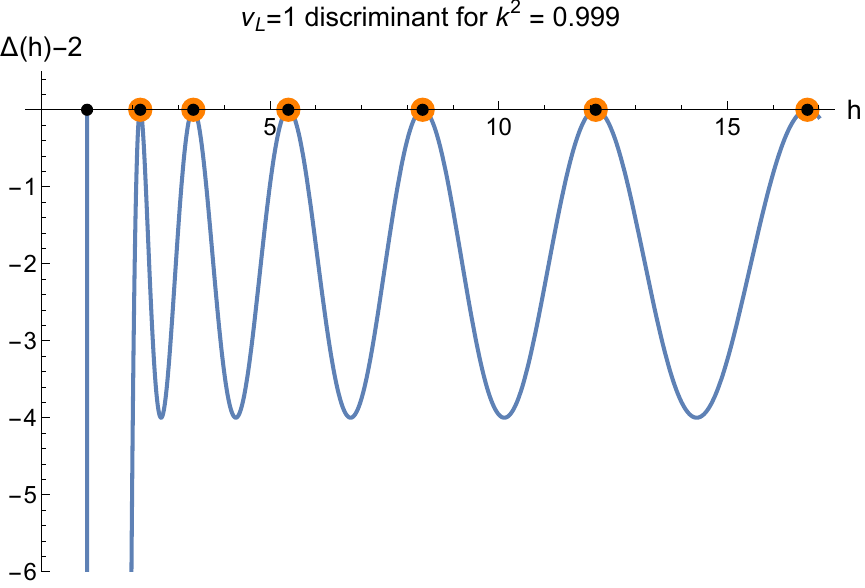}
  \caption{}
  \label{fig:LameEVs2}
\end{subfigure}  \\
\begin{subfigure}{.45\textwidth}
  \centering
  \includegraphics[width=\linewidth]{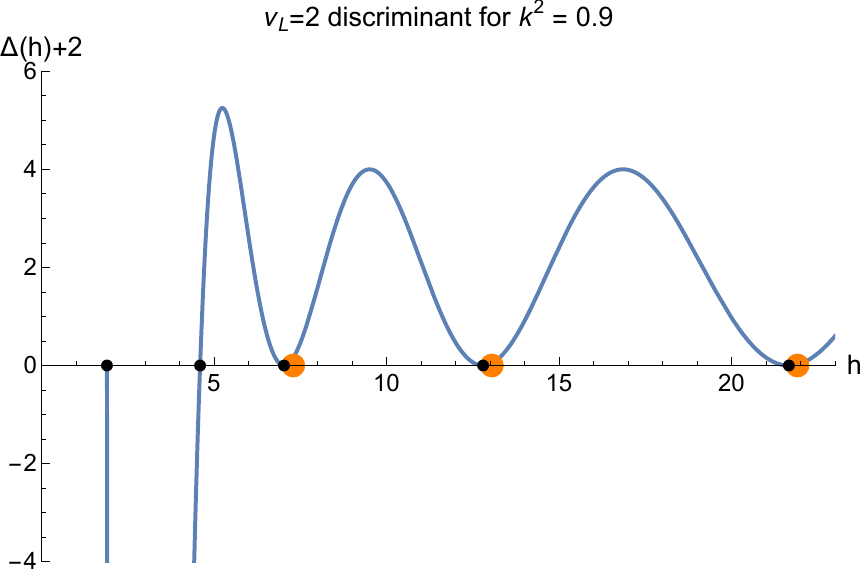}
  \caption{}
  \label{fig:LameEVs3}
\end{subfigure}  \qquad %
\begin{subfigure}{.45\textwidth}
  \centering
  \includegraphics[width=\linewidth]{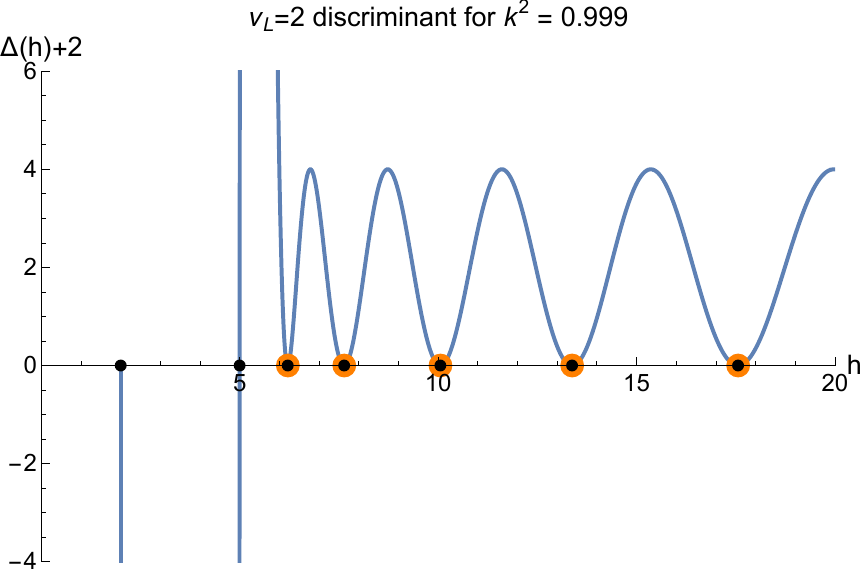}
  \caption{}
  \label{fig:LameEVs4}
\end{subfigure}  
\caption{Graphs of the Hill discriminant plus or minus two for $\nu_{\rm L} = 1,2$ and $\rmk^2 = 0.9, 0.999$.  The zeros of $\Delta - 2$ for $\nu_{\rm L} = 1$ give the eigenvalues $a_{1}^{0}$ and $a_{1}^{2m} = b_{1}^{2m}$, $m \in \mathbbm{N}$, relevant for the fluctuation spectrum around the sine-Gordon kink.  The zeros of $\Delta + 2$ for $\nu_{\rm L} = 2$ give the eigenvalues $b_{2}^1, a_{2}^1$ and $b_{2}^{2m+1} = a_{2}^{2m+1}$, $m \in \mathbbm{N}$, relevant for the fluctuation spectrum around the $\phi^4$-theory kink.  The black dots are the eigenvalues obtained using Mathematica's FindRoot on a numerical implementation of the discriminant.  The orange dots are the large $mL$ approximation obtained from the quantization conditions \eqref{nu1quantcon}, \eqref{nu2quantcon}.  The values $\rmk^2 = 0.9$ and $\rmk^2 = 0.999$ correspond to $mL \approx 4.89, 9.68$ for sine-Gordon and $mL \approx 7.11,13.7$ for $\phi^4$-theory.}
\label{fig:LameEVs}
\end{figure}
%%%%%%%%%%%%%%%%%%

We set $\Delta_2 = -2$ in the case of $\phi^4$-theory.  Imposing the same requirements that $\delta(\kappa) \to 0$ as $\kappa \to \infty$ and $\delta(-\kappa) = -\delta(\kappa)$, the phase is $\delta(\kappa) = \delta_{\phi^4}(\kappa)$ with
\begin{equation}\label{phi4deltaofkappa}
\delta_{\phi^4}(\kappa) = -2 \arctan(\kappa) - 2 \arctan(\kappa/2) + 4\pi \Theta(\kappa) - 2\pi~.
\end{equation}
Meanwhile $2\mathbf{K} = mL/\sqrt{1+\rmk^2} = (mL/\sqrt{2})(1 + O( e^{-mL/\sqrt{2}}))$.  Hence we find
\begin{equation}\label{nu2quantcon}
 \delta_{\phi^4}(\kappa) + \frac{\kappa m L}{\sqrt{2}} = (2n-1)\pi + O\left(|\kappa|^{-1} \cdot \sqrt{mL} e^{-mL/(2\sqrt{2})}\right)~.
 \end{equation}
This agrees with \eqref{staticquantcon} and \eqref{phi4phase} if $\kappa = \sqrt{2} k/m$.  Using $h = 2 (\frac{\upomega^2}{m^2} + 1)( 1 + O(e^{-mL/\sqrt{2}}))$ from \eqref{finiteLevproblem} in \eqref{kappahB}, we see that this identification of $\kappa$ is consistent with $\upomega^2 = k^2 + 2m^2$.

%%%%%%%%%
\begin{table}[t!]
\begin{center}
\begin{tabular}{| c | c | c | c | c | c | c | c |} \hline
   &  $a_{2}^3$  & $a_{2}^5$ & $ a_{2}^7$ & $a_{2}^9$ & $a_{2}^{11}$ & $a_{2}^{13}$     \\ 
\hline \hline
$\Delta + 2$ root & 6.2089 & 7.6310 & 10.0559 & 13.3753 & 17.5589 & 22.5957  \\
\hline 
large $mL$  & 6.2119 & 7.6338 & 10.0587 & 13.3780 & 17.5617 & 22.5984   \\
\hline
LameEigenvalueA(B) &6.2158 & 7.7430 & 10.6004 & 14.9925 & 21.2052 & 29.5370   \\
\hline
\end{tabular}
\end{center}
\caption{Transcendental doubly-degenerate Lam\'e eigenvalues ($a_{2}^{2m+1} = b_{2}^{2m+1}, m \in \mathbbm{N}$) associated to anti-periodic eigenfunctions for $\nu_{\rm L} = 2$ and $\rmk^2 = 0.999$.}
\label{table:LameEVs}
\end{table}
%%%%%%%%%

Our approach to obtaining the transcendental eigenvalues is to find the roots of $\Delta(h) \mp 2$ numerically, using $h_n \approx \kappa_{n}^2 + \nu_{\rm L} (\nu_{\rm L} + 1)$ with $\kappa_n$ the solutions to \eqref{nu1quantcon} or \eqref{nu2quantcon} for $\nu_{\rm L} = 1,2$ respectively, as a starting point.  Our results are illustrated in Figure \ref{fig:LameEVs}.  Mathematica has built-in functions to compute the $a_{\nu_{\rm L}}^j$'s and $b_{\nu_{\rm L}}^{j}$'s, but these are horribly inaccurate for $\rmk$ close to 1, at least in Version 14.1.0.0 of the software.  For comparison, in Table \ref{table:LameEVs} we give the first several eigenvalues for $\nu_{\rm L} = 2$ and $\rmk^2 = 0.999$ using our root finding approach with the Hill discriminant, the approximation \eqref{nu2quantcon}, and Mathematica's built-in LameEigenvalueA.  The first two are nearly indistinguishable at this value of $\rmk^2$, and clearly Mathematica's built-in command has broken down though no error messages are generated.

H.~Volkmer has confirmed our results in Table \ref{table:LameEVs} using a computational routine written in Maple which approximates the three-term recurrence relation from a Fourier series ansatz for the differential equation with a finite-sized matrix eigenvalue problem.  With $32 \times 32$ matrices he gets the same results as Mathematica, but as the matrix size increases his results converge to ours.

\begin{figure}[t!]
 \centering
\begin{subfigure}{.45\textwidth}
  \centering
  \includegraphics[width=\linewidth]{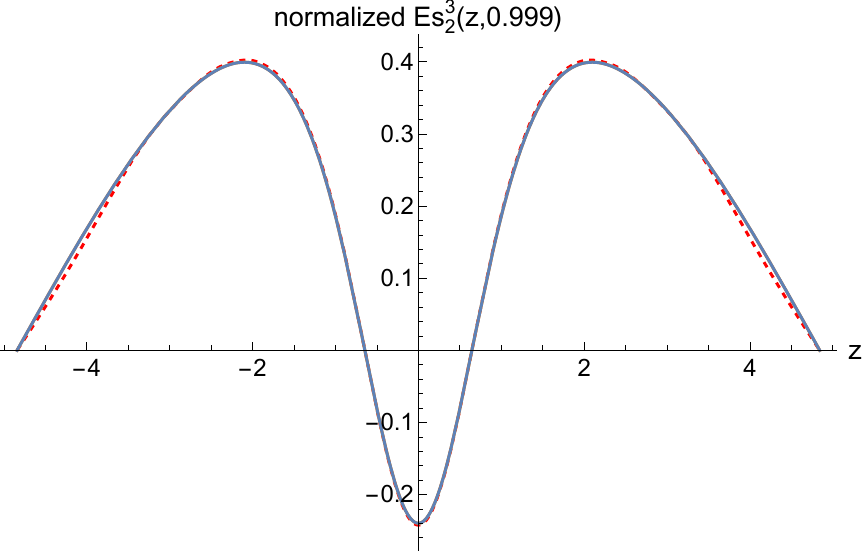}
  \caption{}
  \label{fig:LameEFs1}
\end{subfigure}  \qquad %
\begin{subfigure}{.45\textwidth}
  \centering
  \includegraphics[width=\linewidth]{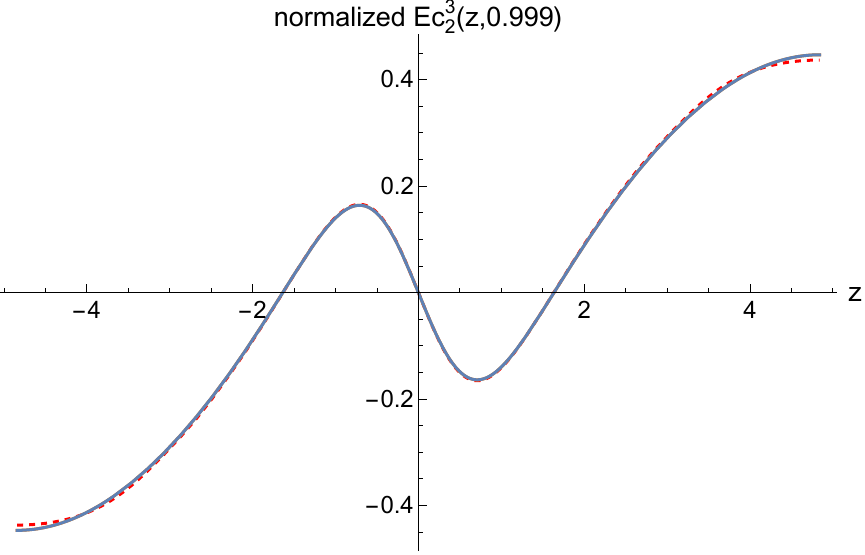}
  \caption{}
  \label{fig:LameEFs2}
\end{subfigure}  \\[1ex]
\begin{subfigure}{.45\textwidth}
  \centering
  \includegraphics[width=\linewidth]{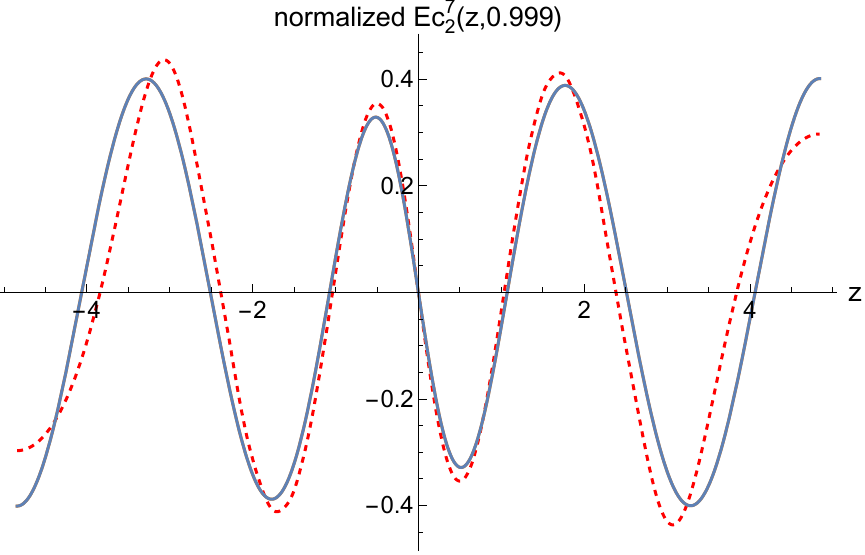}
  \caption{}
  \label{fig:LameEFs3}
\end{subfigure}  \qquad %
\begin{subfigure}{.45\textwidth}
  \centering
  \includegraphics[width=\linewidth]{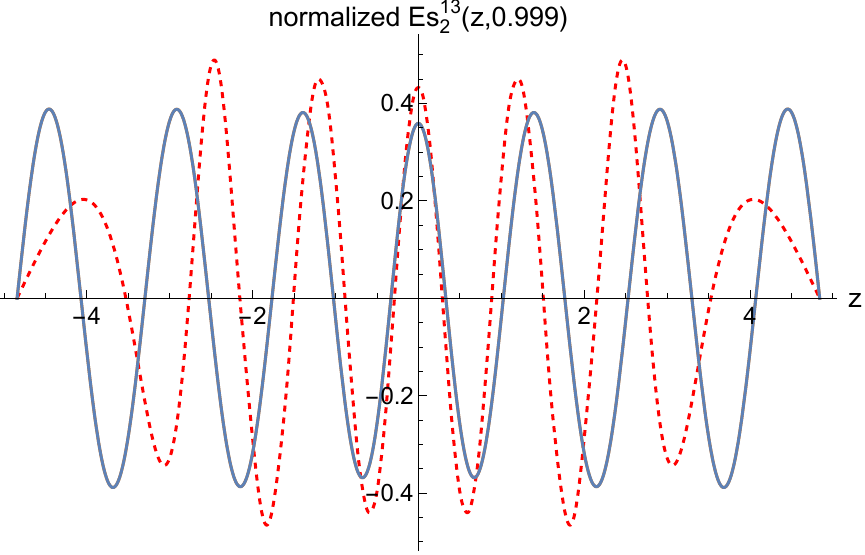}
  \caption{}
  \label{fig:LameEFs4}
\end{subfigure}  
\caption{Anti-periodic eigenfunctions of the $\nu_{\rm L} = 2$ Lam\'e equation with $\rmk^2 = 0.999$, corresponding to transcendental eigenvalues $b_{2}^{3} = a_{2}^{3}$, $a_{2}^7$, and $b_{2}^{13}$.  The orange curve is our custom one obtained by numerical integration with the corresponding eigenvalues obtained from roots of $\Delta + 2$ and normalized to have $L^2$ norm equal to one with respect to $\rho \in [-L/2,L/2]$.  The blue curve is the large $mL$ approximation \eqref{phi4approxmodes}.  These two curves are indistinguishable at this value of $\rmk^2$, and the blue curve is on the top orange curve.  The difference between the two is bounded by $10^{-4}$.  The red dashed curve is Mathematica's built-in Lam\'e eigenfunction, normalized in the same way.}
\label{fig:LameEFs}
\end{figure}

Mathematica's issue with the eigenvalues for $\rmk$ close to one appears to also be affecting its built-in Lam\'e eigenfunctions.  Therefore, with the eigenvalue in hand, we numerically integrate the differential equation \eqref{Lamediffeq} to find the eigenfunctions.  In Figure \ref{fig:LameEFs} we compare these ``custom'' eigenfunctions to Mathematica's built-in ones and to the large $mL$ approximations obtained in \eqref{phi4approxmodes}, \eqref{sGapproxmodes} based on the continuum scattering modes.  At the lowest transcendental mode number, the custom eigenfunction agrees well with Mathematica's built-in one.  This is consistent with the agreement of eigenvalues in Table \ref{table:LameEVs}.  At moderate mode numbers, however, the custom and built-in versions are drastically different.  Meanwhile the agreement between the custom eigenfunction and its large $mL$ approximation is excellent.

%%%%%%%%%%%%%%%%% 
%%%%%%%%%%%%%%%%%%

%%%%%%%%%%%%%%%%%%%%
%%%%%%%%%%%%%%%%%%%%
\section{Boosted Normal Modes}\label{app:boostedmodes}
%%%%%%%%%%%%%%%%%%%%
%%%%%%%%%%%%%%%%%%%%

The form of the quadratic fluctuation Hamiltonian obtained in \cite{Melnikov:2020ret} and restricted to the case of a time-independent background with kink velocity $\beta_i$ is
\begin{align}\label{HHblockform}
\HH_i =&~ \left( \PP_{\uppsi_0}^{\perp} \otimes \mathbbm{1}_2 \right) \left( \begin{array}{c c} \MM & \BB \\ \BB^T & \KK \end{array} \right) \left( \PP_{\uppsi_0}^{\perp} \otimes \mathbbm{1}_2 \right)~,
\end{align}
where
\begin{align}\label{HHblocks}
\MM =&~ \mathbbm{1} + \frac{ | \phi_{\beta_i}' \rangle \langle \phi_{\beta_i}' | }{\langle \uppsi_0 | \phi_{\beta_i}' \rangle^2} ~, \cr
\BB =&~ \beta_i \left( \pd_\rho + \frac{ | \phi_{\beta_i}' \rangle \langle \phi_{\beta_i}'' |}{ \langle \uppsi_0 | \phi_{\beta_i}' \rangle^2} + \frac{ |\phi_{\beta_i}' \rangle \langle \uppsi_0' | }{\langle \uppsi_0| \phi_{\beta_i}' \rangle} \right)~, \cr
\KK =&~  - \pd_{\rho}^2 + V^{(2)}(\phi_{\beta_i}) + \beta_{i}^2 \left( \frac{ |\phi_{\beta_i}'' \rangle \langle  \phi_{\beta_i}'' | }{ \langle \uppsi_0 |\phi_{\beta_i}' \rangle^2} + \frac{ | \phi_{\beta_i}''  \rangle \langle \uppsi_0' | + |\uppsi_0' \rangle \langle \phi_{\beta_i}''  | }{\langle \uppsi_0 |\phi_{\beta_i}' \rangle} \right)~. 
\end{align}
To ease notation in the remainder of this appendix we will set $\beta_i \equiv \beta$.

%%%%%%%%%%%%%%%%%%%%%
\subsection{Reduction to a Twisted Form of the $\beta_i = 0$ Eigenvalue Problem}
%%%%%%%%%%%%%%%%%%%%%

We are interested in the eigenvalue problem
\begin{equation}\label{HHevproblem}
\ii \JJ \HH_i | \eta \rangle  = \nu |\eta \rangle ~,
\end{equation}
for the Hermitian operator $\ii \JJ \HH_i$ with positive-definite symmetric $\HH_i$.  Williamson's theorem tells us that the eigenvalues come in pairs $\pm \nu_a$, $\nu_a > 0$, with complex conjugate eigenvectors $|\eta_a \rangle, |\eta_{a}^\ast \rangle$, where $|\eta_a \rangle$ corresponds to eigenvalue $+\nu_a$.  These eigenvectors can be chosen to satisfy the symplectic orthonormality conditions \eqref{symorth}, which are what ensure a set of canonically normalized normal modes for fluctuations around the boosted kink.  

One can check that
\begin{equation}
\widetilde{\MM} = \mathbbm{1} - \frac{ | \phi_{\beta}' \rangle \langle \phi_{\beta}' | }{\langle \phi_{\beta}' | \phi_{\beta}' \rangle} ~,
\end{equation}
is an inverse to $\MM$ on $( \mathrm{Span}\{ \uppsi_0\})^\perp$:
\begin{equation}
( \PP_{\uppsi_0}^\perp \MM \PP_{\uppsi_0}^\perp)( \PP_{\uppsi_0}^\perp \widetilde{\MM} \PP_{\uppsi_0}^\perp) = \PP_{\uppsi_0}^\perp ~.
\end{equation}
Hence the second row of \eqref{HHevproblem} is solved by taking
\begin{equation}\label{eta1eliminate}
|\eta_{\varpi} \rangle = - \PP_{\uppsi_0}^\perp \widetilde{\MM} \PP_{\uppsi_0}^\perp ( \BB + \ii \nu) | \eta_\varphi \rangle~,
\end{equation}
assuming $\langle \uppsi_0 | \eta_\varphi \rangle = 0$.  Inserting this into the top row equation leads to a second order equation for $|\eta_\varphi\rangle$:
\begin{equation}\label{eta2equation}
0 = \PP_{\uppsi_0}^\perp \left[ \KK - (\BB^T - \ii \nu) \PP_{\uppsi_0}^\perp \widetilde{\MM} \PP_{\uppsi_0}^\perp (\BB + \ii \nu) \right] \PP_{\uppsi_0}^\perp | \eta_2 \rangle \equiv \PP_{\uppsi_0}^\perp \Delta^{(\nu)} \PP_{\uppsi_0}^\perp | \eta_\varphi \rangle ~,
\end{equation}
where in the second step we defined $\Delta^{(\nu)}$ as the operator in the square brackets.

The operator $\Delta^{(\nu)}$ turns out to be a rank-one modification of a local operator $\Delta_{\rm loc}^{(\nu)}$ of the form
\begin{equation}
\Delta^{(\nu)} = \Delta_{\rm loc}^{(\nu)} - \frac{ \Delta_{\rm loc}^{(\nu)} |  \phi_{\beta}' \rangle \langle  \phi_{\beta}' | \Delta_{\rm loc}^{(\nu)} }{ \langle \phi_{\beta}' | \Delta_{\rm loc}^{(\nu)} | \phi_{\beta}' \rangle} ~,
\end{equation}
where
\begin{equation}\label{Deltaloc}
\Delta_{\rm loc}^{(\nu)} = - (1 - \beta^2) \pd_{\rho}^2 + 2 \ii \nu \beta \pd_\rho + V^{(2)}(\phi_{\beta}) - \nu^2 ~,
\end{equation}
and we used that
\begin{equation}\label{Deltaloconbzm}
\Delta_{\rm loc}^{(\nu)} | \phi_{\beta}' \rangle = (\nu^2 - 2\ii \nu \beta \pd_\rho ) | \phi_{\beta}' \rangle ~, \qquad \langle \phi_{\beta}' | \Delta_{\rm loc}^{(\nu)} | \phi_{\beta}' \rangle = \nu^2 \langle \phi_{\beta}' | \phi_{\beta}' \rangle ~.
\end{equation}
We see that $\Delta^{(\nu)}$ annihilates $|\phi_{\beta}' \rangle$ for any $\nu$.  Furthermore if $|\psi_\nu \rangle$ satisfies
\begin{equation}\label{bmodeequation}
\Delta_{\rm loc}^{(\nu)} | \psi_\nu \rangle = 0~,
\end{equation}
it will also be annihilated by $\Delta^{(\nu)}$.  Hence for each such $|\psi_\nu \rangle$ we obtain a solution $|\eta_{\varphi\nu}\rangle$ to \eqref{eta2equation} satisfying $\langle \uppsi_0 | \eta_{\varphi\nu} \rangle = 0$ by taking
\begin{equation}\label{eta2solution}
|\eta_{\varphi\nu} \rangle = |\psi_\nu \rangle - \frac{\langle \uppsi_0 | \psi_\nu \rangle}{\langle \uppsi_0 | \phi_{\beta}' \rangle} |\phi_{\beta}' \rangle ~.
\end{equation}
Back substituting into \eqref{eta1eliminate}, we find after some simplifications,
\begin{equation}
|\eta_{\varpi\nu} \rangle = - \PP_{\uppsi_0}^\perp \left[  (\beta \pd_\rho + \ii \nu) |\psi_\nu \rangle - \beta \frac{\langle \uppsi_0 | \psi_\nu \rangle}{\langle \uppsi_0 | \phi_{\beta}' \rangle} | \phi_{\beta}'' \rangle \right] ~.
\end{equation}

Thus the eigenvalue problem has been reduced to the equation \eqref{bmodeequation}, but what are the boundary conditions we should impose on its solutions?  This is what will determine the spectrum.  To answer this we need to enforce symplectic orthonormality \eqref{symorth}.  We start with the second of these.  Let $|\eta_a \rangle \equiv |\eta_{\nu_a} \rangle$ and $|\eta_b \rangle \equiv |\eta_{\nu_b} \rangle$ be two solutions corresponding to possibly different $\nu_{a,b} > 0$.  Then
\begin{align}\label{orthsym2a}
\langle \eta_{a}^\ast | \ii \JJ | \eta_b \rangle =&~ \ii \int \ed \rho \left\{ \eta_{\varphi a}(\rho) \eta_{\varpi b}(\rho) - \eta_{\varpi a}(\rho) \eta_{\varphi b}(\rho) \right\} \cr
=&~ \ii  \int d\rho \bigg\{ \beta (\psi_{a}' \psi_b - \psi_b' \psi_a) + \ii (\nu_a - \nu_b) \psi_a \psi_b + \cr
&~ \qquad \qquad + \frac{\langle \uppsi_0 | \psi_b \rangle}{\langle \uppsi_0 | \phi_{\beta}'\rangle} \psi_a \left( 2\beta \phi_{\beta}''  - \ii \nu_a \phi_{\beta}' \right) - \frac{\langle \uppsi_0 | \psi_a \rangle}{\langle \uppsi_0 | \phi_{\beta}'\rangle} \psi_b \left( 2 \beta \phi_{\beta}'' - \ii \nu_b \phi_{\beta}' \right) + \cr
&~ \qquad \qquad + \pd_\rho \left[ \frac{\beta \phi_{\beta}'}{\langle \uppsi_0 | \phi_{\beta}' \rangle} ( \langle \uppsi_0 | \psi_a \rangle \psi_b - \langle \uppsi_0 | \psi_b \rangle \psi_a ) \right] \bigg\} ~,
\end{align}
where terms proportional to $\langle \phi_{\beta}' | \phi_{\beta}'' \rangle$ were dropped since this is the integral of the total derivative of a periodic function whether $\phi_{\beta}'$ itself is periodic or anti-periodic.  

To proceed we use two results involving $\Delta_{\rm loc}^{(\nu_a)}$.  First, it follows from \eqref{Deltaloconbzm} and \eqref{bmodeequation} that
\begin{align}
\psi_a (2 \beta \phi_{\beta}'' - \ii \nu_a \phi_{\beta}' ) =&~ \frac{\ii}{\nu_a} \psi_a (\Delta_{\rm loc}^{(\nu_a)} \phi_{\beta}')^\ast  \cr
=&~ \frac{\ii}{\nu_a} \pd_\rho \left[ (1 - \beta^2) (\psi_{a}' \phi_{\beta}' - \psi_a \phi_{\beta}'') - 2\ii \beta \nu_a \psi_a \phi_{\beta}' \right]~.
\end{align}
Second we have, on the one hand,
\begin{equation}
\psi_a (\Delta_{\rm loc}^{(\nu_b)} \psi_b) - \psi_b (\Delta_{\rm loc}^{(\nu_a)} \psi_a) = 0~,
\end{equation}
while on the other
\begin{align}
\psi_a (\Delta_{\rm loc}^{(\nu_b)} \psi_b) - \psi_b (\Delta_{\rm loc}^{(\nu_a)} \psi_a) =&~  (1- \beta_{i}^2) \pd_\rho (\psi_b \psi_a' - \psi_b' \psi_a) - \ii \beta (\nu_a - \nu_b) \pd_\rho (\psi_a \psi_b) + \cr
&~ + \ii (\nu_a + \nu_b) \left[ \beta (\psi_a \psi_b' - \psi_b \psi_a') - \ii (\nu_a - \nu_b) \psi_a \psi_b \right] ~.
\end{align}
Hence
\begin{align}
& \beta (\psi_a' \psi_b - \psi_b' \psi_a) + \ii (\nu_a - \nu_b) \psi_a \psi_b =  \cr
& \qquad \qquad = - \frac{\ii}{(\nu_a + \nu_b)} \pd_\rho \left[ (1-\beta^2) (\psi_a' \psi_b - \psi_b' \psi_a) - \ii \beta (\nu_a - \nu_b) \psi_a \psi_b \right]~.
\end{align}
Thus the entire integrand of \eqref{orthsym2a} is a total derivative.  Explicitly:
\begin{align}
\langle \eta_{a}^\ast | \ii \JJ | \eta_b \rangle =&~ \int d\rho \pd_\rho \bigg\{ \frac{1}{(\nu_a + \nu_b)} \left[ (1-\beta_{i}^2) (\psi_a' \psi_b - \psi_b' \psi_a) - \ii \beta (\nu_a - \nu_b) \psi_a \psi_b \right] + \cr
&~ \qquad \qquad + \frac{\langle \uppsi_0 | \psi_a \rangle}{\nu_b \langle \uppsi_0 | \phi_{\beta}' \rangle} \left[ (1-\beta^2) (\psi_{b}' \phi_{\beta}' - \psi_b \phi_{\beta}'') - \ii \beta \nu_b \psi_b \phi_{\beta}' \right] + \cr
&~ \qquad \qquad -   \frac{\langle \uppsi_0 | \psi_b \rangle}{\nu_a \langle \uppsi_0 | \phi_{\beta}' \rangle} \left[ (1-\beta^2) (\psi_{a}' \phi_{\beta}' - \psi_a \phi_{\beta}'') - \ii \beta \nu_a \psi_a \phi_{\beta}' \right]  \bigg\} ~.
\end{align}

This expression will vanish if $\psi_a$ carries the same periodicity type as $\phi_{\beta}'$ for all $a$.  In particular, for $\phi^4$-theory the $\psi_a$ should be taken \emph{anti-periodic}.  In the past these $\psi_a$ have been taken periodic for $\phi^4$-theory \cite{Jain:1990dq,Melnikov:2020ret}.  Since $\phi_{\beta}'$ is exponentially small at $\rho \to \pm L/2$ for large $mL$, the last two lines vanish automatically, up to exponentially small corrections.  Therefore, we have $\langle \eta_a | \ii \JJ | \eta_b \rangle \approx 0$ up to such corrections if we take $\psi_a$ periodic for all $a$.  However we see from this analysis that anti-periodic conditions are the natural choice for $\phi^4$ theory at finite $mL$, while periodic conditions are the natural choice for sine-Gordon:
\begin{align}\label{bmodeperiodicity}
\textrm{$\phi^4$-theory:} \quad \psi_a(\rho + L) = - \psi_a(\rho)~, \qquad \textrm{sine-Gordon:} \quad \psi_a(\rho + L) = \psi_a(\rho)~,
\end{align}
in which case we have
\begin{equation}
\langle \eta_{a}^\ast | \ii \JJ | \eta_b \rangle = 0~.
\end{equation}

Now it remains to show that, with these periodicity conditions, the $|\eta_a\rangle$ can be chosen to satisfy the first of \eqref{symorth}.  Proceeding similarly, we now have
\begin{align}\label{symorth1a}
\langle \eta_{a} | \ii \JJ | \eta_b \rangle =&~ \ii \int \ed \rho \left\{ \eta_{\varphi a}(\rho)^\ast \eta_{\varpi b}(\rho) - \eta_{\varpi a}(\rho)^\ast \eta_{\varphi b}(\rho) \right\} \cr
=&~   \ii  \int d\rho \bigg\{ \beta (\psi_{a}^{\prime \ast} \psi_b - \psi_b' \psi_{a}^\ast) - \ii (\nu_a + \nu_b) \psi_{a}^\ast \psi_b + \cr
&~ \qquad \qquad + \frac{\langle \uppsi_0 | \psi_b \rangle}{\langle \uppsi_0 | \phi_{\beta}'\rangle} \psi_{a}^\ast \left( 2\beta \phi_{\beta}''  + \ii \nu_a \phi_{\beta}' \right) - \frac{\langle \uppsi_0 | \psi_{a}^\ast \rangle}{\langle \uppsi_0 | \phi_{\beta}'\rangle} \psi_b \left( 2 \beta \phi_{\beta}'' - \ii \nu_b \phi_{\beta}' \right) + \cr
&~ \qquad \qquad + \pd_\rho \left[ \frac{\beta \phi_{\beta}'}{\langle \uppsi_0 | \phi_{\beta}' \rangle} ( \langle \uppsi_0 | \psi_{a}^\ast \rangle \psi_b - \langle \uppsi_0 | \psi_b \rangle \psi_{a}^\ast ) \right] \bigg\}~.
\end{align}
By analogous manipulations we find that the second line can be written as a total derivative, while for the first line we use
\begin{align}
0 =&~ \psi_{a}^\ast (\Delta_{\rm loc}^{(\nu_b)} \psi_b) - \psi_b (\Delta_{\rm loc}^a \psi_{a})^\ast \cr
=&~  (1- \beta^2) \pd_\rho (\psi_b \psi_{a}^{\prime \ast} - \psi_b' \psi_{a}^\ast) + \ii \beta (\nu_a + \nu_b) \pd_\rho (\psi_{a}^\ast \psi_b) + \cr
&~ \qquad + \ii (\nu_a - \nu_b) \left[ \beta (\psi_b \psi_{a}^{\prime \ast} - \psi_{a}^\ast \psi_b') - \ii (\nu_a + \nu_b) \psi_a \psi_b \right] ~.
\end{align}
If $\nu_a \neq \nu_b$ then this shows the first line can be written as a total derivative as well.  Hence, with \eqref{bmodeperiodicity} we conclude
\begin{equation}
\langle \eta_a | \ii \JJ | \eta_b \rangle = 0~, \qquad \textrm{for $a \neq b$}~.
\end{equation}
When $a = b$ the last two lines of \eqref{symorth1a} still vanish so that
\begin{equation}\label{bmodenormcondition}
\langle \eta_a | \ii \JJ | \eta_a \rangle = \int \ed \rho \left\{ \ii \beta (\psi_{a}^{\prime \ast} \psi_a - \psi_{a}' \psi_{a}^\ast ) + 2\nu_a \psi_{a}^\ast \psi_a \right\} ~.
\end{equation}
Requiring this to be equal to one gives the normalization condition for $\psi_a$.

Now let us return to \eqref{bmodeequation}.  Observe that $\Delta_{\rm loc}^{(\nu)}$ can be written in the form 
\begin{align}
\Delta_{\rm loc}^{(\nu)} =&~  e^{\ii \gamma^2 \beta \nu (\rho - \rho_0)}  \left[ - (1-\beta^2) \pd_{\rho}^2 + V^{(2)}(\phi_{\beta}(\rho)) - \gamma^2 \nu^2 \right] e^{-\ii \gamma^2 \beta \nu (\rho - \rho_0)} ~.
\end{align}
Here $\rho_0$ is the same constant that appears in the general relationship between the boosted and static kink profiles as discussed under \eqref{boostedviastatic}:
\begin{equation}
\phi_{\beta}(\rho) = \phi_0(\gamma(\rho - \rho_0); \gamma L) ~.
\end{equation}
This constant happens to vanish for $\phi^4$-theory and sine-Gordon.  The operator in square-brackets has the form of the fluctuation operator around the static kink, but in terms of the Lorentz contracted variable $\gamma (\rho-\rho_0)$, Lorentz contracted parameter $\gamma L$, and eigenvalue $\widetilde{\upomega} = \gamma \nu$.  Hence we can write our solutions to $\Delta_{\rm loc}^{(\nu_a)} |\psi_a \rangle = 0$ in the form
\begin{align}\label{bmodefromsmode}
\psi_{a}(\rho) = e^{\ii \gamma^2 \beta \nu_a (\rho - \rho_0)} \widetilde{\uppsi}_a(\gamma(\rho-\rho_0))~,
\end{align}
where
\begin{equation}\label{nufromomega}
\nu_a = \gamma^{-1} \widetilde{\upomega}_a ~,
\end{equation}
and $\widetilde{\upomega}_a, \widetilde{\uppsi}_a(y)$ solve the eigenvalue problem
\begin{equation}\label{twistedevproblem}
\left[ - \pd_{y}^2 + V^{(2)}(\phi_0(y;\gamma L)) \right] \widetilde{\uppsi}_a(y) = \widetilde{\upomega}_{a}^2 \widetilde{\uppsi}_{a}(y)~,
\end{equation}
where $y \equiv \gamma (\rho - \rho_0)$.  

Although the form of the operator appearing in \eqref{twistedevproblem} is the same as that for fluctuations around the static kink, the eigenvalue problem is not because the boundary conditions on $\widetilde{\uppsi}_a$ are different.  Given \eqref{bmodeperiodicity} and \eqref{bmodefromsmode}, we infer that the $\widetilde{\uppsi}_a$ satisfy twisted periodicity conditions which depend on the eigenvalue and the background kink velocity:
\begin{align}\label{twistedperiodicity}
 \widetilde{\uppsi}_a(y + \gamma L) =&~  e^{\ii \mu_a} \widetilde{\uppsi}_a(y) ~,
 \end{align}
where
\begin{equation}\label{twist}
e^{\ii \mu_a} = \left\{ \begin{array}{l l} -e^{- \ii \beta \gamma \widetilde{\upomega}_a L}~,  & \textrm{$\phi^4$-theory}~, \\ e^{-\ii \beta \gamma  \widetilde{\upomega}_a L}~, & \textrm{sine-Gordon}~. \end{array} \right.
\end{equation}
Equations \eqref{twistedevproblem} and \eqref{twistedperiodicity} with \eqref{twist} give a complete specification of the eigenvalue problem that determines the spectrum and normal modes around the boosted kink.

Returning to the normalization condition \eqref{bmodenormcondition} and using \eqref{bmodefromsmode}, \eqref{nufromomega}, we find
\begin{equation}
\ii \beta (\psi_{a}^{\prime \ast} \psi_a - \psi_{a}^\ast \psi_{a}') = 2 \beta^2\gamma \widetilde{\upomega}_a  \widetilde{\uppsi}_{a}^\ast \widetilde{\uppsi}_a + \ii \beta \left( \widetilde{\uppsi}_{a}^{\prime \ast} \widetilde{\uppsi}_{a} - \widetilde{\uppsi}_{a}^\ast \widetilde{\uppsi}_{a}' \right)~,
\end{equation}
where the prime on $\widetilde{\uppsi}_a$ still denotes differentiation with respect to $\rho$.  This combines with the other term to give
\begin{align}
\langle \eta_a | \ii \JJ | \eta_a \rangle =&~ \int_{-\frac{L}{2} + \rho_0}^{\frac{L}{2} + \rho_0} \ed \rho \left\{ 2 \gamma \widetilde{\upomega}_a \widetilde{\uppsi}_{a}^\ast \widetilde{\uppsi}_a + \ii \beta \left( \widetilde{\uppsi}_{a}^{\prime \ast} \widetilde{\uppsi}_{a} - \widetilde{\uppsi}_{a}^\ast \widetilde{\uppsi}_{a}' \right) \right\} ~,
\end{align}
where we chose a length $L$ domain of integration centered on $\rho_0$ for the periodic integrand.  Now change variables to $y = \gamma (\rho - \rho_0)$.  Then we demand
\begin{align}\label{uppsitildenorm}
1 \stackrel{!}{=}  \langle \eta_a | \ii \JJ | \eta_a \rangle =&~ \int_{-\gamma L/2}^{\gamma L/2} \ed y \left\{ 2 \widetilde{\upomega}_a \widetilde{\uppsi}_{a}^\ast \widetilde{\uppsi}_a + \ii \beta W(\widetilde{\uppsi}_a, \widetilde{\uppsi}_{a}^\ast) \right\}~,
\end{align}
where $W(f,g)(y) = f \pd_y g - g \pd_y f$ is the Wronskian.  

%%%%%%%%%%%%%%%%%%%%%
\subsection{Twisted Lam\'e Eigenvalues}\label{app:twistedEVs}
%%%%%%%%%%%%%%%%%%%%%

The eigenvalue problem we just arrived at is exactly of the form discussed in Appendix \ref{app:Lame}.  Evaluating the potential term in \eqref{twistedevproblem}, we find the Lam\'e equation
\begin{equation}\label{betaLame}
\pd_{z}^2 \widetilde{\uppsi} + \left[ h_{\beta} - \nu_{\rm L} (\nu_{\rm L} + 1) \rmk_{\beta}^2  \sn^2(z,\rmk_{\beta}^2) \right] \widetilde{\uppsi} = 0~,
\end{equation}
where the relationship between parameters and variables is
\begin{align}\label{parametermap}
\textrm{$\phi^4$-theory:}  \quad  z =&~ \frac{m y}{\sqrt{1 + \rmk_{\beta}^2}} = \frac{\gamma m (\rho - \rho_0)}{\sqrt{1 + \rmk_{\beta}^2}} ~, \qquad \nu_{\rm L} = 2~, \cr
h_{\beta} =&~ (1 + \rmk_{\beta}^2) \left( \frac{\widetilde{\upomega}^2}{m^2} + 1 \right)~,  \qquad  \gamma m L  = 2 \sqrt{1 + \rmk_{\beta}^2} \mathbf{K}(\rmk_{\beta}^2)~,   \cr
\textrm{sine-Gordon:} \quad z =&~ \frac{m y}{\rmk_{\beta}} = \frac{ \gamma m  (\rho - \rho_0)}{\rmk_{\beta}}~, \qquad \nu_{\rm L} = 1~,  \cr
h_{\beta} =&~ \rmk_{\beta}^2 \left( \frac{\widetilde{\upomega}^2}{m^2} + 1 \right)~, \qquad  \gamma m L = 2\rmk_{\beta} \mathbf{K}(\rmk_{\beta}^2) ~.
\end{align}
From the discussion in Appendix \ref{app:Lame} we know that the spectrum of the eigenvalue problem with twisted periodicity condition is obtained from the solutions to
\begin{equation}\label{twistedquantcon}
\Delta_{\nu_{\rm L}}(h_{\beta},\rmk_{\beta}^2) = 2 \cos(\mu)~,
\end{equation}
where $\Delta$ is the Hill discriminant.  In our context, $\mu$ itself is a function of the Lam\'e eigenvalue $h_\beta$, elliptic modulus $\rmk_{\beta}$, and $\beta$.  From \eqref{twist} and \eqref{parametermap} we have
\begin{align}
\mu = \mu(h_\beta,\rmk_{\beta}^2,\beta) =&~ \left\{ \begin{array}{l l} 2 \beta \mathbf{K}(\rmk_{\beta}^2) \sqrt{ h_\beta - (1 + \rmk_{\beta}^2) } + \pi ~, & \textrm{$\phi^4$-theory}~, \\ 2 \beta \mathbf{K}(\rmk_{\beta}^2) \sqrt{ h_\beta - \rmk_{\beta}^2} ~, & \textrm{sine-Gordon}~. \end{array} \right.
\end{align}

As in the case of periodic/anti-periodic boundary conditions, we can use the large $mL$ approximation \eqref{Deltaapprox} for the Hill discriminant to find the approximate (twisted version of the) transcendental eigenvalues.  These serve as the initial guess for a numerical approach to the exact transcendental eigenvalues.  In the leading approximation we replace $\Delta$ with $2 \cos(2 \mathbf{K}(\rmk_{\beta}^2)\kappa_\beta + \delta(\kappa_\beta))$, where $\kappa_\beta \equiv \sqrt{h_{\beta} - \nu_{\rm L} (\nu_{\rm L} + 1)}$ is positive real.  This means $h_{\beta}$ is necessarily in the semi-infinite allowed zone, and hence we can only find the transcendental eigenvalues using this approximation.  The quantization condition \eqref{twistedquantcon} then takes the form $\cos(A) = \cos(B)$, up to the error indicated in \eqref{Deltaapprox}.  This has two infinite sequences of solutions: $A = \pm B + 2\pi n$, $n \in \mathbbm{Z}$.  Explicitly,
\begin{align}\label{boostedquantcon1}
 \textrm{$\phi^4$-theory:} \quad  2 \mathbf{K}(\rmk_{\beta}^2) \kappa_\beta + \delta_{\phi^4}(\kappa_\beta) =&~  \pm \left(2\sqrt{2} \mathbf{K}(\rmk_{\beta}^2) \beta \sqrt{ \frac{\kappa_{\beta}^2}{2} + 2} + \pi \right)  + 2\pi n  + \cr
 &~ +  O\left(|\kappa_\beta|^{-1} \sqrt{mL} e^{-mL/(2\sqrt{2})}\right)~,  \cr
 \textrm{sine-Gordon:} \quad  2 \mathbf{K}(\rmk_{\beta}^2) \kappa_\beta + \delta_{\rm sG}(\kappa_\beta) =&~ \pm 2 \mathbf{K}(\rmk_{\beta}^2) \beta \sqrt{\kappa_{\beta}^2 + 1} + 2\pi n + \cr
 &~ +  O\left(|\kappa_\beta|^{-1} mL e^{-mL/2}\right)~,
\end{align}
for $n \in \mathbbm{Z}$.  

The two sequences corresponding to the two choices of sign are distinct for generic $\beta$; the background kink velocity has broken the two-fold degeneracy of the transcendental spectrum around the static kink.  Nevertheless, we can still account for both sequences by extending the allowed $\kappa_\beta$'s to the whole real line so that $|\kappa_\beta| = \sqrt{h_{\beta} - \nu_{\rm L} (\nu_{\rm L} + 1)}$, and then considering just one of the sequences.  In terms of finding the complete set of solutions either sign choice works, since we can get to the other sign choice by sending $(\kappa_\beta,n) \to (-\kappa_\beta,-n)$ and using the fact that we can choose $\delta(\kappa_\beta)$ to be an odd function of $\kappa_\beta$.  If, however, we want to relate $\kappa_\beta$ to a wave number $\widetilde{k}$ of the infinite $mL$ scattering modes as we did in Appendix \ref{app:Lame}, then the correct choice is the bottom sign, as we explain now.

With $\kappa_\beta = \widetilde{k}/m$ and $\kappa_\beta = \sqrt{2} \widetilde{k}/m$ in sine-Gordon and $\phi^4$-theory respectively, we use \eqref{parametermap} to put the quantization conditions \eqref{boostedquantcon1} into the form found in \cite{Melnikov:2020ret}, where the leading order condition was obtained by imposing twisted boundary conditions on the infinite $m L$ continuum spectrum\footnote{See formula (4.101) in \cite{Melnikov:2020ret}, which only considered the $\phi^4$-theory case.  The only difference is a relative shift by $\pi$ due to anti-periodic boundary conditions for the $\psi_a$ (here) versus periodic ones (there).  As we discussed above, anti-periodic conditions are more natural, but the difference between the two choices for the computation of the one-loop kink energy in \cite{Melnikov:2020ret} vanishes exponentially fast in $1/mL$.}:
\begin{equation}\label{boostedquantcon2}
\delta(\widetilde{k}) = y_{n}(\widetilde{k}) + \textrm{exponentially small}
\end{equation}
with the phase shift given in \eqref{phi4phase} or \eqref{sGphase} and $y_{n}(\widetilde{k})$ being one of the two possibilities
\begin{align}\label{boostedyns}
y_{n}^{\pm}(\widetilde{k}) = \left\{ \begin{array}{l l}  2\pi n \pm \pi - L \gamma (\widetilde{k} \mp \beta \sqrt{\widetilde{k}^2 + 2m^2})~, &  \textrm{$\phi^4$-theory}~, \\ 2\pi n - L \gamma (\widetilde{k} \mp \beta \sqrt{ \widetilde{k}^2 + m^2})~, & \textrm{sine-Gordon}~. \end{array} \right.
\end{align}
Now we can decide on the sign by comparing \eqref{boostedquantcon2} with the condition for fluctuations around the static kink, \eqref{nu1quantcon} and \eqref{nu2quantcon}, and requiring the boosted and static results to be Lorentz covariant in the $mL \to \infty$ limit.  Note that \eqref{nu1quantcon} and \eqref{nu2quantcon} have the form \eqref{boostedquantcon2} with $\beta = 0, \widetilde{k} \to k$.  In order to recover the continuum spectrum in the $mL \to \infty$ limit we must simultaneously send $n \to \infty$ such that $2\pi n/(mL) \equiv k_{\rm lab}/m$ is any finite real number.  Then, since the $\delta(k)$ term drops out in the limit, the static quantization condition is $k = k_{\rm lab}$.  Here, the ``lab'' subscript denotes that, when $\beta = 0$, the co-moving frame is the lab frame: $\rho = x$.  The (approximate) frequency $\upomega^{\approx} \equiv \sqrt{k^2 + \mu^2}$ and wave number form an on-shell two-vector $k_{\rm lab}^\mu = (\upomega^{\approx}, k)$.  Now let $k_{\rm kink}^\mu$ be the two-vector of the same fluctuation in an inertial frame moving with velocity $\beta$ relative to the lab frame.  This is the co-moving frame of the kink when the kink has velocity $\beta$.  The two are related by $k_{\rm lab}^\mu = \Lambda^{\mu}_{~\nu} k_{\rm kink}^\mu$ with Lorentz transformation
\begin{equation}
\Lambda = \left( \begin{array}{c c} \gamma &  \beta \gamma \\ \beta \gamma & \gamma \end{array} \right)~,
\end{equation}
implying $k_{\rm lab} = \gamma \left( k_{\rm kink} + \beta \sqrt{ k_{\rm kink}^2 + \mu^2} \right)$.  If we compare this with \eqref{boostedquantcon2} in the limit where $\delta(\widetilde{k})$ is dropped, they agree upon identifying $\widetilde{k} = k_{\rm kink}$ as they should, provided we take the bottom sign in \eqref{boostedyns}.  Hence
\begin{equation}
y_n(\widetilde{k}) = y_{n}^-(\widetilde{k})~.
\end{equation}

As in the $\beta = 0$ case there will be one integer $n_0$ for sine-Gordon and two consecutive integers $n_0, n_1 = n_0 + 1$ for $\phi^4$-theory for which $\delta(\widetilde{k}) = y_n(\widetilde{k})$ has no solution.  In $\phi^4$-theory we find $n_0$ from the condition $ - 2\pi \leq y_{n_0}(0) < 0$, and in sine-Gordon we find it from $-\pi \leq y_{n_0}(k) < \pi$.  The solutions are
\begin{equation}\label{n0values}
\textrm{$\phi^4$-theory:} \quad n_0 = \left\lceil  \frac{\gamma \beta m L}{\sqrt{2} \pi} - \half \right\rceil ~, \qquad \textrm{sine-Gordon:} \quad n_0 = \left\lceil  \frac{\gamma \beta m L}{2\pi} - \half \right\rceil ~,
\end{equation}
where $\lceil x \rceil$ denotes the least integer greater or equal to $x$.  These reduce to $n_0 = 0$ when $\beta = 0$.  For all other $n \in \mathbbm{Z}$, $\delta(\widetilde{k}) = y_n(\widetilde{k})$ determines a $\widetilde{k}(n)$ that gives an approximate root $h_{\beta}^{\approx}(n)$ to \eqref{twistedquantcon} and an approximate frequency $\widetilde{\upomega}^{\approx}(n)$.  We then use Mathematica's FindRoot with a numerical construction of $\Delta$ to obtain the ``exact'' root $h_{\beta}(n)$.  The corresponding frequency $\widetilde{\upomega}(n)$ follows via the relation \eqref{parametermap}.  This determines the transcendental part of the twisted spectrum.

%%%%%%%%%%%%%%%%% 
 \begin{figure}[t!]
 \centering
\begin{subfigure}{.99\textwidth}
  \centering
  \includegraphics[width=\linewidth]{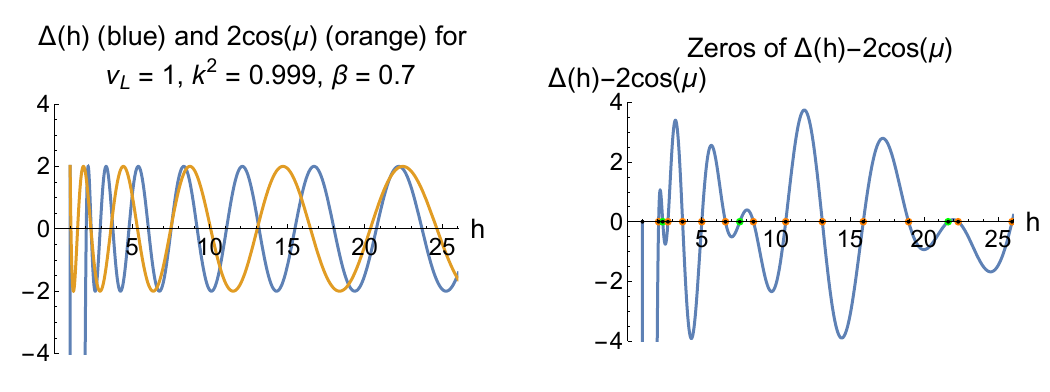}
  \caption{}
  \label{fig:TwistedLameEV1}
\end{subfigure}   \\[1ex]
\begin{subfigure}{.99\textwidth}
  \centering
  \includegraphics[width=\linewidth]{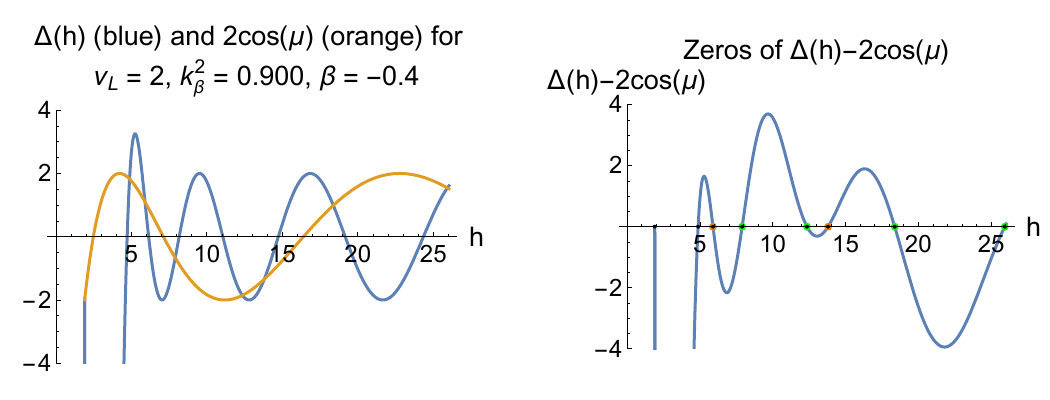}
  \caption{}
  \label{fig:TwistedLameEV2}
\end{subfigure} 
\caption{Roots of $\Delta(h) -2\cos(\mu)$ for two examples.  The black dots indicate the numerically determined root while the larger orange or green dots indicate the large $mL$ approximation used as a starting point for the root search.  The orange dots correspond to increasing mode numbers $n > n_0$ (sine-Gordon) or $n_1$ ($\phi^4$-theory) implying positive values of the wave number $\kappa_n$, while the green dots correspond to decreasing mode numbers $n < n_0$ implying negative values of the wave number $\kappa_n$.}
\label{fig:TwistedLameEVs}
\end{figure}
%%%%%%%%%%%%%%%%%%

For the ``algebraic'' part of the spectrum---that is, those eigenvalues lying in the finite-width allowed zones---the above approach does not apply.  The zero mode eigenvalue $\widetilde{\upomega}_0 = 0$ is still valid since the twist parameter reduces to the $\beta = 0$ periodicity condition for $\widetilde{\upomega}$.  This covers the full spectrum for sine-Gordon.  In $\phi^4$-theory there is one shape mode with an algebraic eigenvalue, and we need to find the twisted version of this eigenvalue.  The plots of the Hill discriminant in Figure \ref{fig:LameEVs} for the $\nu_{\rm L} = 2$ cases show that the width of the band containing this eigenvalue, where we must have $|\Delta| \leq 2$. is extremely small.  On physical grounds we expect that it is in fact exponentially small in $1/mL$.  Hence the $\beta = 0$ version of this eigenvalue should provide a good starting approximation for this root.  Hence we define $h_{\beta}^{\approx}(n_1) = 1 + 4 \rmk_{\beta}^2$.

The first several approximate $h_{\beta,a}^{\approx}$ and exact $h_{\beta,a}$ eigenvalues are compared in Table \ref{table:twistedEVs} for the example shown in Figure \ref{fig:TwistedLameEV1}.

%%%%%%%%%
\begin{table}[t!]
\begin{center}
\begin{tabular}{| c | c | c | c | c | c | c | c |} \hline
 $a = $  &  $1$  & $2$ & $3$ & $4$  & $5$ & $6$     \\ 
\hline \hline
$h_{\beta,a}$ & 2.0903 & 2.3680 & 2.7192 & 3.7081 & 5.0051 & 6.5993  \\
\hline 
$h_{\beta,a}^{\approx}$  & 2.0895 & 2.3664 & 2.7186 & 3.7076 & 5.0045 & 6.5987   \\
\hline
\end{tabular}
\end{center}
\caption{Twisted sine-Gordon eigenvalues for $\beta = 0.7$ and $L = 6.915$ corresponding to $\rmk_{\beta}^2 \approx 0.999$.}
\label{table:twistedEVs}
\end{table}
%%%%%%%%%
  
\begin{figure}[t!]
 \centering
\begin{subfigure}{.45\textwidth}
  \centering
  \includegraphics[width=\linewidth]{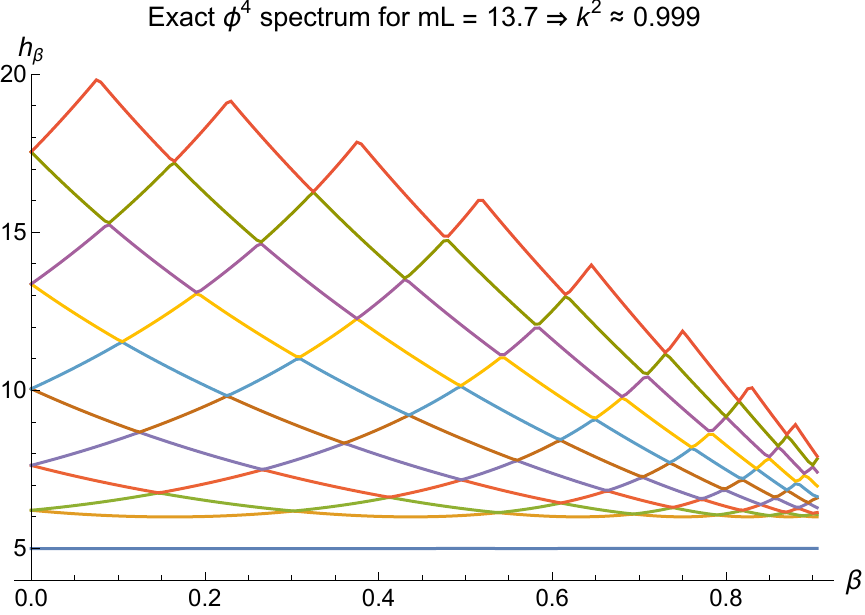}
  \caption{}
  \label{fig:TwistedEVbetadep1}
\end{subfigure}  \qquad %
\begin{subfigure}{.45\textwidth}
  \centering
  \includegraphics[width=\linewidth]{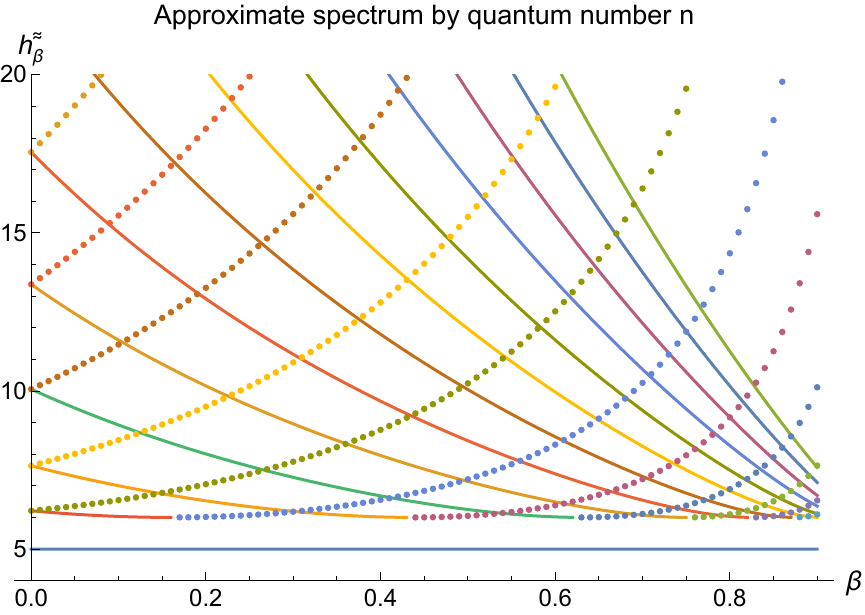}
  \caption{}
  \label{fig:TwistedEVbetadep2}
\end{subfigure} 
\caption{Dependence of the spectrum on $\beta$ for the $\nu_{\rm L} = 2$ case of $\phi^4$ theory.  The graph on the left is for the exact spectrum and the eigenvalues are color coded by increasing energy.  The graph on the right is for the approximate spectrum and the eigenvalues are color coded according to their $n$-value.  Reading from the lower-left corner to the upper-right, the solid curves correspond to $n_a = 2,3,4,\ldots$.  Reading from upper-left to lower-right, the dotted curves correspond to $n_a = -5,-4,-3,\ldots$.  When varying $\beta$ it is important to note that $\rmk_{\beta}^2$ also varies.  The graph is for fixed $\rmk^2 \approx 0.999$, corresponding to $mL = 2 \sqrt{1 + \rmk^2} \mathbf{K}(\rmk^2) = 13.7$.  Then $\rmk_{\beta}^2$ is found by solving $\gamma m L = 2 \sqrt{1 + \rmk_{\beta}^2} \mathbf{K}(\rmk_{\beta}^2)$.  Hence as $\beta$ increases so does $\rmk_{\beta}$, which leads to an overall compression of the transcendental spectrum.  We also plot the algebraic eigenvalue corresponding to the shape mode.  Its variation is too small to notice, but it increases from $1 + 4\rmk^2$ to $5$ as $\beta$ goes from 0 to 1.}
\label{fig:TwistedEVbetadep}
\end{figure}

As in the $\beta = 0$ case, we organize the spectrum $\{ \widetilde{\upomega}_a \}$ (or equivalently $\{ h_{\beta,a} \}$) by energies: $0 = \widetilde{\upomega}_0 < \widetilde{\upomega}_1 < \cdots$.   Figure \ref{fig:TwistedEVbetadep1} illustrates how the low-lying exact spectrum evolves with $\beta$ for fixed $L$.  Nonzero $\beta$ breaks the two-fold degeneracy that was present in the transcendental spectrum for $\beta = 0$.  One eigenvalue from each pair increases while the other initially decreases.  The decreasing eigenvalue appears to cross each increasing eigenvalue from the pairs below it until it bottoms out and begins increasing.  For the exact spectrum these crossings occur when a max or min of $2\cos(\mu(h_{\beta}))$ passes over a max or min of $\Delta(h_{\beta})$.

The detailed behavior of the spectrum is easily understood from the large $mL$ approximation which gives a nearly indistinguishable picture for the $\beta$ dependence of the approximate eigenvalues.  See Figure \ref{fig:TwistedEVbetadep2}.  The initially decreasing (increasing) eigenvalues are associated with wave numbers $k(n)$ that have the same (opposite) sign as $\beta$.  Therefore the smooth falling or rising curves correspond to fixed quantum numbers $n$.  Each initially falling curve becomes a rising curve when its wave number $k(n)$ passes through zero.  At this instant, the $n$-value jumps down by one (sine-Gordon) or two ($\phi^4$-theory) as the phase $\delta(k)$ jumps down by $2\pi$ or $4\pi$.  When $k$ smoothly passes through zero the approximate eigenvalue $\upomega^{\approx}$ touches the threshold of the mass gap.  The kink has caught up with the wave and passed it.  At the same instant the $n_0$ value increases by one according to \eqref{n0values}.

The falling eigenvalues undergo compression while the rising ones undergo rarefaction, leading to a greater and greater asymmetry in the number of $(+\hat{\beta})$- versus $(-\hat{\beta})$-movers as $|\beta|$ increases.  Since the index $a \in \mathbbm{N}_0$ always labels the energy spectrum in increasing order regardless of the value of $\beta$, the crossings mean that the bijection $\mathbbm{N}_0 \to \mathbbm{Z}$: $a \mapsto n_a$ depends nontrivially on $\beta$.  For example, the lowest sixteen eigenvalues shown in Figure \ref{fig:TwistedLameEV1} correspond to $\{n_0 , n_1, \ldots, n_{15}\} = \{1, 2, 0,3,4,5,6,-1,7,8,9,10,11,-2,12,13\}$.  As mentioned above, the quantum number $n$ is proportional to wave number $k$ in the lab frame where the kink is moving, (to the extent that we can ignore the phase shift $\delta(\widetilde{k})$ in the quantization condition \eqref{boostedquantcon2}).  Hence the spreading out of energy values for wave numbers with opposite sign to $\beta$ is due to redshifting.

%%%%%%%%%%%%%%%%% 
%%%%%%%%%%%%%%%%%%

%%%%%%%%%%%%%%%%%%
\subsection{Twisted Lam\'e Eigenfunctions}
%%%%%%%%%%%%%%%%%%

With the eigenvalues in hand we return to the differential equation to construct the eigenfunctions.  Rather than solve \eqref{betaLame} and impose twisted periodicity conditions, it is more convenient to solve \eqref{bmodeequation} and impose (anti-)periodic boundary conditions \eqref{bmodeperiodicity}.  Therefore we rewrite $\Delta_{\rm loc} | \psi \rangle = 0$ in terms of the variable $z$ and parameters $\beta, \rmk_{\beta}^2, h_{\beta}$ using \eqref{parametermap} and obtain the equations
\begin{align}\label{twistedLamediffeq}
0 =&~ \left\{ \pd_{z}^2 - 2 \ii \beta \sqrt{h_{\beta} - (1+ \rmk_{\beta}^2)} \pd_z + \left[ \gamma^{-2} h_{\beta} + \beta^2 (1+\rmk_{\beta}^2) - 6 \rmk_{\beta}^2 \sn^2(z, \rmk_{\beta}^2) \right] \right\} \psi ~, \cr
0 =&~ \left\{ \pd_{z}^2 - 2 \ii \beta \sqrt{h_{\beta} -\rmk_{\beta}^2} \pd_z+ \left[ \gamma^{-2} h_{\beta} + \beta^2 \rmk_{\beta}^2 - 2 \rmk_{\beta}^2 \sn^2(z, \rmk_{\beta}^2) \right] \right\} \psi ~, \raisetag{20pt}
\end{align}
for $\phi^4$-theory and sine-Gordon respectively.  These are one-parameter deformations of the $\nu_{\rm L} = 2$ and $1$ Lam\'e equations and reduce to them when $\beta \to 0$.  

The reflection symmetries of the ordinary Lam\'e equation about $z = \mathbf{K}(\rmk_{\beta}^2)$ and $z = 0$ discussed above \eqref{EcEsbreakdown} are broken by the deformation, due to the linear $\pd_z$ term.  However, if we combine each of these actions with conjugation $\psi \to \psi^\ast$ then the form of the equations is maintained.  Hence, while the two-fold degeneracy of eigenvalues we had previously is broken, most of the discussion about even and odd functions for periodicities $2\mathbf{K}$ and $4\mathbf{K}$ can be carried over provided we replace statements like ``even and $2\mathbf{K}$-periodic'' with ``even real part and odd imaginary part that are $2\mathbf{K}$ periodic.''  

The question of which parity condition to impose for a given eigenvalue $h_{\beta,a}$ is more easily answered after considering the large $mL$ approximation to these modes.  For modes corresponding to transcendental eigenvalues, this approximation is obtained by using \eqref{bmodefromsmode} with $\widetilde{\uppsi}_a$ taken to be the scattering modes in \eqref{flucspectrum} or \eqref{sGstaticmodes} evaluated at the wave number $\widetilde{k}_a \equiv \widetilde{k}(n_a)$ that satisfies the quantization condition $\delta(\widetilde{k}) = y_{n_a}^-(\widetilde{k})$ with \eqref{boostedyns}.  Here the identification $a \mapsto n_a$ is obtained from the ordering of the approximate energy eigenvalues $\widetilde{\upomega}_{a}^{\approx} \equiv \sqrt{\widetilde{k}_{a}^{2} + \mu^2}$, at the given $\beta$.  Thus we write
\begin{align}\label{boostedmodeapprox}
\psi_{a}^{\approx}(\rho) =&~ e^{\ii \gamma \beta \widetilde{\upomega}_{a}^{\approx} \rho} \widetilde{\uppsi}_{a}^{\approx}(\gamma \rho)  = \widetilde{N}_a e^{\ii \gamma \beta \widetilde{\upomega}_{a}^{\approx} \rho} \uppsi_{\underline{\widetilde{k}_{a}}}(\gamma \rho) ~.
\end{align}

We determine the new normalization constant by imposing \eqref{uppsitildenorm}.  The Wronskian in \eqref{uppsitildenorm} for the scattering modes $\uppsi_{\underline{\widetilde{k}_a}}(y)$ evaluates to
\begin{align}
W(\uppsi_{\underline{\widetilde{k}_a}}, \uppsi_{\underline{\widetilde{k}_a}}^\ast) =&~ - \frac{\ii \widetilde{k}_a}{\pi} \widetilde{N}_{a}^2~,
\end{align}
in both cases, and hence \eqref{uppsitildenorm} gives
\begin{equation}\label{Ntildefix1}
1 \approx \widetilde{N}_{a}^2 \left[  2 \gamma \widetilde{\upomega}_{a}^{\approx} \int_{-L/2}^{L/2} \ed \rho \uppsi_{\underline{\widetilde{k}_a}}(\gamma \rho)^\ast \uppsi_{\underline{\widetilde{k}_a}}(\gamma \rho) + \frac{\beta \gamma \widetilde{k}_a L}{\pi} \right] ~,
\end{equation}
up to exponentially small corrections in $1/(mL)$ since we replaced $\widetilde{\upomega}_a \to \widetilde{\upomega}_{a}^{\approx}$.  Using the result discussed around \eqref{phi4approxmodes}, the integral evaluates to $\frac{1}{2\pi\gamma} (\gamma L + \delta'(\widetilde{k}_a))$ up to exponentially small corrections.  Hence the normalization condition is
\begin{align}
1 \approx&~ \frac{\widetilde{N}_{a}^2}{\pi} \left\{ \widetilde{\upomega}_{a}^{\approx} \left[ \gamma L + \delta'(\widetilde{k}_a) \right] + \gamma \beta \widetilde{k}_a L \right\} = \frac{\widetilde{N}_{a}^2}{\pi} \left[ \gamma (\widetilde{\upomega}_{a}^{\approx} + \beta \widetilde{k}_a) L + \widetilde{\upomega}_{a}^\approx \delta'(\widetilde{k}_a) \right]~.
\end{align}
and therefore our approximate $\psi_{a}^{\approx}$ are
\begin{align}\label{boostedapproxmodes}
\psi_{a}^{\approx}(\rho) =&~  \sqrt{\frac{\pi}{\gamma (\widetilde{\upomega}_{a}^{\approx} + \beta \widetilde{k}_a) L + \widetilde{\upomega}_{a}^{\approx} \delta'(\widetilde{k}_a)} } \, e^{\ii \beta \gamma \widetilde{\upomega}_{a}^{\approx} \rho} \uppsi_{\underline{\widetilde{k}_a}}(\gamma \rho) ~. 
\end{align}
Note that $\gamma (\widetilde{\upomega}_{a}^{\approx} + \beta \widetilde{k}_a) = \upomega_{a}^{\approx}$ and $\frac{\ed \widetilde{k}}{\ed k} = \frac{\widetilde{\upomega}_{a}^{\approx}}{\upomega_{a}^{\approx}}$, where $(\upomega_{a}^{\approx}, k_a)$ is the two-momentum in the lab frame.  Hence the normalization constant can be written as
\begin{equation}
\widetilde{N}_{a}^2 = \frac{2\pi}{2\upomega_{a}^{\approx} (L + \frac{\ed \delta}{\ed k}(k_a))}~,
\end{equation}
where we see that it has the same form as $\eqref{phi4approxmodes}$, up to the factor of $(2\upomega_{a}^{\approx})^{-1/2}$.  This is the usual factor associated with the symplectic normalization condition.

The $a = 1$ mode in $\phi^4$-theory is a twisted version of the shape mode corresponding to an algebraic eigenvalue.  There is no approximation based on an approximate expression for the Hill discriminant like in \eqref{boostedquantcon1}.  However, since the width of the finite allowed zones appears to be extremely small at large $mL$ based on numerical investigations, we have used the untwisted value of the eigenvalue, and the corresponding orthonormal mode $\uppsi_1$ in \eqref{boostedmodeapprox} for the approximation (see \eqref{twistedzm}):
\begin{equation}\label{approxshapemode}
\nu_{\rm L} = 2: \qquad \widetilde{\upomega}_{1}^{\approx} \equiv \upomega_{1} =  \sqrt{ \frac{3 \rmk^2 m^2}{1 + \rmk^2}} ~, \qquad \psi_{1}^{\approx}(\rho) = \frac{1}{\sqrt{2\upomega_1}} e^{\ii \gamma \beta \upomega_{1} \rho} \uppsi_1(\gamma \rho)~.
\end{equation}
The normalization arises from \eqref{uppsitildenorm}, where the Wronskian drops out since $\uppsi_1$ is real.

Now let us return to the construction of the exact $\psi_a$.  We take the parity of $\psi_a$ about $z = 0$ to be the same as that of $\psi_{a}^{\approx}$.  This specifies two real initial conditions for the numerical integration of \eqref{twistedLamediffeq}.  In general one requires requires four real initial conditions, but since we have already determined the eigenvalue, finding a third initial condition that guarantees the right periodicity for, say, the real part of $\psi_a$ will automatically imply the right periodicity for the imaginary part.  Hence we use the $z = 0$ value of the approximate solution $\psi_{a}^{\approx}$ as an initial guess for the remaining initial conditions and adjust one free parameter until the appropriate periodicity condition is achieved.  Finally the overall scale is fixed according to the normalization condition that the integral on the right-hand side of \eqref{bmodenormcondition} is one.

%%%%%%%%%%%%%%%%% 
 \begin{figure}[t!]
 \centering
\begin{subfigure}{.45\textwidth}
  \centering
  \includegraphics[width=\linewidth]{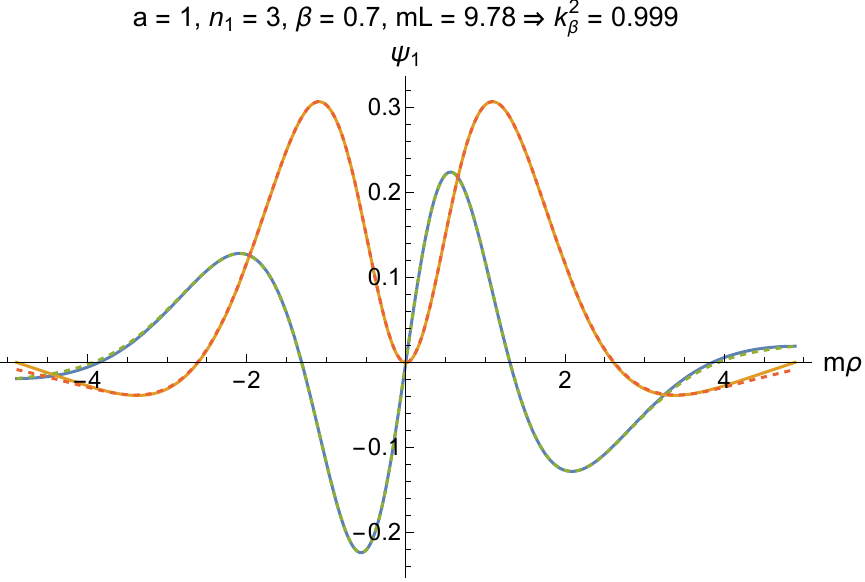}
  \caption{}
  \label{fig:BoostedMode1}
\end{subfigure}  \qquad %
\begin{subfigure}{.45\textwidth}
  \centering
  \includegraphics[width=\linewidth]{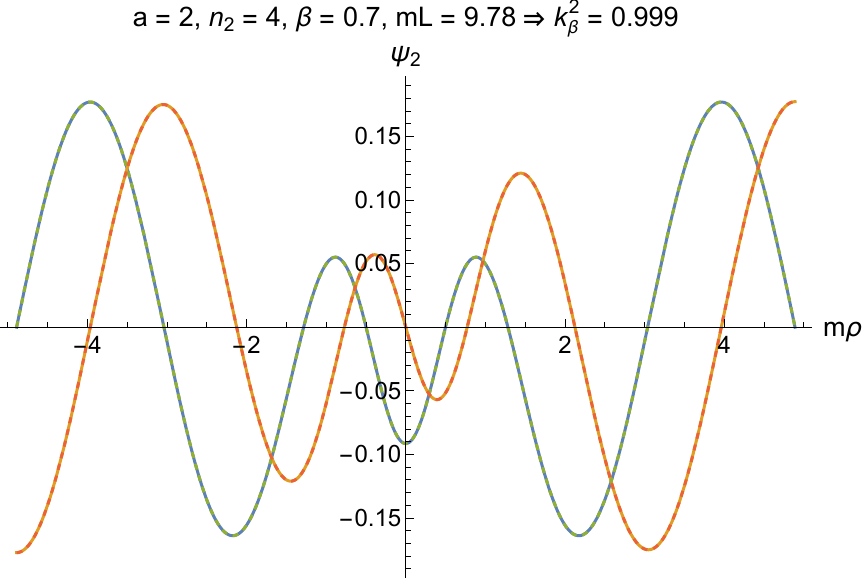}
  \caption{}
  \label{fig:BoostedMode2}
\end{subfigure}  \\[1ex]
\begin{subfigure}{.45\textwidth}
  \centering
  \includegraphics[width=\linewidth]{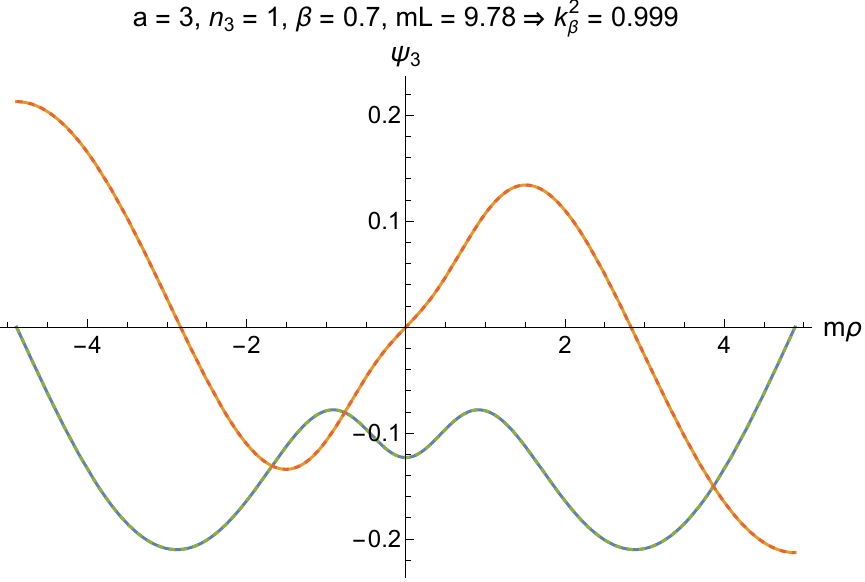}
  \caption{}
  \label{fig:BoostedMode3}
\end{subfigure}  \qquad %
\begin{subfigure}{.45\textwidth}
  \centering
  \includegraphics[width=\linewidth]{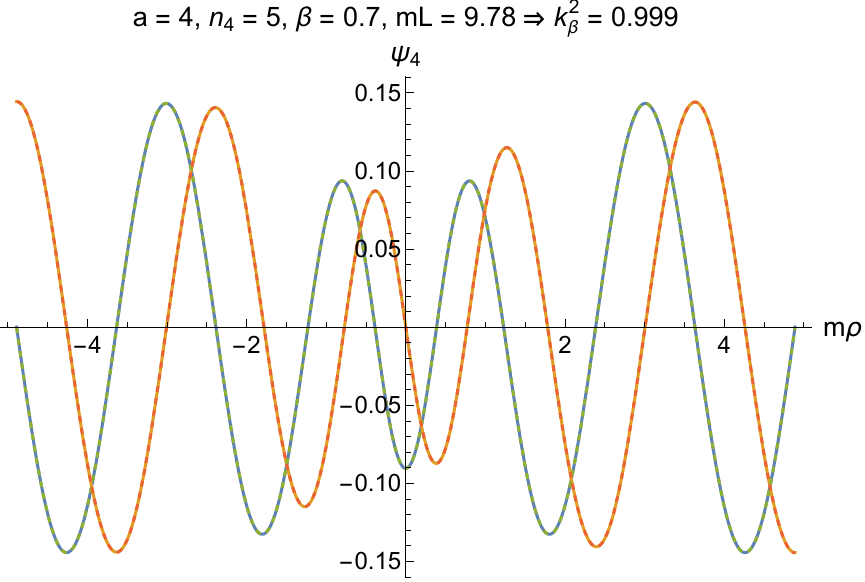}
  \caption{}
  \label{fig:BoostedMode4}
\end{subfigure}  
\caption{The first four anti-periodic boosted modes for $\beta = 0.7$ and $mL = 9.78$, chosen so that $\rmk_{\beta}^2 \approx 0.999$.  The solid blue and orange curves are the real and imaginary parts of the numerically obtained solution $\psi_a$.  The dashed blue and orange curves are the real and imaginary parts of the approximate solution $\psi_{a}^{\approx}$.}
\label{fig:TwistedEFs}
\end{figure}
%%%%%%%%%%%%%%%%%%

In Figure \ref{fig:TwistedEFs} we graph the real and imaginary parts of the $\psi_a$ obtained by numerical integration against their approximate counterparts $\psi_{a}^{\approx}$ for the first few $a$ and a specific choice of $L,\beta$.  Agreement for the transcendental modes $a > 1$ is excellent, while deviations of the approximate shape mode from the numerical solution are noticeable for $\rho$ near $\pm L/2$.  The reason is that the approximate shape mode is not anti-periodic.  Obtaining a properly anti-periodic boosted shape mode requires allowing for a more general linear combination of the Lam\'e polynomial $\uppsi_1$ and a Lam\'e function of the second kind, with relative coefficient adjusted to ensure $\psi_1(\rho + L) = -\psi_1(\rho)$.  A Lam\'e function of the second kind is a non-periodic linearly independent solution to the Lam\'e equation for algebraic eigenvalues.

%%%%%%%%%%%%%%%%%%%
%%%%%%%%%%%%%%%%%%%
\section{Classical Perturbation Theory Details}\label{app:pert}
%%%%%%%%%%%%%%%%%%%
%%%%%%%%%%%%%%%%%%%

In this appendix we compute the source terms appearing in \eqref{secondordersource} for the case $P^{(1)} = \epsilon M_0 \Theta(t-t_0)$ relevant to Subsection \ref{ssec:pertcheck}.  We use the source terms to obtain the first two corrections to the normal modes amplitudes, $A_{a}^{(1,2)}$, and the first correction to the normal mode frequencies, $\nu_{a}^{(1)}$, via \eqref{A1eom} and \eqref{A2eom}.

%%%%%%%%%%%%%%%%%%%
\subsection{First Order Analysis}
%%%%%%%%%%%%%%%%%%%

We start with the construction of the first order source \eqref{firstordersource}.  The first variations are given in \eqref{dHdphi}, which we repeat here for convenience:
 \begin{align}\label{1stordervars}
\frac{\delta H}{\delta \varpi} =&~ \frac{ (P + \langle \varpi | \varphi' \rangle)}{\langle \uppsi_0 | \varphi' \rangle^2} \varphi' + \varpi \cr
\frac{\delta H}{\delta \varphi} =&~ - \frac{ (P + \langle \varpi | \varphi' \rangle )}{\langle \uppsi_0 | \varphi' \rangle^2} \varpi'  + \frac{ (P + \langle \varpi | \varphi' \rangle)^2}{\langle \uppsi_0 |\varphi' \rangle^3} \uppsi_0'  - \varphi'' + V^{(1)}(\varphi) ~.
\end{align}
Taking derivatives with respect to $P$ gives
\begin{align}
\frac{\pd \delta H}{\pd P \delta \varpi} =&~ \frac{\varphi'}{\langle \uppsi_0 | \varphi' \rangle^2} ~, \cr
\frac{\pd \delta H}{\pd P \delta \varphi} =&~ - \frac{ \varpi' }{\langle \uppsi_0 | \varphi' \rangle^2} + \frac{2 (P + \langle \varpi | \varphi' \rangle) \uppsi_{0}' }{\langle \uppsi_0 | \varphi' \rangle^3} ~.
\end{align}
Evaluating these on $(P, \varpi, \varphi) = (P_i, \varpi_{\beta_i}, \phi_{\beta_i})$ with $\varpi_{\beta_i} = - \beta_i (\phi_{\beta_i}' - \langle \uppsi_0 | \phi_{\beta_i}' \rangle \uppsi_0)$ and projecting with $\PP_{\uppsi_0}^\perp$, we obtain the first-order source:
\begin{align}
|\SS^{(1)} \rangle = P^{(1)} \PP_{\uppsi_0}^{\perp} | B_i \rangle~, \qquad \textrm{with} \quad |B_i \rangle =  \frac{1}{\langle \uppsi_0 |\phi_{\beta_i}' \rangle^2} \left( \begin{array}{c} |\phi_{\beta_i}'\rangle \\[1ex] \beta_i ( |\phi_{\beta_i}'' \rangle + \langle \uppsi_0 |\phi_{\beta_i}' \rangle |\uppsi_0' \rangle ) \end{array} \right)  ~.
\end{align}

Since $s_{a}^{(1)} = \langle \eta_a | \SS^{(1)}\rangle$ is time-independent, the $A^{(1)}$ equation of motion \eqref{A1eom} is easily integrated leading to the first order solution \eqref{1storderfieldsol}.  As we noted there, the first order solution oscillates around a nontrivial time-independent configuration
\begin{equation}
|\zeta_{\rm av}^{(1)} \rangle := - P^{(1)} \sum_{a=1}^{\infty} 2 \Re \left\{ \frac{1}{\nu_a} |\eta_a \rangle \langle \eta_a | B_i \rangle \right\} = - P^{(1)} \HH_{i}^{-1} | B_i \rangle = - \HH_{i}^{-1} | \SS^{(1)} \rangle~.
\end{equation}
Our goal is to show that this configuration is the $O(\epsilon)$ term in the expansion of the $\beta_f$ boosted kink profile, where $\beta_f$ is the velocity corresponding to $P_f = P_i + P^{(1)}$.  

The exact $\beta_f$-boosted kink profile has the form
\begin{equation}
|\zeta_{\beta_f} \rangle = \left( \begin{array}{c} - \beta_f \PP_{\uppsi_0}^{\perp} |\phi_{\beta_f}' \rangle \\ |\phi_{\beta_f} \rangle \end{array} \right)~,
\end{equation}
where $\phi_{\beta_f}$ is given explicitly in \eqref{boostedAPkink} with $\beta \to \beta_f$.  The fields have $\epsilon$ dependence through their dependence on $\beta_f$ which is both explicit and implicit through the elliptic modulus $\rmk_{\beta_f}$.  The $\epsilon$ dependence of $\beta_f$ is, in turn, nontrivially determined by the condition $P_f = \langle \phi_{\beta_f}' | \phi_{\beta_f}' \rangle \beta_f$.  We could use these relations to compute the $O(\epsilon)$ terms in the expansion of $\beta_f$ and the field $\phi_{\beta_f}$, but we hold off on doing so for the moment.  Instead, we simply note that $|\zeta_{\beta_f}\rangle$ has an $\epsilon$ expansion with a first order term of the following form:
\begin{align}\label{zetabetaexp}
|\zeta_{\beta_f} \rangle =&~ |\zeta_{\beta_i}\rangle + |\zeta_{\beta_f}^{(1)} \rangle + |\zeta_{\beta_f}^{(2)} \rangle + O(\epsilon^3)~, \qquad \beta_f = \beta_i + \beta_{f}^{(1)} + \beta_{f}^{(2)} + O(\epsilon^3)~, \cr
|\zeta_{\beta_f}^{(1)} \rangle =&~ \left( \begin{array}{c} - \beta_{f}^{(1)} \PP_{\uppsi_0}^\perp |\phi_{\beta_i}' \rangle  - \beta_i \PP_{\uppsi_0}^{\perp} |\phi_{\beta_f}^{(1)\prime} \rangle \\ |\phi_{\beta_f}^{(1)} \rangle \end{array} \right)~.
\end{align}

This motivates us to consider the action of $\HH_i$ on a ket of the form
\begin{equation}
|\zeta \rangle = \left( \begin{array}{c} - c \PP_{\uppsi_0}^\perp | \phi_{\beta_i}' \rangle - \beta_i \PP_{\uppsi_0}^\perp |f' \rangle \\ |f\rangle \end{array} \right)~,
\end{equation}
where $c$ is a constant and the only condition we impose of $|f\rangle$ is that it is orthogonal to the static zero mode, $\langle \uppsi_0 | f \rangle = 0$.  We will find the conditions on $|f\rangle$ and $c$ such that $\HH_i | \zeta \rangle = - |\SS^{(1)} \rangle$, and we will show they are precisely the conditions satisfied by $c = \beta_{f}^{(1)}$ and $|f\rangle = |\phi_{\beta_f}^{(1)}\rangle$.

Using \eqref{HHblockform} with \eqref{HHblocks} a straightforward computation results in the following phase space components for $\HH_i | \zeta \rangle$:
\begin{align}\label{HHonzeta}
( \HH_i | \zeta \rangle )_{\varpi} =&~ -\frac{1}{\langle \uppsi_0 | \phi_{\beta_i}' \rangle^2} \left( c \langle \phi_{\beta_i}' | \phi_{\beta_i}' \rangle + 2\beta_i \langle \phi_{\beta_i}' | f' \rangle \right)   \PP_{\uppsi_0}^\perp |\phi_{\beta_i}' \rangle ~, \cr
 \HH_i | \zeta \rangle )_{\varphi} =&~ \PP_{\uppsi_0}^{\perp} \left\{  \left[ - (1-\beta_{i}^2) \pd_{\rho}^2 + V^{(2)}(\phi_{\beta_i}) \right] |f \rangle  + 2 \beta_i c |\phi_{\beta_i}'' \rangle \right\}  + \cr
 &~ - \frac{\beta_i}{\langle \uppsi_0 |\phi_{\beta_i}' \rangle^2} \left( c \langle \phi_{\beta_i}' | \phi_{\beta_i}' \rangle + 2 \beta_i \langle \phi_{\beta_i}' | f' \rangle \right) \PP_{\uppsi_0}^\perp \left\{ |\phi_{\beta_i}''\rangle + \langle \uppsi_0 |\phi_{\beta_i}' \rangle |\uppsi_0' \rangle \right\} . \qquad
 \end{align}
In order to reach this result we integrated by parts several times in overlaps such as $\langle \uppsi_0 | f' \rangle$.  Thus we require $|f \rangle$ to have the same periodicity properties as the static (or boosted) zero mode so as not to incur any boundary terms.  Comparing \eqref{HHonzeta} with $-|\SS^{(1)}\rangle = -P^{(1)} \PP_{\uppsi_0}^{\perp} | B_i\rangle$, we see that they will agree if the following two conditions hold:
\begin{align}\label{1storderconditions}
c \langle \phi_{\beta_i}' | \phi_{\beta_i}' \rangle + 2 \beta_i \langle \phi_{\beta_i}' | f' \rangle =&~ P^{(1)} ~, \cr
\left[ - (1-\beta_{i}^2) \pd_{\rho}^2 + V^{(2)}(\phi_{\beta_i}) \right] |f \rangle =&~ -2 \beta_i c |\phi_{\beta_i}'' \rangle ~.
\end{align}

Now consider the $\epsilon$ expansion of the condition $P_f = \langle \phi_{\beta_f} | \phi_{\beta_f} \rangle \beta_f$ that determines $\beta_f$:
\begin{align}\label{betaf1condition}
P_i + P^{(1)} =&~ \langle \phi_{\beta_i}' + \phi_{\beta_f}^{(1)\prime} + \cdots | \phi_{\beta_i}' + \phi_{\beta_f}^{(1)\prime} + \cdots \rangle ( \beta_i + \beta_{f}^{(1)} + \cdots )  \cr
\Rightarrow \quad P^{(1)} =&~ 2 \beta_i \langle \phi_{\beta_i}' | \phi_{\beta_f}^{(1)\prime} \rangle + \beta_{f}^{(1)} \langle \phi_{\beta_i}' | \phi_{\beta_i}' \rangle~.
\end{align}
We see that this agrees with the first of \eqref{1storderconditions} if $c = \beta_{f}^{(1)}$ and $|f\rangle = |\phi_{\beta_f}^{(1)} \rangle$.  For the second condition, consider the $\epsilon$ expansion of the nonlinear ODE satisfied by $\phi_{\beta_f}$:
\begin{align}\label{phibetaf1diffeq}
0 =&~ (\beta_{f}^2 - 1) \phi_{\beta_f}'' + V^{(1)}(\phi_{\beta_f}) \cr
\Rightarrow 0 =&~ \left[ (\beta_{i}^2 - 1) + 2 \beta_i \beta_{f}^{(1)} + \cdots \right] \left( \phi_{\beta_i}'' + \phi_{\beta_f}^{(1)\prime\prime} + \cdots \right) + \cr
&~ + V^{(1)}(\phi_{\beta_i}) + V^{(2)}(\phi_{\beta_i}) \phi_{\beta_f}^{(1)} + \cdots \cr
\Rightarrow 0 =&~ \left[ - (1- \beta_{i}^2) \pd_{\rho}^2 + V^{(2)}(\phi_{\beta_i}) \right] \phi_{\beta_f}^{(1)} + 2 \beta_i \beta_{f}^{(1)} \phi_{\beta_i}'' ~. 
\end{align}
We see that this condition is equivalent to the second of \eqref{1storderconditions} if $c = \beta_{f}^{(1)}$ and $|f\rangle = |\phi_{\beta_f}^{(1)} \rangle$.  Hence we have established
\begin{equation}\label{HHzetaresult}
\HH_i | \zeta_{\beta_f}^{(1)} \rangle = - | \SS^{(1)} \rangle \quad \Rightarrow \quad - \HH_{i}^{-1} | \SS^{(1)} \rangle = |\zeta_{\beta_f}^{(1)} \rangle~.
\end{equation}
Therefore the time-averaged first order configuration is $|\zeta_{\rm av}^{(1)} \rangle = | \zeta_{\beta_f}^{(1)} \rangle$ and the full first order solution is
\begin{align}\label{zeta1solwithzetaf}
|\zeta^{(1)} \rangle =&~ \Theta(t-t_\ast) \left( |\zeta_{\beta_f}^{(1)} \rangle + |\zeta_{\rm osc}^{(1)} \rangle \right)~, \qquad \textrm{with} \cr
|\zeta_{\rm osc}^{(1)} \rangle =&~ P^{(1)} \sum_{a=1}^{\infty} 2 \Re\left\{ \frac{\langle \eta_a | B_i \rangle}{\nu_a} e^{-\ii (\nu_a  + \cdots)(t-t_\ast)} | \eta_a \rangle \right\}  ~.
\end{align}

The demonstration we gave above for \eqref{HHzetaresult} is valid for any field theory potential $V$.  When we have an explicit boosted kink profile, $\phi_{\beta_f}$, we can also compute $\phi_{\beta_f}^{(1)}$ and $\beta_{f}^{(1)}$ directly.  We illustrate in the case of sine-Gordon.  First, assuming an expansion for $\beta_f$ of the form in \eqref{zetabetaexp}, we solve the relation \eqref{kbetatoL} perturbatively for the first order term in the $\epsilon$ expansion of $\rmk_{\beta_f}$.  We write $\rmk_{\beta_f} = \rmk_{\beta_i} + \rmk_{\beta_f}^{(1)} + O(\epsilon^2)$, and using the derivative $\mathbf{K}'(x) = \frac{\mathbf{E}(x)}{2x(1-x)}-\frac{1}{2x} \mathbf{K}(x)$ we find
\begin{equation}
\rmk_{\beta_f}^{(1)} = \frac{\gamma_{i}^2 \beta_i \rmk_{\beta_i} (1- \rmk_{\beta_i}^2) \mathbf{K}(\rmk_{\beta_i}^2)}{\mathbf{E}(\rmk_{\beta_i}^2)} \beta_{f}^{(1)} ~.
\end{equation}

Then both arguments of the Jacobi amplitude function in \eqref{boostedAPkink} can be expanded.  For the derivative of $\am(z|\alpha)$ with respect to $\alpha$ we use
\begin{align}
\frac{\pd}{\pd \alpha} \am(z |\alpha) =&~ \frac{z}{2\alpha} \pd_z \am(z|\alpha) + \frac{1}{2(1-\alpha)} \left[ \sn(z|\alpha) \cn(z|\alpha) - \frac{1}{\alpha} \dn(z|\alpha) \varepsilon(z|\alpha) \right]  ~,
\end{align}
where $\varepsilon(z|\alpha)$ is the Jacobi epsilon function.  After a few steps we find the following result for the $O(\epsilon)$ term in the expansion of $\phi_{\beta_f}$:
\begin{align}\label{phibetaf1exp}
\phi_{\beta_f}^{(1)} =&~  \gamma_{i}^2 \beta_i \beta_{f}^{(1)} \left( \rho \pd_\rho \phi_{\beta_i}(\rho) + g_{\rm hom}^{(1)}(\rho) \right) ~,
\end{align}
where
\begin{equation}\label{ghomsG}
g_{\rm hom}^{(1)}(\rho) =  \frac{2 \mathbf{K}(\rmk_{\beta_i}^2)}{\mathbf{E}(\rmk_{\beta_i}^2)} \left( \rmk_{\beta_i}^2 \sn(\tfrac{\gamma_i m\rho}{\rmk_{\beta_i}} \,| \, \rmk_{\beta_i}^2) \cn(\tfrac{\gamma_i m\rho}{\rmk_{\beta_i}} \,| \, \rmk_{\beta_i}^2)  - \dn(\tfrac{\gamma_i m\rho}{\rmk_{\beta_i}} \,| \, \rmk_{\beta_i}^2) \varepsilon(\tfrac{\gamma_i m\rho}{\rmk_{\beta_i}} \,| \, \rmk_{\beta_i}^2) \right) ~.
\end{equation}
One can check that the configuration \eqref{phibetaf1exp} solves the differential equation appearing in the last line of \eqref{phibetaf1diffeq}.  The first term is a particular solution and is familiar from the early work on kinks.  See \emph{e.g.}~\cite{Gervais:1975pa}, where the $\beta_i = 0$ static kink version of this quantity was denoted $\varphi_2$.  It is the only term necessary in the $mL \to \infty$ theory.  The $g_{\rm hom}^{(1)}$ term is a homogeneous solution that corrects the quasi-periodicity of the particular solution so that $\phi_{\beta_f}^{(1)}$ is periodic.  This is achieved with the quasi-periodicity of the Jacobi epsilon function,
\begin{equation}
\varepsilon(z + 2\mathbf{K}(\rmk^2)|\rmk^2) =  \varepsilon(z| \rmk^2) + 2 \mathbf{E}(\rmk^2) ~,
\end{equation}
and the fact that $\pd_\rho \phi_{\beta_i} \propto \dn(\tfrac{\gamma_i m\rho}{\rmk_{\beta_i}} \,| \, \rmk_{\beta_i}^2)$ with just the right proportionality constant.  One can show using asymptotic expansions near $\rmk_{\beta_i} \to 1$ that $|g_{\rm hom}^{(1)}(\rho)|$ is bounded by a quantity that is of order $e^{-\gamma_i m L/2}$ for all $\rho \in [-L/2,L/2]$.

We still need to find $\beta_{f}^{(1)}$, which can be done by inserting \eqref{phibetaf1exp} into \eqref{betaf1condition} and solving.  The result is
\begin{equation}
\beta_{f}^{(1)} = \frac{M_0 \epsilon}{\gamma_{i}^2 \left( \langle \phi_{\beta_i}' | \phi_{\beta_i}' \rangle - 2 \beta_{i}^2 \langle \phi_{\beta_i}'' | g_{\rm hom}^{(1)}\rangle \right) } ~.
\end{equation}
When $mL$ becomes large the second term in the denominator is exponentially small, while $\langle \phi_{\beta_i}' | \phi_{\beta_i}' \rangle \approx \gamma_i M_0$ up to exponentially small corrections.  Hence we recover
\begin{equation}
\lim_{mL \to \infty} \beta_{f}^{(1)}  = \gamma_{i}^{-3} \epsilon~,
\end{equation}
which is the expected result from expansion of the Lorentz-covariant expression
\begin{equation}
\frac{P_f}{\sqrt{P_{f}^2 + M_{0}^2}} = \frac{P_i + M_0 \epsilon}{\sqrt{(P_i + M_0 \epsilon)^2 + M_{0}^2}} = \beta_i + \gamma_{i}^{-3} \epsilon + O(\epsilon^2)~.
\end{equation}
Similarly explicit results can be obtained for $\phi_{\beta_f}^{(1)}$ and $\beta_{f}^{(1)}$ in $\phi^4$ theory, but we will not need them in this paper.

Note that $\langle \eta_a | B_i \rangle = 0$ when $\beta_i = 0$, as the position component of $|B_i \rangle$ is proportional to $\beta_i$ while the momentum component is proportional to $|\uppsi_0\rangle$ in the $\beta_i \to 0$ limit.  Hence the first order source $|\SS^{(1)}\rangle$, and therefore the first order solution $|\zeta^{(1)}\rangle$ vanishes.  The first order velocity is nonzero but constant, being given by $M_0 \epsilon/M_0'$, where $M_0'$ was defined in \eqref{finiteLrelmass} and approaches $M_0$ exponentially fast at large $mL$.

%%%%%%%%%%%%%%%%%%
\subsection{Second Order Source Terms}
%%%%%%%%%%%%%%%%%%

Now we move on to the evaluation of the second order source terms, and we start with the required variations of the Hamiltonian.  The second variations are
\begin{align}\label{2ndordervars}
\frac{\delta^2 H}{\delta\varpi_1 \delta\varpi_2} =&~ \delta_{12} + \frac{ | \varphi' \rangle_1 \langle \varphi' |_2 }{\langle \uppsi_{0} | \varphi' \rangle^2}  ~, \cr
\frac{\delta^2 H}{\delta \varpi_1 \delta \varphi_2} =&~  \frac{ (P + \langle \varpi | \varphi' \rangle)}{\langle \uppsi_0 |\varphi' \rangle^2} \delta_{12} \pd_{2}  - \frac{ |\varphi' \rangle_1 \langle \varpi' |_2 }{\langle \uppsi_{0} |\varphi' \rangle^2} + \frac{2 (P + \langle \varpi | \varphi' \rangle) }{\langle \uppsi_0 |\varphi' \rangle^3} | \varphi' \rangle_{1} \langle \uppsi_{0}' |_2 \cr
\frac{\delta^2 H}{\delta \varphi_1 \delta \varpi_2} =&~ \overleftarrow{\pd}_{1} \delta_{12} \frac{ (P + \langle \varpi | \varphibar \rangle)}{\langle \uppsi_0 |\varphi' \rangle^2} - \frac{ |\varpi' \rangle_1 \langle \varphi' |_2}{\langle \uppsi_0 |\varphi' \rangle^2} + \frac{2 (P + \langle \varpi |\varphi'\rangle)}{\langle \uppsi_0 |\varphi' \rangle^3} |  \uppsi_{0}' \rangle_1 \langle \varphi' |_2 ~,   \cr
\frac{\delta^2 H}{\delta \varphi_1 \delta \varphi_2} =&~  - \delta_{12} \pd_{2}^2 + \delta_{12} V^{(2)}(\varphi_2) + \frac{ |\varpi' \rangle_1 \langle \varpi' |_2 }{\langle \uppsi_{0} |\varphi' \rangle^2} +  \frac{3 (P + \langle \varpi |\varphi' \rangle)^2}{\langle \uppsi_0 |\varphi' \rangle^4} |\uppsi_{0}' \rangle_1 \langle \uppsi_{0}' |_2    + \cr
&~ - \frac{2 (P + \langle \varpi |\varphi' \rangle)}{\langle \uppsi_0 |\varphi' \rangle^3} \left(  |\uppsi_{0}' \rangle_1 \langle \varpi'|_2  +  |\varpi' \rangle_1 \langle \uppsi_{0}' |_2 \right)~.
\end{align}
If we evaluate these on $(P, \varpi, \varphi) \to (P_i , \varpi_{\beta_i}, \phi_{\beta_i})$, and use $\varpi_{\beta_i} = - \beta_i (\phi_{\beta_i}' - \langle \uppsi_0 | \phi_{\beta_i}' \rangle \uppsi_0)$ and $\overleftarrow{\pd}_1 \delta_{12} = - \delta_{12} \pd_2$, we find the blocks $\MM, \BB, \BB^T, \KK$ given in \eqref{HHblocks}.  Clearly the middle two expressions of \eqref{2ndordervars} are related by exchanging $1 \leftrightarrow 2$.  More precisely, we should first make all kets into bras and send $\overleftarrow{\pd}_1 \delta_{12} \to \delta_{21} \pd_1$, so that we view the second variation as a map $L^{2}(\mathbbm{R}) \times L^2(\mathbbm{R}) \to \mathbbm{C}$ rather than as a map $L^{2}(\mathbbm{R}) \to L^{2}(\mathbbm{R})$.  Then the resulting two expressions will be obtained from each other by exchanging $1\leftrightarrow 2$.

The third variation of $H$ with respect to three $\varpi$'s is zero, while
\begin{align}\label{3rdordervars}
\frac{ \delta^3 H}{\delta \varpi_1 \delta \varpi_2 \delta \varphi_3}  =&~ \frac{ \langle \varphi' |_2}{\langle \uppsi_0 | \varphi' \rangle^2}  \delta_{13} \pd_3 + \frac{ | \varphi' \rangle_1 }{\langle \uppsi_0 | \varphi' \rangle^2} \delta_{23} \pd_3 + \frac{2 | \varphi' \rangle_1 \langle \varphi' |_2 \langle \uppsi_0' |_3}{ \langle \uppsi_0 | \varphi' \rangle^3} ~, \cr
\frac{\delta^3 H}{\delta \varpi_1 \delta \varphi_2 \delta \varphi_3} =&~  \left[ \frac{2 (P + \langle \pi | \varphi' \rangle)}{\langle \uppsi_0 | \varphi' \rangle^3} \langle \uppsi_0' |_3   - \frac{ \langle \varpi' |_3}{\langle \uppsi_0 | \varphi' \rangle^2} \right] \delta_{12} \pd_2  + \cr
&~ +  \left[ \frac{2 (P + \langle \varpi | \varphi' \rangle)}{\langle \uppsi_0 | \varphi' \rangle^3} \langle\uppsi_0 '|_2  - \frac{\langle \varpi' |_2}{\langle \uppsi_0 | \varphi' \rangle^2} \right] \delta_{13} \pd_3 + \cr
&~ - \frac{2 | \varphi' \rangle_1}{ \langle \uppsi_0 | \varphi' \rangle^3} \left(  \langle \varpi' |_2 \langle \uppsi_0 ' |_3 +  \langle \uppsi_0' |_2 \langle \varpi' |_3 \right)  + \frac{ 6 (P + \langle \varpi | \varphi' \rangle)}{\langle \uppsi_0 | \varphi' \rangle^4} |\varphi' \rangle_1 \langle \uppsi_0' |_2 \langle \uppsi_0' |_3 \cr
\frac{\delta^3 H}{\delta \varphi_1 \delta \varphi_2 \delta \varphi_3} =&~  \frac{2}{\langle \uppsi_0 | \varphi' \rangle^3} \left( |\varpi' \rangle_1 \langle \varpi' |_2 \langle \uppsi_0' |_3 + | \varpi' \rangle_1 \langle \uppsi_0' |_2 \langle \varpi' |_3 + | \uppsi_0 ' \rangle_1 \langle \varpi' |_2 \langle \varpi' |_3 \right) + \cr
&~ - \frac{6 (P + \langle \varpi | \varphi' \rangle)}{\langle \uppsi_0 | \varphi' \rangle^4} \left( | \varpi' \rangle_1 \langle \uppsi_0 ' |_2 \langle \uppsi_0 ' |_3 + |\uppsi_0 ' \rangle_1 \langle \varpi' |_2 \langle \uppsi_0 ' |_3 + | \uppsi_0 ' \rangle_1 \langle \uppsi_0 ' |_2 \langle \varpi' |_3 \right) + \cr
&~ \delta_{12} \delta_{23} V^{(3)}(\varphi_3) +  \frac{12 (P + \langle \varpi | \varphi' \rangle)^2}{\langle \uppsi_0 | \varphi' \rangle^5 } | \uppsi_0 ' \rangle_1 \langle \uppsi_0 ' |_2 \langle \uppsi_0 ' |_3 ~.
\end{align}
We are representing these variations as maps $L^2(\mathbbm{R}) \times L^2(\mathbbm{R}) \to L^2(\mathbbm{R})$ because this is how they appear in \eqref{secondordersource}.  The other permutations of $\varpi$ and $\varphi$ derivatives can be obtained from \eqref{3rdordervars} with the same type exchange symmetry operation as discussed under \eqref{2ndordervars}.

The second-order source terms in \eqref{secondordersource} receive contributions from three types of $H$ derivatives: those coming from two $P$ derivatives of the first order variations, one $P$ derivative of the second order variations, and no $P$ derivatives of the third order variations.  For the first type the relevant derivatives are
\begin{align}
&  \frac{\pd^2 \delta H}{\pd P^2 \delta \varpi} \bigg|_i  = 0~, \qquad \frac{\pd^2 \delta H}{\pd P^2 \delta \varphi} \bigg|_i  = \frac{2 \uppsi_0'}{\langle \uppsi_0 | \phi_{\beta_i}' \rangle^3} ~.
\end{align}

Next consider the second type.  Taking derivatives of the second variations \eqref{2ndordervars} with respect to $P$ and evaluating on the background gives
\begin{align}
\frac{ \pd \delta^2 H}{\pd P \delta \varpi_1 \delta \varphi_2} \bigg|_i  =&~ \left[ \frac{1}{\langle \uppsi_0 | \varphi' \rangle^2} \delta_{12} \pd_2 + \frac{2 |\varphi' \rangle_1 \langle \uppsi_{0}' |_2 }{\langle \uppsi_0 | \varphi' \rangle^3} \right] \bigg|_i =   \frac{1}{\langle \uppsi_0 | \phi_{\beta_i}' \rangle^2} \delta_{12} \pd_2 + \frac{2 |\phi_{\beta_i}' \rangle_1 \langle \uppsi_0' |_2}{\langle \uppsi_0 |\phi_{\beta_i}' \rangle^3} ~, \cr
\frac{\pd \delta^2 H}{\pd P \delta \varphi_1 \delta \varpi_2} \bigg|_i =&~ \left[ \overleftarrow{\pd}_1 \delta_{12} \frac{1}{\langle \uppsi_0 | \varphi' \rangle^2} + \frac{2 |\uppsi_0' \rangle_1 \langle \varphi' |_2}{\langle \uppsi_0 | \varphi' \rangle^3} \right] \bigg|_i  =  \overleftarrow{\pd}_1 \delta_{12} \frac{1}{\langle \uppsi_0 | \phi_{\beta_i}' \rangle^2} + \frac{2 |\uppsi_0' \rangle_1 \langle \phi_{\beta_i}' |_2 }{\langle \uppsi_0 | \phi_{\beta_i}' \rangle^3} ~, \cr
\frac{\pd \delta^2 H}{\pd P \delta \varphi_1 \delta \varphi_2} \bigg|_i  =&~ \left[ \frac{6 (P + \langle \varpi | \varphi' \rangle )}{\langle \uppsi_0 | \varphi' \rangle^4} |\uppsi_{0}' \rangle_1 \langle \uppsi_0' |_2 - \frac{2}{\langle \uppsi_0 | \varphi' \rangle^3} \left( |\uppsi_0' \rangle_1  \langle \varpi' |_2 + |\varpi' \rangle_1 \langle \uppsi_0' |_2 \right) \right] \bigg|_i  \cr
=&~ \frac{2 \beta_i}{\langle \uppsi_0 |\phi_{\beta_i}' \rangle^3} \left( |\uppsi_0' \rangle_1 \langle \phi_{\beta_i}'' |_2 + |\phi_{\beta_i}'' \rangle_1 \langle \uppsi_0' |_2 + \langle \uppsi_0 | \phi_{\beta_i}' \rangle |\uppsi_0' \rangle_1 \langle \uppsi_0' | \right) ~. \raisetag{24pt}
\end{align}
for the nonzero results.  

For the third type of $H$-derivative that contributes to $|\SS^{(2)}\rangle$ we simply evaluate \eqref{3rdordervars} on the background.  The results are
\begin{align}\label{delta3Heval}
\frac{\delta^3 H}{\delta \varpi_1 \varpi_2 \varphi_3} \bigg|_i =&~ \frac{1}{\langle \uppsi_0 | \phi_{\beta_i}' \rangle^2} \left( |\phi_{\beta_i}' \rangle_1 \delta_{23} \pd_3 + \langle \phi_{\beta_i}' |_2 \delta_{13} \pd_3 \right) + \frac{2 |\phi_{\beta_i}' \rangle_1 \langle \phi_{\beta_i}' |_2 \langle \uppsi_{0}' |_3}{\langle \uppsi_0 |\phi_{\beta_i}' \rangle^3} ~, \cr
\frac{\delta^3 H}{\delta \varpi_1 \varphi_2 \varphi_3} \bigg|_i =&~ \frac{\beta_i}{\langle \uppsi_0 | \phi_{\beta_i}' \rangle^2} \left( \langle \phi_{\beta_i}'' |_3 +  \langle \uppsi_0 | \phi_{\beta_i}' \rangle \langle \uppsi_0' |_3 \right) \delta_{12} \pd_2 + \cr
&~ +  \frac{\beta_i}{\langle \uppsi_0 | \phi_{\beta_i}' \rangle^2} \left( \langle \phi_{\beta_i}'' |_2 + \langle \uppsi_0 | \phi_{\beta_i}' \rangle  \langle \uppsi_0' |_2 \right) \delta_{13} \pd_3 + \cr
&~ + \frac{2 \beta_i | \phi_{\beta_i}' \rangle_1}{\langle \uppsi_0 | \phi_{\beta_i}' \rangle^3} \left( \langle \phi_{\beta_i}'' |_2  \langle \uppsi_0' |_3 + \langle \uppsi_0' |_2 \langle \phi_{\beta_i}'' |_3 + \langle \uppsi_0 | \phi_{\beta_i}' \rangle \langle \uppsi_0' |_2 \langle \uppsi_0' |_3 \right)~, \cr
\frac{\delta^3 H}{\delta \varphi_1 \varphi_2 \varphi_3} \bigg|_i  =&~ \frac{2 \beta_{i}^2}{\langle \uppsi_0 | \phi_{\beta_i}' \rangle^3} \left( |\phi_{\beta_i}'' \rangle_1 \langle \phi_{\beta_i}'' |_2 \langle \uppsi_0' |_3 + |\phi_{\beta_i}'' \rangle_1 \langle \uppsi_0' |_2 \langle \phi_{\beta_i}'' |_3 + |\uppsi_0' \rangle_1 \langle \phi_{\beta_i}'' |_2 \langle \phi_{\beta_i}'' |_3 \right) + \cr
&~ + \frac{2 \beta_{i}^2}{\langle \uppsi_0 | \phi_{\beta_i}' \rangle^2} \left( |\phi_{\beta_i}'' \rangle_1 \langle \uppsi_0' |_2 \langle \uppsi_0' |_3 + |\uppsi_0' \rangle_1 \langle \phi_{\beta_i}'' |_2 \langle \uppsi_0' |_3 + | \uppsi_0' \rangle_1 \langle \uppsi_0' |_2 \langle \phi_{\beta_i}'' |_3 \right) + \cr
&~ + \delta_{12} \delta_{23} V^{(3)}(\phi_{\beta_i 3})  ~.
\end{align}

Now consider the construction of $|\SS^{(2)}\rangle$ via \eqref{secondordersource} using these results.  As we discussed above, the first order solution $|\zeta^{(1)} \rangle$ that appears in $|\SS^{(2)} \rangle$ can be decomposed into a time-independent piece and a piece that oscillates around zero, where the time-independent piece is the $O(\epsilon)$ term in an expansion of the $\beta_f$-boosted kink profile, \eqref{zeta1solwithzetaf}.  It is useful to decompose the contributions to $|\SS^{(2)}\rangle$ according to the power of the oscillating field that appears.

Those terms containing no powers of the $|\zeta_{\rm osc}^{(1)}\rangle$ are denoted $|\SS^{(2,0)}\rangle$.  $|\SS^{(2,0)} \rangle$ receives contributions from all three types of derivative discussed above.  The momentum component of $|\SS^{(2,0)}\rangle$ is
\begin{align}
| \SS_{\varpi}^{(2,0)}\rangle =&~ \PP_{\uppsi_0}^\perp \bigg\{  \left( \frac{\pd \delta^2 H}{\pd P \delta \varpi \delta \varphi_2} \bigg|_i \right)  | \phi_{\beta_f}^{(1)} \rangle_2 + \left( \left. \frac{\delta^3 \HH}{\delta \varpi \delta \varpi_2 \delta \varphi_3} \right|_i \right) | \varpi_{\beta_f}^{(1)} \rangle_2 |\phi_{\beta_f}^{(1)} \rangle_3 + \cr
&~ \qquad +  \half  \left( \left. \frac{\delta^3 \HH}{\delta \varpi \delta \varphi_2 \delta \varphi_3} \right|_i \right) | \phi_{\beta_f}^{(1)} \rangle_2 |\phi_{\beta_f}^{(1)} \rangle_3 \bigg\} ~.
\end{align}
Assembling all the pieces we can bring the result to the form
\begin{align}
& | \SS_{\varpi}^{(2,0)}\rangle = \PP_{\uppsi_0}^\perp \bigg\{  \frac{ \left( P^{(1)} + \langle \phi_{\beta_i}' |\varpi_{\beta_f}^{(1)} \rangle + \beta_i ( \langle \phi_{\beta_i}'' |\phi_{\beta_f}^{(1)} \rangle + \langle \uppsi_0 | \phi_{\beta_i}' \rangle \langle \uppsi_0' |\phi_{\beta_f}^{(1)} \rangle ) \right) }{\langle \uppsi_0 |\phi_{\beta_i}' \rangle^2} |\phi_{\beta_f}^{(1)\prime } \rangle +  \cr
& ~ + \bigg[ \frac{\langle \varpi_{\beta_f}^{(1)} | \phi_{\beta_f}^{(1)\prime} \rangle}{\langle \uppsi_0 |\phi_{\beta_i}' \rangle^2} + \frac{2 \langle \uppsi_0' |\phi_{\beta_f}^{(1)} \rangle}{\langle \uppsi_0 |\phi_{\beta_i}' \rangle}  \frac{ \left( P^{(1)} + \langle \phi_{\beta_i}' |\varpi_{\beta_f}^{(1)} \rangle + \beta_i ( \langle \phi_{\beta_i}'' |\phi_{\beta_f}^{(1)} \rangle + \langle \uppsi_0 | \phi_{\beta_i}' \rangle \langle \uppsi_0' | \phi_{\beta_f}^{(1)} \rangle ) \right) }{\langle \uppsi_0 |\phi_{\beta_i}' \rangle^2}  + \cr
&  \qquad - \frac{\beta_i \langle \uppsi_0' |\phi_{\beta_f}^{(1)} \rangle^2}{\langle \uppsi_0 | \phi_{\beta_i}' \rangle^2} \bigg] | \phi_{\beta_i}' \rangle \bigg\} ~.
\end{align}
Upon inserting $|\varpi_{\beta_f}^{(1)} \rangle = - \beta_{f}^{(1)} | \PP_{\uppsi_0}^\perp |\phi_{\beta_i}' \rangle - \beta_i \PP_{\uppsi_0}^\perp |\phi_{\beta_f}^{(1)\prime} \rangle$ and using \eqref{betaf1condition}, we find that the large fraction appearing in the first and second lines is precisly $\beta_{f}^{(1)}$:
\begin{equation}\label{betaf1nice}
 \frac{ \left( P^{(1)} + \langle \phi_{\beta_i}' |\varpi_{\beta_f}^{(1)} \rangle + \beta_i ( \langle \phi_{\beta_i}'' |\phi_{\beta_f}^{(1)} \rangle + \langle \uppsi_0 | \phi_{\beta_i}' \rangle \langle \uppsi_0' |\phi_{\beta_f}^{(1)} \rangle ) \right) }{\langle \uppsi_0 |\phi_{\beta_i}' \rangle^2} = \beta_{f}^{(1)}~.
 \end{equation}
Hence $|\SS^{(2)}_{\varpi} \rangle$ takes on the more manageable form
\begin{equation}\label{SS20piresult}
| S_{\varpi}^{(2,0)} \rangle =  \PP_{\uppsi_0}^\perp \left\{ \beta_{f}^{(1)} |\phi_{\beta_f}^{(1)\prime}\rangle + \left( \frac{ \langle \varpi_{\beta_f}^{(1)} | \phi_{\beta_f}^{(1)\prime} \rangle - \beta_i \langle \uppsi_0 | \phi_{\beta_f}^{(1)\prime} \rangle^2}{\langle \uppsi_0 | \phi_{\beta_i}' \rangle^2} - 2 \beta_{f}^{(1)} \frac{\langle \uppsi_0 |\phi_{\beta_f}^{(1)\prime} \rangle}{\langle \uppsi_0 |\phi_{\beta_i}' \rangle} \right) |\phi_{\beta_i}' \rangle \right\}~.
\end{equation}

We proceed similarly with the position component of $|\SS^{(2,0)}\rangle$, starting with
\begin{align}\label{SS20def}
|\SS_{\varphi}^{(2,0)}\rangle =&~ \PP_{\uppsi_0}^{\perp} \bigg\{  \frac{P^{(1)2}}{\langle \uppsi_0 |\phi_{\beta_i}' \rangle^3} |\uppsi_0' \rangle +  \left( \frac{\pd \delta^2 H}{\pd P \delta \varphi \delta \varpi_2} \bigg|_i \right) | \varpi_{\beta_f}^{(1)} \rangle_2 + \left( \frac{\pd \delta^2 H}{\pd P \delta \varphi \delta \varphi_2} \bigg|_i \right) |\phi_{\beta_f}^{(1)} \rangle_2 + \ \cr
& \qquad + \half \left( \left. \frac{\delta^3 \HH}{\delta \varphi \delta \varpi_2 \delta \varpi_3} \right|_i \right) | \varpi_{\beta_f}^{(1)} \rangle_2 |\varpi_{\beta_f}^{(1)} \rangle_3  +  \left( \left. \frac{\delta^3 \HH}{\delta \varphi \delta \varpi_2 \delta \varphi_3} \right|_i \right) |\varpi_{\beta_f}^{(1)} \rangle_2 |\phi_{\beta_f}^{(1)} \rangle_3 + \cr
& \qquad  + \half  \left( \left. \frac{\delta^3 \HH}{\delta \varphi \delta \varphi_2 \delta \varphi_3} \right|_i \right) | \phi_{\beta_f}^{(1)} \rangle_2 |\phi_{\beta_f}^{(1)} \rangle_3 \bigg\} ~.
\end{align}
After some work and using \eqref{betaf1nice}, we find the result
\begin{align}\label{SS20phiresult}
| S_{\varphi}^{(2,0)} \rangle =&~ \PP_{\uppsi_0}^\perp \Bigg\{ \beta_i \beta_{f}^{(1)} |\phi_{\beta_f}^{(1)\prime \prime} \rangle + \half | V^{(3)}(\phi_{\beta_i}) \phi_{\beta_f}^{(1)2} \rangle + \beta_{f}^{(1)2} |\phi_{\beta_i}'' \rangle + \beta_i \beta_{f}^{(1)} \langle \uppsi_0 |\phi_{\beta_f}^{(1)\prime} \rangle |\uppsi_0' \rangle   + \cr
&~ \qquad  +  \beta_i  \left( \frac{ \langle \varpi_{\beta_f}^{(1)} | \phi_{\beta_f}^{(1)\prime} \rangle - \beta_i \langle \uppsi_0 | \phi_{\beta_f}^{(1)\prime} \rangle^2}{\langle \uppsi_0 | \phi_{\beta_i}' \rangle^2} - 2 \beta_{f}^{(1)} \frac{\langle \uppsi_0 |\phi_{\beta_f}^{(1)\prime} \rangle}{\langle \uppsi_0 |\phi_{\beta_i}' \rangle} \right) \times \cr
&~ \qquad \qquad \times \left( |\phi_{\beta_i}'' \rangle + \langle \uppsi_0 |\phi_{\beta_i}'\rangle |\uppsi_0' \rangle \right)  \Bigg\} ~. \raisetag{24pt}
\end{align}

The form of the results \eqref{SS20piresult}, \eqref{SS20phiresult} and experience with the first order analysis suggest that this part of the second order source is related to the second order terms in the expansion of $|\zeta_{\beta_f} \rangle$ around $\epsilon = 0$.  To investigate this we consider the action of $\HH_i$ on the new ansatz
\begin{equation}
|\zeta \rangle = \left( \begin{array}{c} - \beta_i \PP_{\uppsi_0}^\perp | f' \rangle - \beta_{f}^{(1)} \PP_{\uppsi_0}^\perp |\phi_{\beta_f}^{(1)\prime} \rangle - c \PP_{\uppsi_0}^\perp | \phi_{\beta_i}' \rangle \\ |f\rangle \end{array} \right)~,
\end{equation}
where $|f\rangle$ is assumed to satisfy $\langle \uppsi_0 | f \rangle = 0$.  If $|f\rangle = |\phi_{\beta_f}^{(2)}\rangle$ and $c = \beta_{f}^{(2)}$, then $|\zeta \rangle = |\zeta_{\beta_f}^{(2)}\rangle$.  Given the form of the ansatz, it is sufficient to compute the action of $\HH_i$ on the ket $\left( - \beta_{f}^{(1)} \PP_{\uppsi_0}^\perp |\phi_{\beta_f}^{(1)\prime} \rangle ~, 0 \right)^T$ and add the result to \eqref{HHonzeta}.  This leads to
\begin{align}\label{Hzetacompare}
(\HH_i | \zeta \rangle)_\varpi =&~ - \beta_{f}^{(1)} | \phi_{\beta_f}^{(1)\prime} \rangle + \cr
&~ - \frac{1}{\langle \uppsi_0 |\phi_{\beta_i}' \rangle^2} \left( c \langle \phi_{\beta_i}' | \phi_{\beta_i}' \rangle + 2\beta_i \langle \phi_{\beta_i}' | f' \rangle  + \beta_{f}^{(1)} \langle \phi_{\beta_i}' | \PP_{\uppsi_0}^\perp | \phi_{\beta_f}^{(1)\prime} \rangle \right) \PP_{\uppsi_0}^\perp |\phi_{\beta_i}' \rangle ~, \cr
 \HH_i | \zeta \rangle )_{\varphi} =&~ \PP_{\uppsi_0}^{\perp} \left[ - (1-\beta_{i}^2) \pd_{\rho}^2 + V^{(2)}(\phi_{\beta_i}) \right] | f  \rangle +  \beta_i  \beta_{f}^{(1)} \PP_{\uppsi_0}^\perp  |\phi_{\beta_f}^{(1)\prime \prime} \rangle  + \cr
 &~  - \beta_i \beta_{f}^{(1)} \langle \uppsi_0 |\phi_{\beta_f}^{(1)\prime} \rangle | \uppsi_0' \rangle  + 2 \beta_i c  \PP_{\uppsi_0}^\perp |\phi_{\beta_i}'' \rangle + \cr
 &~ - \frac{\beta_i}{\langle \uppsi_0 |\phi_{\beta_i}' \rangle^2} \left( c \langle \phi_{\beta_i}' | \phi_{\beta_i}' \rangle + 2 \beta_i \langle \phi_{\beta_i}' | f' \rangle + \beta_{f}^{(1)} \langle \phi_{\beta_i}' | \PP_{\uppsi_0}^\perp |\phi_{\beta_f}^{(1)\prime} \rangle \right) \times \cr
 &~ \qquad \qquad \qquad \times  \PP_{\uppsi_0}^\perp \left(  |\phi_{\beta_i}''\rangle + \langle \uppsi_0 |\phi_{\beta_i}' \rangle |\uppsi_0' \rangle \right) ~. 
\end{align}

Comparing the momentum components of \eqref{Hzetacompare} and $- |\SS^{(2,0)} \rangle$, we find they will agree if the following condition holds:
\begin{align}
& c \langle \phi_{\beta_i}' | \phi_{\beta_i}' \rangle + 2\beta_i \langle \phi_{\beta_i}' | f' \rangle  + \beta_{f}^{(1)} \langle \phi_{\beta_i}' | \PP_{\uppsi_0}^\perp | \phi_{\beta_f}^{(1)\prime} \rangle = \cr
& \qquad \qquad \qquad \qquad   = \langle \varpi_{\beta_f}^{(1)} | \phi_{\beta_f}^{(1)\prime} \rangle - \beta_i \langle \uppsi_0 | \phi_{\beta_f}^{(1)\prime} \rangle^2 - 2 \beta_{f}^{(1)} \langle \uppsi_0 |\phi_{\beta_i}' \rangle \langle \uppsi_0 |\phi_{\beta_f}^{(1)\prime} \rangle ~.
 \end{align}
Inserting $\varpi_{\beta_f}^{(1)} = - \beta_i \PP_{\uppsi_0}^\perp |\phi_{\beta_f}^{(1)\prime}\rangle - \beta_{f}^{(1)} \PP_{\uppsi_0}^\perp | \phi_{\beta_i}' \rangle$, this equation can be rearranged to the form
\begin{equation}\label{2ndccondition}
0 = c \langle \phi_{\beta_i}' |\phi_{\beta_i}' \rangle + 2 \beta_{f}^{(1)} \langle \phi_{\beta_i}' | \phi_{\beta_f}^{(1)\prime} \rangle + \beta_i \langle \phi_{\beta_f}^{(1)\prime} | \phi_{\beta_f}^{(1)\prime} \rangle + 2 \beta_i \langle \phi_{\beta_i}' | f' \rangle ~.
\end{equation}
Now we compare position components, assuming \eqref{2ndccondition} holds.  With this condition we find that the $|\uppsi_0' \rangle$ terms agree on both sides of the equation $(\HH_i | \zeta\rangle )_\varphi = - |\SS_{\varphi}^{(2,0)}\rangle$.  Requiring the remaining terms to agree leads to a condition that can be rearranged to the following form:
\begin{align}\label{2ndfcondition}
 & \left[ - (1-\beta_{i}^2) \pd_{\rho}^2 + V^{(2)}(\phi_{\beta_i}) \right] | f \rangle = \cr
 &  \qquad =   - \left( \beta_{f}^{(1)2} + 2 \beta_i c \right) |\phi_{\beta_i}'' \rangle - 2 \beta_i \beta_{f}^{(1)} | \phi_{\beta_f}^{(1)\prime \prime} \rangle - \half |V^{(3)}(\phi_{\beta_i}) \phi_{\beta_f}^{(1)2} \rangle ~.  \qquad
 \end{align}

The conditions \eqref{2ndccondition} and \eqref{2ndfcondition} are the same form as those determining $\phi_{\beta_f}^{(2)}$ and $\beta_{f}^{(2)}$.  Carrying out \eqref{betaf1condition} one order further leads to
\begin{equation}
0 =  \beta_{f}^{(2)} \langle \phi_{\beta_i}' |\phi_{\beta_i}' \rangle + 2  \beta_{f}^{(1)} \langle \phi_{\beta_i}' | \phi_{\beta_f}^{(1)\prime} \rangle + \beta_i \left( 2 \langle \phi_{\beta_i}' | \phi_{\beta_f}^{(2)\prime} \rangle  + \langle \phi_{\beta_f}^{(1) \prime} | \phi_{\beta_f}^{(1)\prime} \rangle \right)~,
\end{equation}
while carrying out \eqref{phibetaf1diffeq} one order further results in
\begin{align}\label{boosteps2}
\left[ - (1 - \beta_{i}^2) \pd_{\rho}^2 + V^{(2)}(\phi_{\beta_i}) \right] \phi_{\beta_f}^{(2)} =&~  -  \left( 2 \beta_i \beta_{f}^{(2)} + \beta_{f}^{(1)2} \right) \phi_{\beta_i}'' - 2 \beta_i \beta_{f}^{(1)} \phi_{\beta_f}^{(1)\prime\prime}  + \cr
&~ - \half V^{(3)}(\phi_{\beta_i}) \phi_{\beta_f}^{(1)2} ~.
\end{align}
These agree with \eqref{2ndccondition} and \eqref{2ndfcondition} if $c = \beta_{f}^{(2)}$ and $f = \phi_{\beta_f}^{(2)}$, and therefore we have established
\begin{equation}\label{SS20meaning}
|\SS^{(2,0)} \rangle = - \HH_i | \zeta_{\beta_f}^{(2)} \rangle ~.
\end{equation}
This result will be useful in the next subsection for determining the time-independent part of the second order solution.  

For the time-dependent part of the second-order solution we'll instead want an explicit expression for $|\SS^{(2,0)}\rangle$ along the lines of \eqref{SS20phiresult} and \eqref{SS20piresult}.  We can organize these results in terms of phase space quantities as follows.  First, we give an alternative expression for $|\SS_{\varphi}^{(2,0)}\rangle$ using the momentum component of $|\zeta_{\beta_f}^{(1)}\rangle$.  Since $|\varpi_{\beta_f}^{(1)}\rangle = - \beta_{f}^{(1)} \PP_{\uppsi_0}^\perp |\phi_{\beta_i}' \rangle - \beta_i \PP_{\uppsi_0}^\perp |\phi_{\beta_f}^{(1)\prime} \rangle$, we have
\begin{equation}
|\varpi_{\beta_f}^{(1)\prime} \rangle = - \beta_i |\phi_{\beta_f}^{(1)\prime \prime} \rangle - \beta_{f}^{(1)} |\phi_{\beta_i}'' \rangle + \left( \beta_i \langle \uppsi_0 |\phi_{\beta_f}^{(1)\prime} \rangle + \beta_{f}^{(1)} \langle \uppsi_0 |\phi_{\beta_i}' \rangle \right) |\uppsi_0' \rangle~,
\end{equation}
and hence one can write
\begin{align}\label{SS20phialt}
|S_{\varphi}^{(2,0)} \rangle =&~ \PP_{\uppsi_0}^\perp \Bigg\{ - \beta_{f}^{(1)} | \varpi_{\beta_f}^{(1)\prime} \rangle + \half | V^{(3)}(\phi_{\beta_i}) \phi_{\beta_f}^{(1)2} \rangle + \cr
&~ \qquad + \left( \beta_{f}^{(1)2} \langle \uppsi_0 |\phi_{\beta_i}' \rangle  + 2 \beta_i \beta_{f}^{(1)} \langle \uppsi_0 | \phi_{\beta_f}^{(1)\prime} \rangle \right) | \uppsi_0' \rangle + \cr
&~ \qquad + \beta_i \left( \frac{ \langle \varpi_{\beta_f}^{(1)} | \phi_{\beta_f}^{(1)\prime} \rangle - \beta_i \langle \uppsi_0 | \phi_{\beta_f}^{(1)\prime} \rangle^2}{\langle \uppsi_0 | \phi_{\beta_i}' \rangle^2} - 2 \beta_{f}^{(1)} \frac{\langle \uppsi_0 |\phi_{\beta_f}^{(1)\prime} \rangle}{\langle \uppsi_0 |\phi_{\beta_i}' \rangle} \right) \times \cr
&~ \qquad \qquad \times \left( | \phi_{\beta_i}'' \rangle + \langle \uppsi_0 |\phi_{\beta_i}' \rangle |\uppsi_0' \rangle \right) \Bigg\}~.
\end{align}

Now notice that the two terms in \eqref{SS20piresult} can be combined with the first and last terms in \eqref{SS20phialt} to form the phase space kets $-\JJ \pd_\rho |\zeta_{\beta_f}^{(1)}\rangle$ and $|B_i\rangle$.  For the terms in the second line of \eqref{SS20phialt}, it pays to promote $|\uppsi_0' \rangle$ to a phase space ket by defining
\begin{equation}
|\Psi_0 \rangle := \left( \begin{array}{c} 0 \\ |\uppsi_0\rangle \end{array} \right)~.
\end{equation}
It does not appear to be useful to think about the potential term this way, so we just write it out explicitly.  Then we have
\begin{align}\label{SS20final}
|\SS^{(2,0)}\rangle =&~ \PP_{\uppsi_0}^\perp \Bigg\{ - \beta_{f}^{(1)} \JJ |\zeta_{\beta_f}^{(1)\prime} \rangle + \left( \beta_{f}^{(1)2} \langle \uppsi_0 | \phi_{\beta_i}' \rangle - 2 \beta_i \beta_{f}^{(1)} \langle \Psi_0' | \zeta_{\beta_f}^{(1)} \rangle \right) | \Psi_0' \rangle + \cr
&~ \qquad + \left( - \half \langle \zeta_{\beta_f}^{(1)} | \JJ | \zeta_{\beta_f}^{(1)\prime}\rangle - \beta_i \langle \Psi_0' | \zeta_{\beta_f}^{(1)} \rangle^2 + 2 \beta_{f}^{(1)} \langle \uppsi_0 | \phi_{\beta_i}' \rangle \langle \Psi_{0}' | \zeta_{\beta_f}^{(1)} \rangle \right) | B_i \rangle + \cr
&~ \qquad + \half \left( \begin{array}{c} 0 \\ | V^{(3)}(\phi_{\beta_i}) \phi_{\beta_f}^{(1)2} \rangle \end{array} \right) \Bigg\} ~.
\end{align}

Next we consider the contribution $|\SS^{(2,1)}\rangle$ to the second order source, consisting of all terms involving one power of the components of $|\zeta_{\rm osc}^{(1)}\rangle$.  The components of $|\SS^{(2,1)}\rangle$ are
\begin{align}
| \SS_{\varpi}^{(2,1)} \rangle =&~ \PP_{\uppsi_0}^{\perp} \Bigg\{ \left[ \frac{\delta^3 H}{\delta \varpi \delta \varphi_2 \delta \varpi_3} \bigg|_i  | \phi_{\beta_f}^{(1)} \rangle_2 \right] |\varpi_{\rm osc}^{(1)} \rangle_3 + \cr
& \qquad +  \left[  \frac{\pd \delta^2H}{\pd P \delta \varpi \delta \varphi_3} \bigg|_i P^{(1)} + \frac{\delta^3 H }{\delta \varpi \delta \varpi_2 \delta \varphi_3} \bigg|_i  |\varpi_{\beta_f}^{(1)} \rangle_2 +  \frac{\delta^3 H}{\delta \varpi \delta \varphi_2 \delta \varphi_3} \bigg|_i  |\varphi_{\beta_f}^{(1)} \rangle_2 \right] |\varphi_{\rm osc}^{(1)}\rangle \Bigg\}~, \qquad  \cr
| \SS_{\varphi}^{(2,1)} \rangle =&~ \PP_{\uppsi_0}^\perp \Bigg\{ \left[ \frac{\pd \delta^2 H}{\pd P \delta \varphi \delta \varpi_3} \bigg|_i P^{(1)}  +  \frac{\delta^3H}{\delta \varphi \delta \varpi_2 \delta \varpi_3} \bigg|_i  | \varpi_{\beta_i}^{(1)} \rangle_2 +  \frac{\delta^3 H}{\delta \varphi \delta \varphi_2 \delta \varpi_3} \bigg|_i  | \phi_{\beta_f}^{(1)} \rangle_2 \right] | \varpi_{\rm osc}^{(1)}\rangle_3 + \cr
& \qquad  +  \left[  \frac{\pd \delta^2 H}{\pd P \delta \varphi \delta \varphi_3} \bigg|_i P^{(1)} + \frac{\delta^3 H}{\delta \varphi \delta \varpi_2 \delta \varphi_3} \bigg|_i  | \varpi_{\beta_f}^{(1)} \rangle  +  \frac{\delta^3 H}{\delta \varphi \delta \varphi_2 \delta \varphi_3} \bigg|_i  |\phi_{\beta_f}^{(1)} \rangle_2 \right] |\varphi_{\rm osc}^{(1)} \rangle_3 \Bigg\} . \cr
\end{align}
Since $\PP_{\uppsi}^\perp |\varpi_{\beta_f}^{(1)}\rangle = |\varpi_{\beta_f}^{(1)}\rangle$ and similarly for $|\phi_{\beta_f}^{(1)} \rangle$, we recognize the quantities in square brackets as the $O(\epsilon)$ term the expansion of $\HH_f$, the quadratic fluctuation operator around the $\beta_f$ boosted kink.  Hence we can write
\begin{equation}
|\SS^{(2,1)} \rangle = \HH_{f}^{(1)} | \zeta_{\rm osc}^{(1)}\rangle ~.
\end{equation}

We compute the blocks and find
\begin{align}\label{HHf1blocks}
\HH_{f}^{(1)} = \left( \mathbbm{1}_2 \otimes \PP_{\uppsi_0}^\perp \right) \left( \begin{array}{c c} \MM_{f}^{(1)} & \BB_{f}^{(1)} \\ (\BB_{f}^{(1)})^T & \KK_{f}^{(1)} \end{array} \right) \left( \mathbbm{1}_2 \otimes \PP_{\uppsi_0}^\perp \right)~,
\end{align}
with
\begin{align}
\MM_{f}^{(1)} =&~ \frac{ | \phi_{\beta_f}^{(1)\prime} \rangle \langle \phi_{\beta_i}' | + |\phi_{\beta_i}' \rangle \langle \phi_{\beta_f}^{(1)\prime} | }{\langle \uppsi_0 |\phi_{\beta_i}' \rangle^2} - \frac{2 \langle \uppsi_0 |\phi_{\beta_f}^{(1)\prime}\rangle}{\langle \uppsi_0 |\phi_{\beta_i}' \rangle^3} | \phi_{\beta_i}' \rangle \langle \phi_{\beta_i}' | ~, \cr
\BB_{f}^{(1)} =&~ \beta_{f}^{(1)} \left( \mathbbm{1} \pd_\rho + \frac{2 |\phi_{\beta_i}' \rangle \langle \uppsi_0' |}{\langle \uppsi_0 |\phi_{\beta_i}' \rangle} \right) + \frac{\beta_i | \phi_{\beta_f}^{(1)\prime} \rangle ( \langle \phi_{\beta_i}'' + \langle \uppsi_0 |\phi_{\beta_i}' \rangle \langle \uppsi_0' | ) - |\phi_{\beta_i}' \rangle \langle \varpi_{\beta_f}^{(1)\prime} | }{\langle \uppsi_0 |\phi_{\beta_i}' \rangle^2} + \cr
&~ - \frac{2 \beta_i \langle \uppsi_0 |\phi_{\beta_f}^{(1)\prime} \rangle}{\langle \uppsi_0 |\phi_{\beta_i}' \rangle^3} | \phi_{\beta_i}' \rangle \langle \phi_{\beta_i}'' | ~, \cr
\KK_{f}^{(1)} =&~  - \frac{\beta_i}{\langle \uppsi_0 |\phi_{\beta_i}' \rangle^2} \left[ |\varpi_{\beta_f}^{(1)\prime} \rangle (\langle \phi_{\beta_i}''| + \langle \uppsi_0 | \phi_{\beta_i}' \rangle \langle \uppsi_0' | ) + ( |\phi_{\beta_i}'' \rangle + \langle \uppsi_0 |\phi_{\beta_i}' \rangle |\uppsi_0' \rangle) \langle \varpi_{\beta_f}^{(1)\prime } | \right]  + \cr 
&~ + \frac{2 \beta_i \beta_{f}^{(1)}}{\langle \uppsi_0 |\phi_{\beta_i}' \rangle} \left( |\uppsi_0' \rangle \langle \phi_{\beta_i}'' |  + |\phi_{\beta_i}'' \rangle \langle \uppsi_0' | + \langle \uppsi_0 |\phi_{\beta_i}' \rangle |\uppsi_0' \rangle \langle \uppsi_0' | \right) + \cr
&~ -  \frac{2 \beta_{f}^2 \langle \uppsi_0 |\phi_{\beta_f}^{(1)\prime} \rangle}{\langle \uppsi_0 |\phi_{\beta_i}' \rangle^3} |\phi_{\beta_i}'' \rangle \langle \phi_{\beta_i}'' |  + \frac{2 \beta_{i}^2 \langle \uppsi_0 |\phi_{\beta_f}^{(1)\prime}\rangle}{\langle \uppsi_0 |\phi_{\beta_i}' \rangle} |\uppsi_0' \rangle \langle \uppsi_0' | + \mathbbm{1} V^{(3)}(\phi_{\beta_i}) \phi_{\beta_f}^{(1)} ~.
\end{align}

For explicit computation it is useful to have a more manageable phase space representation of the  block matrix in \eqref{HHf1blocks}.  We find that it can be represented as follows in terms of the phase space kets introduced earlier:
\begin{align}\label{HHf1HHtildef1}
\HH_{f}^{(1)} =&~ \left( \mathbbm{1}_2 \otimes \PP_{\uppsi_0}^\perp \right) \widetilde{\HH}_{f}^{(1)} \left( \mathbbm{1}_2 \otimes \PP_{\uppsi_0}^\perp \right)~,
\end{align}
with
\begin{align}\label{HHtildef1}
\widetilde{\HH}_{f}^{(1)} =&~  - \beta_{f}^{(1)} \JJ \pd_\rho - \JJ | \zeta_{\beta_f}^{(1)\prime} \rangle \langle B_i | + |B_i \rangle \langle \zeta_{\beta_f}^{(1)\prime} | \JJ  +  2 \beta_{f}^{(1)} \langle \uppsi_0 |\phi_{\beta_i}' \rangle \left( |B_i \rangle \langle \Psi_0' | + |\Psi_0' \rangle \langle B_i | \right) + \cr
&~ - 2 \beta_i \beta_{f}^{(1)} |\Psi_{0}' \rangle \langle \Psi_0' | + 2 \langle \uppsi_0 |\phi_{\beta_i}' \rangle \langle \Psi_0' |\zeta_{\beta_f}^{(1)} \rangle  |B_i \rangle \langle B_i |  + \cr
&~ -2 \beta_i \langle \Psi_0' | \zeta_{\beta_f}^{(1)} \rangle \left( |B_i \rangle \langle \Psi_0' | + |\Psi_0' \rangle \langle B_i | \right) + \mathbbm{1}_{\varphi} V^{(3)}(\phi_{\beta_i}) \phi_{\beta_f}^{(1)} ~, \raisetag{16pt}
\end{align}
where in the last term $\mathbbm{1}_{\varphi}$ denotes a matrix that is the identity in the lower right (position-position) corner and zero elsewhere.

Finally we consider the contribution $|\SS^{(2,2)}\rangle$ to the second order source consisting of all terms that are quadratic in the components of $|\zeta_{\rm osc}^{(1)}\rangle$.  The momentum component is
\begin{align}
|\SS_{\varpi}^{(2,2)} \rangle =&~ \PP_{\uppsi_0}^\perp \left\{  \frac{\pd^3 H}{\pd \varpi \pd\varpi_2 \pd \varphi_3} \bigg|_i  | \varpi_{\rm osc}^{(1)} \rangle_2 | \varphi_{\rm osc}^{(1)} \rangle_3 +  \half  \frac{\pd^3 H}{\pd \varpi \pd\varphi_2 \pd \varphi_3} \bigg|_i  | \varphi_{\rm osc}^{(1)} \rangle_2 | \varphi_{\rm osc}^{(1)} \rangle_3  \right\} ~,
\end{align}
and straightforward calculation with \eqref{delta3Heval} results in
\begin{align}
| \SS_{\varpi}^{(2,2)} \rangle =&~ \PP_{\uppsi_0}^\perp \Bigg\{ \frac{\langle \varpi_{\rm osc}^{(1)} | \varphi_{\rm osc}^{(1)\prime} \rangle}{\langle \uppsi_0 |\phi_{\beta_i}' \rangle^2} |\phi_{\beta_i}' \rangle + \frac{\langle \phi_{\beta_i}' |\varpi_{\rm osc}^{(1)} \rangle}{\langle \uppsi_0 |\phi_{\beta_i}' \rangle^2} |\varphi_{\rm osc}^{(1)\prime} \rangle + \frac{2 \langle \phi_{\beta_i}' |\varpi_{\rm osc}^{(1)} \rangle \langle \uppsi_0' |\varphi_{\rm osc}^{(1)} \rangle}{\langle \uppsi_0 |\phi_{\beta_i}' \rangle^3} |\phi_{\beta_i}' \rangle + \cr
&~ \qquad + \half \bigg[ \frac{\beta_i}{\langle \uppsi_0 |\phi_{\beta_i}' \rangle^2} \left( 2 \langle \phi_{\beta_i}'' | \varphi_{\rm osc}^{(1)} \rangle |\varphi_{\rm osc}^{(1)\prime} \rangle + 2 \langle \uppsi_0 |\phi_{\beta_i}' \rangle \langle \uppsi_0' |\varphi_{\rm osc}^{(1)} \rangle |\varphi_{\rm osc}^{(1)\prime} \rangle \right) +  \cr
&~ \qquad  + \frac{2 \beta_i}{\langle \uppsi_0 | \phi_{\beta_i}' \rangle^3} \left( 2 \langle \phi_{\beta_i}'' | \varphi_{\rm osc}^{(1)} \rangle \langle \uppsi_0' | \varphi_{\rm osc}^{(1)} \rangle + \langle \uppsi_0 |\phi_{\beta_i}' \rangle \langle \uppsi_0' | \varphi_{\rm osc}^{(1)} \rangle^2 \right) |\phi_{\beta_i}' \rangle \bigg]  \Bigg\} ~. \qquad
\end{align}
Noting that
\begin{equation}\label{Bizetaosc}
\frac{\langle \phi_{\beta_i}' | \varpi_{\rm osc}^{(1)} \rangle + \beta_i ( \langle \phi_{\beta_i}'' | \varphi_{\rm osc}^{(1)} \rangle + \langle \uppsi_0 |\phi_{\beta_i}' \rangle \langle \uppsi_0' | \varphi_{\rm osc}^{(1)} \rangle ) }{\langle \uppsi_0 |\phi_{\beta_i}' \rangle^2} = \langle B_i | \zeta_{\rm osc}^{(1)} \rangle~,
\end{equation}
we can write this result more compactly as
\begin{align}\label{SS22piresult}
| \SS_{\varpi}^{(2,2)} \rangle =&~ \PP_{\uppsi_0}^\perp \Bigg\{ \langle B_i | \zeta_{\rm osc}^{(1)} \rangle | \varphi_{\rm osc}^{(1)\prime} \rangle + \cr
&~ \qquad + \left[ \frac{ \langle \varpi_{\rm osc}^{(1)} | \varphi_{\rm osc}^{(1)\prime}\rangle - \beta_i \langle \uppsi_0 |\varphi_{\rm osc}^{(1)\prime} \rangle^2}{\langle \uppsi_0 | \phi_{\beta_i}' \rangle^2} - \frac{2 \langle B_i |\zeta_{\rm osc}^{(1)} \rangle \langle \uppsi_0 |\varphi_{\rm osc}^{(1)\prime} \rangle}{\langle \uppsi_0 | \phi_{\beta_i}' \rangle} \right] |\phi_{\beta_i}' \rangle \Bigg\}~.\qquad
\end{align}

The position component is
\begin{align}\label{Sdoubpi}
|\SS^{(2,2)}_{\varphi} \rangle =&~ \PP_{\uppsi_0}^{\perp}  \Bigg\{ \half \frac{\delta^3 H}{\delta \varphi \delta \varpi_2 \delta \varpi_3} \bigg|_i | \varpi_{\rm osc}^{(1)} \rangle_2 | \varpi_{\rm osc}^{(1)} \rangle_3 +  \frac{\delta^3 H}{\delta \varphi \delta\varpi_2 \delta \varphi_3} \bigg|_i  | \varpi_{\rm osc}^{(1)} \rangle_2 | \varphi_{\rm osc}^{(1)} \rangle_3 + \cr
&~  \qquad \quad  + \half \frac{\delta^3 H}{\delta \varphi \delta \varphi_2 \delta \varphi_3} \bigg|_i | \varphi_{\rm osc}^{(1)} \rangle_2 | \varphi_{\rm osc}^{(1)} \rangle_3 \Bigg\} ~.
\end{align}
Using \eqref{Bizetaosc} we can bring the result to the form
\begin{align}\label{SS22phiresult}
|\SS^{(2,2)}_{\varphi} \rangle =&~ \PP_{\uppsi_0}^{\perp}  \Bigg\{ -\langle B_i | \zeta_{\rm osc}^{(1)} \rangle | \varpi_{\rm osc}^{(1)\prime} \rangle + \half | V^{(3)}(\phi_{\beta_i}) \varphi_{\rm osc}^{(1)2}\rangle  + \cr
&~ \qquad + \left( \langle \uppsi_0 | \phi_{\beta_i}' \rangle \langle B_i | \zeta_{\rm osc}^{(1)} \rangle^2 + 2 \beta_i \langle B_i | \zeta_{\rm osc}^{(1)} \rangle \langle \uppsi_0 | \varphi_{\rm osc}^{(1)\prime} \rangle \right) |\uppsi_0' \rangle + \cr
&~ \qquad + \beta_i \left[ \frac{ \langle \varpi_{\rm osc}^{(1)} | \varphi_{\rm osc}^{(1)\prime}\rangle - \beta_i \langle \uppsi_0 |\varphi_{\rm osc}^{(1)\prime} \rangle^2}{\langle \uppsi_0 | \phi_{\beta_i}' \rangle^2} - \frac{2 \langle B_i |\zeta_{\rm osc}^{(1)} \rangle \langle \uppsi_0 |\varphi_{\rm osc}^{(1)\prime} \rangle}{\langle \uppsi_0 | \phi_{\beta_i}' \rangle} \right] \times \cr
&~ \qquad \qquad \qquad \times  \left( |\phi_{\beta_i}'' \rangle + \langle \uppsi_0 |\phi_{\beta_i}' \rangle |\uppsi_0' \rangle \right) \Bigg\}~.
\end{align}

Note that \eqref{SS22piresult} and \eqref{SS22phiresult} have the same form as \eqref{SS20piresult} and \eqref{SS20phialt} for the components of $|\SS^{2,0)}\rangle$, if we make the replacement $|\zeta_{\rm osc}^{(1)}\rangle \to |\zeta_{\beta_f}^{(1)}\rangle$ and then send $\langle B_i | \zeta_{\beta_f}^{(1)} \rangle \to \beta_{f}^{(1)}$.  Now, $\langle B_i | \zeta_{\beta_f}^{(1)}\rangle$ is not equal to $\beta_{f}^{(1)}$, but from expanding
\begin{align}
\beta_f =&~ \frac{P_f + \langle \varpi_{\beta_f} | \phi_{\beta_f}' \rangle}{\langle \uppsi_0 | \phi_{\beta_f}' \rangle^2} \cr
=&~ \beta_i + \frac{P^{(1)} + \langle \phi_{\beta_i}' | \varpi_{\beta_f}^{(1)} \rangle - \beta_i \langle \phi_{\beta_i}' | \phi_{\beta_f}^{(1)\prime} \rangle + \beta_i \langle \uppsi_0 | \phi_{\beta_i}' \rangle \langle \uppsi_0 |\phi_{\beta_f}^{(1)\prime} \rangle}{\langle \uppsi_0 |\phi_{\beta_i}' \rangle^2} - \frac{2 \beta_i \langle \uppsi_0 |\phi_{\beta_f}^{(1)\prime} \rangle}{\langle \uppsi_0 | \phi_{\beta_i}' \rangle} + \cr
&~ + O(\epsilon^2)~,
\end{align} 
we see that
\begin{equation}\label{betaf1vsBizetaf1}
\beta_{f}^{(1)} = \frac{P^{(1)}}{\langle \uppsi_0 | \phi_{\beta_i}' \rangle^2} + \langle B_i | \zeta_{\beta_f}^{(1)} \rangle~.
\end{equation}
Hence it must be that the additional contributions to $|\SS^{(2,0)}\rangle$, from the terms involving one and two $P$ derivatives of the second and first variations of $H$ respectively, provide the terms necessary to complete all $\langle B_i | \zeta_{\beta_f}^{(1)} \rangle$'s into $\beta_{f}^{(1)}$'s.

The results \eqref{SS22piresult} and \eqref{SS22phiresult} can be combined into a phase space ket as follows:
\begin{align}
|\SS^{(2,2)} \rangle =&~ \PP_{\uppsi_0}^\perp \Bigg\{ - \langle B_i | \zeta_{\rm osc}^{(1)} \rangle \JJ | \zeta_{\rm osc}^{(1)\prime} \rangle + \langle \zeta_{\rm osc}^{(1)} | \QQ_1 | \zeta_{\rm osc}^{(1)} \rangle |B_i \rangle + \langle \zeta_{\rm osc}^{(1)} | \QQ_2 | \zeta_{\rm osc}^{(1)} \rangle |\Psi_0' \rangle + \cr
&~ \qquad  + \half \left( \begin{array}{c} 0 \\ | V^{(3)}(\phi_{\beta_i}) \varphi_{\rm osc}^{(1)2} \rangle \end{array}\right) \Bigg\}~,
\end{align}
where we've introduced the real symmetric quadratic forms
\begin{align}\label{Q1Q2}
\QQ_1 =&~  - \half \JJ \pd_{\rho} - \beta_i |\Psi_0' \rangle \langle \Psi_0' | + \langle \uppsi_0 | \phi_{\beta_i}' \rangle \left( |B_i \rangle \langle \Psi_0' | + |\Psi_0' \rangle \langle B_i| \right)~, \cr
\QQ_2 =&~ \langle \uppsi_0 | \phi_{\beta_i}' \rangle |B_i \rangle \langle B_i| - \beta_i \left( |\Psi_0' \rangle \langle B_i | + |B_i \rangle \langle \Psi_0' | \right)~.
\end{align}

This form of $|\SS^{(2,2)}\rangle$ makes it easy to read off the phase space kets $|J^{bc} \rangle$ and $|K^{bc}\rangle$ defined in \eqref{JKkets}.  All we need to do is insert the mode sum \eqref{zeta1solwithzetaf} for $|\zeta_{\rm osc}^{(1)} \rangle$ and collect the relevant coefficients.  This leads to
\begin{align}\label{JKexp}
|J^{bc} \rangle =&~ \frac{P^{(1)2} \langle \eta_b | B_i \rangle \langle \eta_c | B_i \rangle}{\nu_b \nu_c} \Bigg\{ - \half ( \langle B_i | \eta_b \rangle \JJ | \eta_{c}' \rangle + \langle B_i | \eta_c \rangle \JJ | \eta_b'\rangle ) + \langle \eta_{b}^\ast | \QQ_1 | \eta_c \rangle  |B_i \rangle  + \cr
&~   \qquad \qquad \qquad \qquad \quad +  \langle \eta_{b}^\ast | \QQ_2 | \eta_c \rangle |\Psi_0' \rangle + \half  \left( \begin{array}{c} 0 \\ | V^{(3)}(\phi_{\beta_i}) \eta_{\varphi b} \eta_{\varphi c} \rangle \end{array} \right) \Bigg\} ~, \cr
| K^{bc} \rangle =&~ \frac{P^{(1)2} \langle \eta_b | B_i \rangle \langle \eta_c^\ast | B_i \rangle}{\nu_b \nu_c} \Bigg\{ - \half ( \langle B_i | \eta_b \rangle \JJ | \eta_{c}^{\ast \prime} \rangle + \langle B_i | \eta_{c}^\ast \rangle \JJ | \eta_b'\rangle ) + \langle \eta_{b}^\ast | \QQ_1 | \eta_{c}^\ast \rangle  |B_i \rangle  + \cr
&~   \qquad \qquad \qquad \qquad \quad +  \langle \eta_{b}^\ast | \QQ_2 | \eta_{c}^\ast \rangle |\Psi_0' \rangle + \half \left( \begin{array}{c} 0 \\ | V^{(3)}(\phi_{\beta_i}) \eta_{\varphi b} \eta_{\varphi c}^\ast \rangle \end{array} \right) \Bigg\} ~. \raisetag{24pt}
\end{align}

We comment on the case $\beta_i = 0$ before concluding this subsection.  The first order configuration $|\zeta^{(1)}\rangle$ is zero in this limit so $|\SS^{(2,1)}\rangle$ and $|\SS^{(2,2)}\rangle$ vanish entirely.  $|\SS_{\varpi}^{(2,0)}\rangle$ vanishes while $|\SS^{(2,0)}_{\varphi} \rangle$ has one nonzero contribution from the very first term in \eqref{SS20def}.  This term shows up in \eqref{SS20phiresult} as the third term in the first line.  Note that $|\uppsi_0' \rangle \propto |\phi_0'' \rangle$. The proportionality constant is correct to make these identified terms in \eqref{SS20def} and \eqref{SS20phiresult} match.  To see this, use that $\langle \uppsi_0 | \phi_{0}' \rangle = \sqrt{M_0'}$ and $\beta_{f}^{(1)} = M_0 \epsilon/M_0'$ in the $\beta_i = 0$ limit.  From either point of view the entire second order source is
\begin{equation}\label{SS20betai0}
\lim_{\beta_i \to 0} |\SS^{(2)}\rangle = \frac{(M_0 \epsilon)^2}{M_{0}^{\prime 2}}\left( \begin{array}{c} 0 \\ |\phi_{0}'' \rangle \end{array} \right)~.
\end{equation}
% 

%%%%%%%%%%%%%%%%%%%
\subsection{Second Order Solution}
%%%%%%%%%%%%%%%%%%%

Now we use the results of the previous subsection to simplify the solution for the second order fields $|\zeta^{(2)}\rangle$ and collective coordinate velocity $\beta^{(2)}$ as much as possible.  We start with the first order correction to the frequencies \eqref{1storderfreqshift} due to the resonant term.  Using \eqref{HHf1HHtildef1} and \eqref{HHtildef1} we get
\begin{align}\label{freqshiftexp}
\nu_{a}^{(1)} =&~ \langle \eta_a | \HH_{f}^{(1)} | \eta_a \rangle = \langle \eta_a |\widetilde{\HH}_{f}^{(1)} | \eta_a \rangle \cr
=&~ \beta_{f}^{(1)} \left( - \langle \eta_a | \JJ | \eta_a' \rangle + 4 \alpha_i \Re \{ b_a  z_{a}^\ast \} - 2 \beta_i |z_a|^2 \right) - 2 \Re \left\{ b_{a}^\ast \langle \eta_a | \JJ | \zeta_{\beta_f}^{(1)\prime} \rangle \right\} + \cr
&~ -2 \alpha_i \alpha_{f}^{(1)} |b_a|^2 + 4 \beta_i \alpha_{f}^{(1)} \Re \{ b_a z_{a}^\ast \} + \left\langle \eta_{\varphi a} \left| V^{(3)}(\phi_{\beta_i}) \phi_{\beta_f}^{(1)} \eta_{\varphi a} \right. \right\rangle ~,
\end{align}
where in addition to $b_a = \langle \eta_a |B_i \rangle$ introduced the following shorthand:
\begin{equation}\label{alphazdef}
\alpha_i = \langle \uppsi_0 |\phi_{\beta_i}' \rangle ~, \qquad \alpha_{f}^{(n)} = \langle \Psi_0 | \zeta_{\beta_f}^{(n)\prime} \rangle = \langle \uppsi_0 | \phi_{\beta_f}^{(n)\prime} \rangle~, \qquad z_a = \langle \eta_a | \Psi_0' \rangle = \langle \eta_{\varphi a} | \uppsi_0' \rangle ~.
\end{equation}

Then for the $A_2$ equation of motion \eqref{A2eomexp} we need the following additional quantities:
\begin{align}\label{A2eompieces}
\langle \eta_a | \SS^{(2,0)} \rangle =&~  \beta_{f}^{(1)2} \alpha_i z_a + \beta_{f}^{(1)} \left( - \langle \eta_a | \JJ | \zeta_{\beta_f}^{(1)\prime} \rangle + 2 \beta_i \alpha_{f}^{(1)} z_a - 2 \alpha_i \alpha_{f}^{(1)} b_a \right) + \cr
&~ - \half b_a \langle \zeta_{\beta_f}^{(1)} | \JJ | \zeta_{\beta_f}^{(1)\prime} \rangle - \beta_i \alpha_{f}^{(1)2} b_a + \half \left\langle \eta_{\varphi a} \left| V^{(3)}(\phi_{\beta_i}) \phi_{\beta_f}^{(1)2} \right. \right\rangle ~, \cr
\langle \eta_a | \HH_{f}^{(1)} | \eta_b \rangle =&~ \beta_{f}^{(1)} \left( - \langle \eta_a | \JJ | \eta_{b}' \rangle + 2 \alpha_i (b_a z_{b}^\ast + z_a b_{b}^\ast) - 2 \beta_i z_{a} z_{b}^\ast \right) + \cr
&~ - \langle \eta_a | \JJ |\zeta_{\beta_f}^{(1)\prime} \rangle b_{b}^\ast - b_a \langle \eta_{b}^\ast | \JJ | \zeta_{\beta_f}^{(1)\prime} \rangle - 2 \alpha_i \alpha_{f}^{(1)} b_a b_{b}^\ast  + \cr
&~ + 2 \beta_i \alpha_{f}^{(1)} (b_a z_{b}^\ast + z_a b_{b}^\ast )+ \left\langle \eta_{\varphi a} \left| V^{(3)}(\phi_{\beta_i}) \phi_{\beta_f}^{(1)} \eta_{\varphi b} \right. \right\rangle ~, \cr
\langle \eta_a | J^{bc} \rangle =&~ \frac{P^{(1)2} b_b b_c}{\nu_b \nu_c} \bigg\{ -\half b_{b}^\ast \langle \eta_a | \JJ | \eta_{c}' \rangle -\half b_{c}^\ast \langle \eta_a | \JJ | \eta_{b}' \rangle - \half b_a \langle \eta_{b}^\ast | \JJ | \eta_{c}' \rangle + \cr
&~ \qquad \qquad + b_a \left[ -\beta_i z_{b}^\ast z_{c}^\ast  + \alpha_i (b_{b}^\ast z_{c}^\ast + z_{b}^\ast b_{c}^\ast) \right] + \cr
&~ \qquad \qquad + z_a \left[ \alpha_i b_{b}^\ast b_{c}^\ast - \beta_i (b_{b}^\ast z_{c}^\ast + z_{b}^\ast b_{c}^\ast) \right] + \half  \left\langle \eta_a \left| V^{(3)}(\phi_{\beta_i}) \eta_{\varphi b} \eta_{\varphi c} \right. \right\rangle \bigg\} ~. \cr
\end{align}
where we used \eqref{SS20final}, \eqref{HHtildef1}, and \eqref{JKexp}.  We can get $\langle \eta_a | \HH_{f}^{(1)} | \eta_{b}^\ast \rangle$ from $\langle \eta_a | \HH_{f}^{(1)} | \eta_b \rangle$ by conjugating everything that carries a $b$ index and $\langle \eta_a |K^{bc} \rangle$ from $\langle \eta_a | J^{bc} \rangle$ by conjugating everything that carries a $c$ index.

In terms of these quantities we can then integrate \eqref{A2eomexp} to obtain $A^{(2)}$ and insert the result into \eqref{modifiedfieldexp} to obtain the second order configuration
\begin{align}\label{zeta2solution}
|\zeta^{(2)} \rangle =&  \sum_{a=1}^{\infty} 2 \Re \left\{ |\eta_a \rangle \left[ \frac{\langle \eta_a |\SS^{(2,0)}\rangle}{\nu_a} + \frac{P^{(1)} b_a \nu_{a}^{(1)}}{\nu_{a}^2} \right] \left( e^{- \ii (\nu_a + \cdots )(t-t_\ast)} - 1 \right) \right\} + \cr
& + \sum_{\substack{a,b=1 \\ a\neq b}}^{\infty} 2 \Re \left\{ |\eta_a \rangle \frac{ P^{(1)} b_{b} \langle \eta_a | \HH_{f}^{(1)} | \eta_b \rangle }{\nu_b (\nu_a - \nu_b)} \left( e^{-\ii (\nu_a + \cdots)(t-t_\ast)} - e^{- \ii (\nu_b + \cdots)(t-t_\ast)} \right) \right\} + \cr
& + \sum_{a,b = 1}^{\infty} 2 \Re \left\{ | \eta_a \rangle \frac{P^{(1)} b_{b}^\ast \langle \eta_a | \HH_{f}^{(1)} | \eta_{b}^\ast \rangle}{\nu_b (\nu_a + \nu_b)} \left( e^{-\ii (\nu_a + \cdots)(t-t_\ast)} - e^{\ii (\nu_b + \cdots)(t-t_\ast)} \right) \right\} + \cr
& + \sum_{a,b,c=1}^{\infty} 2 \Re \left\{ |\eta_a \rangle \frac{ \langle \eta_a | J^{bc} \rangle}{(\nu_a - \nu_b - \nu_c)} \left( e^{-\ii (\nu_a + \cdots)(t-t_\ast)} - e^{-\ii (\nu_b + \nu_c + \cdots)(t-t_\ast)} \right) \right\} + \cr
& + \sum_{a,b,c=1}^{\infty} 2 \Re \left\{ |\eta_a \rangle \frac{ \langle \eta_a | J^{bc\ast} \rangle}{(\nu_a + \nu_b + \nu_c)} \left( e^{-\ii (\nu_a + \cdots)(t-t_\ast)} - e^{\ii (\nu_b + \nu_c + \cdots)(t-t_\ast)} \right) \right\} + \cr
& + \sum_{a,b,c=1}^{\infty} 4 \Re \left\{ |\eta_a \rangle \frac{ \langle \eta_a | K^{bc} \rangle}{(\nu_a - \nu_b + \nu_c)} \left( e^{-\ii (\nu_a + \cdots)(t-t_\ast)} - e^{-\ii (\nu_b - \nu_c + \cdots)(t-t_\ast)} \right) \right\} , \qquad \qquad 
\end{align}
where the entire expression is for $t > t_\ast$.  We have $|\zeta^{(2)}\rangle = 0$ for $t < t_\ast$.  The time-independent terms come from the $-1$ term in the first line and the $e^{-\ii (\nu_b - \nu_c + \cdots)(t-t_\ast)}$ term in the last line when $b = c$.  All remaining terms oscillate around zero.  Hence the second-order time-averaged configuration is
\begin{align}\label{zeta2avapp}
|\zeta_{\rm av}^{(2)} \rangle = - \sum_{a=1}^{\infty} 2 \Re \bigg\{\frac{1}{\nu_a} |\eta_a \rangle \langle \eta_a | \SS^{(2,0)} \rangle + \frac{P^{(1)} b_a  \nu_{a}^{(1)}}{\nu_{a}^2} |\eta_a \rangle  + \frac{2}{\nu_a} |\eta_a \rangle \sum_{b=1}^{\infty} \langle \eta_a | K^{bb} \rangle \bigg\}~.
\end{align}
Note that $|\SS^{(2,0)}\rangle$ and $|K^{bb}\rangle$ are real.  With the result \eqref{SS20meaning} we recognize the first term as $|\zeta_{\beta_f}^{(2)}\rangle$.  Hence, as we noted in text around \eqref{deltazeta2av}, the remaining two terms give the leading deviation of the time-averaged fields from  a $\beta_f$-boosted kink configuration which we denote $|\delta \zeta_{\rm av}^{(2)} \rangle$.  Removing the terms \eqref{zeta2avapp} from \eqref{zeta2solution}, we define the remainder as $|\zeta_{\rm osc}^{(2)}\rangle$.

Our target is the second order collective coordinate velocity which receives contributions from terms linear in the second order solution \eqref{zeta2solution} and terms quadratic in the first order solution \eqref{zeta1solwithzetaf}.  As we explained in the main text, $\beta^{(2)}$ can be brought to the form \eqref{beta2simp1} which we reproduce here:
\begin{equation}\label{beta2fullapp}
\beta^{(2)} = \beta_{f}^{(2)} + \langle B_i | \delta \zeta_{\rm av}^{(2)}\rangle + \langle B_i | \zeta_{\rm osc}^{(2)} \rangle + \langle B_{f}^{(1)} | \zeta_{\rm osc}^{(1)} \rangle + \frac{\langle \zeta_{\rm osc}^{(1)} | \QQ_1 | \zeta_{\rm osc}^{(1)} \rangle}{\langle \uppsi_0 | \phi_{\beta_i}' \rangle^2} ~.
\end{equation}
We note that $|B_{f}^{(1)}\rangle$ can be expressed in terms of the phase space kets introduced earlier via
\begin{equation}\label{Bf1fromknowns}
|B_{f}^{(1)} \rangle = - \frac{1}{\alpha_{i}^2} \left( \JJ | \zeta_{\beta_f}^{(1)\prime} \rangle + 2 \alpha_i \alpha_{f}^{(1)} |B_i \rangle - 2 (\alpha_i \beta_{f}^{(1)} + \beta_i \alpha_{f}^{(1)}) |\Psi_0' \rangle \right)~.
\end{equation}

The time-independent part of \eqref{beta2fullapp} includes the first two terms but also receives a contribution from the last term when we take the same mode from the two factors of $|\zeta_{\rm osc}^{(1)}\rangle$.  Hence we write
\begin{align}\label{deltabeta2}
\beta^{(2)} =&~ \beta_{\rm av}^{(2)} + \beta_{\rm osc}^{(2)} = \beta_{f}^{(2)} + \delta \beta_{\rm av}^{(2)} + \beta_{\rm osc}^{(2)}~, \qquad \textrm{with} \cr
\delta \beta_{\rm av}^{(2)} =&~ \langle B_i | \delta \zeta_{\rm av}^{(2)}\rangle + 2 \sum_{a} \frac{P^{(1)2} |b_{a}|^2}{\nu_{a}^2 \alpha_{i}^2} \langle \eta_a | \QQ_1 | \eta_a \rangle ~.
\end{align}

Let us analyze $\delta \beta_{\rm av}^{(2)}$.  The first term is
\begin{align}
\langle B_i | \delta \zeta_{\rm av}^{(2)}\rangle =&~ -2  \sum_{a=1}^{\infty} \left[ \frac{P^{(1)} |b_{a}|^2 \nu_{a}^{(1)}}{\nu_{a}^2} + \frac{1}{\nu_a} \langle B_i | \left( |\eta_a \rangle \langle \eta_a | + | \eta_{a}^\ast \rangle \langle \eta_{a}^\ast | \right) \sum_{b=1}^{\infty} |K^{bb}\rangle \right] \cr
=&~ 2 \sum_{a=1}^{\infty} \left[ \frac{1}{P^{(1)}} \langle \zeta_{\beta_f}^{(1)} | K^{aa} \rangle - \frac{P^{(1)} |b_{a}|^2 \nu_{a}^{(1)}}{\nu_{a}^2} \right]~,
\end{align}
where in the second step we used the transpose of \eqref{HHzetaresult} and the fact that $|\SS^{(1)} \rangle = P^{(1)} \PP_{\uppsi_0}^\perp |B_i \rangle$.  The inverse of $P^{(1)}$ is not a concern since $|K^{ab}\rangle$ has a factor of $P^{(1)2}$ in it.  In fact, from \eqref{JKexp} we have
\begin{align}
\langle \zeta_{\beta_f}^{(1)} | K^{aa} \rangle =&~ \frac{P^{(1)2} |b_{a}|^2}{\nu_{a}^2} \bigg\{ - \half b_{a}^\ast \langle \eta_{a} | \JJ | \zeta_{\beta_f}^{(1)\prime} \rangle - \half b_a \langle \eta_{a}^\ast | \JJ | \zeta_{\beta_f}^{(1)\prime} \rangle -\half \langle B_i | \zeta_{\beta_f}^{(1)} \rangle \langle \eta_a | \JJ | \eta_{a}' \rangle + \cr
&~ \qquad \qquad \quad + \langle B_i | \zeta_{\beta_f}^{(1)} \rangle \left[ - \beta_i |z_a|^2 + 2 \alpha_i \Re \{ b_a z_{a}^\ast \} \right]  + \cr
&~ \qquad \qquad \quad - \langle \Psi_0 | \zeta_{\beta_f}^{(1)\prime} \rangle \left[ \alpha_i |b_{a}|^2 - 2\beta_i \Re\{ z_a b_{a}^\ast \} \right] + \cr
&~ \qquad \qquad \quad     + \half \left\langle \eta_{\varphi a} \left| V^{(3)}(\phi_{\beta_i}) \phi_{\beta_f}^{(1)} \eta_{\varphi a} \right. \right\rangle \bigg\}  ~.
\end{align}
But recall \eqref{betaf1vsBizetaf1} and \eqref{alphazdef}.   Using these brings us to
\begin{align}
\langle \zeta_{\beta_f}^{(1)} | K^{aa} \rangle =&~ \frac{P^{(1)3} |b_{a}|^2}{\nu_{a}^2 \alpha_{i}^2} \left( \half \langle \eta_a | \JJ | \eta_a' \rangle + \beta_i |z_a|^2 - 2\alpha_i \Re\{ b_a z_{a}^\ast \} \right) + \cr
&~ +   \frac{P^{(1)2} |b_{a}|^2}{\nu_{a}^2} \bigg\{ - \Re \left\{ b_{a}^\ast \langle \eta_{a} | \JJ | \zeta_{\beta_f}^{(1)\prime} \rangle \right\}  -\half \beta_{f}^{(1)} \langle \eta_a | \JJ | \eta_{a}' \rangle + \cr
&~ \qquad \qquad \qquad + \beta_{f}^{(1)} \left[ - \beta_i |z_a|^2 + 2 \alpha_i \Re \{ b_a z_{a}^\ast \} \right]  + \cr
&~ \qquad \qquad \qquad - \alpha_{f}^{(1)} \left[ \alpha_i |b_{a}|^2 - 2\beta_i \Re\{ z_a b_{a}^\ast \} \right] +  \half \left\langle \eta_{\varphi a} \left| V^{(3)}(\phi_{\beta_i}) \phi_{\beta_f}^{(1)} \eta_{\varphi a} \right. \right\rangle \bigg\}   \cr
=&~  \frac{P^{(1)3} |b_{a}|^2}{\nu_{a}^2 \alpha_{i}^2} \left( \half \langle \eta_a | \JJ | \eta_a' \rangle + \beta_i |z_a|^2 - 2\alpha_i \Re\{ b_a z_{a}^\ast \} \right) + \frac{ P^{(1)2} |b_a|^2 \nu_{a}^{(1)}}{2 \nu_{a}^2} ~.
\end{align}
In the last step we noted that the set of terms in the big curly brackets is exactly $\half \nu_{a}^{(1)}$, \eqref{freqshiftexp}.

Meanwhile the summand of the last term of \eqref{deltabeta2} is
\begin{align}
\frac{2 P^{(1)2} |b_{a}|^2}{\nu_{a}^2 \alpha_{i}^2} \langle \eta_a | \QQ_1 | \eta_a \rangle =&~ \frac{2 P^{(1)2} |b_a|^2}{\nu_{a}^2 \alpha_{i}^2} \left[ - \half \langle \eta_a | \JJ | \eta_a' \rangle - \beta_i |z_a|^2 + 2 \alpha_i \Re \{ b_a z_{a}^\ast \} \right] ~.
\end{align}
Hence we see that
\begin{equation}
\frac{2}{P^{(1)}} \langle \zeta_{\beta_f}^{(1)} | K^{aa} \rangle +  \frac{2 P^{(1)2} |b_{a}|^2}{\nu_{a}^2 \alpha_{i}^2} \langle \eta_a | \QQ_1 | \eta_a \rangle = \frac{ P^{(1)2} |b_a|^2 \nu_{a}^{(1)}}{\nu_{a}^2} ~,
\end{equation}
and therefore
\begin{align}
\delta \beta_{\rm av}^{(2)} =&~ - P^{(1)} \sum_{a=1}^{\infty} \frac{|b_a|^2 \nu_{a}^{(1)}}{\nu_{a}^2} ~.
\end{align}

Finally we turn to $\beta_{\rm osc}^{(2)}$.  We note that it generally involves oscillations with frequencies $\nu_a, \nu_a + \nu_b$, and $\nu_a - \nu_b$.  Thus we write
\begin{align}
\beta_{\rm osc}^{(2)} =&~ \sum_{a=1}^{\infty} 2 \Re \left\{ \beta_{a}^{(2)} e^{-\ii (\nu_a + \cdots)(t-t_\ast)} \right\} + \sum_{b,c=1}^{\infty} 2 \Re \left\{ \beta_{\mathbbm{C},bc}^{(2)} e^{-\ii (\nu_b + \nu_c + \cdots)(t-t_\ast)} \right\} + \cr
&~ + \sum_{\substack{b,c = 1 \\ b\neq c}}^{\infty} 2 \Re \left\{ \beta_{\mathbbm{R},bc}^{(2)} e^{-\ii (\nu_b - \nu_c + \cdots)(t-t_\ast)} \right\}~,
\end{align}
where the coefficients $\beta_{a}^{(2)}$, $\beta_{\mathbbm{C},ab}^{(2)}$, and $\beta_{\mathbbm{R},ab}^{(2)}$ are to be found and we can take $\beta_{\mathbbm{C},ab}^{(2)}$ symmetric in $ab$ and $\beta_{\mathbbm{R},ab}^{(2)}$ Hermitian.  We straightforwardly collect all of the relevant terms from \eqref{beta2fullapp} and find
\begin{align}\label{betaa2final}
\beta_{a}^{(2)} =&~ P^{(1)} \frac{b_a \langle B_{f}^{(1)} | \eta_a \rangle}{\nu_a} + P^{(1)} \frac{|b_{a}|^2 \nu_{a}^{(1)}}{\nu_{a}^2} + \frac{b_{a}^\ast \langle \eta_a | \SS^{(2,0)} \rangle}{\nu_a}  + \cr
&~ + P^{(1)} \sum_{\substack{b=1 \\ b\neq a}}^{\infty} \frac{ \nu_a b_{a}^\ast b_b \langle \eta_a | \HH_{f}^{(1)} | \eta_b \rangle + \nu_b b_{a} b_{b}^\ast \langle \eta_{a}^\ast | \HH_{f}^{(1)} | \eta_{b}^\ast \rangle}{\nu_a \nu_b (\nu_a - \nu_b)}  + \cr
&~ +  P^{(1)} \sum_{b=1}^{\infty} \frac{ \nu_a b_{a}^\ast b_{b}^\ast  \langle \eta_a | \HH_{f}^{(1)} | \eta_{b}^\ast \rangle - \nu_b b_a b_b \langle \eta_{a}^\ast | \HH_{f}^{(1)} | \eta_b \rangle }{\nu_a \nu_b (\nu_a + \nu_b)} + \cr
&~ +  \sum_{b,c = 1}^{\infty} \left( \frac{b_{a}^\ast \langle \eta_a |J^{bc} \rangle}{\nu_a - \nu_b - \nu_c} + \frac{b_{a}^\ast \langle \eta_a | J^{bc\ast} \rangle}{\nu_a + \nu_b + \nu_c} + \frac{2 b_{a}^\ast \langle \eta_a | K^{bc} \rangle}{\nu_a - \nu_b + \nu_c} \right)  ~,
\end{align}
\begin{align}\label{betaCab2final}
\beta_{\mathbbm{C},bc}^{(2)} =&~ \frac{P^{(1)2}}{\alpha_{i}^2} \frac{b_b b_c \langle \eta_{b}^\ast | \QQ_1 | \eta_c \rangle}{\nu_b \nu_c}  - \sum_{a=1}^{\infty} \left( \frac{b_{a}^\ast \langle \eta_a | J^{bc} \rangle}{ \nu_a - \nu_b - \nu_c} + \frac{b_a \langle \eta_{a}^\ast |J^{bc} \rangle}{\nu_a + \nu_b + \nu_c} \right)~,
\end{align}
and
\begin{align}\label{betaRab2final}
\beta_{\mathbbm{R},bc}^{(2)} =&~ \frac{P^{(1)2}}{\alpha_{i}^2} \frac{ b_b b_{c}^\ast \langle \eta_{b}^\ast | \QQ_1 | \eta_{c}^\ast \rangle}{\nu_b \nu_c} - \sum_{a=1}^{\infty} \left( \frac{b_{a}^\ast \langle \eta_a | K^{bc}  \rangle}{\nu_a - \nu_b + \nu_c} + \frac{b_a \langle \eta_{a}^\ast | K^{cb} \rangle}{\nu_a + \nu_b - \nu_c} \right) ~.
\end{align}
The relevant quantities going into these coefficients can be obtained from \eqref{A2eompieces}, \eqref{Bf1fromknowns}, \eqref{Q1Q2}, and slight modifications of these.  We have not made any significant attempts to simplify them further.

We conclude with some comments on the $\beta_i \to 0$ limit.  In this case every term of \eqref{zeta2avapp} vanishes except the very first one involving $|\SS^{(2,0)}\rangle$ where we have \eqref{SS20betai0}.  There is a time-independent contribution that is still given by $|\zeta_{\beta_f}^{(2)} \rangle$, but in order to find this solution explicitly we need an analog of the first order analysis we did around \eqref{betaf1condition} and \eqref{phibetaf1diffeq}.  The conditions determining $\beta_{f}^{(n)}$ at the first three orders are
\begin{align}
& (\beta_{f}^{(1)} + \beta_{f}^{(2)} + \beta_{f}^{(3)} + \cdots) \left[ \langle \phi_{0}' | \phi_{0}' \rangle + 2 \langle \phi_{0}' | \phi_{\beta_f}^{(2)\prime} \rangle + \cdots \right] = M_0 \epsilon \cr
 \Rightarrow \quad & \beta_{f}^{(1)} = \frac{M_0 \epsilon}{M_0'} ~, \qquad \beta_{f}^{(2)} = 0~, \qquad \beta_{f}^{(3)} = \frac{2 M_0 \epsilon \langle \phi_{0}'' | \phi_{\beta_f}^{(2)} \rangle}{M_{0}^{\prime 2}}  = 2 \frac{M_0 \epsilon \langle \uppsi_0' | \phi_{\beta_f}^{(2)} \rangle}{(M_0')^{3/2}} ~. \qquad 
\end{align}
The condition for the configuration $\phi_{\beta_f}^{(2)}$ is
\begin{align}
&  (\beta_{f}^{(1)2} - 1 + \cdots) ( \phi_{0}'' + \phi_{\beta_f}^{(2)\prime\prime} + \cdots) + V^{(1)}(\phi_0) + V^{(2)}(\phi_0) \phi_{\beta_f}^{(2)} + \cdots = 0 \cr
\Rightarrow \quad & \left[ - \pd_{\rho}^2 + V^{(2)}(\phi_0) \right] \phi_{\beta_f}^{(2)} = - \frac{(M_0 \epsilon)^2}{M_{0}^{\prime 2}} \phi_{0}'' ~,
\end{align}
and the solution with the appropriate periodicity conditions is
\begin{equation}
 \phi_{\beta_f}^{(2)} = \frac{(M_0 \epsilon)^2}{2 M_{0}^{\prime 2}} \left( \rho \phi_{0}' + g_{\rm hom}^{(1)} \right)~, \qquad (\beta_i = 0)~,
\end{equation}
where $g_{\rm hom}^{(1)}$ is as in \eqref{ghomsG} for sine-Gordon but with $\beta_i = 0$.  Inserting the solution back into $\beta_{f}^{(3)}$ we get
\begin{align}\label{betaf3betai0}
\beta_{f}^{(3)}  = - \frac{M_{0}^3 \epsilon^3}{2 M_{0}^{\prime 3}} \left( 1 + \frac{2}{M_{0}'} \langle \phi_{0}' | g_{\rm hom}^{(1)\prime} \rangle \right)~, \qquad (\beta_i = 0)~.
\end{align}
The overlap with $g_{\rm hom}^{(1)}$ vanishes exponentially fast at large $mL$ resulting in the expected $-\epsilon^3/2$.  Meanwhile the oscillating part of the second order solution is
\begin{align}\label{zeta2oscbetai0}
\lim_{\beta_i \to 0} |\zeta_{\rm osc}^{(2)}\rangle =&~ \sum_{a=1}^{\infty} 2 \Re \left\{ |\eta_a \rangle \frac{\langle \eta_a | \SS^{(2,0)} \rangle}{\nu_a} e^{-\ii (\nu_a+\cdots) (t-t_\ast)} \right\} \cr
=&~ \frac{M_{0}^2 \epsilon^2}{(M_{0}')^{3/2}} \sum_{a=1}^{\infty} 2 \Re \left\{ |\eta_a \rangle \frac{\langle \eta_{\varphi a} | \uppsi_0' \rangle}{\nu_a} e^{-\ii (\nu_a+\cdots) (t-t_\ast)} \right\}~, \qquad (\beta_i = 0)~, \qquad
\end{align}
for $t > t_\ast$.

Returning to \eqref{betapertbetai0}, we recall that the entire time-dependent $\beta^{(2)}$ is zero, and this is consistent with the results above.  The third order velocity is
\begin{equation}
\beta^{(3)} =  \frac{2 M_0 \epsilon}{(M_{0}')^{3/2}} \langle \uppsi_0' | \varphi^{(2)} \rangle ~, \qquad (\beta_i = 0)~,
\end{equation}
and the time-independent part of this is consistent with the $\beta_{f}^{(3)}$ in \eqref{betaf3betai0}.  For the oscillating part we use the bottom component of \eqref{zeta2oscbetai0}.  This leads to
\begin{align}
\beta^{(3)} =&~ \beta_{f}^{(3)} + \beta_{\rm osc}^{(3)} ~, \cr
\beta_{\rm osc}^{(3)} =&~ \frac{4 M_{0}^3 \epsilon^3}{M_{0}^{\prime 3}} \sum_a \frac{ | \langle \uppsi_0' | \eta_{\varphi a} \rangle |^2}{\nu_a} \cos((\nu + \cdots)(t-t_\ast)) ~, \qquad (\beta_i = 0)~, 
\end{align}
for $t > t_\ast$.

%%%%%%%%%%%%%%%%%%%
%%%%%%%%%%%%%%%%%%%
\section{Translation-invariant Quartic Lattice Coupling}\label{app:latquartic}
%%%%%%%%%%%%%%%%%%%
%%%%%%%%%%%%%%%%%%%

Since the quadratic terms in $V_{\rm ti}(\vec{\phi})$ and $V_{\rm loc}(\vec{\phi})$ agree we need only focus on the quartic interaction for $\phi^4$ theory:
\begin{equation}\label{H4}
H_4 :=  g^2 \int_{-L/2}^{L/2} \ed x \phi_{\rm approx}(t,x)^4 = \sum_{j_1,\ldots, j_4} V_{j_1 j_2,j_2,j_4} \phi_{j_1}(t) \phi_{j_2}(t) \phi_{j_3}(t) \phi_{j_4}(t) ~,
\end{equation}
where
\begin{equation}\label{V4lattice}
V_{j_1 j_2j_3j_4} = \frac{g^2}{(2N)^4} \sum_{n_1,\ldots,n_4} \int_{L/2}^{L/2} \ed x \prod_{i=1}^4 e^{2\pi \ii (n_i - \half) (\frac{x}{L} - \frac{j_i}{2N})} ~.
\end{equation}
The $t$-dependence plays no role in the following discussion and will henceforth be suppressed.

The integral over $x$ in \eqref{V4lattice} gives $L \delta_{n_1 + n_2 + n_3 + n_4 - 2,0}$.  The quadruple sum over $n_i \in \{-N + 1, \ldots, N\}$ subject to this constraint can be performed.  Taking at first the $j_i \in \mathbbm{R}$ so that $z_i = 2^{2\pi \ii j_i/(2N)}$ are arbitrary phases, the result takes the form
\begin{align}
V_{j_1j_2j_3j_4} =&~ \frac{L g^2}{(2N)^4 (z_1 \cdots z_4)^{N}} \left\{ \frac{z_{1}^{2N} (z_{2}^{2N} + z_{3}^{2N} + z_{4}^{2N})}{s_{12} s_{13} s_{14}} + \cdots + \frac{ z_{4}^{2N} (z_{1}^{2N} + z_{2}^{2N} + z_{3}^{2N})}{s_{41} s_{42} s_{43}} \right\}~,
\end{align}
with
\begin{align}
s_{k\ell} =&~ \left(\frac{z_k}{z_\ell}\right)^{1/2} - \left( \frac{z_\ell}{z_k} \right)^{1/2} = 2\ii \sin\left( \tfrac{\pi}{2N} (j_k - j_\ell)\right)~.  \raisetag{20pt}
\end{align}
It follows from this that for integer $j_i \in \{-N+1,\ldots, N\}$, all different, $V_{j_1 j_2 j_3 j_4} = 0$.  We obtain the values when some of the indices are equal by taking limits.  Up to symmetric permutations, the full set of results is
\begin{align}\label{V4integer}
V_{ijk\ell} =&~ 0~, \qquad \textrm{$i,j,k,\ell \in \{ -N +1, \ldots, N\}$, all different}~, \cr
V_{jk\ell\ell} =&~ \frac{g^2 L}{(2N)^4} \cdot \frac{ (-1)^{j-k} N}{\sin\left[ \frac{\pi}{2N}(j-\ell)\right] \sin\left[ \frac{\pi}{2N} (k-\ell)\right] } ~, \qquad j \neq k \neq \ell \neq j ~, \cr
V_{jjkk} =&~ \frac{g^2 L}{(2N)^4} \cdot \frac{2N}{\sin^2\left[\frac{\pi}{2N}(j-k)\right] } ~, \qquad j \neq k ~, \cr
V_{jjjk} =&~ \frac{g^2 L}{(2N)^4} \cdot \left( - \frac{(-1)^{j-k} N \cos\left[ \frac{\pi}{2N} (j-k)\right] }{ \sin^2 \left[ \frac{\pi}{2N} (j-k)\right]} \right)~,  \qquad j \neq k~, \cr
V_{jjjj} =&~ \frac{g^2 L}{(2N)^4} \cdot \frac{2}{3} ((2N)^3 + N) ~.
\end{align}

Now consider the quartic coupling \eqref{H4}:
\begin{align}
H_4 =&~ g^2 \sum_{j_1,\ldots,j_4 = -N+1}^{N} V_{j_1 j_2 j_3 j_4} \phi_{j_1} \cdots \phi_{j_4} ~.
\end{align}
We decompose this into the different types of terms we found above:
\begin{align}\label{H4type}
H_4 =&~ g^2 \bigg\{ \sum_{j=-N+1}^{N} V_{jjjj} \phi_{j}^4 + 4 \sum_{\substack{j,k = -N+1 \\ j \neq k}}^{N} V_{jjjk} \phi_{j}^3 \phi_k + 3 \sum_{\substack{j,k = -N+1 \\ j\neq k}}^{N} V_{jjkk} \phi_{j}^2 \phi_{k}^2 + \cr
&~ \qquad  +  6 \sum_{\substack{j,k,\ell = -N+1 \\ j \neq k \neq \ell}}^{N} V_{jk\ell\ell} \phi_{j} \phi_{k} \phi_{\ell}^2  \bigg\}  ~.
\end{align}
Inserting \eqref{V4integer} brings us to
\begin{align}
H_{4} = \frac{g^2 L}{(2N)^{4}} (A + B + C + D)~,
\end{align}
with
\begin{align}
A =&~ \left( \frac{2}{3} (2N)^{3} + \frac{1}{3} (2N) \right) \sum_{j=-N+1}^{N} \phi_{j}^4 ~, \cr
B =&~ - 2 (2N) \sum_{\substack{ j,k = -N+1 \\ j \neq k }}^{N} \frac{ (-1)^{j-k} \cos \left[ \frac{\pi}{2N} (j-k)\right]}{ \sin^2\left[ \frac{\pi}{2N} (j-k)\right]} \phi_{j}^3 \phi_k ~, \cr
C =&~ 3 (2N) \sum_{\substack{j,\ell = -N+1 \\ j \neq \ell}}^{N} \frac{1}{\sin^2\left[ \frac{\pi}{2N} (j - \ell) \right]} \phi_{j}^2 \phi_{\ell}^2 ~, \cr
D =&~ 3 (2N) \sum_{\substack{ j,k,\ell = -N+1 \\j \neq k \neq \ell}}^{N} \frac{(-1)^{j-k}}{\sin \left[ \frac{\pi}{2N} (j-\ell)\right] \sin\left[ \frac{\pi}{2N} (k-\ell)\right] } \phi_j \phi_k \phi_{\ell}^2 ~.
\end{align}

Observe that $C$ fills in the missing terms of $D$ when $j = k$, so that we can write
\begin{align}
C + D =&~ 3 (2N) \sum_{\ell = -N+1}^{N} \phi_{\ell}^2 \sum_{\substack{j=-N+1 \\ j \neq \ell}}^{N} \frac{ (-1)^{j-\ell}}{\sin \left[\frac{\pi}{2N} (\ell - j)\right]} \phi_j \sum_{\substack{k = -N+1 \\ k \neq \ell}}^{N} \frac{(-1)^{k - \ell}}{ \sin\left[ \frac{\pi}{2N} (\ell - k)\right]} \phi_k~.
\end{align}
Comparing the $j$ and $k$ sums to \eqref{DA1}, we find that $C + D$ can be written compactly in terms of $\phi_{\ell}$ and its derivative:
\begin{equation}
C + D = \frac{3 (2N) L^2}{\pi^2} \sum_{\ell = - N+1}^{N} \phi_{\ell}^2 (D \phi)_\ell (D \phi)_\ell ~.
\end{equation}

Now consider $B$ which we write as
\begin{align}
B =&~ 2 (2N) \sum_{j = -N+1}^{N} \phi_{j}^3 \sum_{\substack{ k = -N+1 \\k \neq j}}^{N} \frac{(-1)^{j-k + 1} \cos\left[ \frac{\pi}{2N}(j-k)\right]}{ \sin^2 \left[ \frac{\pi}{2N}(j-k)\right]^2} \phi_k ~.
\end{align}
Comparing with \eqref{DA2}, we see that the sum over $k$ is proportional to the second SLAC derivative, except it is missing the $k = j$ terms.  Adding and subtracting them brings us to
\begin{equation}
B = \frac{(2N) L^2}{\pi^2} \sum_{j = - N+1}^N \phi_{j}^3 (D^{(2)}\phi)_j + \frac{(2N)}{3} ((2N)^2 - 1) \sum_{j=-N+1}^N \phi_{j}^4 ~.
\end{equation}
Now observe how this adds to $A$ perfectly to cancel out the linear term in $N$ and make the coefficient of the $(2N)^3$ term equal to one:
\begin{align}
A + B =&~ (2N)^3 \sum_{j=-N+1}^{N} \phi_{j}^4 + \frac{(2N) L^2}{\pi^2} \sum_{j=-N+1}^{N} \phi_{j}^3 (D^{(2)} \phi)_j ~.
\end{align}

Thus, putting the pieces together, we have
\begin{align}\label{H4result}
H_4 =&~ \frac{g^2 L}{(2N)^{4}} \bigg\{ (2N)^3  \sum_{j=-N+1}^{N} \phi_{j}^4 + \frac{(2N) L^2}{\pi^2} \sum_{j=-N+1}^{N} \left[ \phi_{j}^3 (D^{(2)}\phi)_j + 3 \phi_{j}^2 (D \phi)_j (D \phi)_j \right] \bigg\} ~.
\end{align}
%

%%%%%%%%%%%%%%%%%%%%%%%%%%%%%%%%%%%%%%%%%%%%%%%%%%%%

% Bibliography

%\bibliographystyle{utphys}
%\bibliography{AcceleratingSolitonsBiblio}

\begin{thebibliography}{10}

\bibitem{Dashen:1974ci}
R.~F. Dashen, B.~Hasslacher, and A.~Neveu, ``{Nonperturbative Methods and
  Extended Hadron Models in Field Theory 1. Semiclassical Functional
  Methods},'' \href{http://dx.doi.org/10.1103/PhysRevD.10.4114}{{\em Phys. Rev.
  D} {\bf 10} (1974)  4114}.

\bibitem{Dashen:1974cj}
R.~F. Dashen, B.~Hasslacher, and A.~Neveu, ``{Nonperturbative Methods and
  Extended Hadron Models in Field Theory 2. Two-Dimensional Models and Extended
  Hadrons},'' \href{http://dx.doi.org/10.1103/PhysRevD.10.4130}{{\em Phys. Rev.
  D} {\bf 10} (1974)  4130--4138}.

\bibitem{Goldstone:1974gf}
J.~Goldstone and R.~Jackiw, ``{Quantization of Nonlinear Waves},''
  \href{http://dx.doi.org/10.1103/PhysRevD.11.1486}{{\em Phys. Rev. D} {\bf 11}
  (1975)  1486--1498}.

\bibitem{Gervais:1974dc}
J.-L. Gervais and B.~Sakita, ``{Extended Particles in Quantum Field
  Theories},'' \href{http://dx.doi.org/10.1103/PhysRevD.11.2943}{{\em Phys.
  Rev. D} {\bf 11} (1975)  2943}.

\bibitem{Callan:1975yy}
C.~G. Callan, Jr. and D.~J. Gross, ``{Quantum Perturbation Theory of
  Solitons},''
\href{http://dx.doi.org/10.1016/0550-3213(75)90150-9}{{\em Nucl. Phys.} {\bf
  B93} (1975)  29--55}.
%%CITATION = NUPHA,B93,29;%%.

\bibitem{Christ:1975wt}
N.~Christ and T.~Lee, ``{Quantum Expansion of Soliton Solutions},''
  \href{http://dx.doi.org/10.1103/PhysRevD.12.1606}{{\em Phys. Rev. D} {\bf 12}
  (1975)  1606}.

\bibitem{Gervais:1975pa}
J.-L. Gervais, A.~Jevicki, and B.~Sakita, ``{Perturbation Expansion Around
  Extended Particle States in Quantum Field Theory. 1.},''
  \href{http://dx.doi.org/10.1103/PhysRevD.12.1038}{{\em Phys. Rev. D} {\bf 12}
  (1975)  1038}.

\bibitem{Tomboulis:1975gf}
E.~Tomboulis, ``{Canonical Quantization of Nonlinear Waves},''
\href{http://dx.doi.org/10.1103/PhysRevD.12.1678}{{\em Phys. Rev.} {\bf D12}
  (1975)  1678}.
%%CITATION = PHRVA,D12,1678;%%.

\bibitem{Weisz:1977ii}
P.~H. Weisz, ``{Exact Quantum Sine-Gordon Soliton Form-Factors},''
  \href{http://dx.doi.org/10.1016/0370-2693(77)90097-1}{{\em Phys. Lett. B}
  {\bf 67} (1977)  179--182}.

\bibitem{Karowski:1978vz}
M.~Karowski and P.~Weisz, ``{Exact Form-Factors in (1+1)-Dimensional Field
  Theoretic Models with Soliton Behavior},''
  \href{http://dx.doi.org/10.1016/0550-3213(78)90362-0}{{\em Nucl. Phys. B}
  {\bf 139} (1978)  455--476}.

\bibitem{Babujian:1998uw}
H.~M. Babujian, A.~Fring, M.~Karowski, and A.~Zapletal, ``{Exact form-factors
  in integrable quantum field theories: The Sine-Gordon model},''
  \href{http://dx.doi.org/10.1016/S0550-3213(98)00737-8}{{\em Nucl. Phys. B}
  {\bf 538} (1999)  535--586}, \href{http://arxiv.org/abs/hep-th/9805185}{{\tt
  arXiv:hep-th/9805185}}.

\bibitem{Evslin:2019xte}
J.~Evslin, ``{Manifestly Finite Derivation of the Quantum Kink Mass},''
  \href{http://dx.doi.org/10.1007/JHEP11(2019)161}{{\em JHEP} {\bf 11} (2019)
  161}, \href{http://arxiv.org/abs/1908.06710}{{\tt arXiv:1908.06710
  [hep-th]}}.

\bibitem{Evslin:2020qow}
J.~Evslin, ``{The Ground State of the Sine-Gordon Soliton},''
  \href{http://arxiv.org/abs/2003.11384}{{\tt arXiv:2003.11384 [hep-th]}}.

\bibitem{Evslin:2020azr}
J.~Evslin and H.~Guo, ``{Two-Loop Scalar Kinks},''
  \href{http://dx.doi.org/10.1103/PhysRevD.103.125011}{{\em Phys. Rev. D} {\bf
  103} (2021) no.~12, 125011}, \href{http://arxiv.org/abs/2012.04912}{{\tt
  arXiv:2012.04912 [hep-th]}}.

\bibitem{Evslin:2021gxs}
J.~Evslin, ``{\ensuremath{\phi}4 kink mass at two loops},''
  \href{http://dx.doi.org/10.1103/PhysRevD.104.085013}{{\em Phys. Rev. D} {\bf
  104} (2021) no.~8, 085013}, \href{http://arxiv.org/abs/2104.07991}{{\tt
  arXiv:2104.07991 [hep-th]}}.

\bibitem{Evslin:2022xvs}
J.~Evslin, ``{Form factors for meson-kink scattering},''
  \href{http://dx.doi.org/10.1016/j.physletb.2022.137177}{{\em Phys. Lett. B}
  {\bf 830} (2022)  137177}, \href{http://arxiv.org/abs/2204.06194}{{\tt
  arXiv:2204.06194 [hep-th]}}.

\bibitem{Melnikov:2020ret}
I.~V. Melnikov, C.~Papageorgakis, and A.~B. Royston, ``{Accelerating
  solitons},'' \href{http://dx.doi.org/10.1103/PhysRevD.102.125002}{{\em Phys.
  Rev. D} {\bf 102} (2020) no.~12, 125002},
  \href{http://arxiv.org/abs/2007.11028}{{\tt arXiv:2007.11028 [hep-th]}}.

\bibitem{Melnikov:2020iol}
I.~V. Melnikov, C.~Papageorgakis, and A.~B. Royston, ``{Forced Soliton Equation
  and Semiclassical Soliton Form Factors},''
  \href{http://dx.doi.org/10.1103/PhysRevLett.125.231601}{{\em Phys. Rev.
  Lett.} {\bf 125} (2020) no.~23, 231601},
  \href{http://arxiv.org/abs/2010.10381}{{\tt arXiv:2010.10381 [hep-th]}}.

\bibitem{MR4717595}
H.~Volkmer, ``On the {H}ill discriminant of {L}am\'e's differential equation,''
  \href{http://dx.doi.org/10.3842/SIGMA.2024.021}{{\em SIGMA Symmetry
  Integrability Geom. Methods Appl.} {\bf 20} (2024)  Paper No. 021, 9}.

\bibitem{Drell:1976bq}
S.~D. Drell, M.~Weinstein, and S.~Yankielowicz, ``{Variational Approach to
  Strong Coupling Field Theory. 1. Phi**4 Theory},''
  \href{http://dx.doi.org/10.1103/PhysRevD.14.487}{{\em Phys. Rev. D} {\bf 14}
  (1976)  487}.

\bibitem{Drell:1976mj}
S.~D. Drell, M.~Weinstein, and S.~Yankielowicz, ``{Strong Coupling Field
  Theories. 2. Fermions and Gauge Fields on a Lattice},''
  \href{http://dx.doi.org/10.1103/PhysRevD.14.1627}{{\em Phys. Rev. D} {\bf 14}
  (1976)  1627}.

\bibitem{Pearsonthesis}
R.~B. Pearson, {\em Variational Methods and Bounds for Lattice Field Theories}.
\newblock PhD thesis, Stanford University, 1976.

\bibitem{Guo:2019hiu}
H.~Guo and J.~Evslin, ``{Finite derivation of the one-loop sine-Gordon soliton
  mass},'' \href{http://dx.doi.org/10.1007/JHEP02(2020)140}{{\em JHEP} {\bf 02}
  (2020)  140}, \href{http://arxiv.org/abs/1912.08507}{{\tt arXiv:1912.08507
  [hep-th]}}.

\bibitem{Evslin:2022opz}
J.~Evslin, A.~B. Royston, and B.~Zhang, ``{Cut-off kinks},''
  \href{http://dx.doi.org/10.1007/JHEP01(2023)073}{{\em JHEP} {\bf 01} (2023)
  073}, \href{http://arxiv.org/abs/2210.16523}{{\tt arXiv:2210.16523
  [hep-th]}}.

\bibitem{Rajaraman:1982is}
R.~Rajaraman, {\em Solitons and instantons}.
\newblock North-Holland Publishing Co., Amsterdam, 1982.

\bibitem{Faddeev:1977rm}
L.~Faddeev and V.~Korepin, ``{Quantum Theory of Solitons: Preliminary
  Version},'' \href{http://dx.doi.org/10.1016/0370-1573(78)90058-3}{{\em Phys.
  Rept.} {\bf 42} (1978)  1--87}.

\bibitem{ZAMOLODCHIKOV1978525}
A.~B. Zamolodchikov and A.~B. Zamolodchikov, ``Relativistic factorized s-matrix
  in two dimensions having o(n) isotopic symmetry,''
  \href{http://dx.doi.org/https://doi.org/10.1016/0550-3213(78)90239-0}{{\em
  Nuclear Physics B} {\bf 133} (1978) no.~3, 525--535}.

\bibitem{PhysRevD.11.2088}
S.~Coleman, \href{http://dx.doi.org/10.1103/PhysRevD.11.2088}{``Quantum
  sine-gordon equation as the massive thirring model,''{\em Phys. Rev. D} {\bf
  11} (Apr, 1975)  2088--2097}.

\bibitem{Derrick:1964ww}
G.~H. Derrick, ``{Comments on nonlinear wave equations as models for elementary
  particles},'' \href{http://dx.doi.org/10.1063/1.1704233}{{\em J. Math. Phys.}
  {\bf 5} (1964)  1252--1254}.

\bibitem{Sakamoto:1999yk}
M.~Sakamoto, M.~Tachibana, and K.~Takenaga, ``{Spontaneously broken
  translational invariance of compactified space},''
  \href{http://dx.doi.org/10.1016/S0370-2693(99)00555-9}{{\em Phys. Lett. B}
  {\bf 457} (1999)  33--38}, \href{http://arxiv.org/abs/hep-th/9902069}{{\tt
  arXiv:hep-th/9902069}}.

\bibitem{Mussardo:2004zn}
G.~Mussardo, V.~Riva, and G.~Sotkov, ``{Semiclassical scaling functions of
  sine-Gordon model},''
  \href{http://dx.doi.org/10.1016/j.nuclphysb.2004.08.004}{{\em Nucl. Phys. B}
  {\bf 699} (2004)  545--574}, \href{http://arxiv.org/abs/hep-th/0405139}{{\tt
  arXiv:hep-th/0405139}}.

\bibitem{Mussardo:2005dx}
G.~Mussardo, V.~Riva, G.~Sotkov, and G.~Delfino, ``{Kink scaling functions in
  2-D non-integrable quantum field theories},''
  \href{http://dx.doi.org/10.1016/j.nuclphysb.2005.12.008}{{\em Nucl. Phys. B}
  {\bf 736} (2006)  259--287}, \href{http://arxiv.org/abs/hep-th/0510102}{{\tt
  arXiv:hep-th/0510102}}.

\bibitem{Pawellek:2008st}
M.~Pawellek, ``{Quantum mass correction for the twisted kink},''
  \href{http://dx.doi.org/10.1088/1751-8113/42/4/045404}{{\em J. Math. Phys.}
  {\bf 42} (2009)  045404}, \href{http://arxiv.org/abs/0802.0710}{{\tt
  arXiv:0802.0710 [hep-th]}}.

\bibitem{Pawellek:2008gs}
M.~Pawellek, ``{Quantization of Sine-Gordon solitons on the circle:
  Semiclassical versus exact results},''
  \href{http://dx.doi.org/10.1016/j.nuclphysb.2008.10.001}{{\em Nucl. Phys. B}
  {\bf 810} (2009)  527--541}, \href{http://arxiv.org/abs/0808.0696}{{\tt
  arXiv:0808.0696 [hep-th]}}.

\bibitem{Gervais:1976ws}
J.-L. Gervais and A.~Jevicki, ``{Point Canonical Transformations in Path
  Integral},'' \href{http://dx.doi.org/10.1016/0550-3213(76)90422-3}{{\em Nucl.
  Phys. B} {\bf 110} (1976)  93--112}.

\bibitem{Fernandez:1986bt}
J.~C. Fernandez, M.~J. Coupil, O.~Legrand, and G.~Reinisch, ``{Relativistic
  Dynamics of Sine-Gordon Solitons Trapped in Confining Potentials},''
  \href{http://dx.doi.org/10.1103/PhysRevB.34.6207}{{\em Phys. Rev. B} {\bf 34}
  (1986)  6207--6213}.

\bibitem{Pryce:1948pf}
M.~H.~L. Pryce, ``{The Mass center in the restricted theory of relativity and
  its connection with the quantum theory of elementary particles},''
  \href{http://dx.doi.org/10.1098/rspa.1948.0103}{{\em Proc. Roy. Soc. Lond. A}
  {\bf 195} (1948)  62--81}.

\bibitem{MR1433936}
M.~J. Ablowitz, B.~M. Herbst, and C.~M. Schober, ``On the numerical solution of
  the sine-{G}ordon equation. {II}. {P}erformance of numerical schemes,'' {\em
  J. Comput. Phys.} {\bf 131} (1997) no.~2, 354--367.

\bibitem{MR3245858}
L.~N. Trefethen and J.~A.~C. Weideman, ``The exponentially convergent
  trapezoidal rule,'' \href{http://dx.doi.org/10.1137/130932132}{{\em SIAM
  Rev.} {\bf 56} (2014) no.~3, 385--458}.

\bibitem{Costella:2002js}
J.~P. Costella, ``{A New proposal for the fermion doubling problem. 2.
  Improving the operators for finite lattices},''
  \href{http://arxiv.org/abs/hep-lat/0207015}{{\tt arXiv:hep-lat/0207015}}.

\bibitem{Costella:2004re}
J.~P. Costella, ``{A Strange property of lattices with an even number of
  sites},'' \href{http://arxiv.org/abs/hep-lat/0404009}{{\tt
  arXiv:hep-lat/0404009}}.

\bibitem{sanz-serna_1992}
J.~M. Sanz-Serna, ``Symplectic integrators for hamiltonian problems: an
  overview,'' \href{http://dx.doi.org/10.1017/S0962492900002282}{{\em Acta
  Numerica} {\bf 1} (1992)  243–286}.

\bibitem{Stuart1994ModelPI}
A.~M. Stuart and A.~R. Humphries, ``Model problems in numerical stability
  theory for initial value problems,'' {\em SIAM Rev.} {\bf 36} (1994)
  226--257.

\bibitem{MR2655369}
H.~Volkmer, ``Lam\'e{} functions,'' in {\em N{IST} handbook of mathematical
  functions}, pp.~683--695.
\newblock U.S. Dept. Commerce, Washington, DC, 2010.

\bibitem{NIST:DLMF}
``{\it NIST Digital Library of Mathematical Functions}.''
  \url{https://dlmf.nist.gov/}, release 1.2.2 of 2024-09-15.
\newblock \url{https://dlmf.nist.gov/}. F.~W.~J. Olver, A.~B. {Olde Daalhuis},
  D.~W. Lozier, B.~I. Schneider, R.~F. Boisvert, C.~W. Clark, B.~R. Miller,
  B.~V. Saunders, H.~S. Cohl, and M.~A. McClain, eds.

\bibitem{MR698781}
A.~Erd\'elyi, W.~Magnus, F.~Oberhettinger, and F.~G. Tricomi, {\em Higher
  transcendental functions. {V}ol. {III}}.
\newblock Robert E. Krieger Publishing Co., Inc., Melbourne, FL, 1981.
\newblock Based on notes left by Harry Bateman, Reprint of the 1955 original.

\bibitem{MR2377687}
R.~S. Maier, ``Lam\'e{} polynomials, hyperelliptic reductions and {L}am\'e{}
  band structure,'' \href{http://dx.doi.org/10.1098/rsta.2007.2063}{{\em
  Philos. Trans. R. Soc. Lond. Ser. A Math. Phys. Eng. Sci.} {\bf 366} (2008)
  no.~1867, 1115--1153}, \href{http://arxiv.org/abs/math-ph/0309005}{{\tt
  arXiv:math-ph/0309005 [math-ph]}}.

\bibitem{MR559928}
W.~Magnus and S.~Winkler, {\em Hill's equation}.
\newblock Dover Publications, Inc., New York, 1979.
\newblock Corrected reprint of the 1966 edition.

\bibitem{MR2400}
E.~L. Ince, ``Further investigations into the periodic {L}am\'e{} functions,''
  {\em Proc. Roy. Soc. Edinburgh} {\bf 60} (1940)  83--99.

\bibitem{MR1361127}
F.~Gesztesy and R.~Weikard, ``Lam\'e{} potentials and the stationary (m){K}d{V}
  hierarchy,'' \href{http://dx.doi.org/10.1002/mana.19951760107}{{\em Math.
  Nachr.} {\bf 176} (1995)  73--91}.

\bibitem{MR156022}
H.~Hochstadt, ``Functiontheoretic properties of the discriminant of {H}ill's
  equation,'' \href{http://dx.doi.org/10.1007/BF01111426}{{\em Math. Z.} {\bf
  82} (1963)  237--242}.

\bibitem{MR3075381}
M.~S.~P. Eastham, {\em The spectral theory of periodic differential equations}.
\newblock Texts in Mathematics (Edinburgh). Scottish Academic Press, Edinburgh;
  Hafner Press, New York, 1973.

\bibitem{Jain:1990dq}
P.~Jain, ``{Relativistic energy momentum relationship for a soliton},''
  \href{http://dx.doi.org/10.1103/PhysRevD.41.3273}{{\em Phys. Rev. D} {\bf 41}
  (1990)  3273--3276}.

\end{thebibliography}

\providecommand{\href}[2]{#2}\begingroup\raggedright\endgroup

\end{document}